%% file: main.tex
\documentclass[sigconf]{acmart}
\AtBeginDocument{%
  }

\setcopyright{acmlicensed}
\copyrightyear{2026}
\acmYear{2026}
\acmDOI{XXXXXXX.XXXXXXX}
\acmConference[KDD '26]{32nd SIGKDD Conference on Knowledge Discovery and Data Mining}{August 09--13, 2026}{Jeju, Korea}
\acmISBN{978-1-4503-XXXX-X/2026/06}

\usepackage[utf8]{inputenc} 
\usepackage[T1]{fontenc}    
\usepackage{CJKutf8}        

\usepackage{amsmath, amsthm, amsfonts}
\usepackage{graphicx}
\usepackage{subcaption}
\usepackage{booktabs}
\usepackage{array, makecell, multirow, colortbl}
\usepackage{xcolor} 
\usepackage{tikz}
\usetikzlibrary{trees}
\usepackage{url}
\usepackage{hyperref}
\usepackage{cleveref}
\usepackage{nicefrac, microtype, enumitem, appendix, pifont}
\usepackage{listings, tcolorbox}
\usepackage{changepage}
\usepackage[ruled,vlined]{algorithm2e}
\usepackage{balance}
\usepackage{wrapfig}
\usepackage{tcolorbox}
\tcbuselibrary{breakable}

\copyrightyear{2026}
\acmYear{2026}
\setcopyright{cc}
\setcctype{by}
\acmConference[KDD '26]{Proceedings of the 32nd ACM SIGKDD Conference on Knowledge Discovery and Data Mining V.2}{August 09--13, 2026}{Jeju Island, Republic of Korea}
\acmBooktitle{Proceedings of the 32nd ACM SIGKDD Conference on Knowledge Discovery and Data Mining V.2 (KDD '26), August 09--13, 2026, Jeju Island, Republic of Korea}
\acmDOI{10.1145/3770855.3817500}
\acmISBN{979-8-4007-2259-2/2026/08}

\acmConference[KDD'26]{KDD 2026}{August 9 to 13, 2026}{Jeju, Korea}

\newcommand{\projectname}{\textbf{\textsc{AlphaForgeBench}}\xspace}

\definecolor{tabblue}{HTML}{1F77B4}

\begin{document}

\title{AlphaForgeBench: Benchmarking End-to-End Trading Strategy Design with Large Language Models}

\author{Wentao Zhang}
\authornote{These authors contributed equally to this work.}
\affiliation{%
  \institution{Nanyang Technological University}
  \country{Singapore}}
\email{zhangwent963@gmail.com}

\author{Mingxuan Zhao}
\authornotemark[1]
\affiliation{%
  \institution{The Hong Kong University of Science and Technology (Guangzhou)}
  \country{China}}
\email{mzhao085@connect.hkust-gz.edu.cn}

\author{Jincheng Gao}
\authornotemark[1]
\affiliation{%
  \institution{The Hong Kong University of Science and Technology (Guangzhou)}
  \country{China}}
\email{jinchenggao@hkust-gz.edu.cn}

\author{Jieshun You}
\affiliation{%
  \institution{Hong Kong Polytechnic University}
  \country{Hong Kong}}
\email{24041917g@connect.polyu.hk}

\author{Huaiyu Jia}
\affiliation{%
  \institution{The Hong Kong University of Science and Technology (Guangzhou)}
  \country{China}}
\email{hjia351@connect.hkust-gz.edu.cn}

\author{Yilei Zhao}
\affiliation{%
  \institution{Nanyang Technological University}
  \country{Singapore}}
\email{YILEI002@e.ntu.edu.sg}

\author{Bo An}
\affiliation{%
  \institution{Nanyang Technological University}
  \country{Singapore}}
\email{boan@ntu.edu.sg}

\author{Shuo Sun}
\authornote{Corresponding author.}
\affiliation{%
  \institution{The Hong Kong University of Science and Technology (Guangzhou)}
  \country{China}}
\email{shuosun@hkust-gz.edu.cn}

\renewcommand{\shortauthors}{Wentao Zhang et al.}

\begin{abstract}
The rapid advancement of Large Language Models (LLMs) has catalyzed the proliferation of diverse financial benchmarks, progressively evolving from static knowledge evaluation to increasingly sophisticated interactive trading simulations. Nevertheless, existing frameworks that assess real-time trading performance largely overlook a fundamental failure mode: the severe behavioral instability exhibited by LLMs in sequential decision-making under financial uncertainty. Through extensive empirical investigation, we demonstrate that when deployed as direct trading agents, LLMs manifest extreme run-to-run variance, produce inconsistent action sequences even under strictly deterministic decoding configurations, and exhibit irrational action flipping across temporally adjacent decision steps. We systematically attribute these pathological behaviors to the models' fundamentally stateless autoregressive architectures, which lack persistent memory of prior actions, and their pronounced sensitivity to continuous-to-discrete action mappings inherent in portfolio allocation tasks. These deficiencies collectively undermine the validity and trustworthiness of numerous existing online and offline financial trading benchmarks, rendering their evaluations unreliable, non-reproducible, and uninformative for meaningful model comparison. To address these limitations, we introduce \projectname, a principled evaluation framework that reconceptualizes the role of LLMs from stochastic execution agents to quantitative researchers capable of systematic financial reasoning. Rather than requiring models to emit discrete trading actions, \projectname tasks LLMs with generating executable alpha factors and composing factor-based trading strategies grounded in financial domain knowledge. This paradigm shift decouples reasoning from execution mechanics, enabling fully deterministic and reproducible evaluation while maintaining close alignment with real-world quantitative research workflows. Extensive experiments across multiple state-of-the-art LLMs demonstrate that \projectname effectively eliminates execution-induced instability, yields highly reproducible outcomes, and provides a rigorous and discriminative benchmark for assessing LLMs' capacity for financial reasoning, strategy formulation, and alpha discovery. Webpage at \url{https://finbrain-lab-hkustgz.github.io/AlphaForgeBench}.
\end{abstract}




\maketitle

\section{Introduction}

The rapid advancement of large language models (LLMs) has spurred the development of numerous benchmarks to assess model capabilities across diverse domains. In the financial domain, early benchmarks focused on evaluating general financial knowledge through question-answering (QA) tasks, including numerical reasoning over financial reports (e.g., FinQA~\cite{chen2021finqa}, TAT-QA~\cite{zhu2021tat}), conversational finance QA (e.g., ConvFinQA~\cite{chen2022convfinqa}), and comprehensive multi-task evaluation frameworks (e.g., BloombergGPT~\cite{wu2023bloomberggpt}, FinGPT~\cite{liu2023fingpt}, PIXIU~\cite{xie2023pixiu}, FinBen~\cite{xie2024finben}). However, these benchmarks primarily measure \emph{encyclopedic knowledge} and \emph{static reasoning} over historical snapshots, which fail to capture an LLM's ability to make sequential trading decisions under non-stationary market conditions. As the field evolved, researchers shifted toward evaluating LLMs' real-time trading capabilities, with benchmarks such as Alpha Arena~\cite{alpaha_arena_2026} establishing live trading evaluation paradigms that assess LLM agents' adaptability and alpha-seeking capabilities directly within dynamic, executing financial markets. While these trading benchmarks have expanded the scope of financial LLMs evaluation, they have largely overlooked a critical issue: LLMs exhibit extreme \textit{instability} in their performance on financial trading tasks.

Specifically, when LLMs or LLM-based agents directly emit trading decisions (e.g., \textsc{buy}/\textsc{hold}/\textsc{sell} signals), their outputs demonstrate severe instability across multiple dimensions. \textbf{(1) Run-to-run variance in performance metrics.} Under identical settings, the same LLM produces dramatically different trading trajectories across multiple runs on the same financial data (e.g., OHLCV time series with technical indicators), resulting in substantial variance in returns, drawdowns, and other performance metrics. \textbf{(2) Inconsistent action sequences even under deterministic decoding.} Even with \texttt{temperature=0} (deterministic decoding), LLMs generate completely different trading action sequences across runs on identical market data, exhibiting no consistency in decision-making patterns. \textbf{(3) Rapid action flipping persists despite hard constraints.} LLMs exhibit a tendency to rapidly flip trading actions (e.g., buying immediately after selling, or selling immediately after buying) when performing trading tasks. Even when explicit behavioral guardrails are incorporated into prompts (e.g., minimum holding periods, cooldown windows, or historical action sequences), these prompt-based constraints cannot fully eliminate such rapid flipping behavior. Detailed analyses are provided in Appx.~\ref{appx_sec:instability_analysis}.

Fundamentally, this phenomenon can be primarily attributed to the following three aspects. First, LLMs are inherently stateless, instantaneous decision-making models. In trading tasks, each action output is an independent re-evaluation based on a "current input snapshot." The model possesses no natural memory of its recent buy or sell executions, nor does it view position holding as a long-term state requiring consistency. Consequently, even slight variations in market features in the subsequent step can lead the model to generate divergent or even diametrically opposite actions. Second, LLMs are highly sensitive to the mapping of continuous market signals to discrete actions. While inputs such as prices and technical indicators vary continuously, the outputs are discrete actions. This continuous-to-discrete mapping amplifies the impact of minor fluctuations, causing the model to abruptly shift its stance upon slight indicator reversals or semantic shifts. Crucially, it lacks the mechanisms of inertia, tolerance intervals, and strategic waiting typically inherent in professional trading. Third, LLMs fundamentally perform classification rather than strategy optimization. Directly prompting an LLM to output an action essentially tasks it with assigning a "most reasonable action" label to the current market state, rather than maximizing long-term returns within a sequential decision-making framework subject to costs, positions, and time constraints. Because the model is agnostic to transaction fees, slippage, and penalties for excessive or insufficient trading, it is highly prone to collapsing into extreme patterns of either over-trading or complete inactivity.

To address these limitations, we propose a new evaluation paradigm that shifts from black-box trading to white-box logic generation. We introduce \textbf{\textsc{AlphaForgeBench}}, a comprehensive benchmark designed to assess LLMs on their ability to generate executable alpha factors and trading strategy code, effectively positioning the LLM as a quantitative researcher rather than a stochastic execution agent. This paradigm shift offers three critical advantages that directly mitigate the aforementioned instability. First, it fundamentally solves the continuous-to-discrete sensitivity problem by decoupling reasoning from execution. By compelling the LLM to formalize its decision boundaries into explicit algorithmic rules, the stochastic nature of the model is confined to the generation phase, rendering the subsequent execution strictly deterministic and immune to the random action flipping observed in direct-trading tasks. Second, it resolves the state management issue. Unlike stateless models that struggle to maintain continuity, the generated code inherently preserves internal states and logic across the entire time series to ensure consistent decision-making. Third, this setup mirrors the real-world workflow of quantitative finance where researchers synthesize strategies and engines execute them. This alignment enables transparent logic verification and provides a rigorous metric to assess whether the model has truly learned financial reasoning or is merely overfitting to market noise. Our contributions are summarized as follows:

\begin{itemize}[leftmargin=*, nosep]
\item We systematically analyze the critical flaws in existing financial trading benchmarks. We demonstrate that the inherent instability of LLMs in direct-trading tasks renders traditional performance metrics unreliable, preventing an accurate assessment of true financial reasoning capabilities.

\item We introduce \projectname, a novel benchmark that repositions LLMs from stochastic agents to quantitative researchers. By evaluating the generation of executable alpha factors and strategy code, our framework provides a deterministic and robust metric for measuring financial logic synthesis.

\item We conduct extensive experiments across state-of-the-art LLMs to validate the effectiveness of our approach. The results confirm that \projectname offers a significantly more stable and discriminatory assessment of financial capabilities compared to direct-trading baselines.
\end{itemize}

\begin{figure*}[t]
  \centering
  \includegraphics[width=0.9\textwidth]{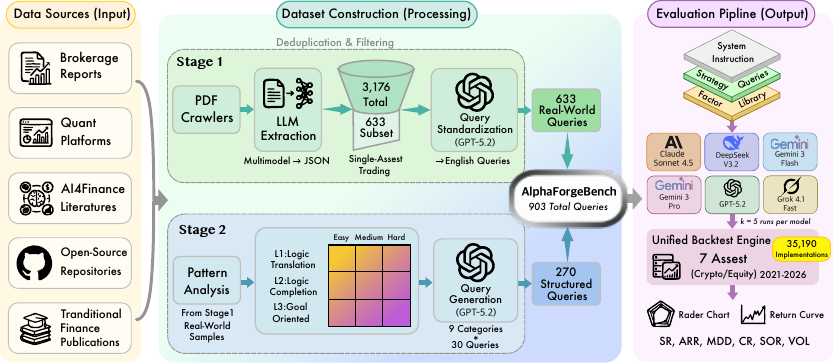}
  \vspace{-0.25cm}
  \caption{The framework of \projectname.}
  \label{fig:alphaforgebench}
  \vspace{-0.25cm}
\end{figure*}

\section{Related Work}

\subsection{Financial Knowledge Benchmark}
Recent advancements in financial question answering have driven the development of diverse benchmarks ranging from specific reasoning tasks to holistic system evaluations. Early efforts focused on numerical and hybrid reasoning, with TAT-QA \cite{zhu2021tat} addressing multi-step reasoning over tabular-textual data, FinQA \cite{chen2021finqa} targeting numerical reasoning on financial reports, and ConvFinQA \cite{chen2022convfinqa} extending this to conversational contexts. As LLMs evolved, researchers introduced comprehensive frameworks to evaluate broader capabilities: BloombergGPT \cite{wu2023bloomberggpt} validated the efficacy of domain-specific pretraining, FinGPT \cite{liu2023fingpt} democratized internet-scale financial data for open-source training, PIXIU \cite{xie2023pixiu} established a multi-task instruction tuning benchmark, and FinBen \cite{xie2024finben} offered a holistic evaluation across seven dimensions of financial intelligence. addressing the complexity of long-form generation, FinTextQA \cite{chen2024fintextqa} utilized RAG-based metrics for extensive textbook-level queries. Furthermore, recent works have tailored benchmarks to specific linguistic and user needs, with CFinBench\cite{nie2025cfinbench} and Fin-Eva \cite{fin_eva_2025} constructing fine-grained evaluation systems for Chinese financial knowledge, while UCFE \cite{yang2025ucfe} pioneered a user-centric framework to align model performance with dynamic human preferences across diverse roles. However, these existing frameworks remain largely confined to passive knowledge retrieval and phenomenon explication, failing to bridge the critical gap between theoretical financial understanding and actionable, strategy-driven trading execution in practice.

\vspace{-0.2cm}
\subsection{Financial Trading Benchmark}

The evolution of financial LLMs has transitioned from passive analysis to autonomous decision-making agents, necessitating rigorous benchmarks for trading and forecasting. In terms of agentic architecture, FINCON \cite{yu2024fincon} introduces a hierarchical multi-agent system with conceptual verbal reinforcement for risk-aware strategies, while AlphaFin \cite{chen2025stockbench} establishes a retrieval-augmented pipeline to evaluate end-to-end fundamental reasoning and alpha generation. \cite{lopez2023can,xie2023wall} analyze the performance of the large model in terms of price prediction and returns. To standardize decision-making assessments, INVESTORBENCH \cite{li2025investorbench} provides a multi-asset simulation environment, revealing that agents struggle to outperform buy-and-hold baselines. Critically, addressing the "time-travel" bias in historical backtesting, recent works have pivoted toward live-market evaluation: DeepFund \cite{li2025time} demonstrates that even SOTA models incur losses in real-time fund management. FutureX \cite{zeng2025futurex, prophetarena2025} pioneers a live, anti-contamination forecasting framework to evaluate the real-time predictive intelligence of LLM agents on unfolding global events. To eliminate look-ahead bias and bridge the gap between simulation and reality, Alpha Arena \cite{alpaha_arena_2026}, RockAlpha \cite{rockalpha2025}, and LiveTradeBench \cite{yu2025livetradebench} collectively establish a live trading benchmark paradigm, rigorously evaluating the real-time adaptability and alpha-seeking capabilities of LLM agents directly within dynamic, executing financial markets.

However, these live evaluation paradigms introduce a fundamental \emph{reproducibility crisis}: evaluation results are bound to the specific execution time window and cannot be independently replicated, a problem further compounded by the inherent stochasticity of LLMs, which produce drastically different trading actions across runs even under deterministic decoding (\Cref{appx_sec:instability_analysis}), rendering single-run live evaluations statistically unreliable. While existing benchmarks have advanced the evaluation of autonomous trading agents, they overlook the necessity of a holistic alpha mining pipeline that integrates strategy formulation, executable code generation, and rigorous backtesting, and largely neglect quantifying the inherent stochasticity and instability of LLM-driven financial decision-making.

\vspace{-0.3cm}
\section{AlphaForgeBench}

In this section we present \projectname, a benchmark that evaluates LLMs as quantitative researchers who synthesize executable trading strategies rather than emit point-wise trading actions. The overall framework is illustrated in \Cref{fig:alphaforgebench} and is organized along three axes. First, we describe the \emph{dataset construction} process (\Cref{sec:dataset}), which proceeds in two stages: Stage~1 collects natural-language queries together with their ground-truth alpha factors and trading strategies from diverse real-world sources; Stage~2 draws on the patterns and complexity profiles observed in these real-world samples to systematically generate augmented queries across a $3\times 3$ level--grade difficulty taxonomy via LLMs, combining authenticity with controlled diagnostic granularity. Second, we detail the \emph{evaluation pipeline} (\Cref{sec:pipeline}), in which each query is fed to the evaluated LLM to produce executable factor and strategy code, and the generated code is then executed within a standardized backtest engine across multiple assets and market regimes to yield quantitative performance profiles.
Third, we introduce the \emph{evaluation methodology} (\Cref{sec:eval_method}), comprising financial performance metrics and statistical protocols that assess both the absolute quality and the cross-run stability of LLM-generated strategies.

\vspace{-0.4cm}
\subsection{Dataset Construction}
\label{sec:dataset}

The construction of \projectname follows a two-stage pipeline that combines the ecological validity of real-world strategies with the diagnostic precision of synthetically structured queries.

\textbf{Stage 1: Real-world Strategy Collection.}
We curate a diverse corpus of alpha factors and factor-based trading strategies from five complementary source categories: brokerage research reports, quantitative investment platforms (WorldQuant~\citep{worldquant_platform}, JoinQuant~\citep{joinquant_platform}), AI-in-finance literature, open-source repositories (Qlib~\citep{yang2020qlib}, OpenFE~\citep{zhang2023openfe}), and traditional finance publications. We build an automated extraction agent, powered by \textit{gemini-3-flash-preview}, that ingests each collected document and produces structured records comprising factor names, mathematical definitions, trading logic, and financial rationale. After deduplication and quality filtering, Stage~1 yields \textbf{3,176} factor-strategy entries spanning three strategy types: single-asset trading (633), portfolio management (2,172), and multi-asset trading (371). In this work, we restrict evaluation to the \textbf{633} single-asset subset to isolate the LLM's signal-generation capability from confounding portfolio-construction effects, thereby serving as an ecologically valid baseline for gauging benchmark difficulty. This restriction is a deliberate design choice rather than a limitation of scope: the single-asset setting provides a cleaner, more controlled evaluation environment that targets core signal-generation and rule-construction capabilities without the additional complexity introduced by asset allocation, cross-asset dependencies, and portfolio-level risk constraints. The portfolio management (2,172) and multi-asset trading (371) subsets are reserved for systematic evaluation in future work, along with the corresponding backtesting infrastructure already developed. Full extraction prompts and dataset statistics are provided in \Cref{appx_sec:construction_alphaforgebench}.

\textbf{Stage 2: LLM-augmented Structured Query Generation.}
While Stage~1 offers ecologically valid test cases, its query distribution is non-uniform across difficulty and is not tailored for controlled diagnosis of specific cognitive demands. Stage~2 therefore constructs \textbf{270} additional benchmark queries under a $3\times 3$ \emph{level--grade} taxonomy, grounded in the strategy patterns and complexity profiles observed in the real-world collection. The three \emph{levels} isolate distinct strategy-generation skills: \textbf{Level~1} (Logic Translation) provides fully specified if--then rules to test faithful code translation; \textbf{Level~2} (Logic Completion) supplies strategic skeletons with critical parameters omitted, requiring domain-grounded inference; and \textbf{Level~3} (Goal-Oriented Generation) specifies only high-level investment objectives, demanding end-to-end strategy design from first principles. Orthogonally, three \emph{grades} (Easy, Medium, Hard) modulate complexity via the number of conditions, the degree of underspecification, and the depth of state-dependent control flow, yielding nine fine-grained difficulty cells (Appx.~\ref{appx_sec:construction_alphaforgebench}).

In summary, the \projectname query set comprises two complementary components: \textbf{633} real-world queries from Stage~1 that ensure ecological validity, and \textbf{270} difficulty-specialized queries from Stage~2 that enable controlled, fine-grained diagnostic evaluation across the full $3\times 3$ level--grade grid, providing balanced coverage and facilitating stratified analyses of model failure modes under varying cognitive demands.

\subsection{Evaluation Pipeline}
\label{sec:pipeline}

Given the curated query set described above (633 real-world queries from Stage~1 and 270 difficulty-specialized queries from Stage~2), \projectname evaluates each model via a standardized generate-and-backtest pipeline (\Cref{fig:alphaforgebench}), comprising prompt instantiation, code synthesis, and backtest-based assessment.

\textbf{Step 1: Prompt construction.}
Each benchmark query is assembled into a standardized prompt comprising three semantically distinct components (see the \emph{Evaluation Pipeline} panel of \Cref{fig:alphaforgebench}): (i)~a \emph{system instruction} that defines the code-generation task, specifies the available data schema (open-high-low-close-volume (OHLCV) columns augmented with precomputed technical-indicator factors), and prescribes the expected output interface; (ii)~the \emph{strategy query}, i.e., the natural-language description of the target trading strategy drawn from either Stage~1 or Stage~2; and (iii)~a \emph{factor-library reference} that enumerates all supported indicator names together with their formal mathematical definitions, providing the model with a complete and unambiguous specification of the default feature space. Notably, the model is not restricted to this predefined indicator set: if a strategy requires novel factors, the model may generate the corresponding factor-computation code, which our backtest engine dynamically registers and incorporates into the evaluation, thereby granting models the flexibility to extend the feature space on the fly. The prompt template is held strictly identical across all evaluated models, ensuring that any observed performance differences can be attributed solely to model capabilities rather than prompt-engineering artifacts.
 
\textbf{Step 2: Code generation.}
The assembled prompt is dispatched to each evaluated LLM through its official API. The model must return a self-contained Python function (\texttt{generate\_signal}) that consumes a dataframe of OHLCV columns and precomputed indicator factors and produces a trading-signal series dictating position actions (e.g., invest versus hold cash). We enforce strict conformance to the backtest engine's interface contract (function signature, permitted column references, and output format), enabling fully automated execution without manual intervention.

\textbf{Step 3: Backtest-based assessment.}
Each generated implementation is executed within a unified, deterministic backtest engine on historical daily price data spanning seven assets across two market regimes: cryptocurrency and US equity. The engine computes a suite of standard financial metrics covering return generation, risk exposure, and risk-adjusted efficiency, producing fully reproducible quantitative profiles that support systematic comparison across models, query sources, and the nine Stage~2 difficulty cells. This deterministic design ensures that any observed cross-run variance originates exclusively from the inherent stochasticity of LLM generation rather than evaluation-side randomness.

\subsection{Evaluation Methodology}
\label{sec:eval_method}

We evaluate LLM-generated strategies along three complementary dimensions. First, each backtest run yields a suite of standard financial metrics spanning return generation, risk exposure, and risk-adjusted efficiency; to account for the stochasticity of LLM generation, all metrics are reported as mean $\pm$ standard deviation over multiple independent runs, capturing both expected performance and generation stability. Second, results are organized into stratified tables along three axes: (i)~overall model ranking aggregated across all queries and assets, (ii)~per-asset decomposition over the 7 backtest assets to quantify cross-market generalization from cryptocurrency to US equity, and (iii)~per-level decomposition (Stage~2 only) by the three difficulty levels of the $3\times 3$ taxonomy to reveal how capabilities degrade under increasing cognitive demands. Third, tabular results are complemented by radar charts that render each model's multi-metric profile for intuitive risk-return comparison, grouped bar charts and box plots that expose cross-asset robustness and inter-model dispersion, and cumulative return curves that surface temporal dynamics such as divergence during market stress and convergence in calm regimes.

\vspace{-0.1cm}
\section{Experiments}
\label{sec:experiments}

We validate \projectname along two complementary evaluation tracks that mirror the two-stage dataset construction. \textbf{Track~1} (real-world queries) evaluates all 633 single-asset queries from Stage~1 to establish ecological validity, baseline difficulty calibration, and a consistency check against the structured track. \textbf{Track~2} (structured queries) evaluates the 270 queries from Stage~2 organized by the $3\times 3$ level--grade taxonomy, enabling fine-grained diagnosis of model strengths, weaknesses, and specific failure modes across difficulty dimensions. This dual-track design also serves as an empirical bias-mitigation mechanism: concordance between model rankings on Stage~1 (real-world) and Stage~2 (LLM-augmented) queries validates that the structured generation process preserves genuine capability differences rather than introducing systematic distributional bias favoring particular models.

\vspace{-0.2cm}
\subsection{Experimental Settings}
\textbf{Evaluated models.}
We benchmark six frontier LLMs spanning five providers: \textit{claude-sonnet-4.5}, \textit{deepseek-v3.2}, \textit{gemini-3-flash-preview}, \textit{gemini-3-pro-preview}, \textit{\textit{gpt-5.2}}, and \textit{grok-4.1-fast}. All models are queried through their official APIs under identical prompt templates with no model-specific tuning.

\textbf{Generation protocol.}
Every experiment runs $k{=}5$ independent generations per query to quantify run-to-run variability. Stage~1 uses $\tau{=}0.7$ ($633~queries \times 6~models \times 5~runs = 18{,}990~samples$); Stage~2 is evaluated at both $\tau{=}0.7$ and $\tau{=}0$ (greedy) for a temperature ablation ($270~queries \times 6~models \times 2~temperatures \times 5~runs = 16{,}200~samples$). The grand total is \textbf{35,190} generated strategy implementations. All metrics are reported as mean $\pm$ std.

\textbf{Backtest configuration.}
Each generated strategy is executed within the unified backtest engine across \textbf{7 assets} spanning two distinct market regimes: 2 cryptocurrencies (BTCUSDT, ETHUSDT; sourced from Binance) and 5 US equities (AAPL, GOOGL, MSFT, NVDA, TSLA; sourced from Yahoo Finance). The evaluation window spans \textbf{5 years} (2021-01-01 to 2026-01-01), deliberately selected to cover heterogeneous market conditions including bull rallies, bear corrections, AI-driven recoveries, and prolonged consolidation phases, ensuring that no single strategy style is systematically favored. We adopt a controlled execution protocol with daily data frequency, a 300-day lookback window, long-only single-asset semantics (binary invest/cash signal), and a fixed one-way transaction cost of $10^{-3}$, thereby isolating intrinsic signal quality from confounding portfolio-construction effects. The long-only constraint is adopted for cross-market consistency, as short-selling is subject to asymmetric restrictions and costs across equity and cryptocurrency markets. Slippage and liquidity constraints are intentionally excluded: these factors are highly dependent on market microstructure (e.g., order-book depth and execution mechanisms) and vary substantially across assets and trading frequencies; introducing a unified slippage model would add evaluation-side uncertainty without improving the fairness of cross-model comparisons. Full parameter justifications are provided in Appx.~\ref{appx_sec:details_of_alphaforgebench_experiments}.

\textbf{Evaluation metrics.}
We assess each strategy using six standard financial metrics: Annual Rate of Return (\textbf{ARR}), Sharpe Ratio (\textbf{SR}), Maximum Drawdown (\textbf{MDD}), Calmar Ratio (\textbf{CR}), Sortino Ratio (\textbf{SoR}), and Volatility (\textbf{VOL}). These jointly capture return generation (ARR), risk exposure (MDD, VOL), and risk-adjusted efficiency (SR, CR, SoR), providing a multi-dimensional profile of strategy quality. Formal definitions are given in Appx.~\ref{appx_sec:details_of_alphaforgebench_experiments}.

\vspace{-0.1cm}
\subsection{Results on Real-world Queries}
\label{sec:stage1_real_world_results}

\begin{table}[htb]
  \vspace{-0.2cm}
  \centering
  \caption{Results on real-world queries (mean $\pm$ std, 633 queries), stratified by overall (green) and asset (gray). Best per block in \textbf{bold}; $\uparrow$/$\downarrow$: higher/lower is better.}
  \label{tab:stage1_main_results}
  \vspace{-0.3cm}
  
  \scriptsize
  \renewcommand{\arraystretch}{0.4}
  \setlength{\tabcolsep}{0.9pt}
  \begin{tabular}{lcccccc}
  \toprule
  \textbf{Model} & \textbf{SR}$\uparrow$ & \textbf{ARR}$\uparrow$ & \textbf{MDD}$\downarrow$ & \textbf{CR}$\uparrow$ & \textbf{SoR}$\uparrow$ & \textbf{VOL}$\downarrow$ \\
  
  \midrule
  \rowcolor{green!5} \multicolumn{7}{c}{\textbf{Overall}} \\
  \textit{claude-sonnet-4.5} & 0.378{\scriptsize$\pm$0.268} & 0.138{\scriptsize$\pm$0.122} & 0.138{\scriptsize$\pm$0.122} & 1.456{\scriptsize$\pm$1.106} & 0.636{\scriptsize$\pm$0.488} & 0.187{\scriptsize$\pm$0.165} \\
  \textit{deepseek-v3.2} & 0.329{\scriptsize$\pm$0.272} & 0.116{\scriptsize$\pm$0.122} & \textbf{0.114}{\scriptsize$\pm$0.120} & \textbf{1.575}{\scriptsize$\pm$1.227} & 0.548{\scriptsize$\pm$0.494} & \textbf{0.155}{\scriptsize$\pm$0.163} \\
  \textit{gemini-3-flash-preview} & 0.388{\scriptsize$\pm$0.268} & 0.142{\scriptsize$\pm$0.122} & 0.138{\scriptsize$\pm$0.119} & 1.504{\scriptsize$\pm$1.131} & 0.648{\scriptsize$\pm$0.488} & 0.189{\scriptsize$\pm$0.161} \\
  \textbf{\textit{gemini-3-pro-preview}} & \textbf{0.449}{\scriptsize$\pm$0.262} & \textbf{0.171}{\scriptsize$\pm$0.123} & 0.174{\scriptsize$\pm$0.119} & 1.411{\scriptsize$\pm$1.165} & \textbf{0.767}{\scriptsize$\pm$0.493} & 0.237{\scriptsize$\pm$0.162} \\
  \textit{gpt-5.2} & 0.342{\scriptsize$\pm$0.279} & 0.123{\scriptsize$\pm$0.119} & 0.122{\scriptsize$\pm$0.118} & 1.534{\scriptsize$\pm$1.386} & 0.575{\scriptsize$\pm$0.503} & 0.166{\scriptsize$\pm$0.161} \\
  \textit{grok-4.1-fast} & 0.366{\scriptsize$\pm$0.276} & 0.135{\scriptsize$\pm$0.122} & 0.142{\scriptsize$\pm$0.124} & 1.396{\scriptsize$\pm$1.038} & 0.618{\scriptsize$\pm$0.500} & 0.192{\scriptsize$\pm$0.168} \\

  \midrule
  \rowcolor{gray!15} \multicolumn{7}{c}{\textbf{BTCUSDT} (Cryptocurrency)} \\
  \textit{claude-sonnet-4.5} & 0.279{\scriptsize$\pm$0.234} & 0.110{\scriptsize$\pm$0.099} & 0.189{\scriptsize$\pm$0.154} & 0.732{\scriptsize$\pm$0.792} & 0.490{\scriptsize$\pm$0.398} & 0.213{\scriptsize$\pm$0.175} \\
  \textit{deepseek-v3.2} & 0.239{\scriptsize$\pm$0.231} & 0.094{\scriptsize$\pm$0.097} & \textbf{0.155}{\scriptsize$\pm$0.155} & 0.861{\scriptsize$\pm$0.967} & 0.421{\scriptsize$\pm$0.393} & 0.175{\scriptsize$\pm$0.175} \\
  \textit{gemini-3-flash-preview} & 0.294{\scriptsize$\pm$0.225} & 0.115{\scriptsize$\pm$0.096} & 0.193{\scriptsize$\pm$0.152} & 0.800{\scriptsize$\pm$0.865} & 0.511{\scriptsize$\pm$0.380} & 0.218{\scriptsize$\pm$0.171} \\
  \textbf{\textit{gemini-3-pro-preview}} & \textbf{0.338}{\scriptsize$\pm$0.223} & \textbf{0.131}{\scriptsize$\pm$0.097} & 0.241{\scriptsize$\pm$0.149} & 0.628{\scriptsize$\pm$0.552} & \textbf{0.592}{\scriptsize$\pm$0.376} & 0.272{\scriptsize$\pm$0.167} \\
  \textit{gpt-5.2} & 0.255{\scriptsize$\pm$0.230} & 0.098{\scriptsize$\pm$0.097} & 0.170{\scriptsize$\pm$0.152} & 0.719{\scriptsize$\pm$0.713} & 0.446{\scriptsize$\pm$0.394} & 0.192{\scriptsize$\pm$0.172} \\
  \textit{grok-4.1-fast} & 0.273{\scriptsize$\pm$0.231} & 0.104{\scriptsize$\pm$0.096} & 0.196{\scriptsize$\pm$0.160} & 0.701{\scriptsize$\pm$0.782} & 0.481{\scriptsize$\pm$0.389} & 0.220{\scriptsize$\pm$0.179} \\
  \midrule
  \rowcolor{gray!15} \multicolumn{7}{c}{\textbf{ETHUSDT} (Cryptocurrency)} \\
  \textit{claude-sonnet-4.5} & 0.260{\scriptsize$\pm$0.232} & 0.119{\scriptsize$\pm$0.149} & 0.205{\scriptsize$\pm$0.208} & 0.778{\scriptsize$\pm$0.626} & 0.398{\scriptsize$\pm$0.359} & 0.242{\scriptsize$\pm$0.244} \\
  \textit{deepseek-v3.2} & 0.220{\scriptsize$\pm$0.233} & 0.103{\scriptsize$\pm$0.147} & \textbf{0.169}{\scriptsize$\pm$0.205} & \textbf{0.851}{\scriptsize$\pm$0.669} & 0.336{\scriptsize$\pm$0.361} & \textbf{0.200}{\scriptsize$\pm$0.241} \\
  \textit{gemini-3-flash-preview} & 0.260{\scriptsize$\pm$0.236} & 0.121{\scriptsize$\pm$0.151} & 0.206{\scriptsize$\pm$0.203} & 0.737{\scriptsize$\pm$0.606} & 0.400{\scriptsize$\pm$0.363} & 0.242{\scriptsize$\pm$0.239} \\
  \textbf{\textit{gemini-3-pro-preview}} & \textbf{0.306}{\scriptsize$\pm$0.224} & \textbf{0.137}{\scriptsize$\pm$0.152} & 0.263{\scriptsize$\pm$0.198} & 0.633{\scriptsize$\pm$0.558} & \textbf{0.475}{\scriptsize$\pm$0.344} & 0.307{\scriptsize$\pm$0.232} \\
  \textit{gpt-5.2} & 0.219{\scriptsize$\pm$0.218} & 0.084{\scriptsize$\pm$0.106} & 0.180{\scriptsize$\pm$0.198} & 0.675{\scriptsize$\pm$0.641} & 0.336{\scriptsize$\pm$0.333} & 0.211{\scriptsize$\pm$0.232} \\
  \textit{grok-4.1-fast} & 0.255{\scriptsize$\pm$0.233} & 0.112{\scriptsize$\pm$0.145} & 0.218{\scriptsize$\pm$0.209} & 0.645{\scriptsize$\pm$0.564} & 0.392{\scriptsize$\pm$0.358} & 0.254{\scriptsize$\pm$0.244} \\
  \midrule
  \rowcolor{gray!15} \multicolumn{7}{c}{\textbf{AAPL} (US Equity)} \\
  \textit{claude-sonnet-4.5} & 0.757{\scriptsize$\pm$0.533} & 0.151{\scriptsize$\pm$0.118} & 0.079{\scriptsize$\pm$0.059} & 2.282{\scriptsize$\pm$1.784} & 1.068{\scriptsize$\pm$0.771} & 0.123{\scriptsize$\pm$0.089} \\
  \textit{deepseek-v3.2} & 0.704{\scriptsize$\pm$0.546} & 0.138{\scriptsize$\pm$0.118} & \textbf{0.069}{\scriptsize$\pm$0.060} & \textbf{2.529}{\scriptsize$\pm$2.036} & 0.994{\scriptsize$\pm$0.782} & \textbf{0.108}{\scriptsize$\pm$0.089} \\
  \textit{gemini-3-flash-preview} & 0.756{\scriptsize$\pm$0.522} & 0.151{\scriptsize$\pm$0.116} & 0.081{\scriptsize$\pm$0.058} & 2.252{\scriptsize$\pm$1.910} & 1.067{\scriptsize$\pm$0.756} & 0.126{\scriptsize$\pm$0.087} \\
  \textbf{\textit{gemini-3-pro-preview}} & \textbf{0.805}{\scriptsize$\pm$0.500} & \textbf{0.162}{\scriptsize$\pm$0.113} & 0.094{\scriptsize$\pm$0.059} & 1.954{\scriptsize$\pm$1.431} & \textbf{1.138}{\scriptsize$\pm$0.738} & 0.143{\scriptsize$\pm$0.087} \\
  \textit{gpt-5.2} & 0.668{\scriptsize$\pm$0.547} & 0.132{\scriptsize$\pm$0.120} & 0.071{\scriptsize$\pm$0.061} & 2.273{\scriptsize$\pm$1.880} & 0.944{\scriptsize$\pm$0.794} & 0.110{\scriptsize$\pm$0.092} \\
  \textit{grok-4.1-fast} & 0.686{\scriptsize$\pm$0.545} & 0.138{\scriptsize$\pm$0.119} & 0.077{\scriptsize$\pm$0.062} & 2.095{\scriptsize$\pm$1.709} & 0.973{\scriptsize$\pm$0.786} & 0.119{\scriptsize$\pm$0.092} \\
  \midrule
  \rowcolor{gray!15} \multicolumn{7}{c}{\textbf{GOOGL} (US Equity)} \\
  \textit{claude-sonnet-4.5} & 0.657{\scriptsize$\pm$0.539} & 0.190{\scriptsize$\pm$0.203} & 0.121{\scriptsize$\pm$0.121} & 2.374{\scriptsize$\pm$2.209} & 1.177{\scriptsize$\pm$1.024} & 0.184{\scriptsize$\pm$0.174} \\
  \textit{deepseek-v3.2} & 0.541{\scriptsize$\pm$0.532} & 0.151{\scriptsize$\pm$0.191} & \textbf{0.099}{\scriptsize$\pm$0.116} & 2.204{\scriptsize$\pm$1.988} & 0.959{\scriptsize$\pm$1.008} & \textbf{0.150}{\scriptsize$\pm$0.169} \\
  \textit{gemini-3-flash-preview} & 0.656{\scriptsize$\pm$0.539} & 0.188{\scriptsize$\pm$0.205} & 0.124{\scriptsize$\pm$0.123} & 2.335{\scriptsize$\pm$2.279} & 1.170{\scriptsize$\pm$1.023} & 0.190{\scriptsize$\pm$0.176} \\
  \textbf{\textit{gemini-3-pro-preview}} & \textbf{0.752}{\scriptsize$\pm$0.525} & \textbf{0.225}{\scriptsize$\pm$0.209} & 0.150{\scriptsize$\pm$0.129} & 2.341{\scriptsize$\pm$2.210} & \textbf{1.390}{\scriptsize$\pm$1.011} & 0.229{\scriptsize$\pm$0.183} \\
  \textit{gpt-5.2} & 0.602{\scriptsize$\pm$0.556} & 0.179{\scriptsize$\pm$0.206} & 0.109{\scriptsize$\pm$0.121} & \textbf{2.502}{\scriptsize$\pm$2.422} & 1.089{\scriptsize$\pm$1.056} & 0.167{\scriptsize$\pm$0.175} \\
  \textit{grok-4.1-fast} & 0.617{\scriptsize$\pm$0.542} & 0.175{\scriptsize$\pm$0.203} & 0.121{\scriptsize$\pm$0.128} & 2.331{\scriptsize$\pm$2.231} & 1.113{\scriptsize$\pm$1.028} & 0.183{\scriptsize$\pm$0.184} \\
  \midrule
  \rowcolor{gray!15} \multicolumn{7}{c}{\textbf{MSFT} (US Equity)} \\
  \textit{claude-sonnet-4.5} & 0.142{\scriptsize$\pm$0.248} & 0.020{\scriptsize$\pm$0.036} & 0.091{\scriptsize$\pm$0.068} & 0.319{\scriptsize$\pm$0.703} & 0.231{\scriptsize$\pm$0.285} & 0.121{\scriptsize$\pm$0.088} \\
  \textit{deepseek-v3.2} & 0.133{\scriptsize$\pm$0.242} & 0.020{\scriptsize$\pm$0.034} & \textbf{0.079}{\scriptsize$\pm$0.067} & \textbf{0.352}{\scriptsize$\pm$0.776} & 0.214{\scriptsize$\pm$0.280} & \textbf{0.105}{\scriptsize$\pm$0.088} \\
  \textit{gemini-3-flash-preview} & 0.147{\scriptsize$\pm$0.250} & 0.021{\scriptsize$\pm$0.036} & 0.091{\scriptsize$\pm$0.066} & 0.289{\scriptsize$\pm$0.662} & 0.235{\scriptsize$\pm$0.288} & 0.121{\scriptsize$\pm$0.086} \\
  \textbf{\textit{gemini-3-pro-preview}} & \textbf{0.176}{\scriptsize$\pm$0.253} & \textbf{0.024}{\scriptsize$\pm$0.037} & 0.102{\scriptsize$\pm$0.067} & 0.316{\scriptsize$\pm$0.620} & \textbf{0.268}{\scriptsize$\pm$0.290} & 0.137{\scriptsize$\pm$0.086} \\
  \textit{gpt-5.2} & 0.125{\scriptsize$\pm$0.245} & 0.018{\scriptsize$\pm$0.034} & 0.081{\scriptsize$\pm$0.070} & 0.293{\scriptsize$\pm$0.608} & 0.204{\scriptsize$\pm$0.284} & 0.108{\scriptsize$\pm$0.091} \\
  \textit{grok-4.1-fast} & 0.138{\scriptsize$\pm$0.232} & 0.019{\scriptsize$\pm$0.034} & 0.086{\scriptsize$\pm$0.069} & 0.306{\scriptsize$\pm$0.665} & 0.217{\scriptsize$\pm$0.273} & 0.115{\scriptsize$\pm$0.090} \\
  \midrule
  \rowcolor{gray!15} \multicolumn{7}{c}{\textbf{NVDA} (US Equity)} \\
  \textit{claude-sonnet-4.5} & 0.289{\scriptsize$\pm$0.324} & 0.090{\scriptsize$\pm$0.135} & 0.133{\scriptsize$\pm$0.153} & 1.071{\scriptsize$\pm$1.506} & 0.558{\scriptsize$\pm$0.635} & 0.207{\scriptsize$\pm$0.233} \\
  \textit{deepseek-v3.2} & 0.242{\scriptsize$\pm$0.315} & 0.074{\scriptsize$\pm$0.128} & \textbf{0.110}{\scriptsize$\pm$0.149} & 1.090{\scriptsize$\pm$1.522} & 0.469{\scriptsize$\pm$0.618} & \textbf{0.171}{\scriptsize$\pm$0.227} \\
  \textit{gemini-3-flash-preview} & 0.304{\scriptsize$\pm$0.320} & 0.096{\scriptsize$\pm$0.136} & 0.134{\scriptsize$\pm$0.152} & \textbf{1.189}{\scriptsize$\pm$1.555} & 0.586{\scriptsize$\pm$0.633} & 0.209{\scriptsize$\pm$0.231} \\
  \textbf{\textit{gemini-3-pro-preview}} & \textbf{0.372}{\scriptsize$\pm$0.318} & \textbf{0.112}{\scriptsize$\pm$0.138} & 0.184{\scriptsize$\pm$0.162} & 0.939{\scriptsize$\pm$1.314} & \textbf{0.739}{\scriptsize$\pm$0.650} & 0.284{\scriptsize$\pm$0.244} \\
  \textit{gpt-5.2} & 0.272{\scriptsize$\pm$0.315} & 0.086{\scriptsize$\pm$0.132} & 0.122{\scriptsize$\pm$0.151} & 1.094{\scriptsize$\pm$1.355} & 0.532{\scriptsize$\pm$0.634} & 0.190{\scriptsize$\pm$0.229} \\
  \textit{grok-4.1-fast} & 0.293{\scriptsize$\pm$0.327} & 0.088{\scriptsize$\pm$0.130} & 0.139{\scriptsize$\pm$0.155} & 0.952{\scriptsize$\pm$1.262} & 0.563{\scriptsize$\pm$0.634} & 0.214{\scriptsize$\pm$0.235} \\
  \midrule
  \rowcolor{gray!15} \multicolumn{7}{c}{\textbf{TSLA} (US Equity)} \\
  \textit{claude-sonnet-4.5} & 0.274{\scriptsize$\pm$0.420} & 0.289{\scriptsize$\pm$0.381} & 0.140{\scriptsize$\pm$0.174} & 2.611{\scriptsize$\pm$2.772} & 0.545{\scriptsize$\pm$0.713} & 0.216{\scriptsize$\pm$0.264} \\
  \textit{deepseek-v3.2} & 0.240{\scriptsize$\pm$0.393} & 0.239{\scriptsize$\pm$0.359} & \textbf{0.115}{\scriptsize$\pm$0.164} & 2.629{\scriptsize$\pm$2.745} & 0.466{\scriptsize$\pm$0.680} & \textbf{0.179}{\scriptsize$\pm$0.251} \\
  \textit{gemini-3-flash-preview} & 0.300{\scriptsize$\pm$0.425} & 0.304{\scriptsize$\pm$0.383} & 0.138{\scriptsize$\pm$0.168} & 2.781{\scriptsize$\pm$2.794} & 0.572{\scriptsize$\pm$0.713} & 0.215{\scriptsize$\pm$0.256} \\
  \textbf{\textit{gemini-3-pro-preview}} & \textbf{0.396}{\scriptsize$\pm$0.444} & \textbf{0.409}{\scriptsize$\pm$0.400} & 0.186{\scriptsize$\pm$0.175} & 2.756{\scriptsize$\pm$2.738} & \textbf{0.770}{\scriptsize$\pm$0.738} & 0.290{\scriptsize$\pm$0.264} \\
  \textit{gpt-5.2} & 0.275{\scriptsize$\pm$0.431} & 0.272{\scriptsize$\pm$0.379} & 0.120{\scriptsize$\pm$0.163} & \textbf{2.972}{\scriptsize$\pm$3.080} & 0.517{\scriptsize$\pm$0.710} & 0.189{\scriptsize$\pm$0.251} \\
  \textit{grok-4.1-fast} & 0.303{\scriptsize$\pm$0.422} & 0.315{\scriptsize$\pm$0.383} & 0.153{\scriptsize$\pm$0.173} & 2.659{\scriptsize$\pm$2.676} & 0.597{\scriptsize$\pm$0.713} & 0.237{\scriptsize$\pm$0.262} \\
  \bottomrule
  \end{tabular}
  \vspace{-0.3cm}
\end{table}

\subsubsection{Overall model comparison.}
As shown in the Overall block of \Cref{tab:stage1_main_results}, three salient observations emerge. \textbf{(1)~A clear performance hierarchy exists on return-oriented metrics.} Models separate into three tiers: \textit{gemini-3-pro-preview} leads decisively, followed by a middle tier (\textit{gemini-3-flash-preview}, \textit{claude-sonnet-4.5}), with the remaining models trailing. The 5.5\% ARR spread between the best and worst models translates to a 47\% relative improvement, which is economically meaningful over the 5-year evaluation horizon. \textbf{(2)~Return and risk rankings are inversely correlated}, suggesting that LLMs encode distinct implicit risk preferences in their generated code. Notably, \textit{gemini-3-pro-preview} consistently favors aggressive, high-conviction signal logic (highest SR and ARR, but also highest MDD and VOL), whereas \textit{deepseek-v3.2} converges on conservative, risk-controlled strategies (lowest MDD and VOL, yet highest CR). This return-risk inversion underscores the necessity of multi-metric evaluation: a single metric cannot capture the full spectrum of model behavior. \textbf{(3)~Observed performance gaps are statistically reliable.} The intra-query standard deviation across 5 runs is typically an order of magnitude smaller than the inter-query standard deviation, confirming that ranking differences reflect genuine capability gaps rather than generation noise. Detailed distributional and visual analyses are provided in Appx.~\ref{appx_sec:appendix_results_stage_1}.

\subsubsection{Per-asset analysis.}
Disaggregating by asset in \Cref{tab:stage1_main_results} yields three further observations. \textbf{(1)~A consistent difficulty gradient emerges across assets.} Trend-rich US large-caps (AAPL, GOOGL) are the easiest targets for all models, cryptocurrencies (BTCUSDT, ETHUSDT) occupy an intermediate tier with higher volatility and deeper drawdowns, and MSFT proves the hardest asset, likely due to its narrower trading ranges during the evaluation period. \textbf{(2)~High-volatility assets amplify inter-model dispersion.} TSLA exhibits both the highest absolute returns and the widest cross-model variance, indicating that extreme and highly non-stationary price dynamics effectively magnify differences in the signal logic generated by each LLM. \textbf{(3)~Model rankings are robust across asset classes.} The relative ordering identified in the aggregate analysis is preserved across all seven assets: return-oriented leaders and risk-oriented leaders maintain their respective positions regardless of whether the market is cryptocurrency or US equity. This cross-asset stability strongly suggests that the implicit risk preferences encoded by each LLM are intrinsic to its own generation behavior rather than artifacts of any particular market environment.

\subsubsection{Findings and conclusions.}
Synthesizing the above analyses, the Stage~1 evaluation yields four principal findings.
\textbf{(1)~High code-generation reliability}: all six frontier LLMs produce executable strategy code with $>$96\% backtest pass rates, establishing a solid foundation for the code-generation evaluation paradigm.
\textbf{(2)~Stable and reproducible evaluation}: the intra-query (run-to-run) variance is an order of magnitude smaller than the inter-query variance, confirming that the code-generation paradigm confines LLM stochasticity to a single generation step while guaranteeing deterministic execution thereafter, in stark contrast to direct-trading benchmarks where run-to-run variance routinely exceeds inter-model variance.
\textbf{(3)~Implicit risk preferences}: as revealed by the return-risk inversion in the overall comparison, different LLMs encode distinct and stable risk personalities in their generated code, and these characteristic profiles persist across all runs, assets, and metrics.
\textbf{(4)~Cross-asset robustness}: model rankings and a shared asset difficulty gradient are consistent across all 7 assets spanning cryptocurrency and US equity markets, indicating that the observed capability hierarchy is environment-agnostic.
Taken together, these findings validate the strategy code-generation paradigm as a stable, reproducible, and discriminative framework for benchmarking LLM capabilities in quantitative finance. They also demonstrate that real-world queries alone can reveal meaningful capability differences among frontier models, thereby providing a solid empirical foundation and directly motivating the more fine-grained, difficulty-stratified diagnostic evaluation in Stage~2.

\subsection{Results on LLM-augmented Queries}
\label{sec:stage2_results}

\begin{table*}[htb]
  \centering
  \caption{Results on LLM-augmented queries (mean $\pm$ std, 5 runs) at $T{=}0$ and $T{=}0.7$, stratified by overall (green), level (blue), and asser (gray). Best per block in \textbf{bold}; $\uparrow$/$\downarrow$: higher/lower is better.}
  \vspace{-0.3cm}
  \label{tab:stage2_main_results}
  \scriptsize
  \renewcommand{\arraystretch}{0.4}
  \setlength{\tabcolsep}{2pt}
  \begin{tabular}{lcccccccccccc}
  \toprule
  \multirow{2}{*}{\textbf{Model}} & \multicolumn{2}{c}{\textbf{SR}$\uparrow$} & \multicolumn{2}{c}{\textbf{ARR}$\uparrow$} & \multicolumn{2}{c}{\textbf{MDD}$\downarrow$} & \multicolumn{2}{c}{\textbf{CR}$\uparrow$} & \multicolumn{2}{c}{\textbf{SOR}$\uparrow$} & \multicolumn{2}{c}{\textbf{VOL}$\downarrow$} \\
  & T=0 & T=0.7 & T=0 & T=0.7 & T=0 & T=0.7 & T=0 & T=0.7 & T=0 & T=0.7 & T=0 & T=0.7 \\

  \midrule
  \rowcolor{green!5} \multicolumn{13}{c}{\textbf{Overall}} \\
  \textit{claude-sonnet-4.5} & 0.513{\scriptsize$\pm$0.270} & 0.508{\scriptsize$\pm$0.266} & 0.164{\scriptsize$\pm$0.114} & 0.162{\scriptsize$\pm$0.113} & 0.150{\scriptsize$\pm$0.111} & 0.147{\scriptsize$\pm$0.110} & \textbf{1.650{\scriptsize$\pm$0.864}} & 1.634{\scriptsize$\pm$0.620} & 0.806{\scriptsize$\pm$0.478} & 0.795{\scriptsize$\pm$0.471} & 0.205{\scriptsize$\pm$0.153} & 0.202{\scriptsize$\pm$0.152} \\
  \textit{deepseek-v3.2} & 0.430{\scriptsize$\pm$0.280} & 0.424{\scriptsize$\pm$0.289} & 0.132{\scriptsize$\pm$0.110} & 0.130{\scriptsize$\pm$0.112} & 0.127{\scriptsize$\pm$0.108} & 0.123{\scriptsize$\pm$0.109} & 1.586{\scriptsize$\pm$0.800} & 1.570{\scriptsize$\pm$0.761} & 0.672{\scriptsize$\pm$0.477} & 0.660{\scriptsize$\pm$0.489} & 0.173{\scriptsize$\pm$0.149} & 0.168{\scriptsize$\pm$0.150} \\
  \textit{gpt-5.2} & 0.415{\scriptsize$\pm$0.307} & 0.417{\scriptsize$\pm$0.308} & 0.130{\scriptsize$\pm$0.122} & 0.130{\scriptsize$\pm$0.121} & \textbf{0.119{\scriptsize$\pm$0.116}} & \textbf{0.117{\scriptsize$\pm$0.114}} & 1.599{\scriptsize$\pm$0.794} & 1.660{\scriptsize$\pm$0.821} & 0.645{\scriptsize$\pm$0.525} & 0.643{\scriptsize$\pm$0.522} & \textbf{0.163{\scriptsize$\pm$0.160}} & \textbf{0.160{\scriptsize$\pm$0.157}} \\
  \textit{gemini-3-flash-preview} & 0.523{\scriptsize$\pm$0.289} & 0.530{\scriptsize$\pm$0.264} & 0.162{\scriptsize$\pm$0.122} & 0.165{\scriptsize$\pm$0.113} & 0.148{\scriptsize$\pm$0.117} & 0.151{\scriptsize$\pm$0.111} & 1.618{\scriptsize$\pm$0.904} & 1.658{\scriptsize$\pm$1.384} & 0.807{\scriptsize$\pm$0.506} & 0.820{\scriptsize$\pm$0.468} & 0.204{\scriptsize$\pm$0.162} & 0.206{\scriptsize$\pm$0.153} \\
  \textbf{\textit{gemini-3-pro-preview}} & \textbf{0.628{\scriptsize$\pm$0.255}} & \textbf{0.627{\scriptsize$\pm$0.242}} & \textbf{0.208{\scriptsize$\pm$0.112}} & \textbf{0.209{\scriptsize$\pm$0.107}} & 0.191{\scriptsize$\pm$0.109} & 0.188{\scriptsize$\pm$0.104} & 1.586{\scriptsize$\pm$0.665} & 1.639{\scriptsize$\pm$0.651} & \textbf{1.004{\scriptsize$\pm$0.465}} & \textbf{0.999{\scriptsize$\pm$0.439}} & 0.262{\scriptsize$\pm$0.150} & 0.259{\scriptsize$\pm$0.143} \\
  \textit{grok-4.1-fast} & 0.421{\scriptsize$\pm$0.269} & 0.429{\scriptsize$\pm$0.267} & 0.134{\scriptsize$\pm$0.108} & 0.135{\scriptsize$\pm$0.107} & 0.125{\scriptsize$\pm$0.102} & 0.127{\scriptsize$\pm$0.102} & 1.629{\scriptsize$\pm$0.741} & \textbf{1.692{\scriptsize$\pm$0.819}} & 0.658{\scriptsize$\pm$0.461} & 0.668{\scriptsize$\pm$0.457} & 0.171{\scriptsize$\pm$0.140} & 0.173{\scriptsize$\pm$0.140} \\

    \midrule
    \rowcolor{blue!5} \multicolumn{13}{c}{\textbf{Level 1} (Logic Translation)} \\
    \textit{claude-sonnet-4.5} & 0.549{\scriptsize$\pm$0.306} & 0.551{\scriptsize$\pm$0.305} & 0.176{\scriptsize$\pm$0.131} & 0.177{\scriptsize$\pm$0.131} & 0.157{\scriptsize$\pm$0.119} & 0.158{\scriptsize$\pm$0.119} & \textbf{1.676{\scriptsize$\pm$0.746}} & \textbf{1.690{\scriptsize$\pm$0.719}} & 0.852{\scriptsize$\pm$0.544} & 0.856{\scriptsize$\pm$0.543} & 0.219{\scriptsize$\pm$0.167} & 0.219{\scriptsize$\pm$0.166} \\
    \textbf{\textit{deepseek-v3.2}} & \textbf{0.561{\scriptsize$\pm$0.296}} & \textbf{0.559{\scriptsize$\pm$0.296}} & \textbf{0.180{\scriptsize$\pm$0.128}} & \textbf{0.181{\scriptsize$\pm$0.128}} & 0.163{\scriptsize$\pm$0.117} & 0.162{\scriptsize$\pm$0.117} & 1.664{\scriptsize$\pm$0.706} & 1.675{\scriptsize$\pm$0.711} & \textbf{0.873{\scriptsize$\pm$0.529}} & \textbf{0.870{\scriptsize$\pm$0.530}} & 0.226{\scriptsize$\pm$0.164} & 0.224{\scriptsize$\pm$0.163} \\
    \textit{gpt-5.2} & 0.544{\scriptsize$\pm$0.313} & 0.545{\scriptsize$\pm$0.315} & 0.175{\scriptsize$\pm$0.132} & 0.175{\scriptsize$\pm$0.133} & 0.156{\scriptsize$\pm$0.121} & 0.156{\scriptsize$\pm$0.121} & 1.637{\scriptsize$\pm$0.769} & 1.642{\scriptsize$\pm$0.773} & 0.846{\scriptsize$\pm$0.552} & 0.849{\scriptsize$\pm$0.554} & 0.217{\scriptsize$\pm$0.169} & 0.217{\scriptsize$\pm$0.169} \\
    \textit{gemini-3-flash-preview} & 0.543{\scriptsize$\pm$0.316} & 0.543{\scriptsize$\pm$0.313} & 0.176{\scriptsize$\pm$0.133} & 0.175{\scriptsize$\pm$0.132} & 0.157{\scriptsize$\pm$0.122} & 0.156{\scriptsize$\pm$0.121} & 1.644{\scriptsize$\pm$0.774} & 1.672{\scriptsize$\pm$0.763} & 0.847{\scriptsize$\pm$0.556} & 0.845{\scriptsize$\pm$0.552} & 0.218{\scriptsize$\pm$0.170} & 0.217{\scriptsize$\pm$0.169} \\
    \textit{gemini-3-pro-preview} & 0.545{\scriptsize$\pm$0.314} & 0.543{\scriptsize$\pm$0.314} & 0.176{\scriptsize$\pm$0.132} & 0.175{\scriptsize$\pm$0.133} & 0.157{\scriptsize$\pm$0.121} & \textbf{0.156{\scriptsize$\pm$0.121}} & 1.647{\scriptsize$\pm$0.771} & 1.651{\scriptsize$\pm$0.772} & 0.850{\scriptsize$\pm$0.554} & 0.846{\scriptsize$\pm$0.554} & 0.218{\scriptsize$\pm$0.169} & 0.217{\scriptsize$\pm$0.169} \\
    \textit{grok-4.1-fast} & 0.532{\scriptsize$\pm$0.302} & 0.532{\scriptsize$\pm$0.304} & 0.172{\scriptsize$\pm$0.130} & 0.172{\scriptsize$\pm$0.130} & \textbf{0.155{\scriptsize$\pm$0.119}} & 0.156{\scriptsize$\pm$0.119} & 1.675{\scriptsize$\pm$0.714} & 1.674{\scriptsize$\pm$0.734} & 0.827{\scriptsize$\pm$0.537} & 0.829{\scriptsize$\pm$0.539} & \textbf{0.215{\scriptsize$\pm$0.166}} & \textbf{0.216{\scriptsize$\pm$0.166}} \\
    \midrule
    \rowcolor{blue!5} \multicolumn{13}{c}{\textbf{Level 2} (Parameter Inference)} \\
    \textit{claude-sonnet-4.5} & 0.482{\scriptsize$\pm$0.249} & 0.482{\scriptsize$\pm$0.249} & 0.151{\scriptsize$\pm$0.104} & 0.149{\scriptsize$\pm$0.102} & 0.136{\scriptsize$\pm$0.109} & 0.136{\scriptsize$\pm$0.106} & \textbf{1.819{\scriptsize$\pm$1.217}} & 1.659{\scriptsize$\pm$0.673} & 0.746{\scriptsize$\pm$0.442} & 0.739{\scriptsize$\pm$0.433} & 0.185{\scriptsize$\pm$0.148} & 0.185{\scriptsize$\pm$0.145} \\
    \textit{deepseek-v3.2} & 0.401{\scriptsize$\pm$0.251} & 0.406{\scriptsize$\pm$0.273} & 0.117{\scriptsize$\pm$0.095} & 0.119{\scriptsize$\pm$0.100} & 0.110{\scriptsize$\pm$0.103} & 0.112{\scriptsize$\pm$0.108} & 1.564{\scriptsize$\pm$0.805} & 1.568{\scriptsize$\pm$0.832} & 0.612{\scriptsize$\pm$0.417} & 0.625{\scriptsize$\pm$0.454} & 0.149{\scriptsize$\pm$0.139} & 0.152{\scriptsize$\pm$0.146} \\
    \textit{gpt-5.2} & 0.366{\scriptsize$\pm$0.262} & 0.384{\scriptsize$\pm$0.266} & 0.104{\scriptsize$\pm$0.097} & 0.109{\scriptsize$\pm$0.095} & \textbf{0.095{\scriptsize$\pm$0.096}} & \textbf{0.095{\scriptsize$\pm$0.093}} & 1.553{\scriptsize$\pm$0.894} & 1.798{\scriptsize$\pm$0.963} & 0.546{\scriptsize$\pm$0.431} & 0.564{\scriptsize$\pm$0.432} & \textbf{0.129{\scriptsize$\pm$0.130}} & \textbf{0.130{\scriptsize$\pm$0.127}} \\
    \textit{gemini-3-flash-preview} & 0.493{\scriptsize$\pm$0.271} & 0.515{\scriptsize$\pm$0.243} & 0.144{\scriptsize$\pm$0.112} & 0.151{\scriptsize$\pm$0.100} & 0.130{\scriptsize$\pm$0.111} & 0.138{\scriptsize$\pm$0.105} & 1.698{\scriptsize$\pm$1.239} & 1.792{\scriptsize$\pm$2.233} & 0.736{\scriptsize$\pm$0.462} & 0.778{\scriptsize$\pm$0.420} & 0.177{\scriptsize$\pm$0.152} & 0.188{\scriptsize$\pm$0.142} \\
    \textbf{\textit{gemini-3-pro-preview}} & \textbf{0.604{\scriptsize$\pm$0.224}} & \textbf{0.632{\scriptsize$\pm$0.197}} & \textbf{0.196{\scriptsize$\pm$0.096}} & \textbf{0.209{\scriptsize$\pm$0.089}} & 0.171{\scriptsize$\pm$0.095} & 0.184{\scriptsize$\pm$0.091} & 1.684{\scriptsize$\pm$0.761} & 1.740{\scriptsize$\pm$0.753} & \textbf{0.942{\scriptsize$\pm$0.394}} & \textbf{0.991{\scriptsize$\pm$0.357}} & 0.234{\scriptsize$\pm$0.129} & 0.252{\scriptsize$\pm$0.123} \\
    \textit{grok-4.1-fast} & 0.401{\scriptsize$\pm$0.240} & 0.422{\scriptsize$\pm$0.239} & 0.127{\scriptsize$\pm$0.090} & 0.133{\scriptsize$\pm$0.089} & 0.117{\scriptsize$\pm$0.088} & 0.122{\scriptsize$\pm$0.089} & 1.639{\scriptsize$\pm$0.803} & \textbf{1.820{\scriptsize$\pm$1.084}} & 0.622{\scriptsize$\pm$0.402} & 0.653{\scriptsize$\pm$0.395} & 0.158{\scriptsize$\pm$0.119} & 0.165{\scriptsize$\pm$0.120} \\
    \midrule
    \rowcolor{blue!5} \multicolumn{13}{c}{\textbf{Level 3} (Goal-Oriented Generation)} \\
    \textit{claude-sonnet-4.5} & 0.507{\scriptsize$\pm$0.247} & 0.492{\scriptsize$\pm$0.234} & 0.165{\scriptsize$\pm$0.104} & 0.160{\scriptsize$\pm$0.102} & 0.156{\scriptsize$\pm$0.102} & 0.148{\scriptsize$\pm$0.104} & 1.452{\scriptsize$\pm$0.353} & 1.553{\scriptsize$\pm$0.414} & 0.820{\scriptsize$\pm$0.434} & 0.792{\scriptsize$\pm$0.420} & 0.212{\scriptsize$\pm$0.140} & 0.202{\scriptsize$\pm$0.142} \\
    \textit{deepseek-v3.2} & 0.329{\scriptsize$\pm$0.240} & 0.304{\scriptsize$\pm$0.235} & 0.099{\scriptsize$\pm$0.086} & 0.090{\scriptsize$\pm$0.084} & 0.107{\scriptsize$\pm$0.094} & \textbf{0.095{\scriptsize$\pm$0.087}} & 1.530{\scriptsize$\pm$0.876} & 1.458{\scriptsize$\pm$0.723} & 0.531{\scriptsize$\pm$0.407} & 0.483{\scriptsize$\pm$0.391} & 0.142{\scriptsize$\pm$0.126} & \textbf{0.126{\scriptsize$\pm$0.118}} \\
    \textit{gpt-5.2} & 0.336{\scriptsize$\pm$0.303} & 0.322{\scriptsize$\pm$0.297} & 0.112{\scriptsize$\pm$0.120} & 0.106{\scriptsize$\pm$0.118} & 0.105{\scriptsize$\pm$0.120} & 0.099{\scriptsize$\pm$0.114} & \textbf{1.608{\scriptsize$\pm$0.702}} & 1.542{\scriptsize$\pm$0.682} & 0.542{\scriptsize$\pm$0.524} & 0.516{\scriptsize$\pm$0.510} & 0.143{\scriptsize$\pm$0.163} & 0.135{\scriptsize$\pm$0.156} \\
    \textit{gemini-3-flash-preview} & 0.532{\scriptsize$\pm$0.276} & 0.531{\scriptsize$\pm$0.226} & 0.167{\scriptsize$\pm$0.119} & 0.169{\scriptsize$\pm$0.102} & 0.159{\scriptsize$\pm$0.116} & 0.158{\scriptsize$\pm$0.105} & 1.511{\scriptsize$\pm$0.541} & 1.509{\scriptsize$\pm$0.347} & 0.838{\scriptsize$\pm$0.487} & 0.837{\scriptsize$\pm$0.416} & 0.216{\scriptsize$\pm$0.159} & 0.214{\scriptsize$\pm$0.144} \\
    \textbf{\textit{gemini-3-pro-preview}} & \textbf{0.734{\scriptsize$\pm$0.167}} & \textbf{0.705{\scriptsize$\pm$0.157}} & \textbf{0.252{\scriptsize$\pm$0.087}} & \textbf{0.244{\scriptsize$\pm$0.081}} & 0.245{\scriptsize$\pm$0.086} & 0.225{\scriptsize$\pm$0.082} & 1.429{\scriptsize$\pm$0.341} & 1.526{\scriptsize$\pm$0.293} & \textbf{1.221{\scriptsize$\pm$0.335}} & \textbf{1.159{\scriptsize$\pm$0.309}} & 0.334{\scriptsize$\pm$0.118} & 0.309{\scriptsize$\pm$0.114} \\
    \textit{grok-4.1-fast} & 0.331{\scriptsize$\pm$0.219} & 0.332{\scriptsize$\pm$0.211} & 0.102{\scriptsize$\pm$0.085} & 0.100{\scriptsize$\pm$0.082} & \textbf{0.105{\scriptsize$\pm$0.088}} & 0.102{\scriptsize$\pm$0.086} & 1.573{\scriptsize$\pm$0.701} & \textbf{1.578{\scriptsize$\pm$0.510}} & 0.526{\scriptsize$\pm$0.375} & 0.521{\scriptsize$\pm$0.362} & \textbf{0.140{\scriptsize$\pm$0.119}} & 0.137{\scriptsize$\pm$0.117} \\
  
  \midrule
  \rowcolor{gray!15} \multicolumn{13}{c}{\textbf{BTCUSDT} (Cryptocurrency)} \\
  \textit{claude-sonnet-4.5} & 0.353{\scriptsize$\pm$0.186} & 0.354{\scriptsize$\pm$0.186} & 0.134{\scriptsize$\pm$0.085} & 0.133{\scriptsize$\pm$0.085} & 0.203{\scriptsize$\pm$0.128} & 0.199{\scriptsize$\pm$0.128} & 0.817{\scriptsize$\pm$0.655} & 0.833{\scriptsize$\pm$0.676} & 0.601{\scriptsize$\pm$0.330} & 0.601{\scriptsize$\pm$0.327} & 0.233{\scriptsize$\pm$0.149} & 0.230{\scriptsize$\pm$0.149} \\
  \textit{deepseek-v3.2} & 0.307{\scriptsize$\pm$0.202} & 0.297{\scriptsize$\pm$0.205} & 0.114{\scriptsize$\pm$0.087} & 0.110{\scriptsize$\pm$0.088} & 0.173{\scriptsize$\pm$0.130} & 0.169{\scriptsize$\pm$0.131} & 0.821{\scriptsize$\pm$0.687} & 0.854{\scriptsize$\pm$0.758} & 0.526{\scriptsize$\pm$0.348} & 0.513{\scriptsize$\pm$0.348} & 0.198{\scriptsize$\pm$0.150} & 0.194{\scriptsize$\pm$0.152} \\
  \textit{gpt-5.2} & 0.301{\scriptsize$\pm$0.212} & 0.302{\scriptsize$\pm$0.211} & 0.111{\scriptsize$\pm$0.094} & 0.111{\scriptsize$\pm$0.093} & \textbf{0.158{\scriptsize$\pm$0.139}} & \textbf{0.159{\scriptsize$\pm$0.137}} & \textbf{0.958{\scriptsize$\pm$0.781}} & \textbf{0.954{\scriptsize$\pm$0.789}} & 0.497{\scriptsize$\pm$0.373} & 0.499{\scriptsize$\pm$0.370} & \textbf{0.184{\scriptsize$\pm$0.161}} & \textbf{0.185{\scriptsize$\pm$0.159}} \\
  \textit{gemini-3-flash-preview} & 0.373{\scriptsize$\pm$0.205} & 0.371{\scriptsize$\pm$0.176} & 0.137{\scriptsize$\pm$0.094} & 0.138{\scriptsize$\pm$0.085} & 0.203{\scriptsize$\pm$0.139} & 0.205{\scriptsize$\pm$0.130} & 0.876{\scriptsize$\pm$0.723} & 0.855{\scriptsize$\pm$0.683} & 0.621{\scriptsize$\pm$0.357} & 0.624{\scriptsize$\pm$0.315} & 0.235{\scriptsize$\pm$0.160} & 0.238{\scriptsize$\pm$0.150} \\
  \textbf{\textit{gemini-3-pro-preview}} & \textbf{0.420{\scriptsize$\pm$0.162}} & \textbf{0.425{\scriptsize$\pm$0.143}} & \textbf{0.164{\scriptsize$\pm$0.078}} & \textbf{0.167{\scriptsize$\pm$0.071}} & 0.254{\scriptsize$\pm$0.125} & 0.251{\scriptsize$\pm$0.118} & 0.797{\scriptsize$\pm$0.643} & 0.837{\scriptsize$\pm$0.623} & \textbf{0.739{\scriptsize$\pm$0.313}} & \textbf{0.736{\scriptsize$\pm$0.258}} & 0.294{\scriptsize$\pm$0.142} & 0.291{\scriptsize$\pm$0.134} \\
  \textit{grok-4.1-fast} & 0.299{\scriptsize$\pm$0.194} & 0.313{\scriptsize$\pm$0.187} & 0.111{\scriptsize$\pm$0.086} & 0.114{\scriptsize$\pm$0.084} & 0.174{\scriptsize$\pm$0.124} & 0.177{\scriptsize$\pm$0.124} & 0.797{\scriptsize$\pm$0.715} & 0.869{\scriptsize$\pm$0.710} & 0.508{\scriptsize$\pm$0.334} & 0.524{\scriptsize$\pm$0.325} & 0.199{\scriptsize$\pm$0.142} & 0.203{\scriptsize$\pm$0.142} \\
  \midrule
  \rowcolor{gray!15} \multicolumn{13}{c}{\textbf{ETHUSDT} (Cryptocurrency)} \\
  \textit{claude-sonnet-4.5} & 0.297{\scriptsize$\pm$0.225} & 0.290{\scriptsize$\pm$0.226} & 0.151{\scriptsize$\pm$0.150} & 0.147{\scriptsize$\pm$0.150} & 0.245{\scriptsize$\pm$0.184} & 0.243{\scriptsize$\pm$0.183} & \textbf{0.592{\scriptsize$\pm$0.376}} & \textbf{0.563{\scriptsize$\pm$0.346}} & 0.465{\scriptsize$\pm$0.348} & 0.455{\scriptsize$\pm$0.349} & 0.285{\scriptsize$\pm$0.220} & 0.282{\scriptsize$\pm$0.220} \\
  \textit{deepseek-v3.2} & 0.252{\scriptsize$\pm$0.218} & 0.244{\scriptsize$\pm$0.217} & 0.123{\scriptsize$\pm$0.138} & 0.118{\scriptsize$\pm$0.137} & 0.216{\scriptsize$\pm$0.182} & 0.211{\scriptsize$\pm$0.184} & 0.545{\scriptsize$\pm$0.370} & 0.560{\scriptsize$\pm$0.440} & 0.396{\scriptsize$\pm$0.336} & 0.382{\scriptsize$\pm$0.335} & 0.249{\scriptsize$\pm$0.216} & 0.243{\scriptsize$\pm$0.218} \\
  \textit{gpt-5.2} & 0.233{\scriptsize$\pm$0.241} & 0.229{\scriptsize$\pm$0.236} & 0.121{\scriptsize$\pm$0.153} & 0.118{\scriptsize$\pm$0.152} & \textbf{0.197{\scriptsize$\pm$0.193}} & \textbf{0.194{\scriptsize$\pm$0.191}} & 0.564{\scriptsize$\pm$0.405} & 0.562{\scriptsize$\pm$0.365} & 0.369{\scriptsize$\pm$0.370} & 0.361{\scriptsize$\pm$0.363} & \textbf{0.228{\scriptsize$\pm$0.230}} & \textbf{0.224{\scriptsize$\pm$0.227}} \\
  \textit{gemini-3-flash-preview} & 0.276{\scriptsize$\pm$0.260} & 0.292{\scriptsize$\pm$0.230} & 0.141{\scriptsize$\pm$0.162} & 0.145{\scriptsize$\pm$0.152} & 0.246{\scriptsize$\pm$0.195} & 0.249{\scriptsize$\pm$0.184} & 0.507{\scriptsize$\pm$0.416} & 0.544{\scriptsize$\pm$0.354} & 0.444{\scriptsize$\pm$0.385} & 0.460{\scriptsize$\pm$0.352} & 0.284{\scriptsize$\pm$0.235} & 0.288{\scriptsize$\pm$0.222} \\
  \textbf{\textit{gemini-3-pro-preview}} & \textbf{0.358{\scriptsize$\pm$0.224}} & \textbf{0.355{\scriptsize$\pm$0.215}} & \textbf{0.181{\scriptsize$\pm$0.156}} & \textbf{0.179{\scriptsize$\pm$0.149}} & 0.311{\scriptsize$\pm$0.174} & 0.305{\scriptsize$\pm$0.166} & 0.543{\scriptsize$\pm$0.360} & 0.559{\scriptsize$\pm$0.330} & \textbf{0.569{\scriptsize$\pm$0.341}} & \textbf{0.564{\scriptsize$\pm$0.326}} & 0.360{\scriptsize$\pm$0.211} & 0.354{\scriptsize$\pm$0.202} \\
  \textit{grok-4.1-fast} & 0.245{\scriptsize$\pm$0.213} & 0.248{\scriptsize$\pm$0.211} & 0.121{\scriptsize$\pm$0.135} & 0.122{\scriptsize$\pm$0.135} & 0.214{\scriptsize$\pm$0.174} & 0.218{\scriptsize$\pm$0.172} & 0.529{\scriptsize$\pm$0.363} & 0.515{\scriptsize$\pm$0.340} & 0.387{\scriptsize$\pm$0.327} & 0.393{\scriptsize$\pm$0.322} & 0.247{\scriptsize$\pm$0.207} & 0.251{\scriptsize$\pm$0.205} \\
  \midrule
  \rowcolor{gray!15} \multicolumn{13}{c}{\textbf{AAPL} (US Equity)} \\
  \textit{claude-sonnet-4.5} & 0.941{\scriptsize$\pm$0.493} & 0.924{\scriptsize$\pm$0.488} & 0.180{\scriptsize$\pm$0.119} & 0.175{\scriptsize$\pm$0.117} & 0.073{\scriptsize$\pm$0.048} & 0.071{\scriptsize$\pm$0.047} & 2.799{\scriptsize$\pm$1.303} & \textbf{2.779{\scriptsize$\pm$1.330}} & 1.262{\scriptsize$\pm$0.726} & 1.240{\scriptsize$\pm$0.721} & 0.116{\scriptsize$\pm$0.074} & 0.113{\scriptsize$\pm$0.072} \\
  \textit{deepseek-v3.2} & 0.803{\scriptsize$\pm$0.515} & 0.784{\scriptsize$\pm$0.526} & 0.145{\scriptsize$\pm$0.116} & 0.142{\scriptsize$\pm$0.119} & 0.059{\scriptsize$\pm$0.047} & 0.058{\scriptsize$\pm$0.048} & 2.774{\scriptsize$\pm$1.499} & 2.722{\scriptsize$\pm$1.400} & 1.074{\scriptsize$\pm$0.732} & 1.047{\scriptsize$\pm$0.750} & 0.095{\scriptsize$\pm$0.072} & 0.093{\scriptsize$\pm$0.074} \\
  \textit{gpt-5.2} & 0.745{\scriptsize$\pm$0.549} & 0.751{\scriptsize$\pm$0.548} & 0.139{\scriptsize$\pm$0.125} & 0.139{\scriptsize$\pm$0.125} & \textbf{0.058{\scriptsize$\pm$0.052}} & \textbf{0.058{\scriptsize$\pm$0.052}} & 2.621{\scriptsize$\pm$1.493} & 2.745{\scriptsize$\pm$1.577} & 0.977{\scriptsize$\pm$0.780} & 0.982{\scriptsize$\pm$0.778} & \textbf{0.092{\scriptsize$\pm$0.079}} & \textbf{0.091{\scriptsize$\pm$0.079}} \\
  \textit{gemini-3-flash-preview} & 0.946{\scriptsize$\pm$0.505} & 0.953{\scriptsize$\pm$0.472} & 0.173{\scriptsize$\pm$0.125} & 0.177{\scriptsize$\pm$0.117} & 0.073{\scriptsize$\pm$0.051} & 0.075{\scriptsize$\pm$0.048} & \textbf{2.861{\scriptsize$\pm$2.392}} & 2.672{\scriptsize$\pm$1.327} & 1.237{\scriptsize$\pm$0.740} & 1.257{\scriptsize$\pm$0.698} & 0.116{\scriptsize$\pm$0.078} & 0.118{\scriptsize$\pm$0.073} \\
  \textbf{\textit{gemini-3-pro-preview}} & \textbf{1.067{\scriptsize$\pm$0.480}} & \textbf{1.075{\scriptsize$\pm$0.446}} & \textbf{0.214{\scriptsize$\pm$0.123}} & \textbf{0.215{\scriptsize$\pm$0.115}} & 0.093{\scriptsize$\pm$0.049} & 0.091{\scriptsize$\pm$0.046} & 2.539{\scriptsize$\pm$1.472} & 2.661{\scriptsize$\pm$1.312} & \textbf{1.456{\scriptsize$\pm$0.729}} & \textbf{1.462{\scriptsize$\pm$0.679}} & 0.146{\scriptsize$\pm$0.073} & 0.144{\scriptsize$\pm$0.070} \\
  \textit{grok-4.1-fast} & 0.744{\scriptsize$\pm$0.494} & 0.751{\scriptsize$\pm$0.485} & 0.139{\scriptsize$\pm$0.114} & 0.139{\scriptsize$\pm$0.112} & 0.059{\scriptsize$\pm$0.046} & 0.059{\scriptsize$\pm$0.046} & 2.613{\scriptsize$\pm$1.453} & 2.615{\scriptsize$\pm$1.386} & 0.986{\scriptsize$\pm$0.711} & 0.993{\scriptsize$\pm$0.697} & 0.094{\scriptsize$\pm$0.071} & 0.094{\scriptsize$\pm$0.070} \\
  \midrule
  \rowcolor{gray!15} \multicolumn{13}{c}{\textbf{GOOGL} (US Equity)} \\
  \textit{claude-sonnet-4.5} & 0.792{\scriptsize$\pm$0.445} & 0.789{\scriptsize$\pm$0.434} & 0.201{\scriptsize$\pm$0.146} & 0.195{\scriptsize$\pm$0.140} & 0.103{\scriptsize$\pm$0.089} & 0.103{\scriptsize$\pm$0.089} & 2.575{\scriptsize$\pm$1.566} & 2.578{\scriptsize$\pm$1.596} & 1.316{\scriptsize$\pm$0.823} & 1.308{\scriptsize$\pm$0.799} & 0.168{\scriptsize$\pm$0.132} & 0.168{\scriptsize$\pm$0.132} \\
  \textit{deepseek-v3.2} & 0.628{\scriptsize$\pm$0.455} & 0.627{\scriptsize$\pm$0.470} & 0.146{\scriptsize$\pm$0.138} & 0.146{\scriptsize$\pm$0.142} & 0.087{\scriptsize$\pm$0.088} & 0.084{\scriptsize$\pm$0.088} & 2.261{\scriptsize$\pm$1.689} & 2.350{\scriptsize$\pm$1.720} & 1.044{\scriptsize$\pm$0.815} & 1.036{\scriptsize$\pm$0.837} & 0.139{\scriptsize$\pm$0.131} & 0.136{\scriptsize$\pm$0.132} \\
  \textit{gpt-5.2} & 0.635{\scriptsize$\pm$0.502} & 0.639{\scriptsize$\pm$0.502} & 0.160{\scriptsize$\pm$0.154} & 0.161{\scriptsize$\pm$0.154} & \textbf{0.080{\scriptsize$\pm$0.089}} & \textbf{0.077{\scriptsize$\pm$0.086}} & 2.636{\scriptsize$\pm$1.946} & 2.648{\scriptsize$\pm$1.621} & 1.055{\scriptsize$\pm$0.897} & 1.051{\scriptsize$\pm$0.890} & \textbf{0.131{\scriptsize$\pm$0.134}} & \textbf{0.127{\scriptsize$\pm$0.131}} \\
  \textit{gemini-3-flash-preview} & 0.802{\scriptsize$\pm$0.492} & 0.808{\scriptsize$\pm$0.442} & 0.195{\scriptsize$\pm$0.157} & 0.199{\scriptsize$\pm$0.146} & 0.104{\scriptsize$\pm$0.097} & 0.105{\scriptsize$\pm$0.089} & 2.435{\scriptsize$\pm$1.751} & 2.487{\scriptsize$\pm$1.605} & 1.325{\scriptsize$\pm$0.902} & 1.336{\scriptsize$\pm$0.822} & 0.170{\scriptsize$\pm$0.143} & 0.170{\scriptsize$\pm$0.132} \\
  \textbf{\textit{gemini-3-pro-preview}} & \textbf{0.960{\scriptsize$\pm$0.446}} & \textbf{0.968{\scriptsize$\pm$0.423}} & \textbf{0.252{\scriptsize$\pm$0.153}} & \textbf{0.256{\scriptsize$\pm$0.143}} & 0.131{\scriptsize$\pm$0.097} & 0.130{\scriptsize$\pm$0.090} & 2.633{\scriptsize$\pm$1.765} & 2.700{\scriptsize$\pm$1.552} & \textbf{1.627{\scriptsize$\pm$0.834}} & \textbf{1.644{\scriptsize$\pm$0.789}} & 0.212{\scriptsize$\pm$0.140} & 0.211{\scriptsize$\pm$0.131} \\
  \textit{grok-4.1-fast} & 0.639{\scriptsize$\pm$0.446} & 0.644{\scriptsize$\pm$0.440} & 0.157{\scriptsize$\pm$0.138} & 0.157{\scriptsize$\pm$0.137} & 0.082{\scriptsize$\pm$0.082} & 0.084{\scriptsize$\pm$0.084} & \textbf{2.646{\scriptsize$\pm$1.915}} & \textbf{2.830{\scriptsize$\pm$2.808}} & 1.065{\scriptsize$\pm$0.798} & 1.079{\scriptsize$\pm$0.785} & 0.133{\scriptsize$\pm$0.123} & 0.137{\scriptsize$\pm$0.125} \\
  \midrule
  \rowcolor{gray!15} \multicolumn{13}{c}{\textbf{MSFT} (US Equity)} \\
  \textit{claude-sonnet-4.5} & 0.438{\scriptsize$\pm$0.312} & 0.433{\scriptsize$\pm$0.308} & 0.084{\scriptsize$\pm$0.062} & 0.083{\scriptsize$\pm$0.061} & 0.078{\scriptsize$\pm$0.047} & 0.077{\scriptsize$\pm$0.046} & 1.145{\scriptsize$\pm$1.180} & 1.097{\scriptsize$\pm$0.971} & 0.657{\scriptsize$\pm$0.450} & 0.644{\scriptsize$\pm$0.443} & 0.113{\scriptsize$\pm$0.067} & 0.111{\scriptsize$\pm$0.065} \\
  \textit{deepseek-v3.2} & 0.362{\scriptsize$\pm$0.314} & 0.360{\scriptsize$\pm$0.321} & 0.066{\scriptsize$\pm$0.062} & 0.065{\scriptsize$\pm$0.064} & 0.062{\scriptsize$\pm$0.046} & \textbf{0.061{\scriptsize$\pm$0.047}} & 1.035{\scriptsize$\pm$0.975} & 1.037{\scriptsize$\pm$0.951} & 0.529{\scriptsize$\pm$0.455} & 0.525{\scriptsize$\pm$0.469} & 0.091{\scriptsize$\pm$0.066} & \textbf{0.088{\scriptsize$\pm$0.068}} \\
  \textit{gpt-5.2} & 0.376{\scriptsize$\pm$0.326} & 0.376{\scriptsize$\pm$0.333} & 0.070{\scriptsize$\pm$0.064} & 0.070{\scriptsize$\pm$0.066} & 0.063{\scriptsize$\pm$0.050} & 0.062{\scriptsize$\pm$0.050} & 1.130{\scriptsize$\pm$0.937} & 1.154{\scriptsize$\pm$1.010} & 0.546{\scriptsize$\pm$0.469} & 0.544{\scriptsize$\pm$0.480} & 0.092{\scriptsize$\pm$0.071} & 0.091{\scriptsize$\pm$0.072} \\
  \textit{gemini-3-flash-preview} & 0.474{\scriptsize$\pm$0.339} & 0.475{\scriptsize$\pm$0.306} & 0.089{\scriptsize$\pm$0.067} & 0.090{\scriptsize$\pm$0.062} & 0.079{\scriptsize$\pm$0.050} & 0.080{\scriptsize$\pm$0.046} & 1.147{\scriptsize$\pm$0.918} & \textbf{1.323{\scriptsize$\pm$2.709}} & 0.692{\scriptsize$\pm$0.498} & 0.696{\scriptsize$\pm$0.439} & 0.115{\scriptsize$\pm$0.070} & 0.117{\scriptsize$\pm$0.064} \\
  \textbf{\textit{gemini-3-pro-preview}} & \textbf{0.576{\scriptsize$\pm$0.318}} & \textbf{0.567{\scriptsize$\pm$0.289}} & \textbf{0.113{\scriptsize$\pm$0.064}} & \textbf{0.111{\scriptsize$\pm$0.059}} & 0.097{\scriptsize$\pm$0.047} & 0.097{\scriptsize$\pm$0.044} & \textbf{1.197{\scriptsize$\pm$0.763}} & 1.201{\scriptsize$\pm$0.787} & \textbf{0.858{\scriptsize$\pm$0.461}} & \textbf{0.844{\scriptsize$\pm$0.423}} & 0.140{\scriptsize$\pm$0.065} & 0.141{\scriptsize$\pm$0.061} \\
  \textit{grok-4.1-fast} & 0.352{\scriptsize$\pm$0.307} & 0.357{\scriptsize$\pm$0.310} & 0.066{\scriptsize$\pm$0.061} & 0.066{\scriptsize$\pm$0.061} & \textbf{0.062{\scriptsize$\pm$0.045}} & 0.063{\scriptsize$\pm$0.045} & 1.067{\scriptsize$\pm$0.897} & 1.035{\scriptsize$\pm$0.861} & 0.515{\scriptsize$\pm$0.442} & 0.517{\scriptsize$\pm$0.444} & \textbf{0.090{\scriptsize$\pm$0.065}} & 0.091{\scriptsize$\pm$0.064} \\
  \midrule
  \rowcolor{gray!15} \multicolumn{13}{c}{\textbf{NVDA} (US Equity)} \\
  \textit{claude-sonnet-4.5} & 0.223{\scriptsize$\pm$0.210} & 0.220{\scriptsize$\pm$0.208} & 0.005{\scriptsize$\pm$0.050} & 0.005{\scriptsize$\pm$0.054} & 0.188{\scriptsize$\pm$0.179} & 0.183{\scriptsize$\pm$0.178} & 0.449{\scriptsize$\pm$0.817} & 0.462{\scriptsize$\pm$0.896} & 0.480{\scriptsize$\pm$0.417} & 0.463{\scriptsize$\pm$0.412} & 0.273{\scriptsize$\pm$0.261} & 0.266{\scriptsize$\pm$0.260} \\
  \textit{deepseek-v3.2} & 0.192{\scriptsize$\pm$0.209} & 0.191{\scriptsize$\pm$0.208} & 0.002{\scriptsize$\pm$0.053} & 0.005{\scriptsize$\pm$0.054} & 0.155{\scriptsize$\pm$0.167} & 0.149{\scriptsize$\pm$0.166} & 0.460{\scriptsize$\pm$0.942} & 0.502{\scriptsize$\pm$0.885} & 0.395{\scriptsize$\pm$0.394} & 0.390{\scriptsize$\pm$0.397} & 0.225{\scriptsize$\pm$0.243} & 0.216{\scriptsize$\pm$0.242} \\
  \textit{gpt-5.2} & 0.165{\scriptsize$\pm$0.203} & 0.164{\scriptsize$\pm$0.205} & 0.001{\scriptsize$\pm$0.052} & 0.002{\scriptsize$\pm$0.050} & \textbf{0.150{\scriptsize$\pm$0.180}} & \textbf{0.147{\scriptsize$\pm$0.177}} & \textbf{0.497{\scriptsize$\pm$1.078}} & \textbf{0.544{\scriptsize$\pm$1.542}} & 0.378{\scriptsize$\pm$0.421} & 0.370{\scriptsize$\pm$0.421} & \textbf{0.218{\scriptsize$\pm$0.262}} & \textbf{0.213{\scriptsize$\pm$0.258}} \\
  \textit{gemini-3-flash-preview} & 0.234{\scriptsize$\pm$0.242} & 0.231{\scriptsize$\pm$0.210} & 0.009{\scriptsize$\pm$0.076} & 0.006{\scriptsize$\pm$0.060} & 0.182{\scriptsize$\pm$0.189} & 0.184{\scriptsize$\pm$0.179} & 0.435{\scriptsize$\pm$1.013} & 0.437{\scriptsize$\pm$0.826} & 0.482{\scriptsize$\pm$0.466} & 0.476{\scriptsize$\pm$0.417} & 0.265{\scriptsize$\pm$0.276} & 0.268{\scriptsize$\pm$0.260} \\
  \textbf{\textit{gemini-3-pro-preview}} & \textbf{0.302{\scriptsize$\pm$0.221}} & \textbf{0.299{\scriptsize$\pm$0.206}} & \textbf{0.011{\scriptsize$\pm$0.073}} & \textbf{0.013{\scriptsize$\pm$0.065}} & 0.247{\scriptsize$\pm$0.183} & 0.242{\scriptsize$\pm$0.172} & 0.493{\scriptsize$\pm$1.063} & 0.535{\scriptsize$\pm$1.285} & \textbf{0.653{\scriptsize$\pm$0.506}} & \textbf{0.635{\scriptsize$\pm$0.411}} & 0.359{\scriptsize$\pm$0.266} & 0.352{\scriptsize$\pm$0.251} \\
  \textit{grok-4.1-fast} & 0.183{\scriptsize$\pm$0.196} & 0.188{\scriptsize$\pm$0.193} & 0.006{\scriptsize$\pm$0.055} & 0.005{\scriptsize$\pm$0.052} & 0.153{\scriptsize$\pm$0.156} & 0.153{\scriptsize$\pm$0.156} & 0.465{\scriptsize$\pm$1.004} & 0.508{\scriptsize$\pm$0.896} & 0.391{\scriptsize$\pm$0.374} & 0.396{\scriptsize$\pm$0.373} & 0.222{\scriptsize$\pm$0.228} & 0.222{\scriptsize$\pm$0.228} \\
  \midrule
  \rowcolor{gray!15} \multicolumn{13}{c}{\textbf{TSLA} (US Equity)} \\
  \textit{claude-sonnet-4.5} & 0.547{\scriptsize$\pm$0.385} & 0.545{\scriptsize$\pm$0.386} & 0.395{\scriptsize$\pm$0.341} & 0.397{\scriptsize$\pm$0.343} & 0.158{\scriptsize$\pm$0.143} & 0.155{\scriptsize$\pm$0.142} & 3.076{\scriptsize$\pm$1.846} & 3.259{\scriptsize$\pm$2.031} & 0.860{\scriptsize$\pm$0.658} & 0.858{\scriptsize$\pm$0.661} & 0.249{\scriptsize$\pm$0.220} & 0.245{\scriptsize$\pm$0.219} \\
  \textit{deepseek-v3.2} & 0.470{\scriptsize$\pm$0.377} & 0.464{\scriptsize$\pm$0.394} & 0.329{\scriptsize$\pm$0.324} & 0.323{\scriptsize$\pm$0.328} & 0.134{\scriptsize$\pm$0.136} & 0.130{\scriptsize$\pm$0.136} & 3.057{\scriptsize$\pm$2.042} & 2.989{\scriptsize$\pm$2.109} & 0.741{\scriptsize$\pm$0.635} & 0.726{\scriptsize$\pm$0.649} & 0.210{\scriptsize$\pm$0.210} & 0.205{\scriptsize$\pm$0.210} \\
  \textit{gpt-5.2} & 0.451{\scriptsize$\pm$0.412} & 0.460{\scriptsize$\pm$0.413} & 0.309{\scriptsize$\pm$0.346} & 0.308{\scriptsize$\pm$0.338} & \textbf{0.124{\scriptsize$\pm$0.144}} & \textbf{0.121{\scriptsize$\pm$0.141}} & 2.894{\scriptsize$\pm$1.629} & 2.991{\scriptsize$\pm$1.740} & 0.691{\scriptsize$\pm$0.685} & 0.692{\scriptsize$\pm$0.676} & \textbf{0.196{\scriptsize$\pm$0.223}} & \textbf{0.193{\scriptsize$\pm$0.218}} \\
  \textit{gemini-3-flash-preview} & 0.554{\scriptsize$\pm$0.451} & 0.580{\scriptsize$\pm$0.398} & 0.391{\scriptsize$\pm$0.372} & 0.399{\scriptsize$\pm$0.339} & 0.152{\scriptsize$\pm$0.153} & 0.157{\scriptsize$\pm$0.144} & 2.921{\scriptsize$\pm$1.918} & 3.065{\scriptsize$\pm$1.772} & 0.851{\scriptsize$\pm$0.735} & 0.891{\scriptsize$\pm$0.668} & 0.241{\scriptsize$\pm$0.235} & 0.247{\scriptsize$\pm$0.221} \\
  \textbf{\textit{gemini-3-pro-preview}} & \textbf{0.712{\scriptsize$\pm$0.383}} & \textbf{0.699{\scriptsize$\pm$0.366}} & \textbf{0.523{\scriptsize$\pm$0.346}} & \textbf{0.522{\scriptsize$\pm$0.331}} & 0.204{\scriptsize$\pm$0.143} & 0.202{\scriptsize$\pm$0.139} & 2.983{\scriptsize$\pm$1.688} & 3.015{\scriptsize$\pm$1.463} & \textbf{1.127{\scriptsize$\pm$0.650}} & \textbf{1.106{\scriptsize$\pm$0.620}} & 0.324{\scriptsize$\pm$0.219} & 0.320{\scriptsize$\pm$0.212} \\
  \textit{grok-4.1-fast} & 0.485{\scriptsize$\pm$0.369} & 0.500{\scriptsize$\pm$0.372} & 0.337{\scriptsize$\pm$0.313} & 0.344{\scriptsize$\pm$0.314} & 0.133{\scriptsize$\pm$0.129} & 0.132{\scriptsize$\pm$0.129} & \textbf{3.244{\scriptsize$\pm$2.087}} & \textbf{3.418{\scriptsize$\pm$2.322}} & 0.756{\scriptsize$\pm$0.608} & 0.772{\scriptsize$\pm$0.613} & 0.210{\scriptsize$\pm$0.199} & 0.211{\scriptsize$\pm$0.200} \\
  \bottomrule
  \end{tabular}
  \vspace{-0.5cm}
\end{table*}

\begin{figure}[h]
  \centering
  \begin{subfigure}[t]{0.23\textwidth}
  \includegraphics[width=\textwidth]{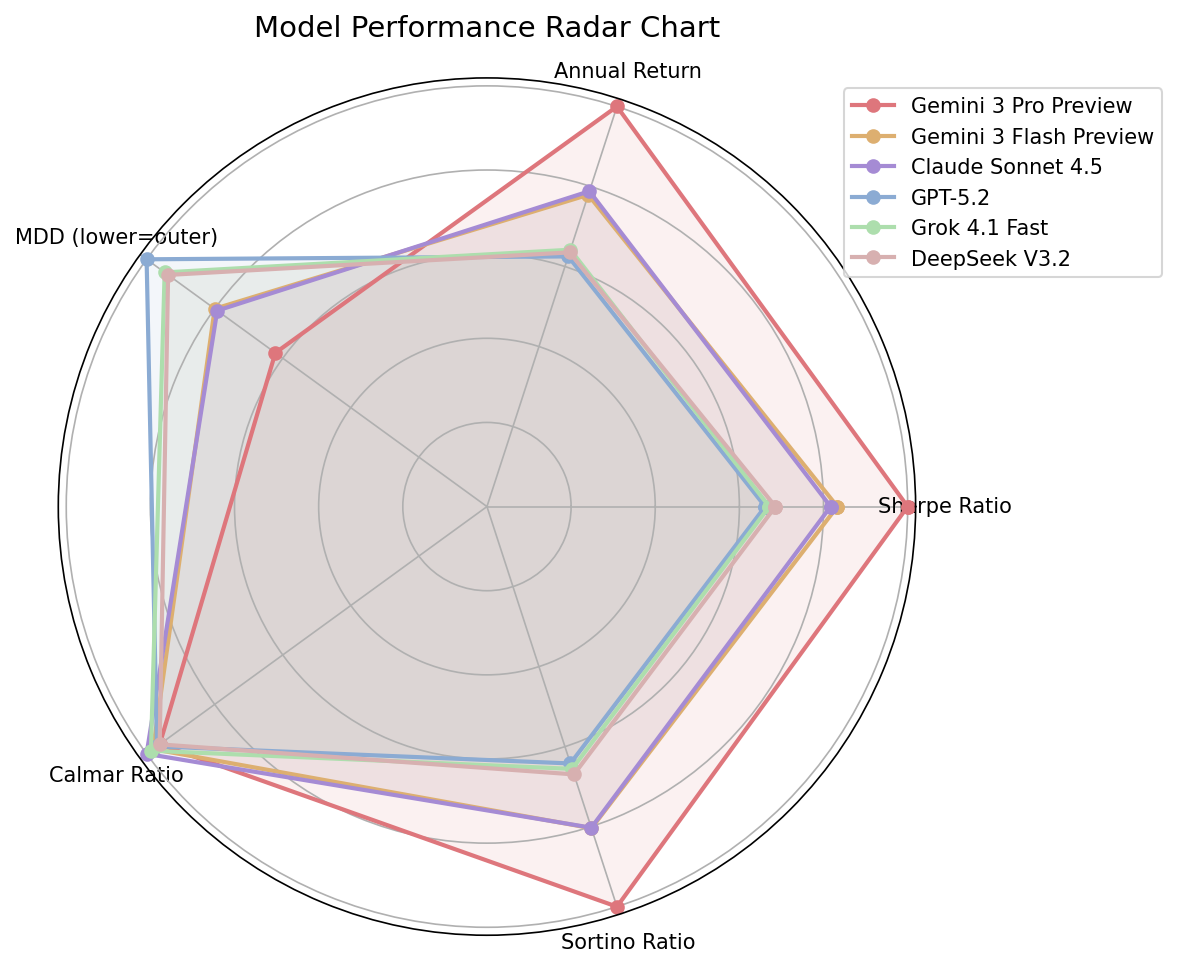}
  \caption{T=0.0}
  \end{subfigure}
  \hfill
  \begin{subfigure}[t]{0.23\textwidth}
  \includegraphics[width=\textwidth]{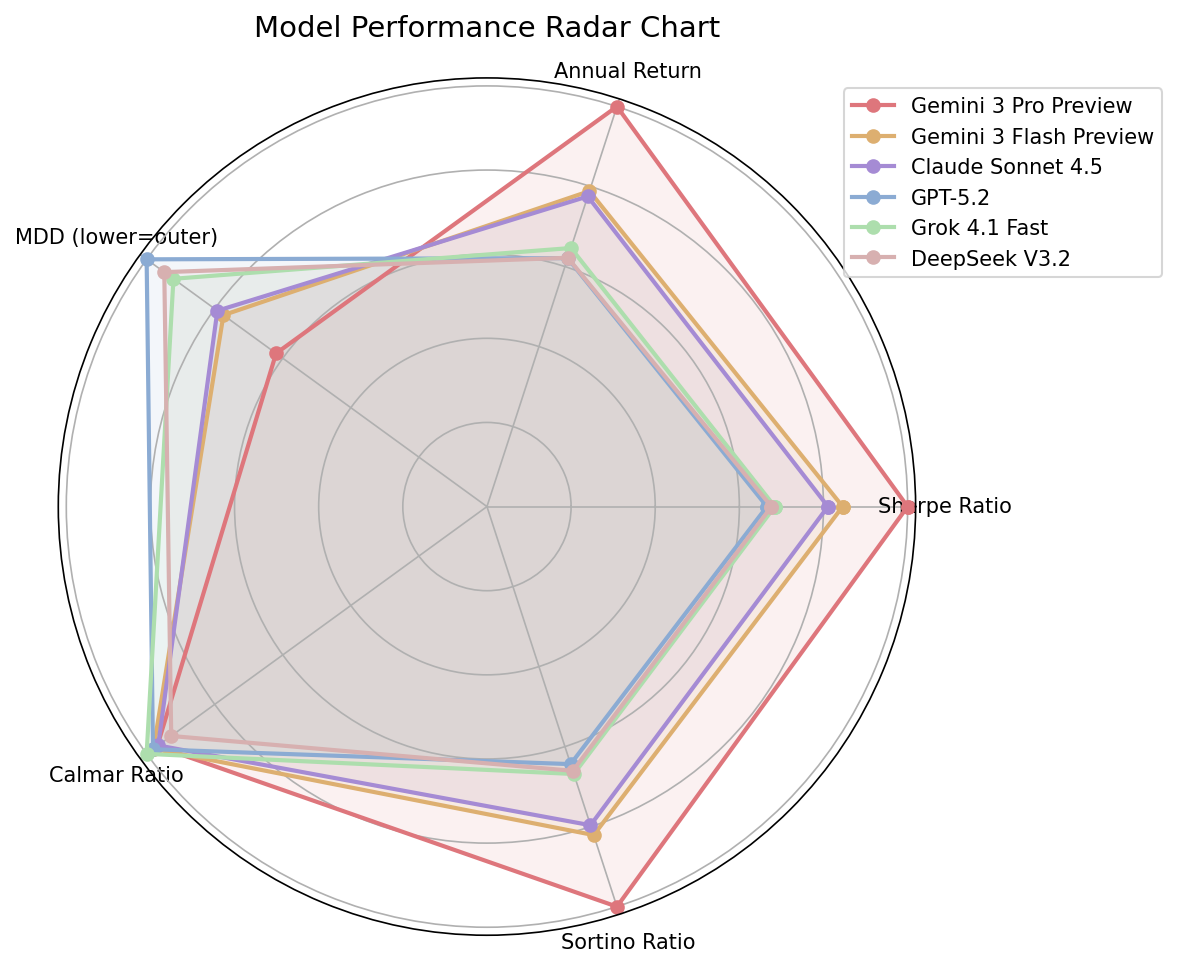}
  \caption{T=0.7}
  \end{subfigure}
  \vspace{-0.3cm}
  \caption{Multi-metric radar profiles on LLM-augmented queries (five metrics, $\tau{=}0$ vs.\ $\tau{=}0.7$). Near-identical polygons confirm temperature-invariant model behavior.}
  \vspace{-0.6cm}
  \label{fig:model_radar}
\end{figure}

\subsubsection{Overall model comparison.}
The Overall block of \Cref{tab:stage2_main_results} reveals four key observations that extend and sharpen the Stage~1 findings.
\textbf{(1)~The three-tier hierarchy is preserved and amplified.} The same model ranking observed in Stage~1 recurs: \textit{gemini-3-pro-preview} leads (SR~=~0.628 at $T{=}0$), a middle tier follows (\textit{gemini-3-flash-preview}, \textit{claude-sonnet-4.5}), and the remaining models trail. Notably, the 7.8\% ARR gap between the best and worst models exceeds the 5.5\% gap in Stage~1, confirming that the controlled difficulty design of Stage~2 effectively amplifies latent capability differences among models.
\textbf{(2)~The return-risk inversion persists.} Consistent with Stage~1, return and risk rankings remain inversely correlated: the top return-generating model (\textit{gemini-3-pro-preview}) incurs the highest drawdown and volatility, whereas the most conservative model (\textit{\textit{gpt-5.2}}, MDD~=~0.119) trails on return metrics. \textit{claude-sonnet-4.5} achieves the best Calmar Ratio (CR~=~1.650 at $T{=}0$), exhibiting the most favorable return-to-drawdown trade-off, further reinforcing the necessity of multi-metric evaluation.
\textbf{(3)~Evaluation is temperature-invariant.} The $T{=}0$ and $T{=}0.7$ columns yield near-identical values across all models and metrics (max SR difference~$<$~0.008). This invariance is a distinctive advantage of the code-generation paradigm: because downstream execution is deterministic, decoding temperature affects only surface-level code variation without altering the resulting trading logic, a property absent in direct-trading benchmarks where temperature directly perturbs each decision.
\textbf{(4)~Radar charts reveal distinct and stable risk profiles.} As shown in \Cref{fig:model_radar}, \textit{gemini-3-pro-preview} spans the largest polygon, dominating on the return and Sortino axes while receding on MDD, visually encoding its aggressive strategy preference. \textit{\textit{gpt-5.2}} and \textit{grok-4.1-fast} form compact, risk-averse polygons that extend furthest on the MDD axis, while \textit{claude-sonnet-4.5} achieves the most balanced shape across all five axes. The near-identical polygon geometry between the $\tau{=}0$ and $\tau{=}0.7$ panels provides strong visual confirmation that these model risk personalities are intrinsic and temperature-invariant.

\begin{figure}[h]
  \centering
  \begin{subfigure}[t]{0.23\textwidth}
  \includegraphics[width=\textwidth]{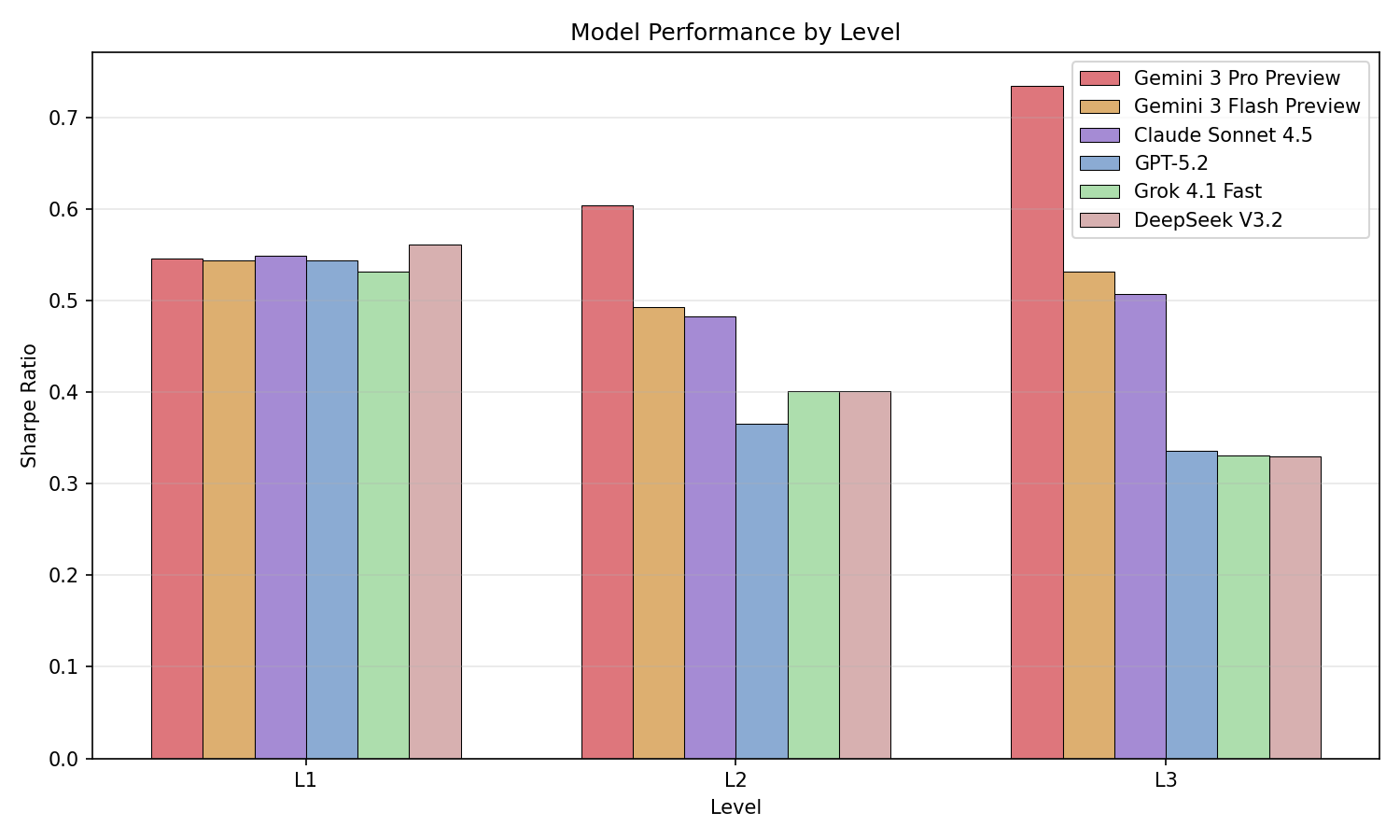}
  \caption{T=0.0}
  \end{subfigure}
  \hfill
  \begin{subfigure}[t]{0.23\textwidth}
  \includegraphics[width=\textwidth]{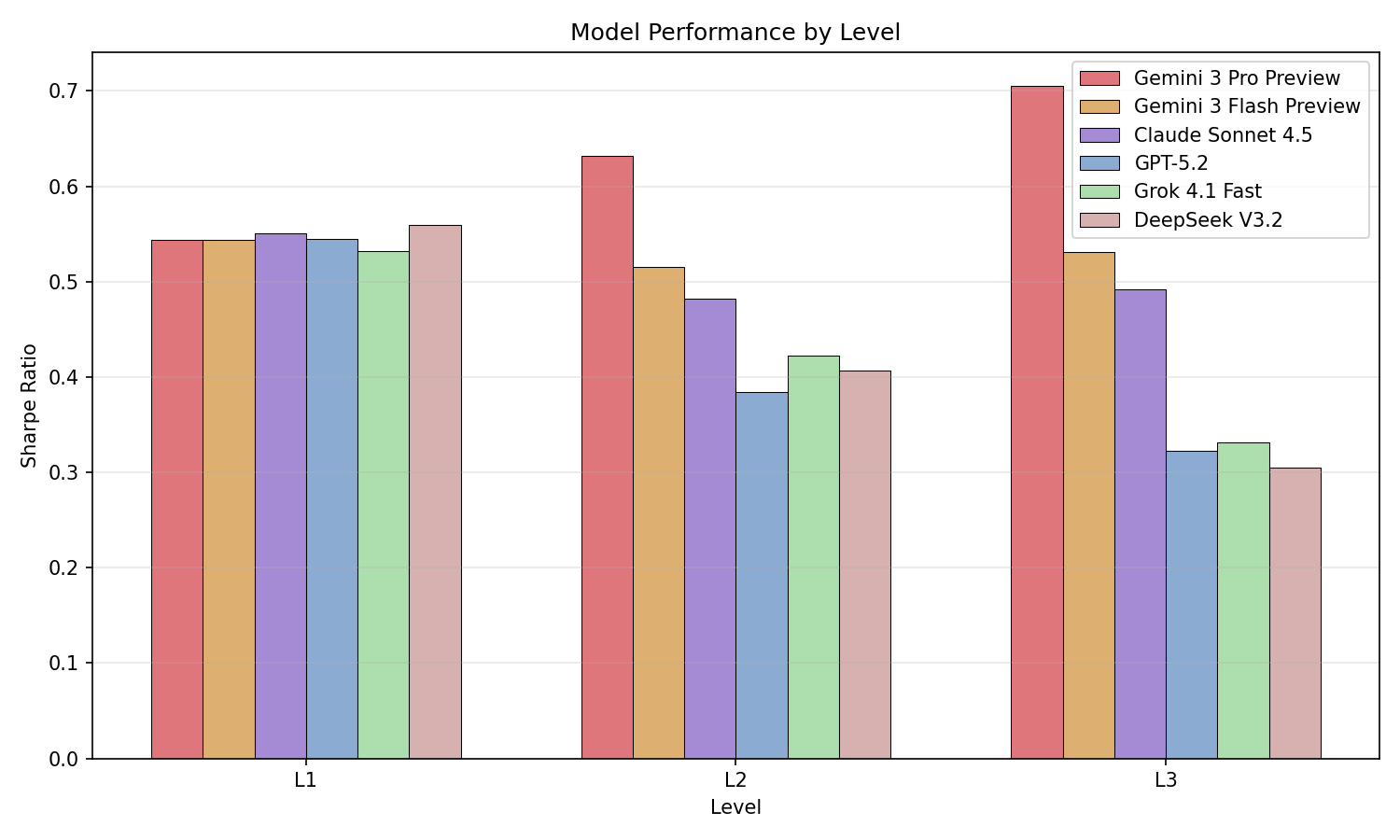}
  \caption{T=0.7}
  \end{subfigure}
  \vspace{-0.3cm}
  \caption{Sharpe Ratio by difficulty level on LLM-augmented queries ($\tau{=}0$ vs.\ $\tau{=}0.7$). Inter-model spread widens monotonically from Level~1 to Level~3.}
  \label{fig:level_grouped_bar}
  \vspace{-0.5cm}
\end{figure}

\subsubsection{Per-level analysis.}
The Level blocks of \Cref{tab:stage2_main_results} and the grouped bar charts in \Cref{fig:level_grouped_bar} reveal a monotonically widening inter-model spread across the three difficulty levels, validating the $3\times 3$ taxonomy as an effective diagnostic tool. Three principal findings emerge. \textbf{(1)~Logic translation (Level~1) is a near-saturated capability}: all models achieve tightly clustered SR values (range~=~0.029), indicating that faithful code translation from fully specified rules presents minimal challenge for frontier LLMs; notably, \textit{deepseek-v3.2} leads at this level despite ranking among the weakest overall, suggesting that code-translation competence is a distinct skill from strategic reasoning ability. \textbf{(2)~Parameter inference (Level~2) exposes domain-knowledge gaps}: the inter-model SR spread widens to $8\!\times$ the Level~1 range once models must supply missing thresholds, lookback windows, and indicator parameters, indicating that grounding underspecified strategy skeletons in reasonable financial parameters effectively separates models with strong domain knowledge from those that lack it; stochastic sampling ($\tau{=}0.7$) marginally benefits the top model at this level, suggesting that sampling diversity can occasionally discover better parameter configurations. \textbf{(3)~Goal-oriented generation (Level~3) produces the most discriminative separation}: the spread reaches $14\!\times$ the Level~1 range, with a striking crossover where models that excel at constrained translation (e.g., \textit{deepseek-v3.2}) decline monotonically from L1 to L3, whereas \textit{gemini-3-pro-preview} actually \emph{improves}, indicating superior open-ended strategy design capabilities. The primary cognitive leap occurs between Level~1 and Level~2 (a 16\% SR decline in model-averaged performance), while the L2-to-L3 transition introduces additional inter-model variance rather than a sharp further mean decline. This crossover demonstrates that the three levels probe fundamentally different cognitive capabilities, and no single model dominates across all levels.

\begin{figure}[h]
  \vspace{-0.2cm}
  \centering
  \begin{subfigure}[t]{0.23\textwidth}
  \includegraphics[width=\textwidth]{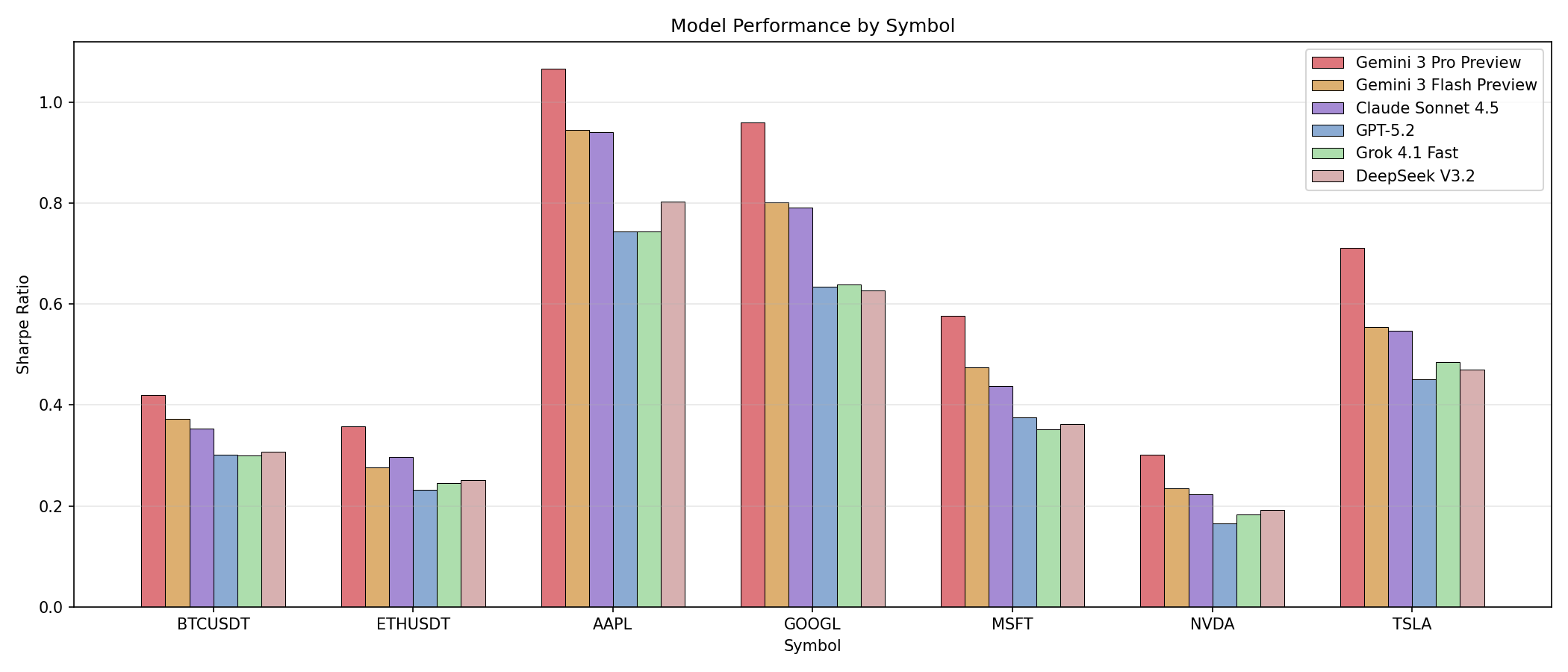}
  \caption{T=0.0}
  \end{subfigure}
  \hfill
  \begin{subfigure}[t]{0.23\textwidth}
  \includegraphics[width=\textwidth]{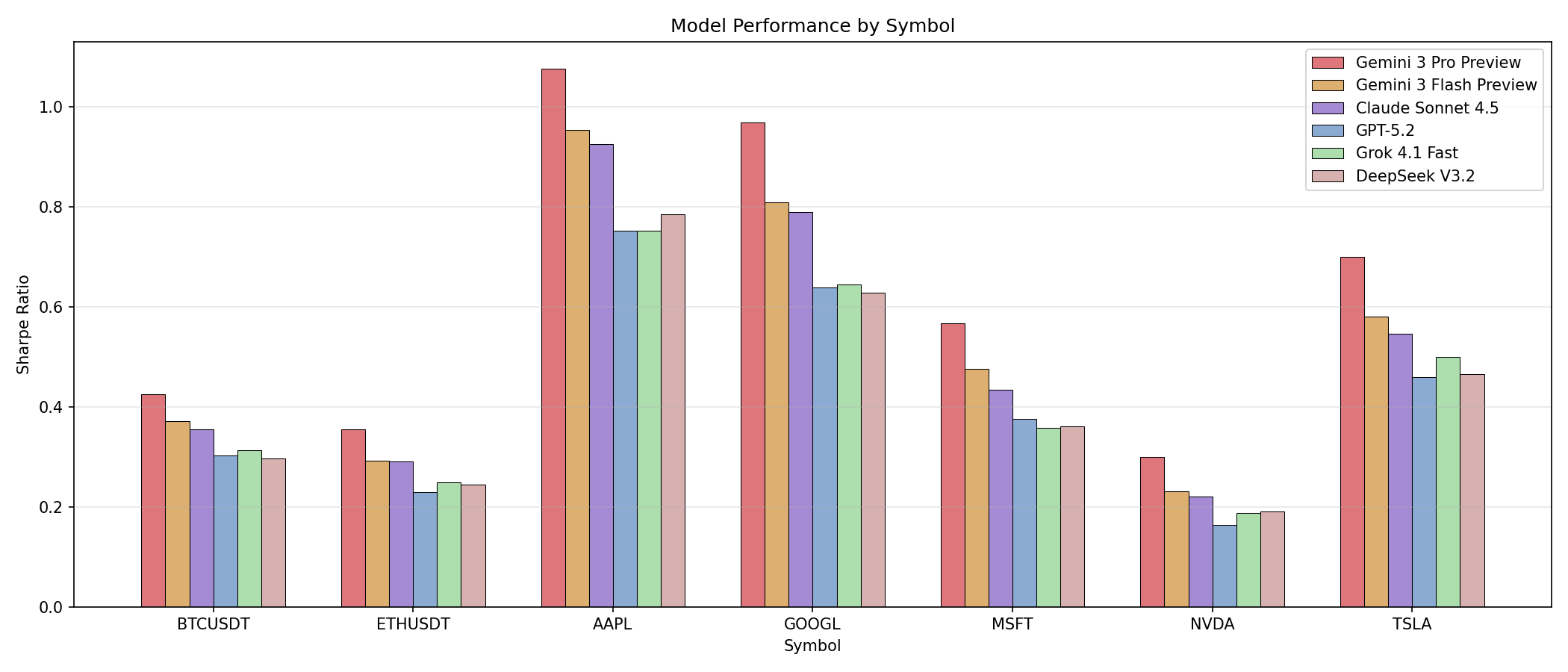}
  \caption{T=0.7}
  \end{subfigure}
  \vspace{-0.3cm}
  \caption{Sharpe Ratio by asset on LLM-augmented queries ($\tau{=}0$ vs.\ $\tau{=}0.7$). A shared difficulty gradient emerges, with AAPL/GOOGL easiest and crypto/NVDA hardest.}
  \label{fig:symbol_grouped_bar}
  \vspace{-0.4cm}
\end{figure}

\subsubsection{Per-asset analysis.}
The asset blocks of \Cref{tab:stage2_main_results} and \Cref{fig:symbol_grouped_bar} yield four observations that mirror and extend the Stage~1 per-asset findings. \textbf{(1)~A systematic difficulty gradient persists across assets.} Trend-rich large-caps (AAPL, GOOGL) remain the easiest targets for all models, MSFT and NVDA constitute the hardest equity environments primarily due to narrower trading ranges and rapid regime shifts respectively, and cryptocurrencies (BTCUSDT, ETHUSDT) rank as the most challenging overall, reflecting higher volatility and fundamentally different 24/7 market dynamics. TSLA again occupies a unique position: it yields the highest absolute returns yet the widest cross-model variance, amplifying the advantage of aggressive signal logic. \textbf{(2)~The Stage~2 difficulty design sharpens inter-model dispersion within each asset.} Compared to Stage~1, the controlled query design produces wider SR spreads on every asset, confirming that the structured queries effectively magnify latent capability differences that real-world queries alone partially obscure. \textbf{(3)~Distinct model specialization patterns emerge.} The top return-generating model excels particularly on high-volatility assets (TSLA, NVDA, cryptocurrencies), suggesting robust handling of challenging market conditions; conversely, \textit{claude-sonnet-4.5} exhibits the most uniform performance across assets, indicating balanced, asset-agnostic strategy generation. \textbf{(4)~Rankings and difficulty gradients are temperature-invariant.} The bar patterns visualized in \Cref{fig:symbol_grouped_bar} remain virtually identical between the $\tau{=}0$ and $\tau{=}0.7$ panels, confirming that both model specialization profiles and the asset-difficulty hierarchy are intrinsic properties fundamentally unaffected by decoding temperature. Detailed per-asset tables and additional visual analyses are provided in Appx.~\ref{appx_sec:appendix_results_stage_2}.

\subsubsection{Aligned return curve analysis.}
\begin{figure}[h]
  \vspace{-0.1cm}
  \centering
  \begin{subfigure}[t]{0.23\textwidth}
  \includegraphics[width=\textwidth]{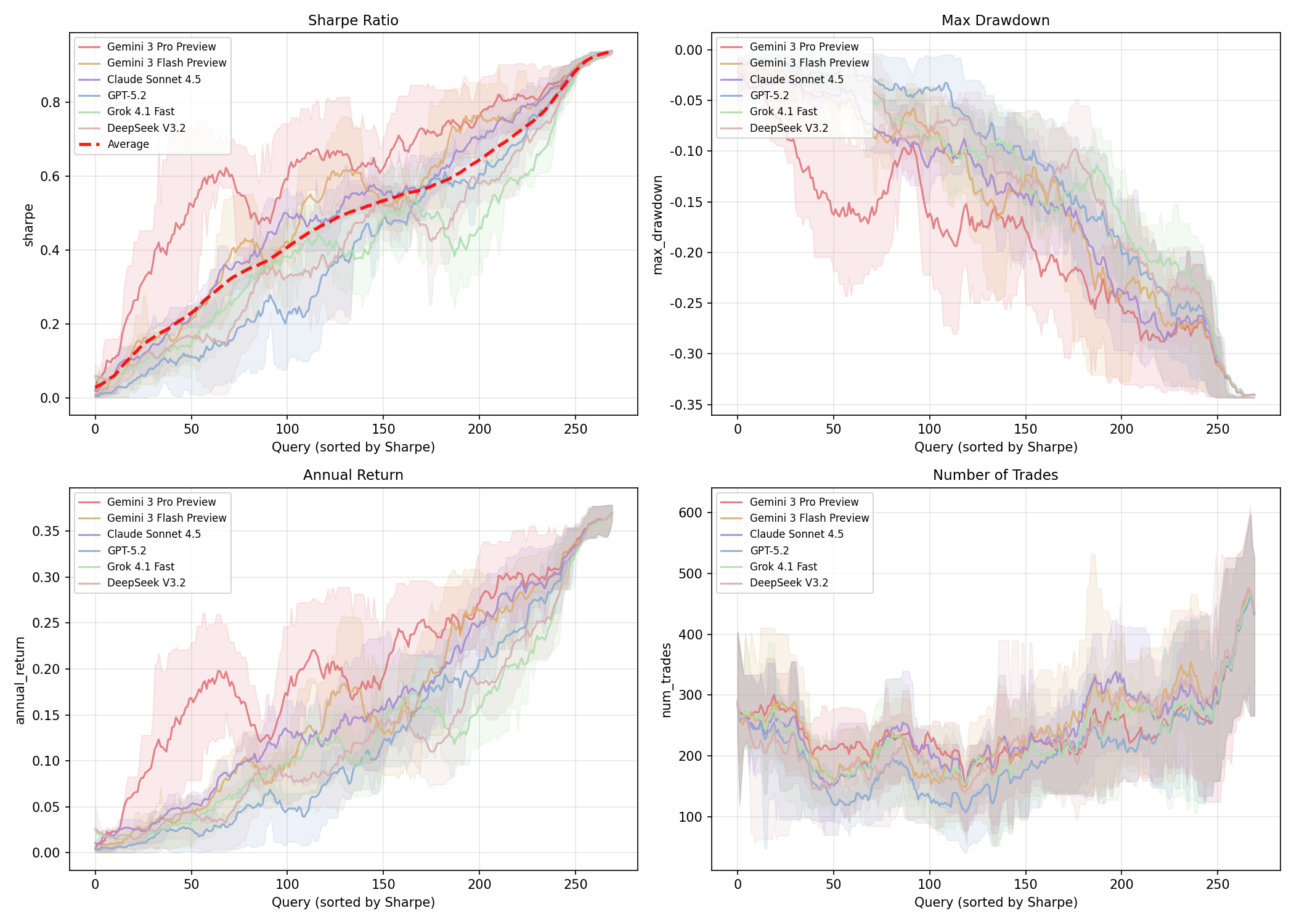}
  \caption{T=0.0}
  \end{subfigure}
  \hfill
  \begin{subfigure}[t]{0.23\textwidth}
  \includegraphics[width=\textwidth]{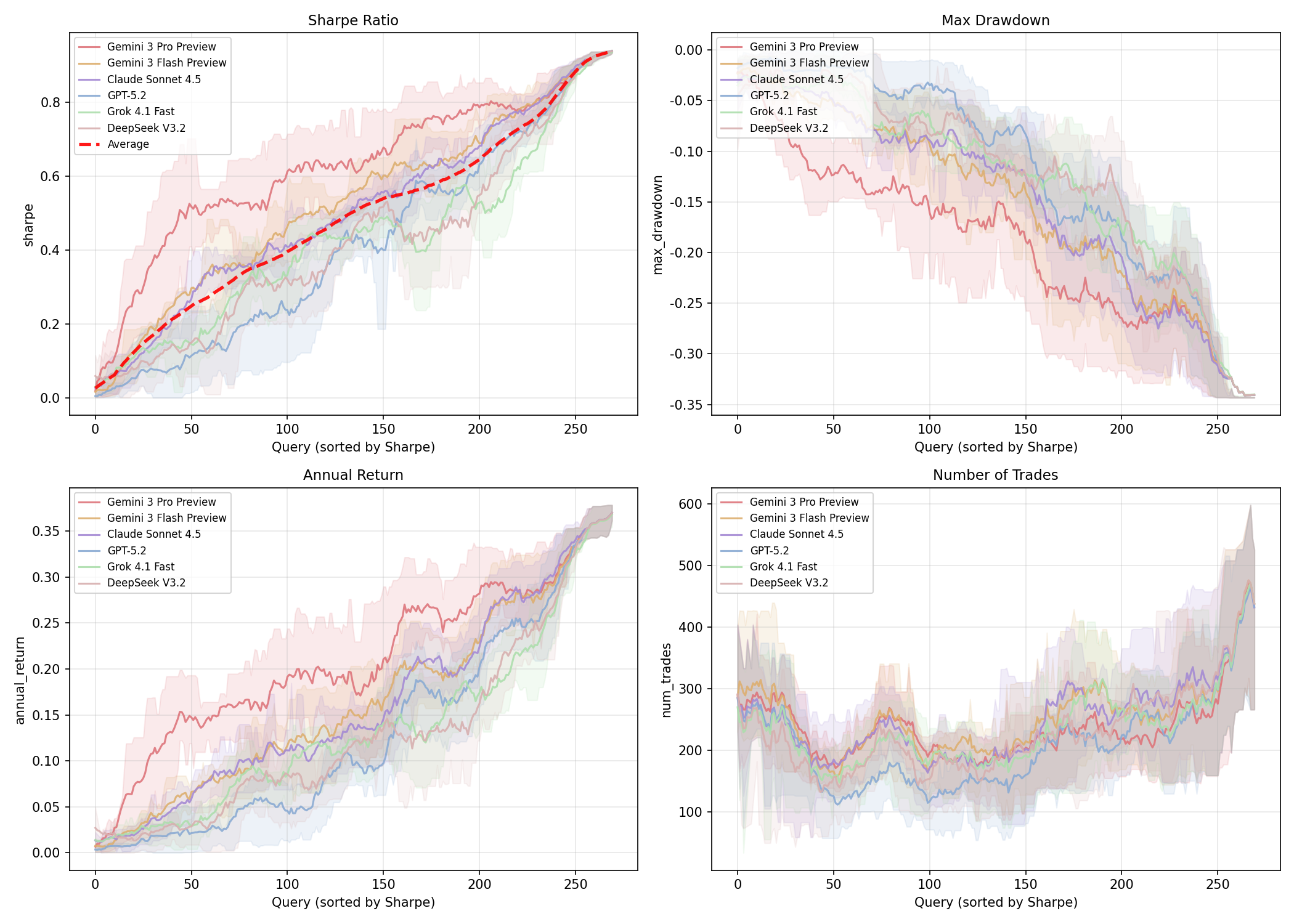}
  \caption{T=0.7}
  \end{subfigure}
  \vspace{-0.2cm}
  \caption{Aligned cumulative return curves on LLM-augmented queries ($\tau{=}0$ vs.\ $\tau{=}0.7$). Shaded bands denote the 25th--75th percentile range over 5 runs.}
  \label{fig:model_robustness}
  \vspace{-0.2cm}
\end{figure}

\Cref{fig:model_robustness} overlays the cumulative return trajectories of all six models across the full query spectrum, with shaded bands denoting the 25th--75th percentile range over 5 independent runs. Four observations emerge. \textbf{(1)~Inter-model separation is persistent and substantial.} The vertical gap between the highest and lowest trajectories far exceeds any individual model's confidence band throughout the entire query range, providing direct visual evidence that performance differences reflect genuine capability gaps rather than generation noise. \textbf{(2)~Run-to-run stability is consistently high.} The narrow shaded bands confirm that independently generated strategies produce tightly clustered outcomes even at $\tau{=}0.7$; notably, \textit{claude-sonnet-4.5} exhibits the narrowest bands (most consistent generation), while \textit{grok-4.1-fast} displays the widest, corroborating its high-variance profile observed in the boxplot analysis. \textbf{(3)~Temperature invariance is visually confirmed.} The $\tau{=}0$ and $\tau{=}0.7$ panels yield virtually identical curve shapes and separation patterns, reinforcing that the code-generation paradigm confines LLM stochasticity to a single generation step while ensuring deterministic downstream execution. \textbf{(4)~Difficulty modulates inter-model divergence.} The separation widens progressively for harder queries and narrows for easier ones, consistent with the per-level finding that Level~3 tasks maximally amplify capability differences across models.

\subsubsection{Per-model stability profiles.}
Appx.~\ref{appx_sec:appendix_results_stage_2} provides detailed per-model analyses of multi-run stability, difficulty-level performance curves, and best-case strategy returns. Three archetypal behavioral profiles emerge from these results. \textbf{(1)~Aggressive-creative profile} (\textit{gemini-3-pro-preview}): uniquely, this model's SR \emph{increases} from Level~1 to Level~3, indicating that open-ended creative freedom amplifies its strengths; however, this comes at the cost of the widest confidence bands among top-tier models and the highest drawdown/volatility. \textbf{(2)~Balanced-stable profile} (\textit{claude-sonnet-4.5}, \textit{gemini-3-flash-preview}): these models exhibit the narrowest run-to-run confidence bands and the mildest Level~1-to-Level~3 degradation, with \textit{claude-sonnet-4.5} achieving the best Calmar Ratio (CR~=~1.650) and \textit{gemini-3-flash-preview} serving as a cost-effective alternative with a nearly flat difficulty profile. \textbf{(3)~Conservative-rigid profile} (\textit{\textit{gpt-5.2}}, \textit{deepseek-v3.2}, \textit{grok-4.1-fast}): these models favor low-risk signal logic (lowest MDD and VOL) but suffer the steepest performance degradation from structured to open-ended tasks; \textit{deepseek-v3.2} exemplifies this pattern most starkly, leading at Level~1 yet dropping to the bottom tier at Level~3, while \textit{grok-4.1-fast} additionally exhibits the highest run-to-run variance, making it the least predictable model in the benchmark. These archetypal profiles suggest that the return-risk trade-off observed in aggregate metrics stems from fundamentally different strategy-generation strategies encoded by each LLM, and that no single model simultaneously optimizes for return, stability, and robustness across difficulty levels.

\subsubsection{Findings and conclusions.}
Synthesizing the above analyses, the Stage~2 evaluation yields five principal findings. \textbf{(1)~Temperature invariance}: the $\tau{=}0$ and $\tau{=}0.7$ results are near-identical across models, metrics, and difficulty levels, with a maximum SR difference below $0.008$. \textbf{(2)~Systematic difficulty progression}: the inter-model SR spread widens from Level~1 ($0.029$) to Level~2 ($8\!\times$) and Level~3 ($14\!\times$), confirming that the $3\times 3$ taxonomy separates models along a controlled cognitive-demand axis. \textbf{(3)~Cross-level ranking reversals}: Level~1 code translation and Level~3 open-ended strategy design rely on distinct capabilities, as shown by the strong crossover in model rankings that would be obscured by aggregate metrics and single-score summaries. \textbf{(4)~Reproducible risk profiles}: the aggressive-creative, balanced-stable, and conservative-rigid patterns persist across stages, assets, and temperatures, indicating stable model-specific strategy behavior under both structured and open-ended query settings. \textbf{(5)~Cross-asset robustness}: model rankings and a shared asset-difficulty gradient are preserved across all 7 assets spanning cryptocurrency and US equity markets, with Stage~2 further sharpening inter-model dispersion within each asset. Together, these findings validate the $3\times 3$ taxonomy as an effective diagnostic tool while complementing the ecological validity of Stage~1, and reinforce the code-generation paradigm as a stable, reproducible, and discriminative benchmark for LLM capability evaluation in finance across both real-world and structured queries.

\section{Discussion}
\label{sec:discussion}

The results suggest that strategy code generation changes both the evaluation interface and the capability under study. Here, end to end denotes the full pipeline from natural language to complete rule based strategy generation and deterministic backtesting, rather than agent style systems that directly emit \texttt{BUY}, \texttt{HOLD}, or \texttt{SELL} actions. The two settings therefore probe different abilities. Agent based trading emphasizes step wise action selection, whereas \projectname focuses on whether a model can synthesize factors, rules, and decision logic into a coherent strategy. The near invariance between $\tau{=}0$ and $\tau{=}0.7$ and the low run to run variance across both stages indicate that, once execution randomness is removed, performance differences more cleanly reflect strategy design ability. This also explains why agent style financial trading is a weak benchmark here, since Appendix C shows that such systems can remain unstable even under identical configurations.

The results further show that financial strategy generation is not a single capability. The widening spread across difficulty levels and the ranking reversals between Level~1 and Level~3 indicate that logic translation, logic completion, and open ended synthesis rely on distinct strengths. The recurring risk profiles across stages and assets likewise suggest that frontier LLMs encode stable preferences in how they trade off return, drawdown, and volatility. \projectname is therefore useful not only as a leaderboard but also as a diagnostic benchmark that separates conservative from aggressive models, structured translators from creative synthesizers, and robust models from those that degrade as task ambiguity increases. This also motivates the current focus on single asset evaluation, which isolates signal generation and strategy logic construction without the added variability of asset allocation, cross asset dependencies, and risk control.

The benchmark should be interpreted within a controlled scope. The current evaluation targets single asset long only strategy design under a standardized backtesting environment with a fixed transaction cost of $10^{-3}$ and without explicit slippage or liquidity modeling. These choices improve comparability and reproducibility, but they also mean that the reported results are better read as controlled measures of strategy design quality than as direct estimates of deployable trading performance. The same logic applies to Stage~2, whose queries are generated through controlled augmentation of Stage~1 to provide diagnostic structure rather than broad ecological coverage. Because Stage~2 remains anchored to real world sources and the predefined $3\times 3$ taxonomy, and model rankings remain broadly aligned between Stage~1 and Stage~2, the structured track appears to preserve genuine capability differences without strong model specific bias. Under this interpretation, \projectname provides a stable and interpretable foundation for evaluating financial strategy design with LLMs.

\section{Conclusion}

We presented \projectname, a benchmark that reframes LLM evaluation in quantitative finance from black box action emission to white box strategy code generation. Across 903 queries, six frontier LLMs, seven assets, and 35,190 total implementations, the results show that this code generation paradigm is temperature invariant, highly reproducible, and more discriminative than direct trading baselines. The benchmark further reveals widening capability gaps across difficulty levels, ranking reversals that separate distinct cognitive skills, and stable model specific risk profiles, establishing \projectname as a rigorous framework for evaluating financial strategy design with LLMs.

\section{Acknowledgments}
Shuo Sun is supported by Guangdong Provincial Key Lab of Integrated Communication, Sensing and Computation for Ubiquitous Internet of Things (No.2023B1212010007). Bo An is supported by the National Research Foundation Singapore and DSO National Laboratories under the AI Singapore Programme (AISG Award No: AISG2-GC-2023-009-1B).

\clearpage
\bibliographystyle{ACM-Reference-Format}
\balance
\bibliography{main}

\clearpage
\onecolumn

\appendix
\input{appendix}

\end{document}

%% file: appendix.tex
\section{Code and Data Availability}
All code and data will be publicly available upon acceptance of this paper, including the benchmark query set, evaluation pipeline, backtest engine, and supplementary scripts for reproducing the reported results.

\vspace{-0.5cm}
\section{Motivation}
Recent progress on financial large language models (FinLLMs) has triggered a surge of benchmarks aimed at measuring their financial capabilities. These benchmarks fall into two main categories: \emph{financial QA benchmarks} that test knowledge and reasoning on static inputs, and \emph{financial trading benchmarks} that ask an LLM to directly emit trading actions. However, neither category can reliably assess an LLM's true financial capability. QA benchmarks are prone to data staleness and memorization, while trading benchmarks suffer from severe \emph{decision instability}: the same model under the same settings can produce drastically different action sequences across runs, rendering single-run evaluations unreproducible and benchmark rankings fragile. This fundamental lack of \emph{stability}, \emph{robustness}, and \emph{reproducibility} motivates us to establish a new evaluation paradigm. Rather than benchmarking LLMs on static QA or unstable action generation, we propose to evaluate two core intermediate artifacts in systematic trading that are inherently more stable and auditable: \textbf{alpha factors} and \textbf{factor-based trading strategies}.
Below we first revisit the limitations of financial QA benchmarks, then discuss the instability pitfalls of action-based trading benchmarks, and finally motivate why factor and factor-strategy generation provides a more principled and comparable evaluation target.

\textbf{Financial QA benchmarks.}
A dominant line of work evaluates FinLLMs via financial question answering (QA) and reasoning tasks, e.g., numerical table QA (e.g., FinQA~\cite{chen2021finqa}, TAT-QA~\cite{zhu2021tat}), conversational finance QA (e.g., ConvFinQA~\cite{chen2022convfinqa}), and sentiment/interpretation style datasets (e.g., FiQA~\cite{maia201818}).
While useful, these QA benchmarks have several limitations:
\textbf{(1) Surface-level \textit{encyclopedic} competence and unfair comparability.}
Many QA datasets are static collections of text/tables (sometimes with charts), which primarily measure recall and short-horizon reasoning over a snapshot. As model scale and pretraining coverage grow, improvements can come from memorization/contamination rather than better \emph{financial decision-making}, making fair comparison difficult.
\textbf{(2) Staleness and temporal leakage under rapidly evolving finance.}
Financial concepts, events, regulations, and market narratives drift quickly. Static QA test sets become outdated, and train--test contamination is hard to rule out, which undermines reliability of benchmark conclusions in real-world settings.
\textbf{(3) Weak robustness and reliability assessment.}
QA metrics typically focus on answer matching and do not stress-test stability under noisy/contradictory multi-source signals, uncertainty calibration, or the cost of hallucinations. In practice, these properties are crucial for downstream trading pipelines.
Most importantly, \textbf{QA tasks cannot measure an LLM's trading capability}: they do not require sequential decision-making with positions, transaction costs, and long-horizon objectives. This gap motivates a second line of benchmarks, namely \textbf{financial trading benchmarks} that aim to evaluate trading decisions.

\textbf{Financial trading benchmarks.}
These benchmarks (e.g., Alpha Arena~\cite{alphaarena2025}) typically provide an LLM with multi-source market information (such as OHLCV time series, technical indicators, fundamentals, and news) and ask it to directly output a \emph{trading action} (e.g., \textsc{buy}/\textsc{sell}/\textsc{hold}, or a target position). Despite their appeal, we argue that current action-emitting trading benchmarks remain insufficient for measuring an LLM's \emph{financial capability} in a principled way:
\textbf{(1) Decision instability undermines reproducibility and can invalidate the benchmark.}
In practice, LLM trading decisions can be highly unstable. Under an identical setting (same model, same prompt template, same market and time period), the generated action sequence may vary substantially across runs due to decoding randomness or minor input perturbations, leading to dramatically different PnL and drawdown statistics. When such variance dominates, benchmark rankings become fragile and hard to reproduce.
\textbf{(2) Capability is confounded with system constraints and evaluation design.}
Reported performance is heavily affected by backtest and execution assumptions (e.g., costs, slippage, position sizing, and rebalancing rules) and by whether additional guardrails are imposed to curb over-trading. Therefore, improvements may reflect better constraint engineering rather than better financial reasoning.
\textbf{(3) Per-step action emission encourages myopic labeling instead of a consistent policy.}
Directly emitting actions resembles short-horizon classification for the current state, whereas profitable trading requires state-consistent, cost-aware, long-horizon policy optimization. As a result, these benchmarks may overestimate competence without capturing stable decision rules.

A key reason behind these limitations is the \textbf{instability} of LLMs action outputs in trading tasks, which can manifest as rapid flipping (e.g., buy then immediately sell or sell then immediately buy). Through comprehensive experiments, we attribute this instability to four major factors:
\textbf{(1) Stateless next-step inference.}
Vanilla LLMs make each action from the current input snapshot, without an inherent notion of persistent portfolio state. Even when recent actions are included in the prompt, small changes in inputs can still trigger inconsistent reversals.
\textbf{(2) Sensitivity in continuous-to-discrete mapping.}
Market signals vary continuously, whereas actions are discrete. This mismatch amplifies minor fluctuations (or minor prompt/wording differences) into action switches, lacking the inertia and tolerance bands commonly used in trading.
\textbf{(3) Action classification rather than policy optimization.}
Asking an LLM to output an action is closer to labeling the ``most reasonable'' action for the current state, instead of optimizing long-term, cost-aware returns in a constrained sequential decision problem. Without explicit optimization pressure, the model has little internal incentive to suppress over-trading.
\textbf{(4) Lack of hard behavioral constraints.}
Without system-level guardrails (e.g., minimum holding periods, cooldown windows, or a portfolio state machine), the LLM's natural linguistic uncertainty is directly executed as trades, magnifying churn and instability.

\textbf{Our motivation.}
Instead of benchmarking LLMs by \emph{direct action emission}, we argue for benchmarking their ability to produce \emph{auditable intermediate artifacts} used in quant research: \textbf{alpha factors} and \textbf{factor-based trading strategies}.
Such artifacts are (i) explicitly stateful when executed in backtests, (ii) naturally robustified via standard evaluation protocols (e.g., out-of-sample tests and turnover/cost analyses), and (iii) more comparable across models because performance is measured on a shared, reproducible pipeline. This motivates AlphaForgeBench as a benchmark for LLM-generated factors and factor strategies.

\input{appendix/appendix_llm_trading}

\input{appendix/appendix_alphaforgebench}

\input{appendix/appendix_real_result}

\input{appendix/appendix_bench_result}

%% file: appendix/appendix_llm_trading.tex
\section{Detailed Analysis of LLMs for Financial Trading}
\label{appx_sec:instability_analysis}

To quantify decision instability in action-emitting trading setups, we evaluate seven mainstream closed-source models (\textit{gemini-3-flash-preview}, \textit{gemini-3-pro-preview}, \textit{grok-4.1-fast}, \textit{deepseek-v3.2}, \textit{gpt-5.2}, \textit{claude-sonnet-4.5}) on a BTC interday trading task over 01/01/2025--01/01/2026 and, using a controlled-variable protocol with all other settings fixed, compare their trading trajectories along two axes: \textbf{(1) run-to-run variability (5 runs per identical setting)} and \textbf{(2) decoding temperature} (\texttt{temperature=0} vs.\ \texttt{temperature=0.7}).

\subsection{Experimental Setup}

\textbf{Trading task design.} We design a controlled single-asset trading environment on BTC (Bitcoin) with daily granularity. At each trading step $t$, the model receives the daily OHLCV (Open, High, Low, Close, Volume) data for BTC up to day $t$, along with a set of commonly used technical indicators (e.g., moving averages, RSI, MACD, Bollinger Bands). The recent price history (past 30 days) and indicator values are formatted into a structured prompt that provides sufficient context for decision-making. The model is prompted to output exactly one discrete action from $\{$\textsc{buy}, \textsc{hold}, \textsc{sell}$\}$, representing a fully invested long position, no change, or fully exiting the position, respectively. We adopt a simple position-sizing rule: each \textsc{buy} invests 100\% of available capital, and each \textsc{sell} liquidates the entire position. No leverage, short-selling, or partial positions are allowed. Transaction costs are set to zero to isolate decision quality from execution assumptions. Importantly, the prompt template, data preprocessing pipeline, and agent configuration are held strictly identical across all models, runs, and temperature settings, ensuring that any observed differences are attributable solely to the model's own decision-making process. This controlled design allows us to attribute instability to the LLM itself rather than to confounding factors in the trading environment.

\textbf{Why these two dimensions.} We focus on two complementary dimensions of instability. \textbf{(1) Run-to-run variability} probes the model's \emph{intrinsic} stochasticity: even at $T=0$ (nominally deterministic decoding), modern LLMs can produce different outputs across runs due to floating-point non-determinism or Mixture-of-Experts (MoE) routing noise. In a sequential trading task, a single divergent action early on can cascade into entirely different subsequent decisions, amplifying the perturbation into dramatically different financial outcomes. \textbf{(2) Decoding temperature} probes the model's sensitivity to a \emph{user-controlled} hyperparameter. If a small change from $T=0$ to $T=0.7$ causes the trading behavior to shift dramatically, the model's decision logic is fragile and over-reliant on sampling noise rather than grounded financial reasoning. Together, these two dimensions span the spectrum from uncontrollable internal randomness to controllable external randomness. If instability is observed on \emph{both} dimensions, it strongly suggests that the model lacks a coherent trading policy and is instead performing noisy per-step classification.

\textbf{Models.} We evaluate six mainstream closed-source models (\textit{gemini-3-flash-preview}, \textit{gemini-3-pro-preview}, \textit{grok-4.1-fast}, \textit{deepseek-v3.2}, \textit{gpt-5.2}, \textit{claude-sonnet-4.5}) on a BTC daily OHLCV trading task over 01/01/2025--01/01/2026. These models are among the strongest, frontier LLMs currently available; they top public benchmarks (e.g., MMLU, HumanEval) and are widely used in production. We select them to (i) span the major commercial providers (Google, Anthropic, OpenAI, xAI, DeepSeek), ensuring our findings are not tied to a single vendor; and (ii) include both flagship (\textit{pro}, \textit{sonnet}) and efficient (\textit{flash}, \textit{fast}) families, so we can assess whether decision instability differs systematically by model scale and intended use case. Notably, we choose the window 01/01/2025--01/01/2026 to minimize data leakage. Some models may have been trained on market data from earlier periods, which would introduce leakage risk; we therefore use the most recent available span.

\textbf{Run-to-run variability.} Under identical settings (fixed decoding temperature, agent configuration, data window, and model), we run the agent trading task 5 times and compare the resulting trajectories along five dimensions. We choose 5 runs as a balance between computational cost and statistical coverage, which is sufficient to reveal systematic instability patterns. \textbf{(1) Pairwise agreement rate} (heatmap) measures how well actions match between each pair of the 5 runs, providing a direct quantification of action-level reproducibility. \textbf{(2) Step-wise agreement} is the Jaccard similarity of actions at each time step across the 5 runs, revealing whether instability is uniformly distributed or concentrated in specific trading periods. \textbf{(3) Action distribution} is a bar chart of the fraction of BUY, HOLD, and SELL actions in each of the 5 runs, showing whether the overall trading strategy remains structurally consistent even when individual actions differ. \textbf{(4) Metric variance} uses box plots to show the spread of Annualized Rolling Return (ARR), Sharpe ratio, and Maximum Drawdown (MDD) across the 5 runs, quantifying how action-level instability translates into financial outcome variance. \textbf{(5) Cumulative return} is a line chart of cumulative return versus step for each of the 5 runs, visualizing the trajectory-level divergence over time. We run 5 times at \texttt{temperature=0.0} and 5 times at \texttt{temperature=0.7}, then produce detailed comparison figures of the five-run results under each fixed temperature.

\textbf{Decoding temperature.} As above, we run each model 5 times at \texttt{temperature=0.0} and 5 times at \texttt{temperature=0.7}. For each temperature, we first aggregate results by averaging over the 5 runs, then compare the two temperatures along the same five dimensions. By averaging, we smooth out run-to-run noise and isolate the systematic effect of temperature on trading behavior. \textbf{(1) Action distribution} is a bar chart of the fraction of BUY, HOLD, and SELL actions at each temperature, averaged over the 5 runs, showing whether temperature shifts the overall trading stance. \textbf{(2) Temperature agreement rate} measures how well actions match between \texttt{temperature=0.0} and \texttt{temperature=0.7}, computed after averaging each temperature's decisions over its 5 runs. \textbf{(3) Metric variance} reports ARR, Sharpe ratio, and MDD for each temperature, each averaged over the 5 runs, indicating whether one setting systematically outperforms. \textbf{(4) Step-wise agreement} is the Jaccard similarity of actions at each time step between the two temperature settings, computed on per-step actions aggregated over the 5 runs at each temperature. \textbf{(5) Cumulative return} is a line chart of cumulative return versus trading step for each temperature, averaged over the 5 runs.

\clearpage
\subsection{Model: \textit{gemini-3-pro-preview}}

\begingroup\centering
\includegraphics[width=0.49\linewidth,height=0.32\textheight,keepaspectratio]{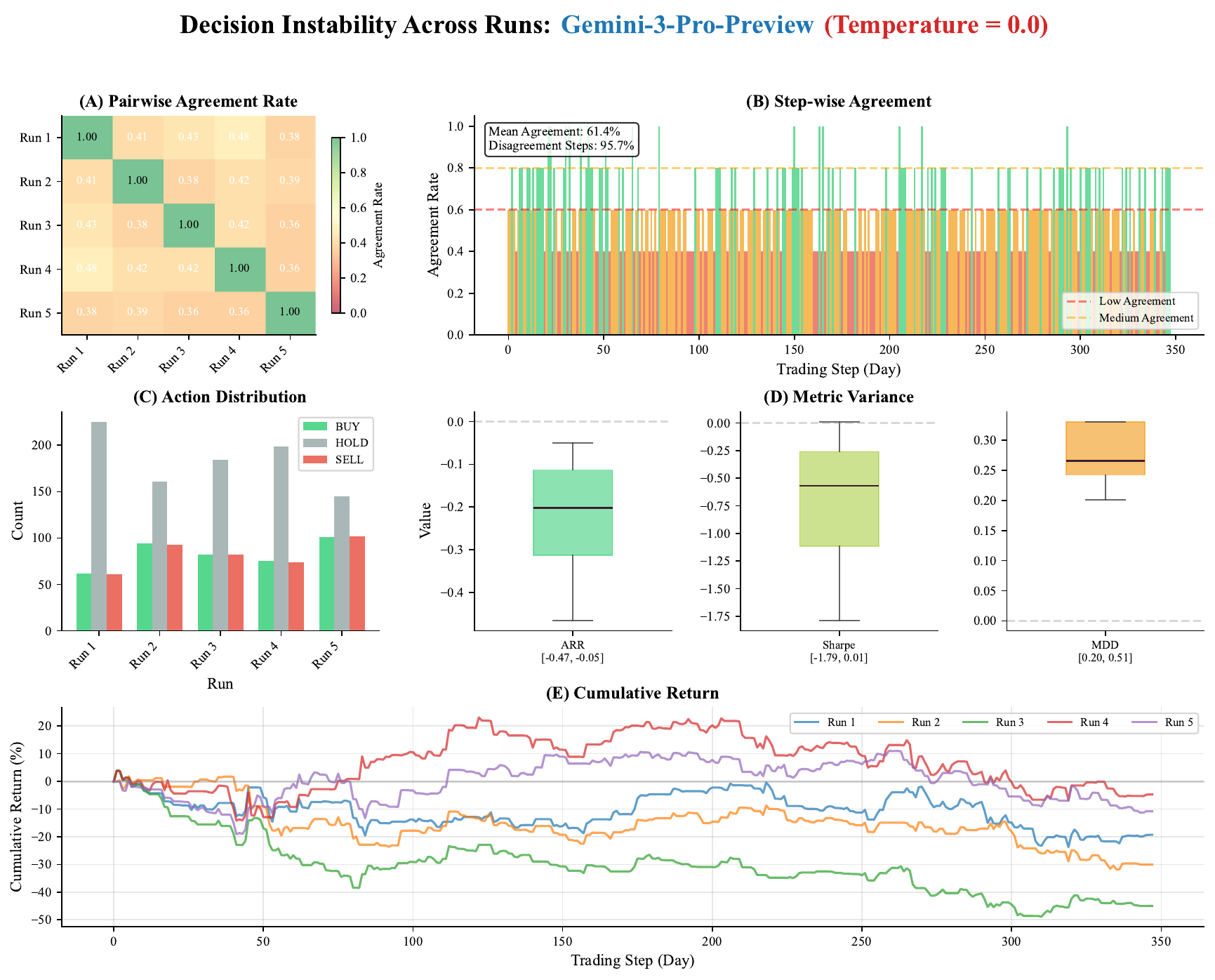}
\hfill
\includegraphics[width=0.49\linewidth,height=0.32\textheight,keepaspectratio]{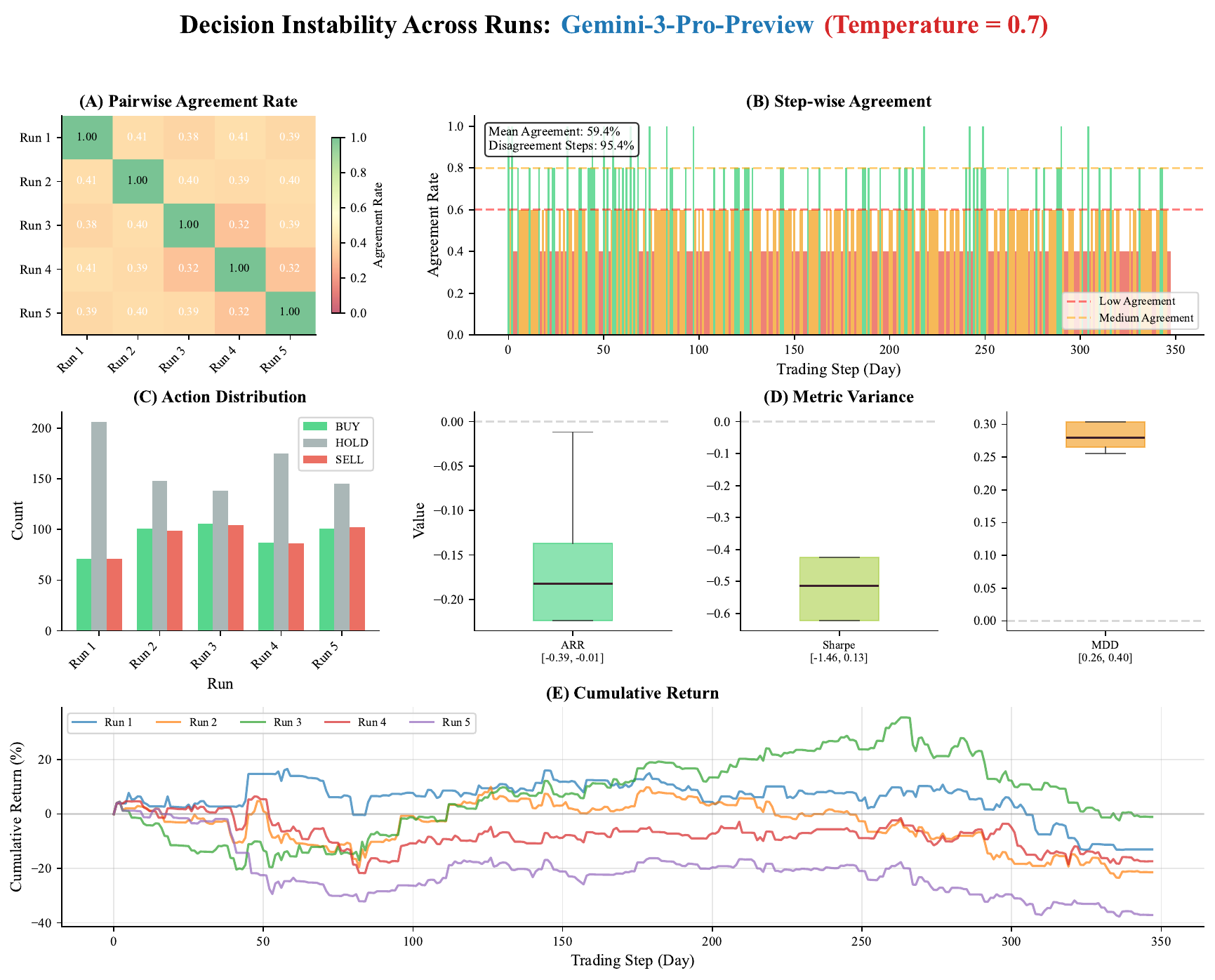}
\vspace{-1.5em}
\captionof{figure}{Cross-run instability of \textit{gemini-3-pro-preview} under different decoding temperatures.}
\label{appx_fig:cross_run_instability_gemini_3_pro_preview}
\par\endgroup

\textbf{Run-to-run Variability.} \Cref{appx_fig:cross_run_instability_gemini_3_pro_preview} summarizes five-run variability from five complementary views. \textbf{(1) Pairwise agreement rate} (Panel A) is remarkably low even at $T=0$, ranging only 0.36--0.48, and further drops to 0.32--0.41 at $T=0.7$, indicating that two runs rarely choose the same actions. \textbf{(2) Step-wise agreement} (Panel B) is similarly unstable: the mean agreement is 61.4\% with a 95.7\% step-level disagreement rate, implying the model almost never reproduces a complete action sequence. \textbf{(3) Action distribution} (Panel C) reveals that this instability is not merely cosmetic; runs differ materially in the frequency of \textsc{buy}/\textsc{sell} events rather than only in rare edge cases. \textbf{(4) Metric variance} (Panel D) shows large dispersion of ARR/Sharpe/MDD across runs, consistent with the above decision volatility. \textbf{(5) Cumulative return} trajectories (Panel E) therefore diverge sharply; e.g., at $T=0$ one run collapses to nearly $-50\%$ while others hover near break-even. Overall, such extreme variance in a nominally greedy setup undermines the reliability of single-run backtests for \textit{gemini-3-pro-preview}.

\begin{wrapfigure}{r}{0.52\linewidth}
    \vspace{-1.em}
    \centering
    \includegraphics[width=\linewidth]{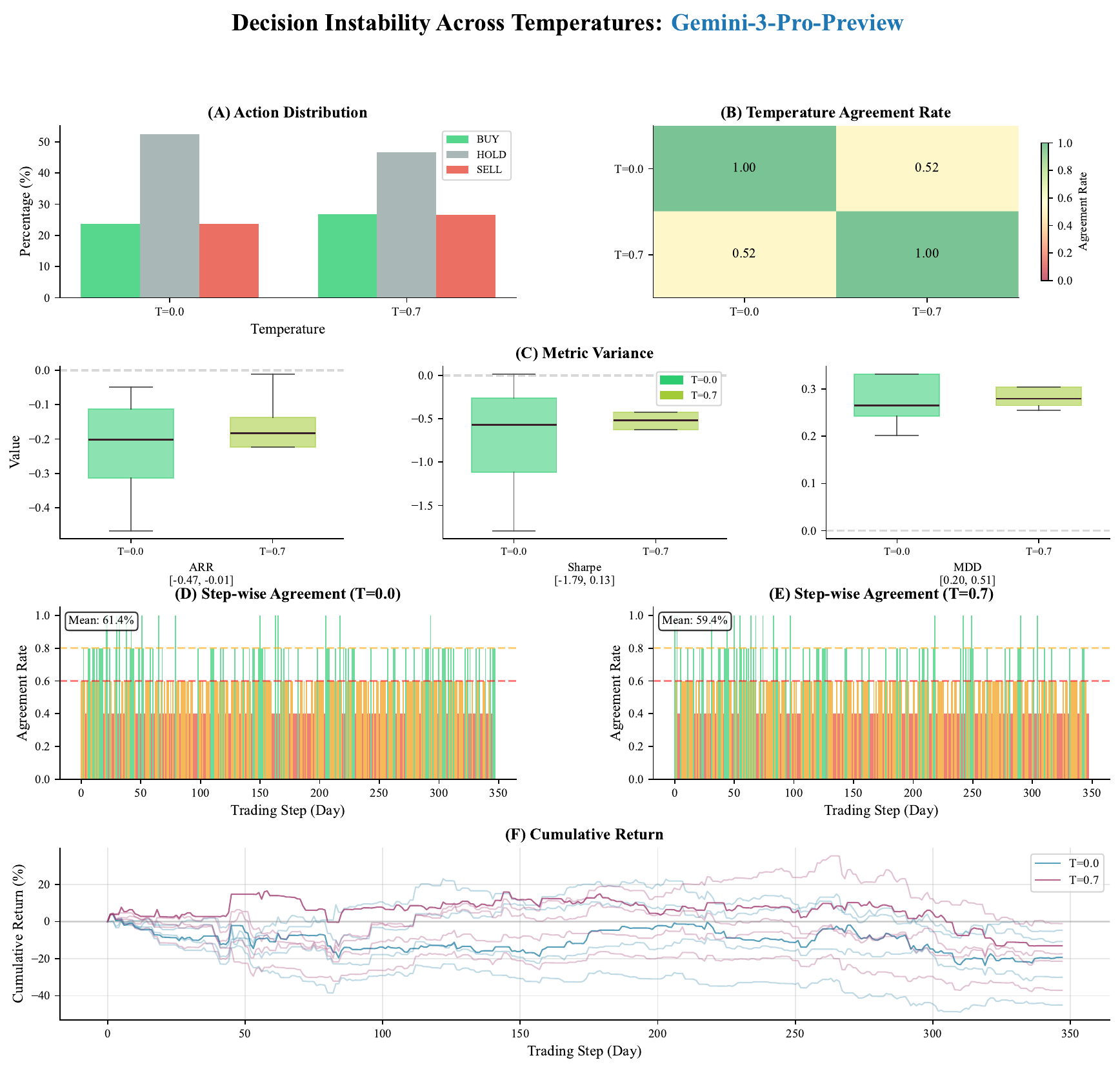}
    \vspace{-1.5em}
    \caption{Decision instability across decoding temperatures (\textcolor{tabblue}{\textit{gemini-3-pro-preview}})}
    \label{appx_fig:cross_temp_instability_gemini_3_pro_preview}
\end{wrapfigure}

\textbf{Decoding Temperature.} \Cref{appx_fig:cross_temp_instability_gemini_3_pro_preview} compares $T=0$ and $T=0.7$ from five perspectives. \textbf{(1) Action distribution} (Panel A) is dominated by \textsc{hold} under both temperatures, with only modest shifts in \textsc{buy}/\textsc{sell} frequency. \textbf{(2) Temperature agreement rate} (Panel B) is only 0.55, showing that switching from greedy decoding to stochastic sampling flips nearly half of the aggregated trading decisions. \textbf{(3) Metric variance} (Panel C) remains consistently poor across temperatures, with negative ARR and Sharpe in both cases and similar MDD ranges. \textbf{(4) Step-wise agreement} (Panels D--E) is low and noisy across the trading horizon, with mean agreement around 61.4\% at $T=0$ and 59.4\% at $T=0.7$, indicating unstable decision logic even within each temperature setting. \textbf{(5) Cumulative return} (Panel F) trajectories largely overlap and trend downward for both temperatures, suggesting that while temperature changes the specific actions, it does not improve the overall profitability profile for \textit{gemini-3-pro-preview}.

\textbf{Summary.} We draw three key conclusions:
\begin{itemize}[leftmargin=*]
    \item Both $T=0.0$ and $T=0.7$ fail to guarantee stable, consistent action sequences across repeated runs under the same setting.
    \item Overall, the stochastic setting ($T=0.7$) tends to yield better cumulative return trajectories than deterministic decoding ($T=0.0$), although performance remains volatile.
    \item Even within an identical setting, the action distribution can differ substantially across runs, indicating large run-to-run variability in trading behavior.
\end{itemize}

\clearpage
\subsection{Model: \textit{gemini-3-flash-preview}}

\begingroup\centering
\includegraphics[width=0.49\linewidth,height=0.32\textheight,keepaspectratio]{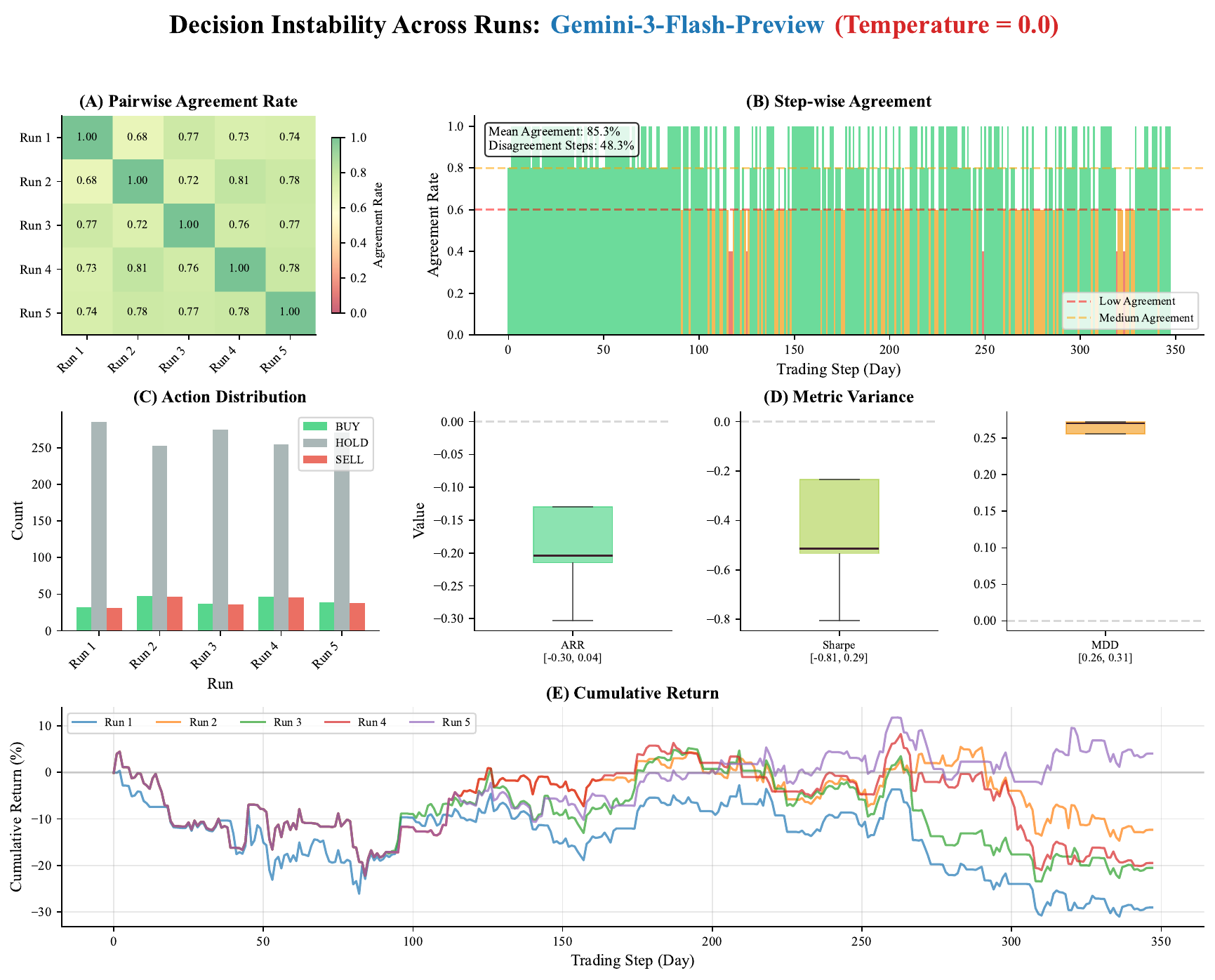}
\hfill
\includegraphics[width=0.49\linewidth,height=0.32\textheight,keepaspectratio]{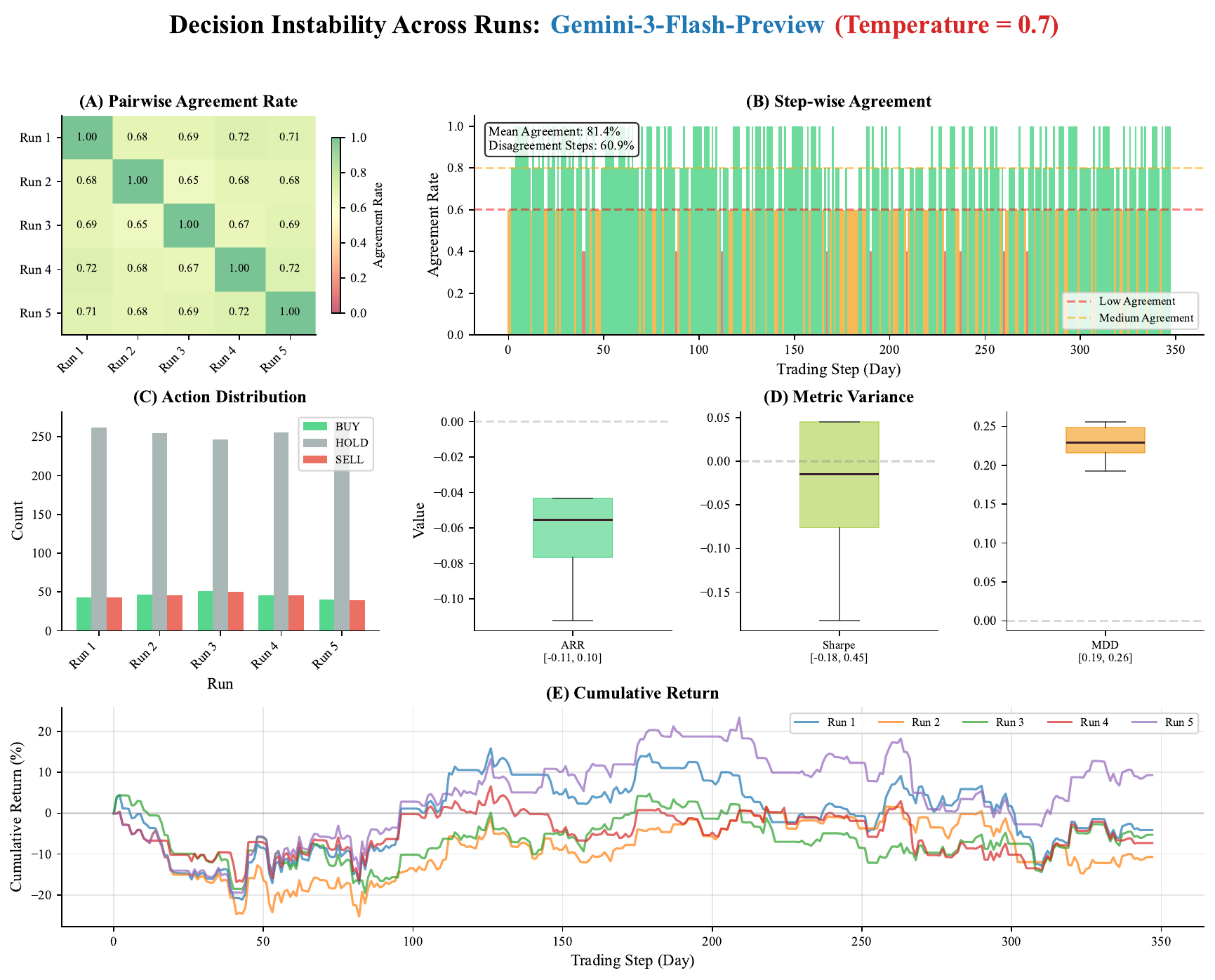}
\vspace{-1.5em}
\captionof{figure}{Cross-run instability of \textit{gemini-3-flash-preview} under different decoding temperatures.}
\label{appx_fig:cross_run_instability_gemini_3_flash_preview}
\par\endgroup

\textbf{Run-to-run Variability.} \Cref{appx_fig:cross_run_instability_gemini_3_flash_preview} summarizes five-run variability from five complementary views. \textbf{(1) Pairwise agreement rate} (Panel A) is relatively high at $T=0$ (mean 85.3\%) and remains high at $T=0.7$ (81.4\%), indicating that \textit{gemini-3-flash-preview} produces more consistent action sequences across runs than its Pro counterpart. \textbf{(2) Step-wise agreement} (Panel B) still fluctuates over the trading horizon and often degrades over time, as small early discrepancies accumulate into divergent actions in later steps. \textbf{(3) Action distribution} (Panel C) is broadly conservative across all runs, but the frequency of rare \textsc{buy}/\textsc{sell} events varies noticeably across runs, suggesting timing instability even when the overall strategy is stable. \textbf{(4) Metric variance} (Panel D) shows noticeable dispersion in ARR, Sharpe, and MDD across runs, confirming that even high agreement rates do not guarantee consistent financial outcomes. \textbf{(5) Cumulative return} trajectories (Panel E) start close but gradually fan out as trading progresses, leading to materially different final outcomes despite similar initial behavior.

\begin{wrapfigure}{r}{0.52\linewidth}
    \vspace{-1.0em}
    \centering
    \includegraphics[width=\linewidth]{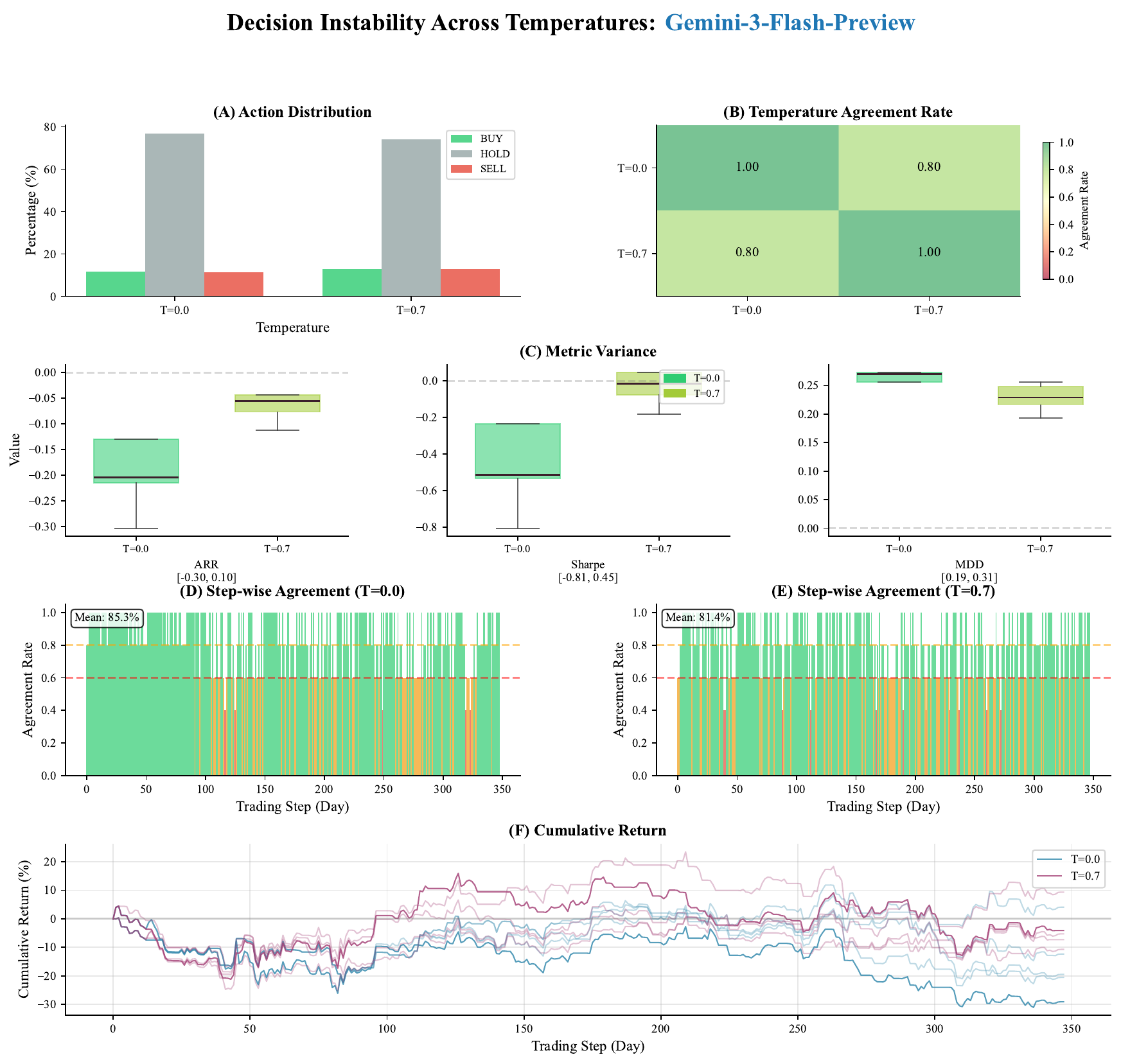}
    \vspace{-1.5em}
    \caption{Decision instability across decoding temperatures (\textcolor{tabblue}{\textit{gemini-3-flash-preview}})}
    \label{appx_fig:cross_temp_instability_gemini_3_flash_preview}
\end{wrapfigure}

\textbf{Decoding Temperature.} \Cref{appx_fig:cross_temp_instability_gemini_3_flash_preview} compares $T=0$ and $T=0.7$ from five perspectives. \textbf{(1) Action distribution} (Panel A) is conservative at both temperatures, with \textsc{hold} dominating around 80\% and only minor shifts in \textsc{buy}/\textsc{sell} frequency. \textbf{(2) Temperature agreement rate} (Panel B) is 0.80, meaning most trading decisions are preserved across temperature settings. \textbf{(3) Metric variance} (Panel C) shows negative ARR and Sharpe at both settings, with $T=0.7$ slightly less negative and marginally better Sharpe. \textbf{(4) Step-wise agreement} (Panels D and E) remains relatively high, with mean agreement about 85.3\% at $T=0$ and 81.4\% at $T=0.7$. \textbf{(5) Cumulative return} (Panel F) trajectories largely overlap, and $T=0.7$ tends to sit slightly above $T=0$.

\textbf{Summary.} For \textit{gemini-3-flash-preview}, we draw three key conclusions:
\begin{itemize}[leftmargin=*]
    \item The consistency between $T=0$ and $T=0.7$ is relatively high (temperature agreement rate $=0.80$), suggesting that decoding temperature has a limited impact on the trading decisions for \textit{gemini-3-flash-preview} compared to other models.
    \item Under an identical setting, run-to-run variability remains substantial; runs exhibit high agreement early on but progressively diverge as small discrepancies accumulate over time, resulting in different final outcomes.
    \item Overall, the stochastic setting ($T=0.7$) tends to yield slightly better cumulative return trajectories than deterministic decoding ($T=0$), although performance remains volatile and both settings underperform.
\end{itemize}

\clearpage
\subsection{Model: \textit{deepseek-v3.2}}

\begingroup\centering
\includegraphics[width=0.49\linewidth,height=0.32\textheight,keepaspectratio]{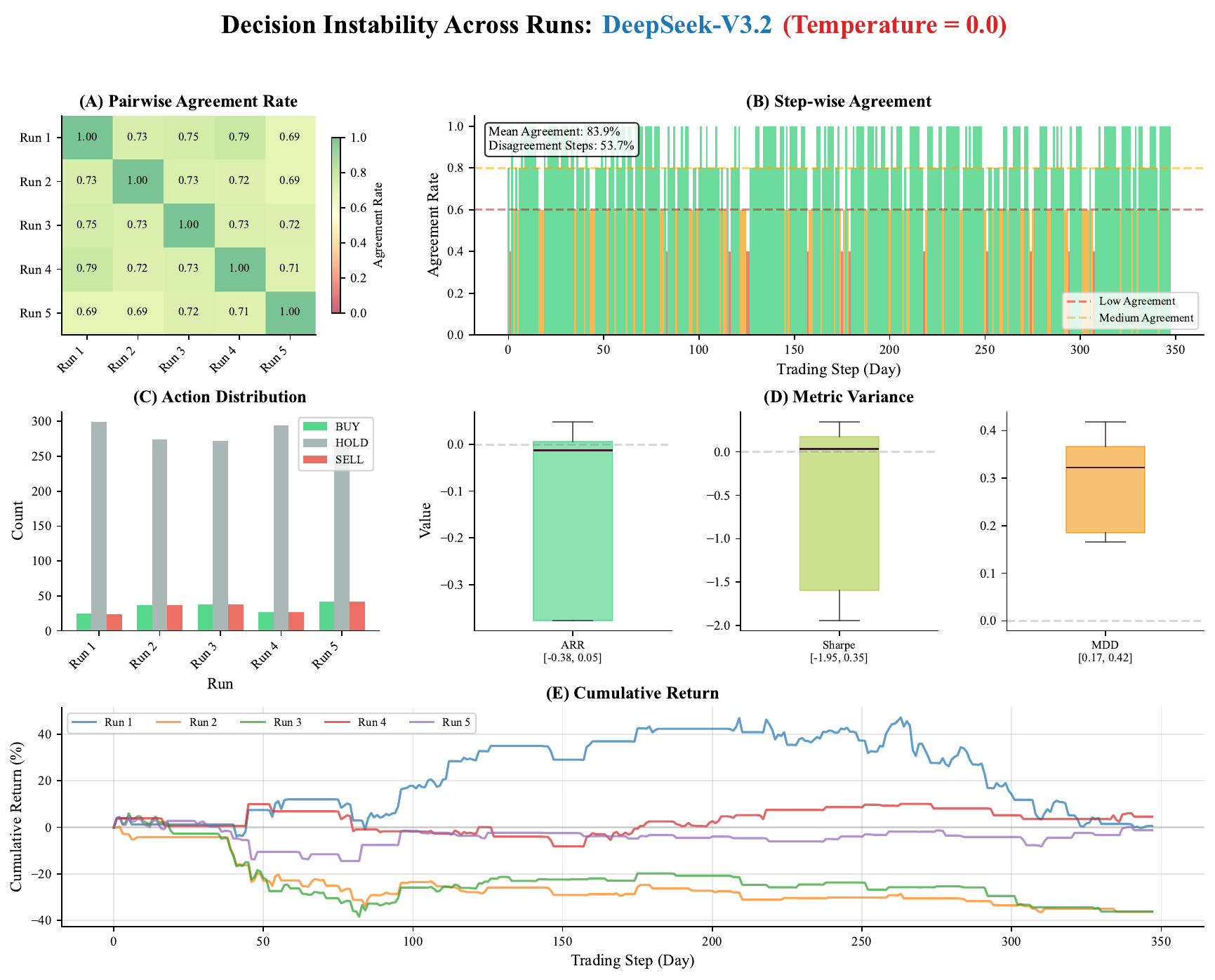}
\hfill
\includegraphics[width=0.49\linewidth,height=0.32\textheight,keepaspectratio]{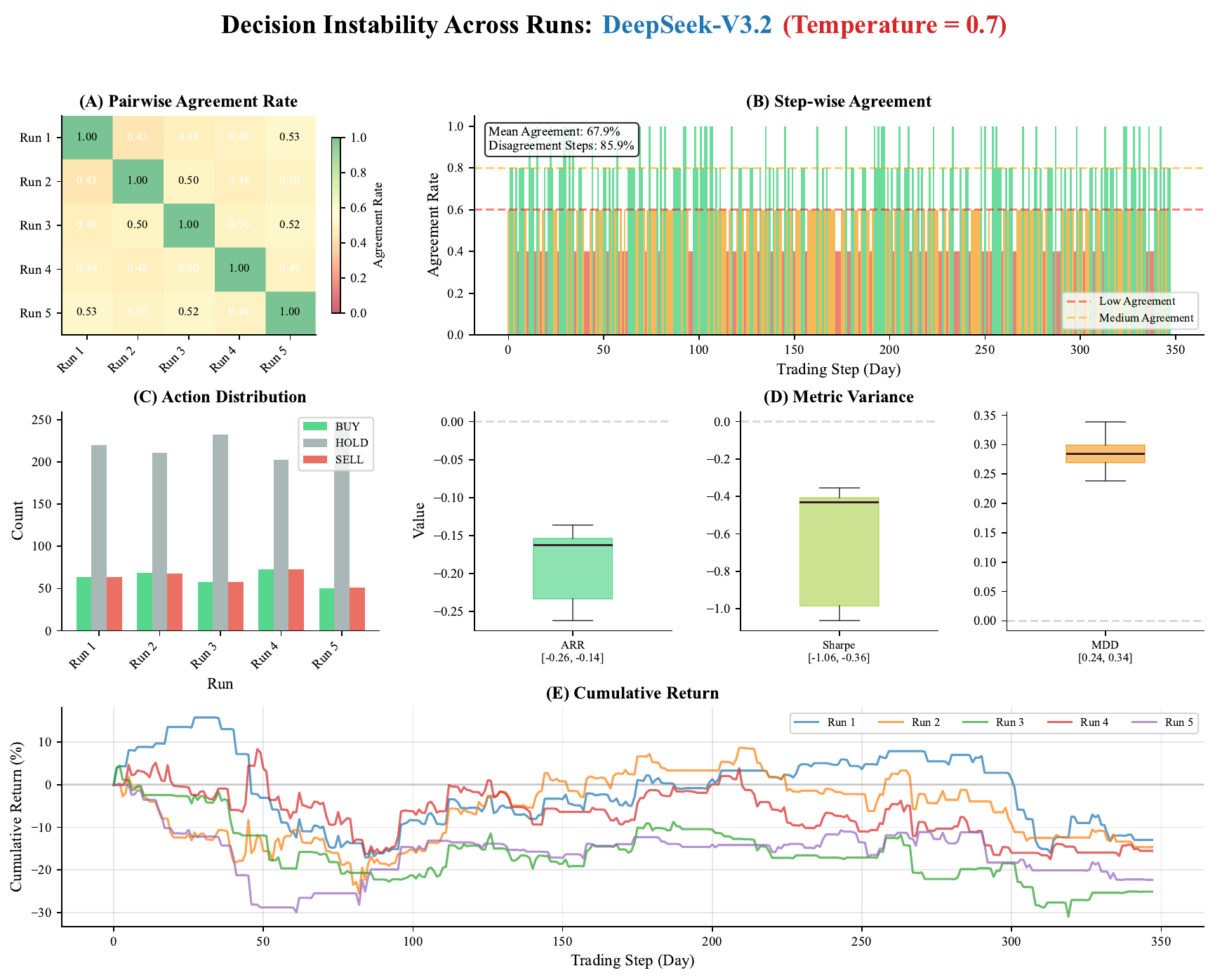}
\vspace{-1.5em}
\captionof{figure}{Cross-run instability of \textit{deepseek-v3.2} under different decoding temperatures.}
\label{appx_fig:cross_run_instability_deepseek_v3p2}
\par\endgroup

\textbf{Run-to-run Variability.} \Cref{appx_fig:cross_run_instability_deepseek_v3p2} summarizes five-run variability from five complementary views. \textbf{(1) Pairwise agreement rate} (Panel A) is relatively high at $T=0$ (mean 83.9\%) but drops substantially at $T=0.7$ (mean 67.9\%), indicating that stochastic sampling significantly disrupts action consistency. \textbf{(2) Step-wise agreement} (Panel B) shows 53.7\% disagreement steps at $T=0$ and 85.9\% at $T=0.7$, reflecting increasing decision instability under stochastic decoding. \textbf{(3) Action distribution} (Panel C) is dominated by \textsc{hold} at $T=0$, but at $T=0.7$ the \textsc{buy}/\textsc{sell} frequency increases and varies substantially across runs. \textbf{(4) Metric variance} (Panel D) shows large dispersion in ARR, Sharpe, and MDD at $T=0$, with one outlier run achieving positive return while others suffer up to $-40\%$ drawdown; at $T=0.7$, the variance is tighter but uniformly negative. \textbf{(5) Cumulative return} trajectories (Panel E) diverge sharply at $T=0$, ranging from $+10\%$ to $-40\%$, whereas at $T=0.7$ trajectories cluster in the $-10\%$ to $-30\%$ loss zone, paradoxically making the stochastic setting more predictable in its failure.

\begin{wrapfigure}{r}{0.52\linewidth}
    \vspace{-1.0em}
    \centering
    \includegraphics[width=\linewidth]{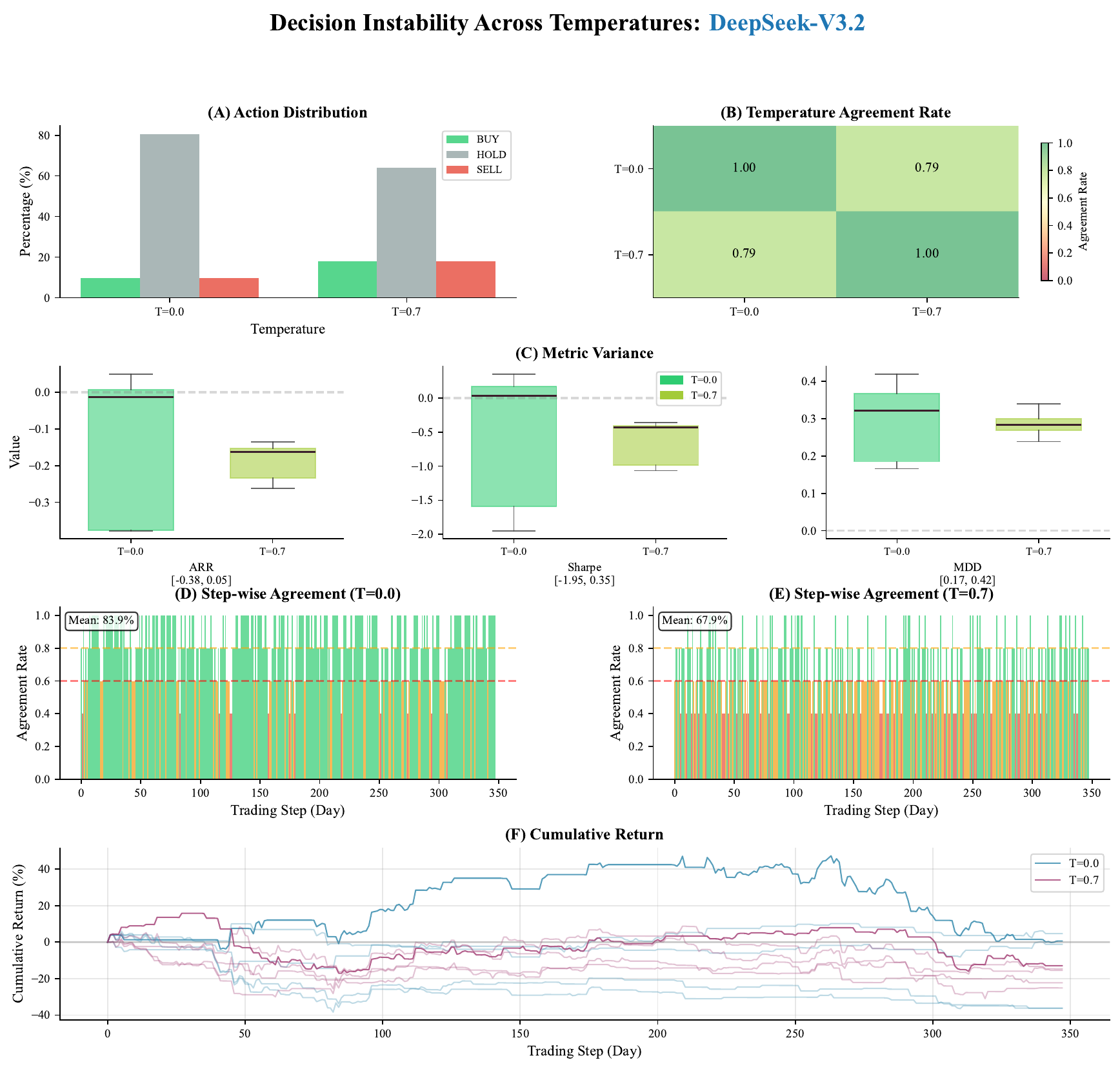}
    \vspace{-1.0em}
    \caption{Decision instability across decoding temperatures (\textcolor{tabblue}{\textit{deepseek-v3.2}})}
    \label{appx_fig:cross_temp_instability_deepseek_v3p2}
\end{wrapfigure}

\textbf{Decoding Temperature.} \Cref{appx_fig:cross_temp_instability_deepseek_v3p2} compares $T=0$ and $T=0.7$ from five perspectives. \textbf{(1) Action distribution} (Panel A) is conservative at both temperatures, with \textsc{hold} dominating 75--80\% and minor shifts in \textsc{buy}/\textsc{sell}. \textbf{(2) Temperature agreement rate} (Panel B) is 0.79, indicating most trading decisions are preserved across temperatures. \textbf{(3) Metric variance} (Panel C) shows negative ARR and Sharpe at both settings; notably, $T=0$ exhibits larger variance than $T=0.7$. \textbf{(4) Step-wise agreement} (Panels D and E) is 83.9\% at $T=0$ and drops to 67.9\% at $T=0.7$. \textbf{(5) Cumulative return} (Panel F) trajectories diverge substantially at $T=0$ ($+40\%$ to $-40\%$) but cluster tightly in the negative zone at $T=0.7$ ($-10\%$ to $-30\%$).

\textbf{Summary.} We draw three key conclusions:
\begin{itemize}[leftmargin=*]
    \item Unlike other models, the deterministic setting ($T=0$) outperforms the stochastic setting ($T=0.7$): at $T=0$, the model adopts a more conservative strategy with fewer trading actions and occasionally achieves positive returns; at $T=0.7$, trading frequency increases but outcomes are uniformly negative.
    \item Stochastic decoding ($T=0.7$) not only degrades profitability but also significantly increases run-to-run inconsistency, making it doubly undesirable for this model.
    \item Despite relatively high agreement rates at $T=0$, runs still diverge substantially in cumulative returns, indicating that even small decision differences can compound into drastically different financial outcomes over time.
\end{itemize}

\clearpage
\subsection{Model: \textit{grok-4.1-fast}}

\begingroup\centering
\includegraphics[width=0.49\linewidth,height=0.32\textheight,keepaspectratio]{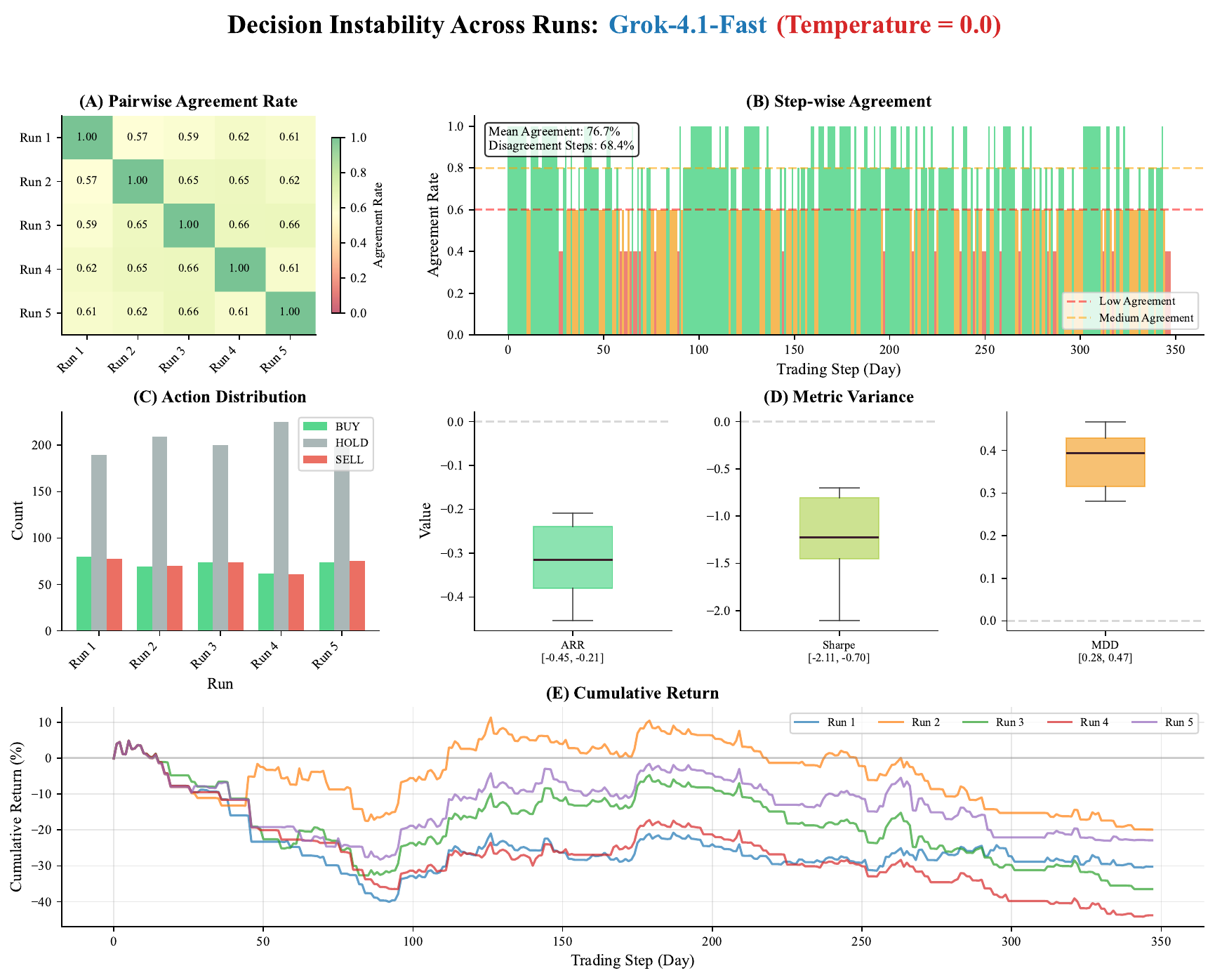}
\hfill
\includegraphics[width=0.49\linewidth,height=0.32\textheight,keepaspectratio]{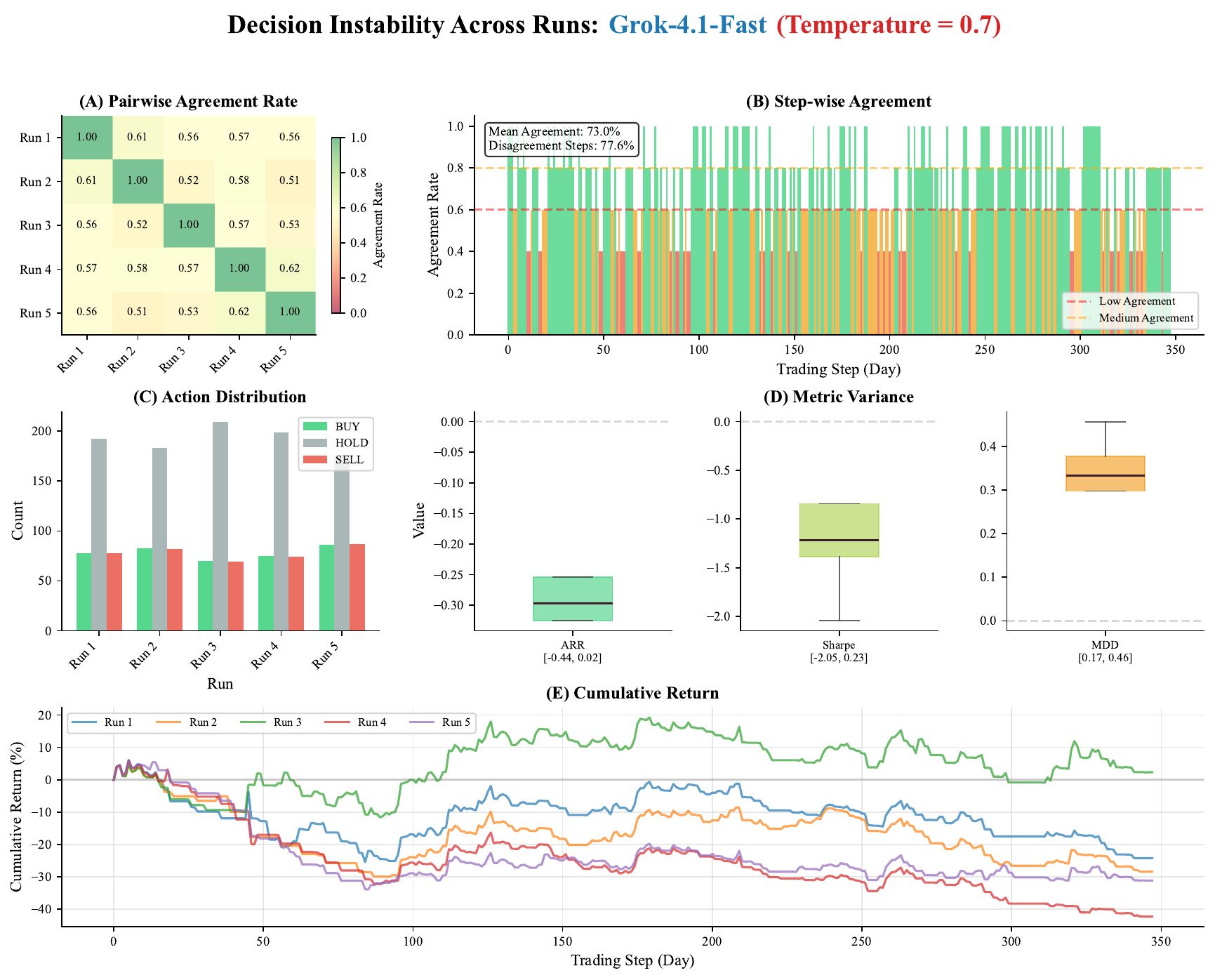}
\vspace{-1.5em}
\captionof{figure}{Cross-run instability of \textit{grok-4.1-fast} under different decoding temperatures.}
\label{appx_fig:cross_run_instability_grok_4p1_fast}
\par\endgroup

\textbf{Run-to-run Variability.} \Cref{appx_fig:cross_run_instability_grok_4p1_fast} summarizes five-run variability from five complementary views. \textbf{(1) Pairwise agreement rate} (Panel A) is moderate at $T=0$ (mean 75.7\%) and slightly lower at $T=0.7$ (mean 73.0\%), indicating substantial decision inconsistency across runs at both temperatures. \textbf{(2) Step-wise agreement} (Panel B) shows 88.4\% disagreement steps at $T=0$ and 77.6\% at $T=0.7$, reflecting persistent instability throughout the trading horizon. \textbf{(3) Action distribution} (Panel C) varies considerably across runs, with some runs dominated by \textsc{hold} while others exhibit more frequent \textsc{buy}/\textsc{sell} actions. \textbf{(4) Metric variance} (Panel D) is substantial at both temperatures: at $T=0$, ARR ranges from $-0.45$ to $-0.21$ with Sharpe from $-2.11$ to $-0.70$; at $T=0.7$, one outlier run achieves near-zero ARR while others remain deeply negative. \textbf{(5) Cumulative return} trajectories (Panel E) diverge significantly at both settings, ranging from $-10\%$ to $-40\%$ at $T=0$ and from $+20\%$ to $-40\%$ at $T=0.7$, highlighting the danger of relying on single-run backtests.

\begin{wrapfigure}{r}{0.52\linewidth}
    \vspace{-1.0em}
    \centering
    \includegraphics[width=\linewidth]{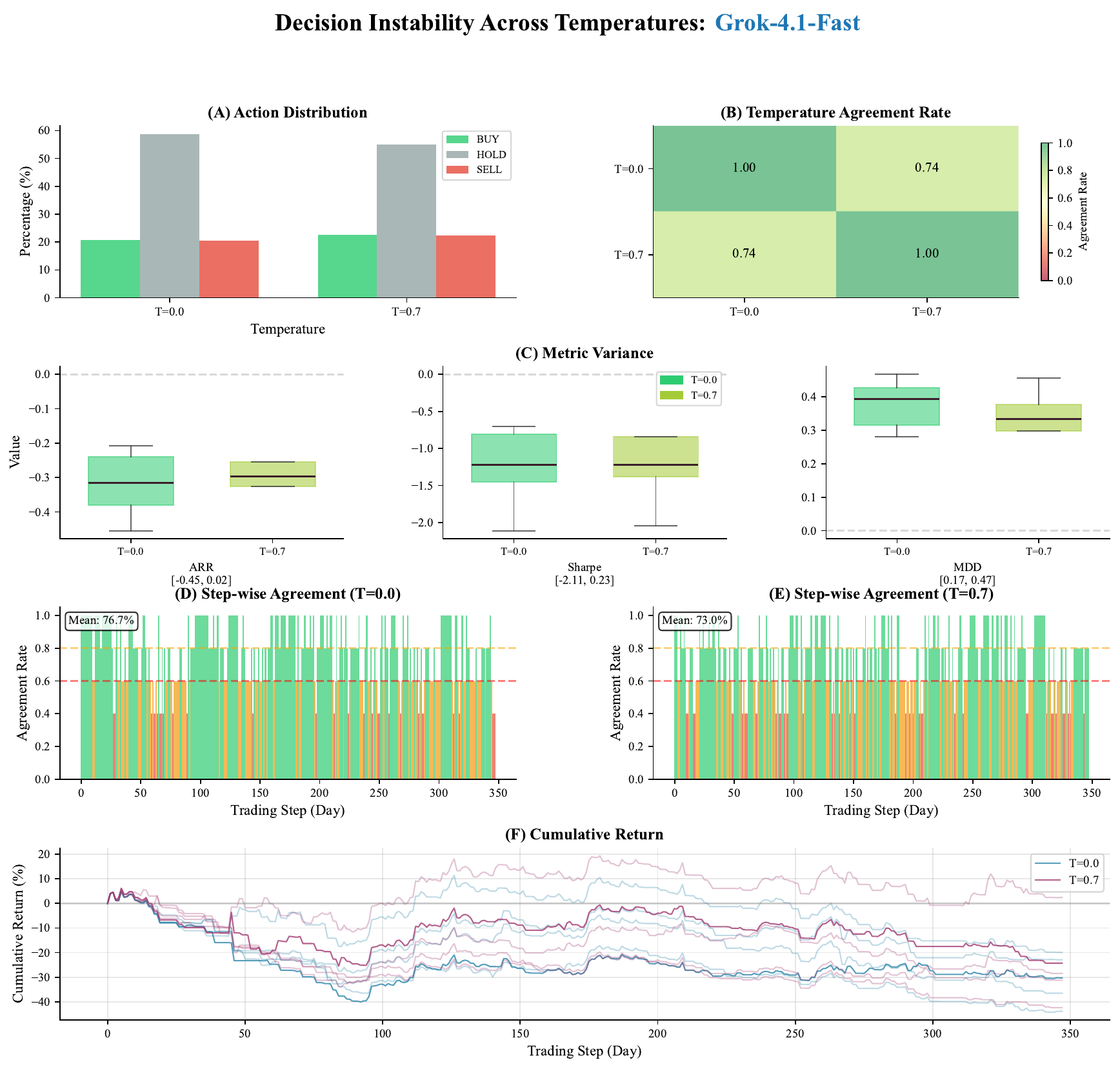}
    \vspace{-1.0em}
    \caption{Decision instability across decoding temperatures (\textcolor{tabblue}{\textit{grok-4.1-fast}})}
    \label{appx_fig:cross_temp_instability_grok_4p1_fast}
\end{wrapfigure}

\textbf{Decoding Temperature.} \Cref{appx_fig:cross_temp_instability_grok_4p1_fast} compares $T=0$ and $T=0.7$ from five perspectives. \textbf{(1) Action distribution} (Panel A) shows a more balanced trading profile than other models, with \textsc{hold} around 55--60\% and notable \textsc{buy}/\textsc{sell} activity (~20\% each) at both temperatures. \textbf{(2) Temperature agreement rate} (Panel B) is 0.74, indicating moderate consistency across temperature settings. \textbf{(3) Metric variance} (Panel C) shows negative ARR and Sharpe at both settings with large variance; ARR ranges from $-0.45$ to $+0.02$ and Sharpe from $-2.11$ to $+0.23$, reflecting high outcome uncertainty. \textbf{(4) Step-wise agreement} (Panels D and E) is 76.7\% at $T=0$ and 73.0\% at $T=0.7$, both lower than other models. \textbf{(5) Cumulative return} (Panel F) trajectories overlap substantially between temperatures, both ranging from $+10\%$ to $-40\%$, suggesting that temperature has limited impact on final outcomes while run-to-run variance dominates.

\textbf{Summary.} We draw three key conclusions:
\begin{itemize}[leftmargin=*]
    \item \textit{grok-4.1-fast} adopts a more aggressive trading profile than other models with more frequent \textsc{buy}/\textsc{sell} actions, yet this does not translate into better performance.
    \item Temperature has limited impact on final outcomes: both $T=0$ and $T=0.7$ produce similarly wide cumulative return ranges with overlapping trajectories.
    \item Run-to-run instability dominates: large metric variance across runs makes single-run backtests highly unreliable for this model.
\end{itemize}

\clearpage
\subsection{Model: \textit{claude-sonnet-4.5}}

\begingroup\centering
\includegraphics[width=0.49\linewidth,height=0.32\textheight,keepaspectratio]{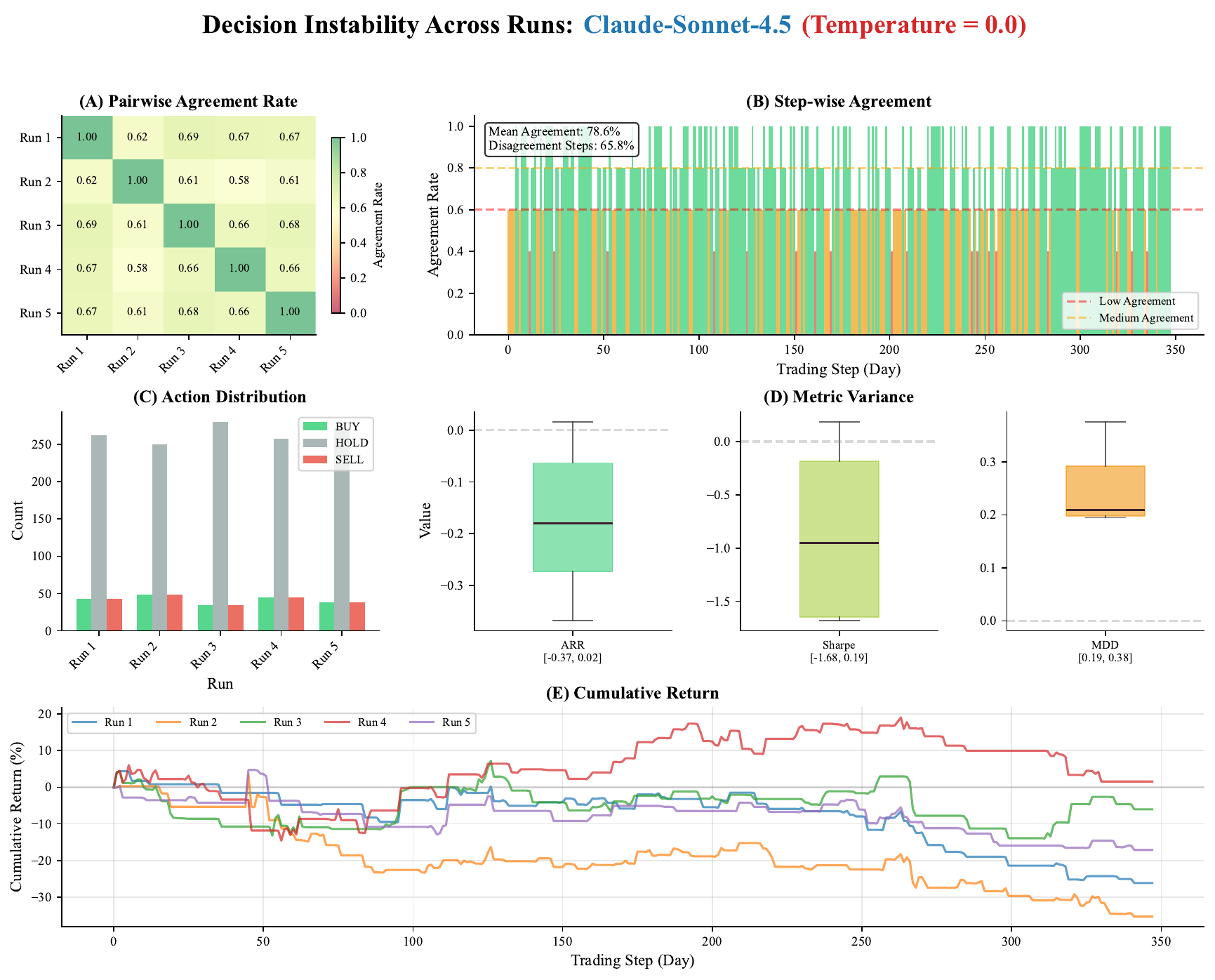}
\hfill
\includegraphics[width=0.49\linewidth,height=0.32\textheight,keepaspectratio]{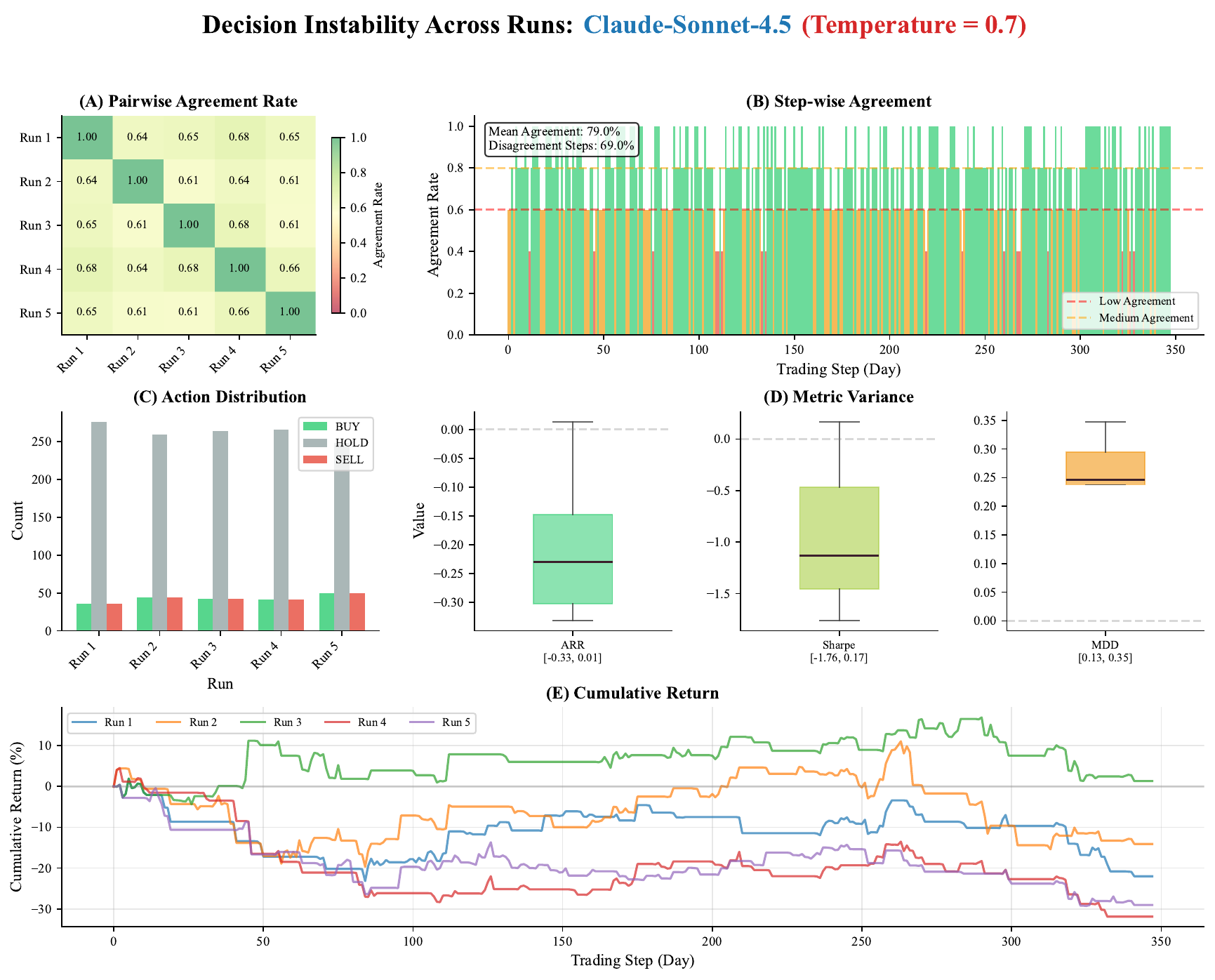}
\vspace{-1.5em}
\captionof{figure}{Cross-run instability of \textit{claude-sonnet-4.5} under different decoding temperatures.}
\label{appx_fig:cross_run_instability_claude_sonnet_4p5}
\par\endgroup

\textbf{Run-to-run Variability.} \Cref{appx_fig:cross_run_instability_claude_sonnet_4p5} summarizes five-run variability from five complementary views. \textbf{(1) Pairwise agreement rate} (Panel A) is moderate at $T=0$ (mean around 79.0\%) and drops slightly at $T=0.7$ (mean around 70.9\%), indicating reasonable but imperfect consistency across runs. \textbf{(2) Step-wise agreement} (Panel B) shows 65.8\% disagreement steps at $T=0$ and 69.0\% at $T=0.7$, reflecting persistent instability throughout the trading horizon. \textbf{(3) Action distribution} (Panel C) is dominated by \textsc{hold} at both temperatures, with relatively consistent \textsc{buy}/\textsc{sell} frequency across runs compared to other models. \textbf{(4) Metric variance} (Panel D) shows moderate dispersion in ARR, Sharpe, and MDD at both temperatures; some runs achieve near-zero or slightly positive returns while others suffer losses up to $-30\%$. \textbf{(5) Cumulative return} trajectories (Panel E) start similarly but gradually diverge, ranging from $+10\%$ to $-30\%$ at both $T=0$ and $T=0.7$, indicating that run-to-run variance dominates temperature effects.

\begin{wrapfigure}{r}{0.52\linewidth}
    \vspace{-1.0em}
    \centering
    \includegraphics[width=\linewidth]{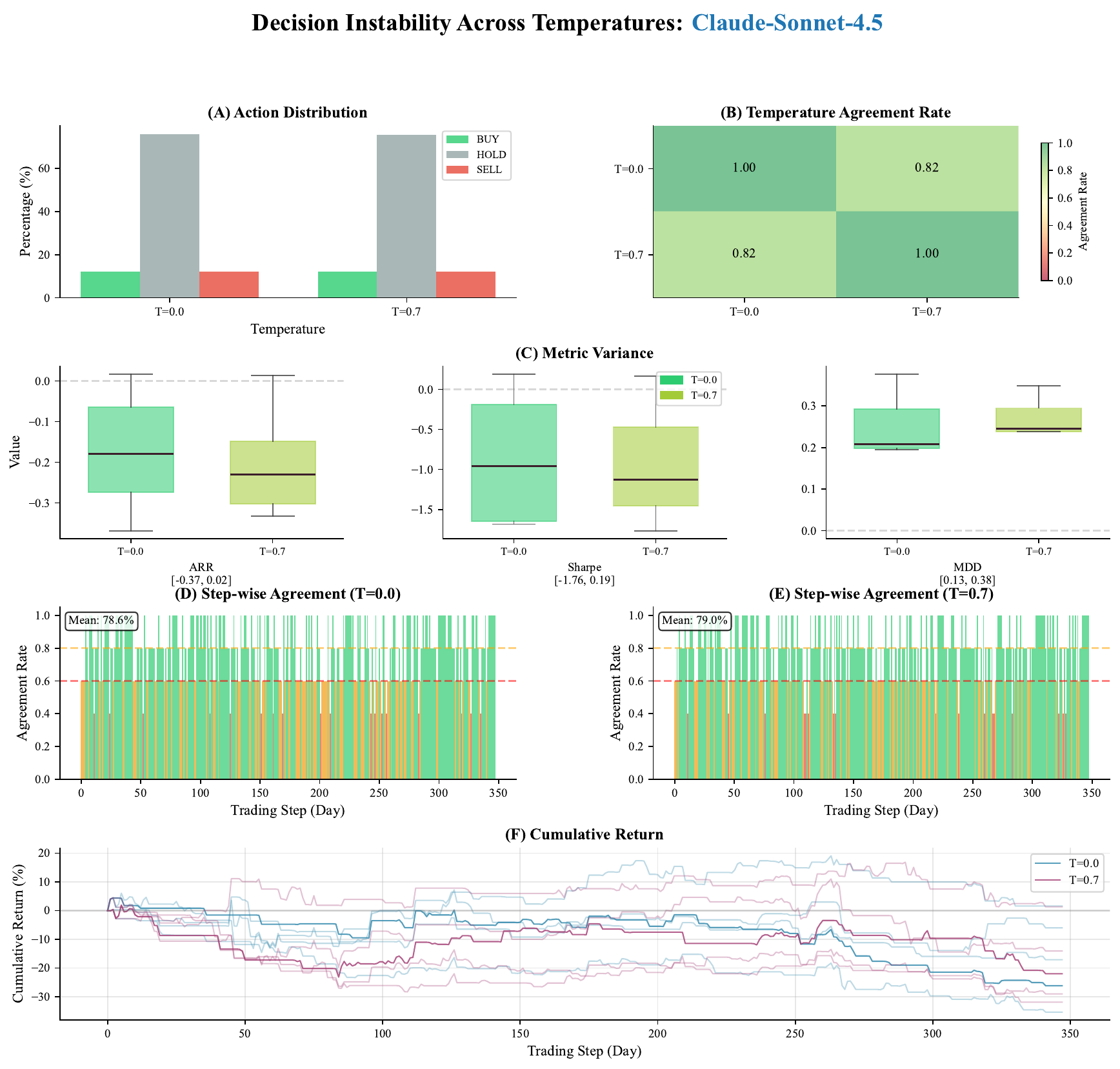}
    \vspace{-1.0em}
    \caption{Decision instability across decoding temperatures (\textcolor{tabblue}{\textit{claude-sonnet-4.5}})}
    \label{appx_fig:cross_temp_instability_claude_sonnet_4p5}
\end{wrapfigure}

\textbf{Decoding Temperature.} \Cref{appx_fig:cross_temp_instability_claude_sonnet_4p5} compares $T=0$ and $T=0.7$ from five perspectives. \textbf{(1) Action distribution} (Panel A) is conservative at both temperatures, with \textsc{hold} dominating around 75\% and balanced \textsc{buy}/\textsc{sell} activity. \textbf{(2) Temperature agreement rate} (Panel B) is 0.82, indicating high consistency across temperature settings, among the highest of all tested models. \textbf{(3) Metric variance} (Panel C) shows negative ARR and Sharpe at both settings with moderate variance; both temperatures exhibit similar performance distributions. \textbf{(4) Step-wise agreement} (Panels D and E) is 78.0\% at $T=0$ and 79.0\% at $T=0.7$, relatively stable across the trading horizon. \textbf{(5) Cumulative return} (Panel F) trajectories overlap substantially between temperatures, both trending downward and ranging from $+10\%$ to $-30\%$, suggesting temperature has limited impact on outcomes.

\textbf{Summary.} For \textit{claude-sonnet-4.5}, we draw three key conclusions:
\vspace{-0.5em}
\begin{itemize}[leftmargin=*]
    \item While the aggregated action distribution is similar across temperatures (agreement rate 0.82), individual runs still produce substantially different action sequences at both $T=0$ and $T=0.7$, with cumulative returns ranging from $+10\%$ to $-30\%$.
    \item The model adopts a conservative \textsc{hold}-dominated strategy with infrequent trading, but this conservatism does not eliminate run-to-run instability; repeated runs under identical settings diverge in financial outcomes.
    \item Changing temperature does not improve stability or profitability; both settings exhibit similar patterns of run-to-run variance and negative ARR/Sharpe ratios.
\end{itemize}

\clearpage
\subsection{Model: \textit{gpt-5.2}}

\begingroup\centering
\includegraphics[width=0.49\linewidth,height=0.32\textheight,keepaspectratio]{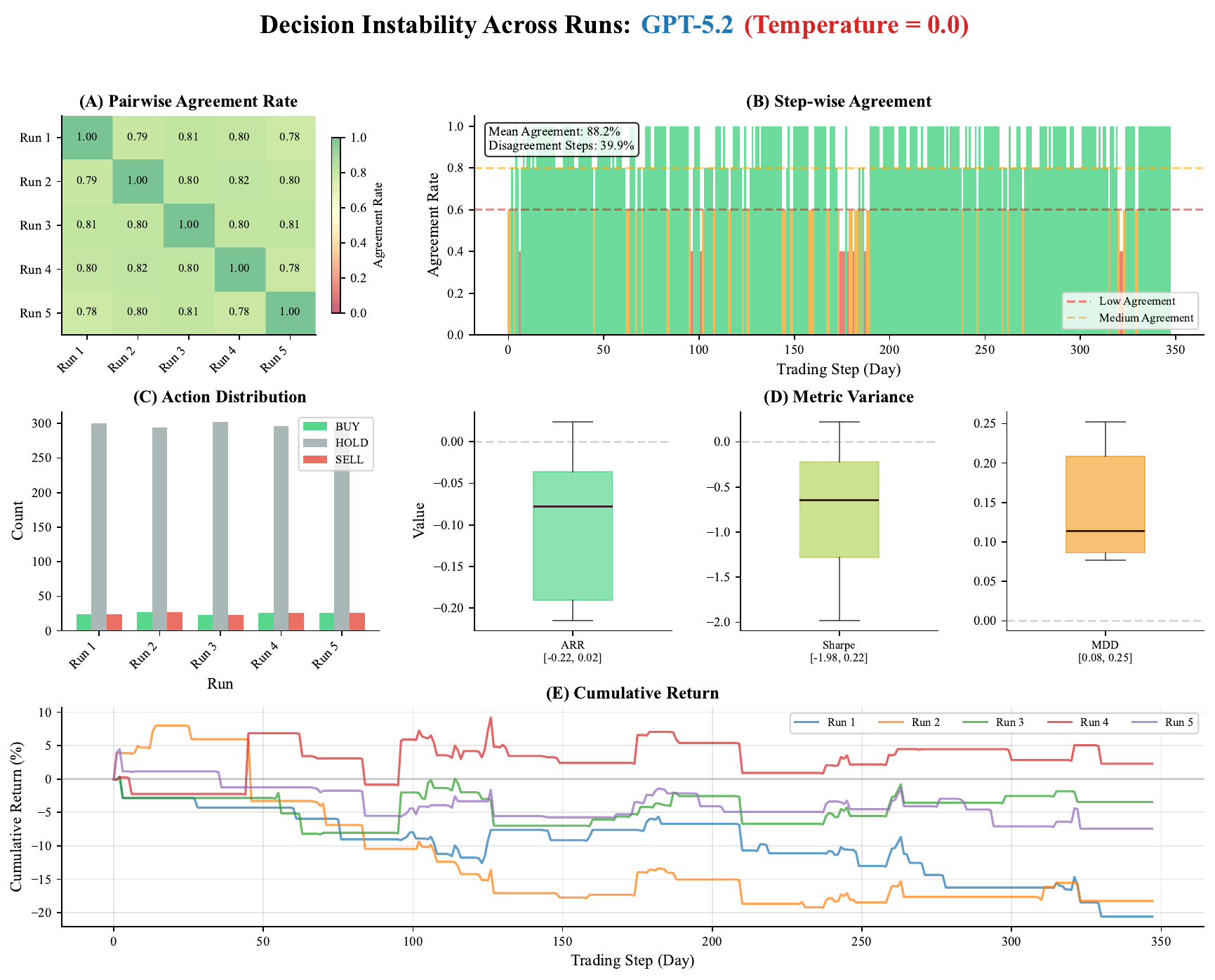}
\hfill
\includegraphics[width=0.49\linewidth,height=0.32\textheight,keepaspectratio]{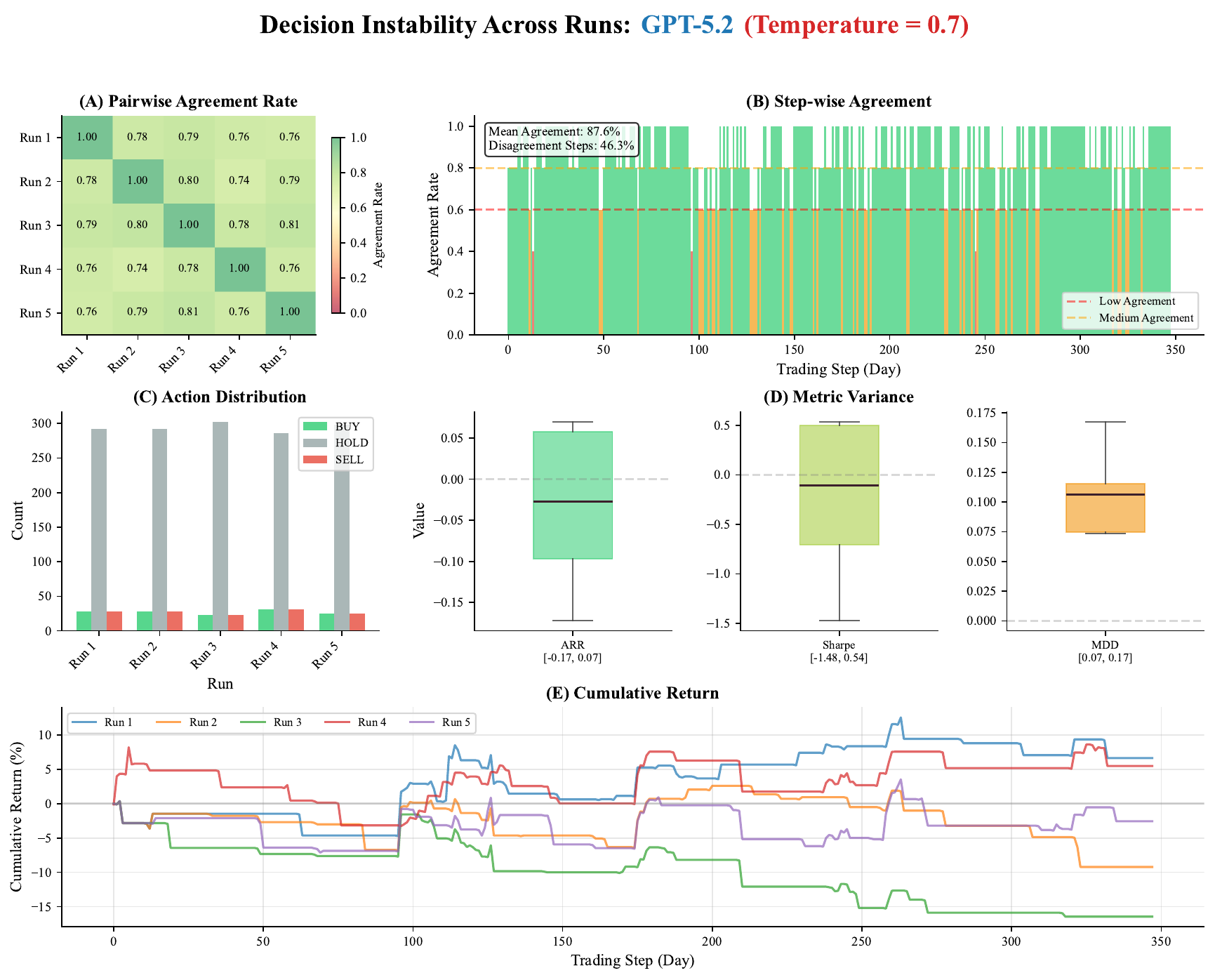}
\vspace{-1.5em}
\captionof{figure}{Cross-run instability of \textit{gpt-5.2} under different decoding temperatures.}
\label{appx_fig:cross_run_instability_gpt_5p2}
\par\endgroup

\textbf{Run-to-run Variability.} \Cref{appx_fig:cross_run_instability_gpt_5p2} summarizes five-run variability from five complementary views. \textbf{(1) Pairwise agreement rate} (Panel A) is high at both temperatures: mean 88.2\% at $T=0$ and 87.6\% at $T=0.7$, indicating strong consistency across runs. \textbf{(2) Step-wise agreement} (Panel B) shows only 39.9\% disagreement steps at $T=0$ and 40.3\% at $T=0.7$, the lowest among all tested models. \textbf{(3) Action distribution} (Panel C) is extremely conservative, with \textsc{hold} dominating over 80\% of decisions and minimal \textsc{buy}/\textsc{sell} activity across all runs. \textbf{(4) Metric variance} (Panel D) shows moderate dispersion; some runs achieve near-zero returns while others suffer modest losses. \textbf{(5) Cumulative return} trajectories (Panel E) remain relatively clustered compared to other models, ranging from $+5\%$ to $-20\%$ at $T=0$ and $+10\%$ to $-15\%$ at $T=0.7$.

\begin{wrapfigure}{r}{0.52\linewidth}
    \vspace{-1.0em}
    \centering
    \includegraphics[width=\linewidth]{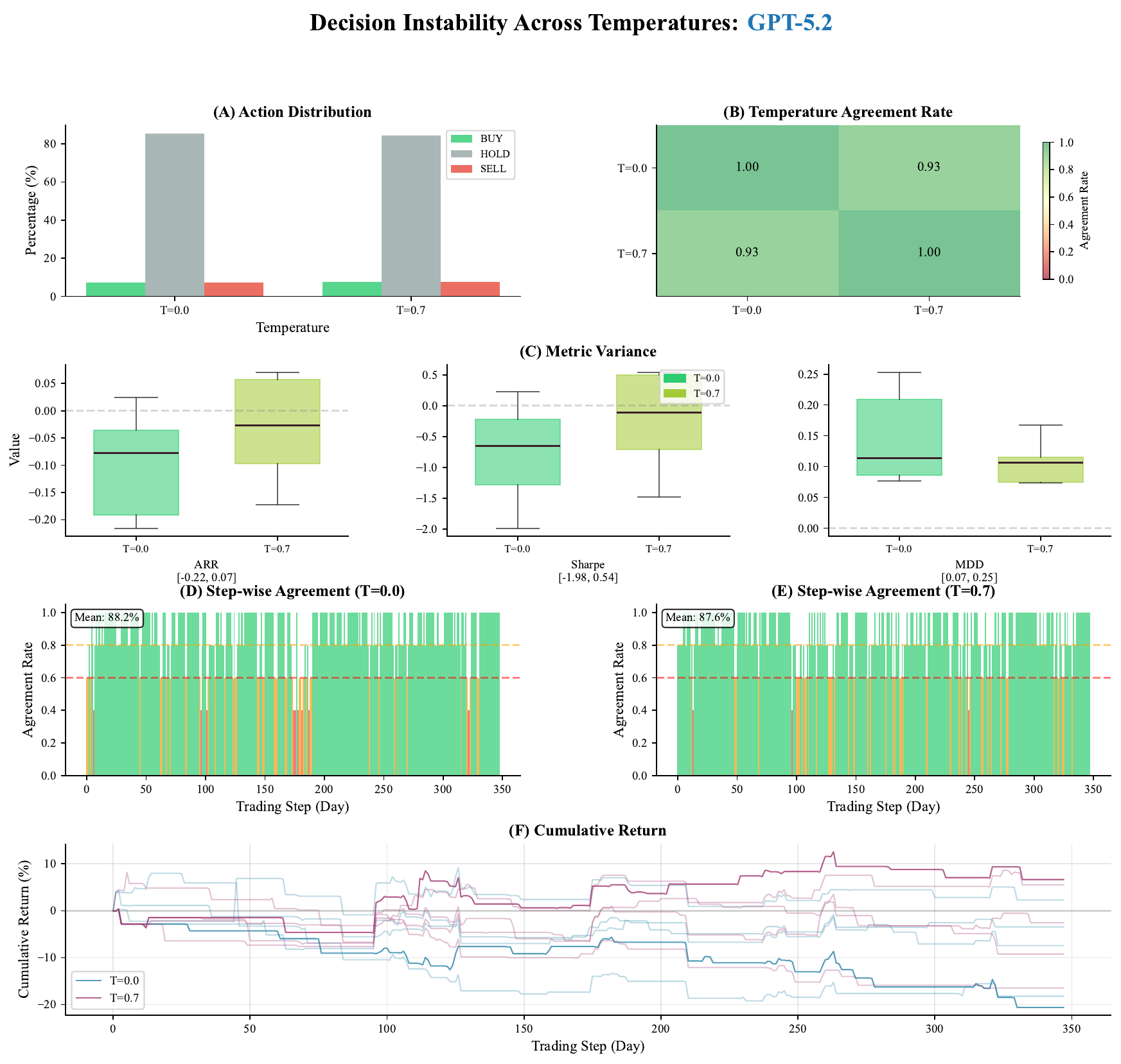}
    \vspace{-1.0em}
    \caption{Decision instability across decoding temperatures (\textcolor{tabblue}{\textit{gpt-5.2}})}
    \label{appx_fig:cross_temp_instability_gpt_5p2}
\end{wrapfigure}

\textbf{Decoding Temperature.} \Cref{appx_fig:cross_temp_instability_gpt_5p2} compares $T=0$ and $T=0.7$ from five perspectives. \textbf{(1) Action distribution} (Panel A) is highly conservative at both temperatures, with \textsc{hold} around 80\% and balanced but infrequent \textsc{buy}/\textsc{sell} actions. \textbf{(2) Temperature agreement rate} (Panel B) is 0.93, the highest among all tested models, indicating near-identical trading decisions regardless of temperature setting. \textbf{(3) Metric variance} (Panel C) shows similar ARR and Sharpe distributions across temperatures, both hovering around break-even to slightly negative. \textbf{(4) Step-wise agreement} (Panels D and E) is 88.2\% at $T=0$ and 87.5\% at $T=0.7$, consistently high throughout the trading horizon. \textbf{(5) Cumulative return} (Panel F) trajectories overlap substantially between temperatures, both ranging from $+10\%$ to $-20\%$, confirming minimal temperature sensitivity.

\textbf{Summary.} For \textit{gpt-5.2}, we draw three key conclusions:
\begin{itemize}[leftmargin=*]
    \item While \textit{gpt-5.2} shows the highest temperature agreement rate (0.93) and relatively higher pairwise agreement than other models, it still exhibits run-to-run variability; individual runs produce different action sequences with cumulative returns ranging from $+10\%$ to $-20\%$.
    \item The model adopts an extremely conservative \textsc{hold}-dominated strategy with over 80\% hold actions, which reduces but does not eliminate run-to-run divergence in financial outcomes.
    \item Changing temperature has minimal impact on aggregated behavior, but neither setting resolves the fundamental instability; performance remains around break-even to slightly negative regardless of configuration.
\end{itemize}

\clearpage

\subsection{Conclusion}

Synthesizing the empirical findings across all six evaluated models, we summarize the key conclusions in \Cref{appx_tab:llm_trading_conclusions}.

\begin{table}[htbp]
\centering
\caption{Summary of key findings on LLM decision instability in trading tasks.}
\label{appx_tab:llm_trading_conclusions}
\small
\begin{tabular}{p{3.2cm}p{10.5cm}}
\toprule
\textbf{Category} & \textbf{Key Findings} \\
\midrule
\multirow{3}{*}{\parbox{3.2cm}{\textbf{Common Characteristics}}} 
& \textbf{Run-to-run instability:} Repeated runs under identical settings produce substantially different action sequences and cumulative returns. \\
& \textbf{Conservative bias without stability:} Most models adopt a \textsc{hold}-dominated strategy, yet this does not ensure consistent performance. \\
& \textbf{Persistent underperformance:} ARR and Sharpe ratios are predominantly negative across all configurations. \\
\midrule
\multirow{5}{*}{\parbox{3.2cm}{\textbf{Temperature Sensitivity}}} 
& \textbf{Gemini Pro/Flash:} Stochastic decoding ($T=0.7$) yields marginally better returns than $T=0$. \\
& \textbf{\textit{deepseek-v3.2}:} Deterministic decoding ($T=0$) outperforms; $T=0.7$ degrades both profitability and consistency. \\
& \textbf{\textit{grok-4.1-fast}:} Insensitive to temperature; run-to-run variance dominates temperature effects. \\
& \textbf{\textit{claude-sonnet-4.5}:} Temperature has limited impact on aggregated behavior, but run-to-run instability persists at both settings. \\
& \textbf{\textit{gpt-5.2}:} Temperature has minimal impact; aggregated actions are similar across temperatures, yet individual runs still vary. \\
\midrule
\multirow{4}{*}{\parbox{3.2cm}{\textbf{Model-Specific}}} 
& \textbf{\textit{gemini-3-pro-preview}:} Most unstable model with lowest pairwise agreement across runs. \\
& \textbf{\textit{grok-4.1-fast}:} More aggressive trading profile, yet no performance improvement. \\
& \textbf{\textit{claude-sonnet-4.5}:} Conservative strategy; moderate run-to-run variance despite similar aggregated behavior across temperatures. \\
& \textbf{\textit{gpt-5.2}:} Relatively higher pairwise agreement than other models, but still exhibits run-to-run divergence in cumulative returns. \\
\midrule
\multirow{4}{*}{\parbox{3.2cm}{\textbf{Instability Dimensions}}} 
& \textbf{Action-level inconsistency:} Same model produces different decisions across runs. \\
& \textbf{Outcome-level divergence:} Small action differences compound into dramatically different returns. \\
& \textbf{Unpredictable temperature sensitivity:} Temperature effects vary across models with no consistent pattern. \\
& \textbf{Persistent underperformance:} All models fail to achieve positive risk-adjusted returns. \\
\bottomrule
\end{tabular}
\end{table}

\textbf{Implications for LLM-based trading.} Our analysis demonstrates that current LLMs are fundamentally unreliable for direct action-emitting trading tasks. While some models (\textit{gpt-5.2} and \textit{claude-sonnet-4.5}) show similar aggregated behavior across temperatures, they still exhibit substantial run-to-run variability in action sequences, leading to divergent financial outcomes under identical settings. These findings underscore the need for evaluation frameworks, such as factor-based benchmarks, that decouple LLM reasoning capability from the inherent noise of direct action generation.

%% file: appendix/appendix_alphaforgebench.tex
\section{Construction of AlphaForgeBench}
\label{appx_sec:construction_alphaforgebench}

The data construction process of \projectname consists of two stages. \textbf{Stage 1: Real-world strategy collection.} We collect natural-language queries and their corresponding ground-truth alpha factors and factor-based trading strategies from diverse real-world sources on the Internet, including brokerage research reports, quantitative investment platforms, AI-in-finance literature, open-source repositories, and traditional finance publications. These real-world samples form the foundation of the benchmark and ensure practical relevance. \textbf{Stage 2: LLM-augmented structured query generation.} Based on the patterns and types observed in the collected real-world strategies, we define three levels of strategy complexity, and within each level we further assign three difficulty grades (easy, medium, and hard). We then use LLMs to systematically generate additional queries at each level-difficulty combination, producing test cases that more precisely target specific aspects of an LLM's strategy generation capability. This two-stage design combines the authenticity of real-world strategies with the controlled granularity of synthetically constructed test cases.

\subsection{Stage 1: Real-world Strategy Collection}

\subsubsection{Data Sources}

In \projectname, we collect a comprehensive set of alpha factors and factor-based trading strategies from five primary sources to ensure diversity and practical relevance.

\begin{itemize}[leftmargin=*, nosep]
    \item \textbf{Brokerage Research Reports}: We analyzed a large collection of quantitative finance research reports from top-tier brokerages (e.g., CITIC Securities, Huatai Securities, and Haitong Securities), which provide both (i) established \emph{alpha factors} and (ii) practical \emph{factor-based trading strategies} for the Chinese stock market or US stock market.
    
    \item \textbf{Quantitative Investment Platforms}: We incorporated both alpha factors and factor-driven strategy formulations from widely used platform libraries, including \textbf{WorldQuant} (e.g., 101 Alphas) and \textbf{JoinQuant}, covering global benchmark alphas and community/industry resources tailored for local markets.
    
    \item \textbf{AI in Finance Literature}: We systematically reviewed recent AI4Finance papers and articles to collect both AI-generated/AI-enhanced factors and factor-based strategy designs (e.g., using deep learning and reinforcement learning).
    
    \item \textbf{Open-Source Repositories}: We leveraged popular open-source projects on GitHub, such as \textbf{OpenFE}, as well as \textbf{Qlib} and its widely used feature set \textbf{Alpha158}, to complement the pool with automated feature engineering patterns and standardized factor/feature libraries, which can be used to construct new factors and further build factor-based trading strategies.
    
    \item \textbf{Traditional Finance Literature}: We queried academic search engines and working-paper repositories of finance research (Google Scholar, SSRN, arXiv and NBER) using curated keyword sets across multi-asset classes. We then performed citation snowballing from cornerstone papers and surveys to expand coverage, prioritizing widely cited and methodologically explicit studies.
\end{itemize}

\subsubsection{Data Extraction Pipeline}

The raw data sources described above are predominantly in PDF format (research reports, academic papers, platform documentation). To systematically extract structured factor definitions and strategy logic from these unstructured documents, we designed an automated extraction pipeline consisting of two steps: document collection followed by LLM-based information extraction.

\textbf{Document collection.}
We used web crawlers to collect PDF documents from the five source categories. For brokerage research reports, we crawled publicly available quantitative research sections from major Chinese and international brokerages. For academic literature, we queried Google Scholar, SSRN, arXiv, and NBER with curated keyword sets (e.g., ``alpha factor,'' ``momentum strategy,'' ``mean reversion,'' ``factor investing'') and downloaded the resulting papers. For quantitative platforms, we scraped strategy descriptions and factor documentation from JoinQuant and WorldQuant community pages.

\textbf{LLM-based extraction.}
Given the multimodal capabilities of recent frontier LLMs, we leverage \textit{gemini-3-flash-preview} to parse each collected PDF document and extract structured information. Specifically, the model receives the full PDF as a multimodal input and is instructed to identify and extract: (i) factor names and their mathematical definitions, (ii) strategy logic and trading rules, (iii) the underlying financial rationale, and (iv) whether the strategy involves deep learning or alternative data sources. The extraction prompt enforces a standardized JSON schema for each identified factor or strategy, ensuring consistent formatting across heterogeneous source documents. The complete prompt template is provided below.

\begin{tcolorbox}[colback=gray!5, colframe=gray!50, title=PDF Extraction Prompt, fonttitle=\bfseries\small, breakable]
\small
\textbf{System:} You are a quantitative strategy extraction expert. Please extract the \textbf{core strategies originally proposed} in this research report PDF.

\medskip
\textbf{Extraction Rules:}
\begin{enumerate}[leftmargin=*, nosep]
    \item Only extract strategies \textbf{originally proposed} in this paper. Do \textbf{not} extract:
    \begin{itemize}[nosep]
        \item Strategies merely mentioned in the abstract
        \item Strategies cited from other literature
        \item Existing strategies used as comparison baselines
    \end{itemize}
    \item A single report may contain multiple core strategies; extract all of them.
    \item If the report does not propose any concrete executable strategy, return an empty list.
\end{enumerate}

\medskip
\textbf{For each strategy, extract:}
\begin{enumerate}[leftmargin=*, nosep]
    \item \textbf{Name}: Strategy name (Chinese and English)
    \item \textbf{Factor}: Core factor name(s) used by the strategy
    \item \textbf{Definition}: Mathematical definition of the factor(s) in LaTeX
    \item \textbf{Strategy Logic}: Complete trading rules, including conditions for long, short, and neutral positions
    \item \textbf{Design Rationale (CoT)}: Detailed explanation of why the strategy works:
    \begin{itemize}[nosep]
        \item What market phenomenon does it capture? (e.g., momentum, mean reversion, microstructure)
        \item Why can this factor/signal predict future returns? What is the financial intuition?
        \item What are the advantages over traditional methods?
        \item Under what market conditions does it perform better or worse?
    \end{itemize}
    \item \textbf{Strategy Type}: \texttt{rule\_based} (directly implementable with technical indicators) or \texttt{model\_based} (requires a pre-trained ML/DL model)
    \item \textbf{Strategy Code}: Python code implementing the strategy
\end{enumerate}

\medskip
\textbf{Code Specification:}
\begin{itemize}[leftmargin=*, nosep]
    \item Function name must be \texttt{generate\_signal}
    \item Input: \texttt{df: pd.DataFrame} with columns \texttt{open, high, low, close, volume}
    \item Output: \texttt{pd.Series} with values $\{1, -1, 0\}$ (long / short / neutral)
    \item For \texttt{rule\_based}: compute signals directly from \texttt{df} columns
    \item For \texttt{model\_based}: assume a pre-computed \texttt{model\_score} column exists in \texttt{df}
\end{itemize}

\medskip
\textbf{Output Schema (JSON):}
\begin{lstlisting}[basicstyle=\ttfamily\scriptsize, breaklines=true]
{
  "title": "Report title",
  "strategies": [
    {
      "name": "Strategy name",
      "factor": "Core factor name",
      "definition": "LaTeX formula",
      "logic": "Trading rules",
      "reason": "Design rationale (CoT)",
      "strategy_type": "rule_based | model_based",
      "code": "Python code"
    }
  ]
}
\end{lstlisting}
\label{appx_prompt:extraction_prompt}
\end{tcolorbox}

\subsubsection{Dataset Statistics and Scope}

After extraction and deduplication, the full \projectname dataset comprises \textbf{3,176} factor-strategy entries spanning three strategy types, as summarized in \Cref{appx_tab:dataset_stats}. We describe each type below.

\textbf{Single-asset Trading} (633 entries) strategies operate on individual assets in isolation. Each strategy takes as input the historical price and indicator data of a single asset (e.g., BTCUSDT, ETHUSDT, AAPL, GOOGL, MSFT, NVDA, TSLA) and produces trading signals (buy/sell/hold) or alpha factors for that asset alone. These strategies are self-contained and asset-agnostic: the same logic can be independently applied to any individual asset without requiring cross-asset coordination. Examples include momentum-based timing strategies, mean-reversion signals, and technical indicator combinations.

\textbf{Portfolio Management} (2,172 entries) strategies involve cross-sectional analysis and portfolio-level allocation across multiple assets simultaneously. Rather than generating signals for a single asset, these strategies rank or score a universe of assets and construct a weighted portfolio. Typical examples include factor-based long-short portfolios (e.g., buying the top decile and shorting the bottom decile by a value or momentum factor), risk-parity allocation, and sector rotation strategies. These strategies require the model to reason about relative asset characteristics and portfolio-level constraints such as diversification and risk budgets.

\textbf{Multi-asset Trading} (371 entries) strategies involve coordinated trading across different asset classes or instruments. Unlike portfolio management strategies that rank within a single universe, multi-asset strategies exploit relationships between heterogeneous assets, such as cross-market arbitrage (e.g., BTC spot vs.\ futures), pair trading between correlated assets, or macro-driven allocation across equities, bonds, and commodities. These strategies demand understanding of inter-market dynamics and cross-asset dependencies.

In this paper, we focus the benchmark evaluation on the \textbf{single-asset trading} subset. Single-asset trading strategies are ideal for controlled evaluation, as they isolate the LLM's ability to generate correct and profitable trading logic from the confounding effects of portfolio construction, asset allocation, and cross-asset coordination. The portfolio management and multi-asset trading subsets are reserved for future work.

\begin{table}[h]
\centering
\caption{\projectname dataset statistics by strategy type.}
\label{appx_tab:dataset_stats}
\begin{tabular}{lrc}
\toprule
\textbf{Strategy Type} & \textbf{Count} & \textbf{Description} \\
\midrule
Single-asset Trading & 633 & Strategies operating on individual assets \\
Portfolio Management & 2,172 & Portfolio-level allocation and cross-sectional ranking strategies \\
Multi-asset Trading & 371 & Cross-asset and inter-market trading strategies \\
\midrule
\textbf{Total} & \textbf{3,176} & \\
\bottomrule
\end{tabular}
\end{table}

\subsubsection{Strategy Query Generation}
\label{appx_sec:strategy_query_generation}

The structured strategy entries extracted in Stage~1 contain rich metadata (strategy name, logic, factors, rationale), but they cannot be directly used as real-world benchmark queries because they may include implementation details, unsupported data sources (e.g., deep learning models, order-book data), or non-English descriptions. To produce clean, implementation-agnostic English queries suitable for evaluating LLM code generation, we employ \textit{gpt-5.2} to transform each extracted strategy entry into a standardized real-world benchmark query.

The generation process takes as input the structured strategy fields (name, logic, reason, factors) and produces a requirements-only English query that preserves the core trading intent while ensuring implementability within the backtest framework's constraints (OHLCV data and precomputed technical indicators only).

\begin{tcolorbox}[colback=gray!5, colframe=gray!50, title=Query Generation System Prompt, fonttitle=\bfseries\small, breakable]
\small
\textbf{System:} You are a benchmark query generator for single-asset trading strategies. Given a SOURCE strategy entry (name/logic/reason/factors), generate ONE English, requirements-only query that a code-generation model can implement.

\medskip
\textbf{Hard Constraints:}
\begin{itemize}[leftmargin=*, nosep]
    \item The query MUST be implementable using ONLY \texttt{df['open','high','low','close','volume']} and the allowed precomputed factor columns.
    \item The strategy MUST be stateless (no position tracking or ``if already in position'' logic).
    \item Long-only semantics are assumed by the system: do NOT mention ``long-only'' or ``no short selling.''
    \item Do NOT mention code, libraries, backtest engines, or implementation details.
    \item If the source strategy relies on deep learning, alternative data, options IV, order-book/OFI, or other unsupported signals, approximate the intent using technical indicators (EMA/SMA/RSI/MACD/BB/ATR/STD/VOL/OBV/etc.).
\end{itemize}

\medskip
\textbf{Output Schema (JSON):}
\begin{lstlisting}[basicstyle=\ttfamily\scriptsize, breaklines=true]
{
  "query": "<3-8 sentence requirement description>",
  "summary": "<10-20 word one-sentence summary>",
  "approximation_notes": "<how unsupported parts were approximated>",
  "used_factors": ["ema_20", "rsi_14", ...]
}
\end{lstlisting}
\end{tcolorbox}

\begin{tcolorbox}[colback=gray!5, colframe=gray!50, title=Query Generation User Prompt Template, fonttitle=\bfseries\small, breakable]
\small
\textbf{SOURCE STRATEGY (for generation only):}
\begin{itemize}[leftmargin=*, nosep]
    \item \texttt{source\_id}: Strategy identifier
    \item \texttt{strategy\_name}: Name of the strategy
    \item \texttt{is\_deep\_learning}: Whether the strategy involves deep learning
    \item \texttt{strategy\_logic}: Complete trading logic description
    \item \texttt{strategy\_reason}: Rationale behind the strategy
    \item \texttt{Source factor hints}: Factor names and definitions (may be unsupported; used as semantic guidance)
\end{itemize}

\medskip
\textbf{TASK:} Generate a benchmark query that is implementable with the allowed DataFrame columns. Keep the core intent (trend-following / mean-reversion / volatility control / breakout / risk management), but translate unsupported signals into technical-indicator proxies.

\medskip
\textbf{OUTPUT:} Return ONLY a single JSON object with keys: \texttt{query}, \texttt{summary}, \texttt{approximation\_notes}, \texttt{used\_factors}.
\end{tcolorbox}

\subsection{Stage 2: LLM-augmented Structured Query Generation}

\subsubsection{Motivation and Design Overview}

While the real-world strategies collected in Stage 1 provide authentic and diverse test cases, they are not systematically organized by difficulty and may not uniformly cover the full spectrum of capabilities required for strategy generation. To address this, Stage 2 introduces a structured query generation framework that complements the real-world samples with synthetically constructed queries designed to probe specific aspects of an LLM's strategy generation ability in a controlled and fine-grained manner.

\subsubsection{Levels and Dimensions of Difficulty}

The framework evaluates three orthogonal dimensions of difficulty: (1)~\textit{Granularity of Strategy Logic}, measuring how much trading logic is left underspecified in the query; (2)~\textit{Semantic--Symbolic Alignment to the Factor Library}, measuring whether the model can map natural-language financial concepts to concrete factor APIs; and (3)~\textit{Complexity of Logical Structure}, characterizing the algorithmic complexity of the target strategy. These three dimensions are jointly organized into three progressive levels (Level~1 through Level~3), where each level defines a characteristic setting along every dimension. Within each level, we further assign three difficulty grades (\textit{Easy}, \textit{Medium}, and \textit{Hard}) that modulate the task complexity through factors such as the number of conditions, the degree of underspecification, and the depth of logical control flow and the degree of autonomous design required. This $3 \times 3$ level--grade taxonomy yields nine distinct difficulty cells, enabling fine-grained and systematic evaluation of an LLM's strategy generation capability as described in \Cref{appx_tab:difficulty_taxonomy}.

\textbf{Level 1: Logic Translation.}
At this level, queries provide fully specified IF--THEN rules with explicit numerical parameters, and the model is evaluated primarily on faithful code translation rather than financial inference.
Along the \emph{Granularity} axis (\textit{Logic Translation}), the query leaves no implementation decisions to the model.
Along the \emph{Alignment} axis (\textit{Explicit Mapping}), indicator names mentioned in the query map directly to factor-library variables (e.g., ``RSI'' $\to$ \texttt{rsi\_14}), so the task reduces to precise retrieval and correct API invocation.
Along the \emph{Complexity} axis (\textit{Sequential Thresholding}), the target strategy is composed of pointwise boolean decisions over threshold comparisons.
Within this level, three difficulty grades are distinguished by the number of conditions, the depth of logical composition, and the coordination effort across factors:

\begin{itemize}[leftmargin=*, nosep]
    \item \textit{Easy.}
    The query specifies a single-indicator threshold rule with one explicit entry condition (e.g., ``buy when RSI drops below~30'').
    The indicator-to-factor mapping is one-to-one and unambiguous, and the implementation requires only a single conditional statement.

    \item \textit{Medium.}
    The query specifies a conjunction of two to three indicator-based conditions with AND/OR connectives, each with explicitly stated thresholds (e.g., ``buy when RSI~$<$~30 and the closing price is above the 20-day EMA'').
    Queries at this grade may also include explicit crossover semantics (e.g., golden cross, death cross) and fixed take-profit/stop-loss thresholds.
    Although every condition maps straightforwardly to a factor variable, the model must correctly compose multiple boolean predicates and handle their logical conjunction.
    All strategies are stateless: entry and exit conditions are evaluated independently on each bar based on current data only.

    \item \textit{Hard.}
    The query specifies a multi-factor strategy with signal-strength-based position sizing and explicit priority rules for resolving conflicting signals (e.g., ``allocate full position if RSI~$<$~20, half position if RSI~$<$~30; combine short-term, medium-term, and long-term indicators; if the risk signal fires, override the entry signal'').
    The strategy involves four or more nested conditions spanning multiple time-window factors, yet all logic remains fully explicit and stateless---position size is determined by current signal strength on each bar, not by tracking previous positions.
\end{itemize}

\textbf{Level 2: Logic Completion.}
At this level, queries provide a strategic skeleton but deliberately leave critical implementation details unspecified, requiring the model to supply plausible defaults grounded in domain knowledge.
Along the \emph{Granularity} axis (\textit{Logic Completion}), key parameters such as thresholds, lookback windows, or the operational definition of qualitative terms (e.g., ``significant deviation'') are omitted and must be inferred by the model.
Along the \emph{Alignment} axis (\textit{Conceptual Mapping}), the query employs financial jargon or qualitative descriptors (e.g., ``oversold''), and the model must identify the underlying quantitative proxy and operationalize it via appropriate factors and thresholds.
Along the \emph{Complexity} axis (\textit{Multi-Factor Composition}), the strategy combines multiple indicators into a coherent trading rule, and the model must infer how to integrate them---including conditional filters, confirmation logic, and risk controls---from incomplete specifications.
Within this level, three difficulty grades are distinguished by the extent of underspecification, the abstractness of semantic alignment, and the sophistication of multi-factor composition:

\begin{itemize}[leftmargin=*, nosep]
    \item \textit{Easy.}
    The query omits a single implementation parameter (e.g., a specific threshold or lookback period) while leaving the remainder of the logic explicit.
    The jargon used is standard and widely recognized (e.g., ``overbought''), mapping in one hop to a well-known indicator with a conventional default (e.g., ``oversold'' $\to$ RSI~$<$~30).

    \item \textit{Medium.}
    The query omits multiple parameters simultaneously (e.g., both the lookback window and the deviation threshold for a mean-reversion strategy), requiring the model to jointly infer coherent defaults.
    The jargon involves domain-specific compound concepts (e.g., ``volume-confirmed breakout with trend confirmation and chop filter''), and the model must decompose the term into constituent factors and determine their interaction.

    \item \textit{Hard.}
    The query provides only a high-level strategic skeleton with most quantitative details left unspecified (e.g., ``implement a mean-reversion strategy with appropriate risk controls''), demanding that the model design a complete parameterization from domain priors.
    The alignment requires interpreting abstract, multi-interpretation jargon (e.g., ``liquidity drying up,'' ``panic selling,'' ``risk appetite shift'') whose operationalization depends on market regime assumptions.
    Risk control logic is also left unspecified and must be inferred by the model.
\end{itemize}

\textbf{Level 3: Goal-Oriented Generation.}
At this level, the query states only a high-level investment objective or describes an abstract pattern, and the model must design an end-to-end strategy architecture from first principles.
Along the \emph{Granularity} axis (\textit{Goal-Oriented Generation}), no concrete trading rules are provided; the model must autonomously formulate the strategy logic, select appropriate indicators, and determine all parameters.
Along the \emph{Alignment} axis (\textit{Intent/Cross-Modal Mapping}), the query expresses abstract intent or visually described patterns (e.g., a hammer candlestick), requiring the model to synthesize multi-factor compositions from primitives such as \texttt{open}, \texttt{high}, \texttt{low}, and \texttt{close} to represent higher-order structures.
Along the \emph{Complexity} axis (\textit{Constraint-Driven Design}), the strategy must satisfy multiple simultaneous constraints and resolve potential conflicts among competing objectives, all based on current-bar data only---no position tracking, regime history, or state machines are permitted.
Within this level, three difficulty grades are distinguished by the ambiguity of the objective, the complexity of cross-modal reasoning, and the depth of constraint-driven design:

\begin{itemize}[leftmargin=*, nosep]
    \item \textit{Easy.}
    The query states a single, well-defined objective with a clear stylistic category (e.g., ``design a trend-following strategy for equity indices'') and a small number of constraints (e.g., a maximum drawdown limit).
    The abstract concept to be operationalized corresponds to a single canonical pattern (e.g., ``golden cross''), and the model must compose two related indicators (e.g., short-term and long-term moving averages).

    \item \textit{Medium.}
    The query specifies a multi-faceted objective with competing sub-goals (e.g., ``capture momentum while limiting drawdown in volatile regimes''), requiring the model to balance return-seeking and risk-controlling components.
    The alignment involves translating a visually or qualitatively described pattern (e.g., ``cup-and-handle formation'') into a conjunction of geometric and volumetric conditions across multiple OHLCV-derived factors.
    The strategy must incorporate current-bar regime detection (e.g., using ATR or standard deviation to classify the current bar as high- or low-volatility) and adapt its behavior accordingly, with all decisions based solely on current data.

    \item \textit{Hard.}
    The query provides only a vague or open-ended investment mandate (e.g., ``generate consistent risk-adjusted returns in sideways markets with a maximum drawdown constraint''), and the model must autonomously select the strategy archetype, define the signal logic, and calibrate all parameters.
    The alignment demands cross-modal synthesis of abstract financial intuitions (e.g., ``a volatility compression preceding a breakout'') into multi-factor composite signals that have no single canonical representation.
    The strategy must resolve conflicting objectives (e.g., maximize returns vs.\ minimize drawdown vs.\ reduce turnover) through a complex priority-based rule system, with multi-layer risk controls that can override entry signals and position sizing that adapts to multiple current-bar factors simultaneously; all decisions must be based on current bar data only, using conditional priority rules rather than state tracking. Performance at this grade distinguishes genuine quantitative reasoning and strategy design capability from mere NL-to-code competence.
\end{itemize}

\begin{table}[ht]
\centering
\caption{Summary of the $3\times 3$ difficulty taxonomy for LLM-augmented query generation. Each cell characterizes the task along three dimensions: \textit{Granularity} (how much logic is specified), \textit{Alignment} (how indicator names relate to factor APIs), and \textit{Complexity} (algorithmic structure of the target strategy).}
\label{appx_tab:difficulty_taxonomy}
\setlength{\tabcolsep}{4pt}
\resizebox{\textwidth}{!}{%
\begin{tabular}{c c l l l}
\toprule
\textbf{Level} & \textbf{Grade} & \textbf{Granularity} & \textbf{Alignment} & \textbf{Complexity} \\
\midrule
\multirow{3}{*}{\shortstack{Level 1\\[2pt]\textit{Logic Transl.}}}
& Easy   & Single rule; all params explicit & One-to-one name match          & Single threshold \\
& Medium & 2--3 conjunctive conditions      & Direct match w/ disambiguation & Multi-condition conjunction \\
& Hard   & All explicit; 4+ nested conditions  & Multiple mixed-connective look-ups & Position sizing + priority override \\
\midrule
\multirow{3}{*}{\shortstack{Level 2\\[2pt]\textit{Logic Compl.}}}
& Easy   & One parameter omitted            & Standard jargon, known default & Simple single-factor rule \\
& Medium & Multiple params omitted jointly  & Compound concept decomposition & Multi-factor combined inference \\
& Hard   & Only skeleton; most details open & Ambiguous descriptors          & Full parameterization + risk logic \\
\midrule
\multirow{3}{*}{\shortstack{Level 3\\[2pt]\textit{Goal-Orient.}}}
& Easy   & Single clear objective           & Single canonical pattern       & Basic end-to-end design \\
& Medium & Competing multi-faceted goals    & Visual pattern $\to$ multi-factor & Current-bar regime detection \\
& Hard   & Vague open-ended mandate         & Cross-modal, no canonical form & Priority-based conflict resolution \\
\bottomrule
\end{tabular}%
}
\end{table}

\subsubsection{Benchmark Query Generation}
\label{appx_sec:benchmark_query_generation}

To systematically populate the $3\times 3$ difficulty taxonomy, we use \textit{gpt-5.2} to generate strategy queries for each of the nine level--grade cells. For every cell, the model produces three queries per batch---one for each trading style (Conservative, Aggressive, and Balanced)---and the process iterates with deduplication until the target count per cell is reached (90 queries per cell, 810 total before filtering to 30 per cell for the final benchmark). The generation prompt is assembled from six modular components, presented below: (1)~a \emph{base system prompt} that defines the generator's role, constraints, and the complete factor library; (2)~a \emph{level overview} summarizing all three difficulty levels; (3)~\emph{category definitions} for all nine cells with the current cell highlighted; (4)~a \emph{current task emphasis} block reinforcing the active cell's constraints; (5)~\emph{trading style definitions}; and (6)~\emph{output format requirements} specifying the JSON schema. When prior queries have already been generated for the same cell, a \emph{deduplication instruction} listing recent strategy summaries is appended to encourage diversity.

\begin{tcolorbox}[colback=gray!5, colframe=gray!50, title=Base System Prompt, fonttitle=\bfseries\small, breakable]
\small
You are an experienced quantitative trading expert generating strategy requirement queries for a benchmark. Write queries in a professional tone that guides models to produce high-quality trading strategies.

\medskip
INTERNAL NOTE (do NOT mention in queries): All strategies are long-only by default. Do not include phrases like ``no short selling'' or ``long-only'' in the generated queries---this constraint is already handled by the system.

\medskip
Each query is requirements-only: describe the trading logic and constraints, but do not ask to write code, do not include implementation details, and do not reference any libraries, APIs, or backtesting engines.

\medskip
\textbf{Available Data \& Allowed Factors (strict)}

\medskip
Assume you trade one single stock using a pandas-like DataFrame \texttt{df} with precomputed columns only. You may reference only the following variables:

\begin{itemize}[leftmargin=*, nosep]
\item Price/volume series: \texttt{df['open']}, \texttt{df['high']}, \texttt{df['low']}, \texttt{df['close']}, \texttt{df['volume']} (assume they exist).
\item Precomputed factors (use exact column names, \texttt{\{n\}} is any positive integer period):
\end{itemize}

\medskip
\textbf{Price-scale factors} (same magnitude as price, compare with price):
\begin{itemize}[leftmargin=*, nosep]
\item SMA: \texttt{df['sma\_\{n\}']} = SMA(close, n) -- simple moving average
\item EMA: \texttt{df['ema\_\{n\}']} = EMA(close, n) -- exponential moving average
\item Bollinger Bands: \texttt{df['bb\_upper\_\{n\}']}, \texttt{df['bb\_middle\_\{n\}']}, \texttt{df['bb\_lower\_\{n\}']}
\item ATR: \texttt{df['atr\_\{n\}']} -- average true range
\end{itemize}

\medskip
\textbf{Fixed-range 0--100:}
\begin{itemize}[leftmargin=*, nosep]
\item RSI: \texttt{df['rsi\_\{n\}']} = 100 - 100/(1 + avg\_gain/avg\_loss) -- relative strength index
\item MFI: \texttt{df['mfi\_\{n\}']} -- money flow index
\item Stochastic: \texttt{df['stoch\_k\_\{n\}']}, \texttt{df['stoch\_d\_\{n\}']} -- KDJ indicator (NO j line)
\end{itemize}

\medskip
\textbf{Normalized 0--1:}
\begin{itemize}[leftmargin=*, nosep]
\item RSV: \texttt{df['rsv\_\{n\}']} = (close - ts\_min(low,n)) / (ts\_max(high,n) - ts\_min(low,n))
\item Count ratios: \texttt{df['cntp\_\{n\}']} = count(ret>0,n)/n, \texttt{df['cntn\_\{n\}']} = count(ret<0,n)/n
\item Sum ratios: \texttt{df['sump\_\{n\}']} = ts\_sum(pos\_ret,n)/ts\_sum(abs\_ret,n), \texttt{df['sumn\_\{n\}']} = 1 - sump
\item Volume sum: \texttt{df['vsump\_\{n\}']} = ts\_sum(pos\_vol\_chg,n)/ts\_sum(abs\_vol\_chg,n), \texttt{df['vsumn\_\{n\}']} = 1 - vsump
\end{itemize}

\medskip
\textbf{Normalized 0 to (n-1)/n} (position in window, NEVER reaches 1.0):
\begin{itemize}[leftmargin=*, nosep]
\item Position: \texttt{df['imax\_\{n\}']} = argmax(high,n)/n, \texttt{df['imin\_\{n\}']} = argmin(low,n)/n (e.g., imax\_20 ranges 0 to 0.95)
\item Rank: \texttt{df['rank\_\{n\}']} = ts\_rank(close,n)/n (e.g., rank\_20 ranges 0 to 0.95)
\end{itemize}

\medskip
\textbf{Normalized -1 to 1:}
\begin{itemize}[leftmargin=*, nosep]
\item Count diff: \texttt{df['cntd\_\{n\}']} = cntp - cntn
\item Sum diff: \texttt{df['sumd\_\{n\}']} = 2*sump - 1
\item Volume diff: \texttt{df['vsumd\_\{n\}']} = 2*vsump - 1
\item Correlation: \texttt{df['corr\_\{n\}']} = ts\_corr(close, log(volume), n), \texttt{df['cord\_\{n\}']} = ts\_corr(delta(close), delta(volume), n)
\end{itemize}

\medskip
\textbf{Normalized -(n-1)/n to (n-1)/n:}
\begin{itemize}[leftmargin=*, nosep]
\item Position diff: \texttt{df['imxd\_\{n\}']} = (argmax(high,n) - argmin(low,n))/n (e.g., imxd\_20 ranges -0.95 to 0.95)
\end{itemize}

\medskip
\textbf{Ratio around 1.0} (compare with 1.0):
\begin{itemize}[leftmargin=*, nosep]
\item MA ratio: \texttt{df['ma\_\{n\}']} = ts\_mean(close,n)/close
\item ROC: \texttt{df['roc\_\{n\}']} = close.shift(n)/close
\item VMA: \texttt{df['vma\_\{n\}']} = ts\_mean(volume,n)/volume ($\geq$0)
\end{itemize}

\medskip
\textbf{Ratio $\geq$1.0} (always >= 1):
\begin{itemize}[leftmargin=*, nosep]
\item Max ratio: \texttt{df['max\_\{n\}']} = ts\_max(close,n)/close (window max >= current close)
\end{itemize}

\medskip
\textbf{Ratio 0--1} (always <= 1):
\begin{itemize}[leftmargin=*, nosep]
\item Min ratio: \texttt{df['min\_\{n\}']} = ts\_min(close,n)/close (window min <= current close)
\end{itemize}

\medskip
\textbf{Ratio $\geq$0} (always non-negative):
\begin{itemize}[leftmargin=*, nosep]
\item Std: \texttt{df['std\_\{n\}']} = ts\_std\_dev(close,n)/close
\item Volume std: \texttt{df['vstd\_\{n\}']} = ts\_std\_dev(volume,n)/volume
\item WVMA: \texttt{df['wvma\_\{n\}']} = ts\_std\_dev(abs(ret)*vol,n)/ts\_mean(abs(ret)*vol,n)
\end{itemize}

\medskip
\textbf{Unbounded} (can be positive or negative):
\begin{itemize}[leftmargin=*, nosep]
\item Quantile: \texttt{df['qtlu\_\{n\}']} = (close - quantile\_80(close,n))/close, \texttt{df['qtld\_\{n\}']} = (close - quantile\_20(close,n))/close
\item Beta: \texttt{df['beta\_\{n\}']} = (close.shift(n) - close)/(n * close)
\end{itemize}

\medskip
\textbf{Candlestick shape:}
\begin{itemize}[leftmargin=*, nosep]
\item \texttt{df['klen']} = (high-low)/open ($\geq$0)
\item \texttt{df['kup']} = (high - max(open,close))/open ($\geq$0), \texttt{df['kup2']} = (high - max(open,close))/(high-low)
\item \texttt{df['klow']} = (min(open,close) - low)/open ($\geq$0), \texttt{df['klow2']} = (min(open,close) - low)/(high-low)
\item \texttt{df['kmid']} = (close - open)/close (unbounded), \texttt{df['kmid2']} = (close - open)/(high-low)
\item \texttt{df['ksft']} = (2*close - high - low)/open (unbounded), \texttt{df['ksft2']} = (2*close - high - low)/(high-low)
\end{itemize}

\medskip
\textbf{Unbounded} (no fixed threshold, use crossover or trend):
\begin{itemize}[leftmargin=*, nosep]
\item MACD: \texttt{df['macd']}, \texttt{df['macd\_signal']}, \texttt{df['macd\_hist']}
\item CCI: \texttt{df['cci\_\{n\}']} -- commodity channel index
\item OBV: \texttt{df['obv']} -- on-balance volume
\item LogVol: \texttt{df['logvol']} = log(volume + 1)
\end{itemize}

\medskip
Do not invent new factor names. Do not reference external data (news, fundamentals, VIX, options, macro). Everything must be expressible using the allowed columns.
\end{tcolorbox}

\begin{tcolorbox}[colback=gray!5, colframe=gray!50, title=Level Overview, fonttitle=\bfseries\small, breakable]
\small
\textbf{Level Overview (Understanding the Hierarchy)}

\medskip
The three levels represent a spectrum from explicit to abstract:

\medskip
\textbf{Level~1: Explicit Translation (White-box)}\\
\textit{Core Idea}: Fully transparent -- all variable names and rules are explicit.\\
Query MUST include exact column names (e.g., \texttt{df['rsi\_14']}) AND specific numeric thresholds (e.g., < 30). No financial knowledge required -- can be directly translated to code.

\medskip
\textbf{Level~2: Domain Inference (Grey-box)}\\
\textit{Core Idea}: Strategy skeleton -- factor types given, values to be inferred.\\
Query uses financial jargon to describe the strategy skeleton. It indicates WHICH factors to use but leaves specific parameter values for the model to infer based on domain knowledge.

\medskip
\textbf{Level~3: Strategic Synthesis (Black-box)}\\
\textit{Core Idea}: Goal-oriented -- neither factors nor values are specified.\\
Query provides only abstract objectives and constraints. The model must independently decide which factors to use and what values to set.
\end{tcolorbox}

\begin{tcolorbox}[colback=gray!5, colframe=gray!50, title=Category Definitions (All 9 Cells), fonttitle=\bfseries\small, breakable]
\small
All nine category definitions are provided to the generator for reference. The current task's cell is highlighted with ``$\leftarrow$ CURRENT TASK''.

\medskip
\textbf{L1\_easy --- Single IF-THEN Rule}\\
Level~1: Explicit Translation \& Syntactic Mapping (white-box).
Logic must be fully explicit: clear IF/THEN rules with verbatim factor column names. No finance knowledge is required; avoid jargon that needs interpretation.
L1-easy specific: A single IF-THEN rule; 1 entry + 1 exit condition; all thresholds/periods explicit.

\medskip
\textbf{L1\_medium --- Multiple Conditions with AND/OR}\\
Level~1: Explicit Translation \& Syntactic Mapping (white-box).
Logic must be fully explicit: clear IF/THEN rules with verbatim factor column names.
L1-medium specific: Multiple conditions with AND/OR; explicit ``cross'' semantics (e.g., golden cross, death cross); fixed take-profit/stop-loss thresholds.
IMPORTANT: Strategies are STATELESS---describe entry/exit conditions based on current bar data only. Do NOT use phrases like ``if already in position'' or ``once I'm long.'' Each bar independently evaluates conditions.

\medskip
\textbf{L1\_hard --- Multi-Factor Position Sizing with Priority Rules}\\
Level~1: Explicit Translation \& Syntactic Mapping (white-box).
Logic must be fully explicit: clear IF/THEN rules with verbatim factor column names.
L1-hard specific: Signal-strength-based position sizing (e.g., full position if RSI < 20, half position if RSI < 30); multiple time window factors combined (e.g., short-term + medium-term + long-term indicators); explicit priority rules when signals conflict (e.g., risk signal overrides entry signal); nested conditional logic with 4+ conditions.
IMPORTANT: Strategies are STATELESS---position size is determined by CURRENT signal strength, not by tracking previous positions. Each bar independently evaluates all conditions.

\medskip
\textbf{L2\_easy --- Simple Jargon, Few Missing Parameters}\\
Level~2: Domain Inference \& Logic Completion (grey-box).
You may use financial jargon, but it must be interpretable using allowed factors. Provide a strategy skeleton and intentionally leave some parameters to be inferred.
L2-easy specific: Jargon maps in one hop to common indicators (e.g., ``oversold'' $\rightarrow$ RSI); few missing thresholds.

\medskip
\textbf{L2\_medium --- Combined Mapping, Multiple Missing Parameters}\\
Level~2: Domain Inference \& Logic Completion (grey-box).
You may use financial jargon, but it must be interpretable using allowed factors.
L2-medium specific: Jargon requires combined mapping and multiple missing parameters. E.g., ``volume-confirmed breakout + trend confirmation + chop filter.''

\medskip
\textbf{L2\_hard --- Abstract Jargon, Risk Logic Inference Required}\\
Level~2: Domain Inference \& Logic Completion (grey-box).
L2-hard specific: Abstract, multi-interpretation jargon (e.g., ``liquidity drying up,'' ``panic selling,'' ``risk appetite shift''); must be made coherent using allowed factors; risk controls also need completion.

\medskip
\textbf{L3\_easy --- Single Objective + Few Constraints}\\
Level~3: Strategic Synthesis \& Goal-Oriented Generation (black-box objective).
Provide only an abstract objective and constraints; the model must design the whole strategy from scratch.
L3-easy specific: Single objective + a small number of constraints (e.g., max drawdown limit).

\medskip
\textbf{L3\_medium --- Multiple Constraints + Current-Bar Regime Detection}\\
Level~3: Strategic Synthesis \& Goal-Oriented Generation (black-box objective).
L3-medium specific: Objective + multiple constraints (e.g., drawdown limit + position size cap + trading frequency); regime detection based on CURRENT bar factors (e.g., use ATR/std to detect high/low volatility regime); different behavior for different market conditions, all determined from current data.
IMPORTANT: Regime detection must use CURRENT bar indicators only (e.g., ``if std\_20 > 0.05, treat as high volatility''). Do NOT track regime history or state transitions.

\medskip
\textbf{L3\_hard --- Conflicting Objectives with Priority-Based Resolution}\\
Level~3: Strategic Synthesis \& Goal-Oriented Generation (black-box objective).
L3-hard specific: Conflicting objectives (e.g., maximize returns vs.\ minimize drawdown vs.\ reduce turnover); complex priority-based rule system to resolve conflicts; multi-layer risk controls that can override entry signals; position sizing that adapts to multiple current-bar factors simultaneously.
IMPORTANT: All decisions must be based on CURRENT bar data only. Do NOT describe state machines, regime tracking, or any form of historical state memory. Use conditional priority rules instead (e.g., ``if risk condition A, ignore entry signal B'').
\end{tcolorbox}

\begin{tcolorbox}[colback=gray!5, colframe=gray!50, title=Trading Style Definitions, fonttitle=\bfseries\small, breakable]
\small
Each batch generates one query per style:

\begin{enumerate}[leftmargin=*, nosep]
    \item \textbf{Conservative}: Focus on capital preservation and risk control. Prefer confirmed signals with multiple validations. Tighter stop-loss, smaller position sizes, lower trading frequency.
    \item \textbf{Aggressive}: Focus on capturing large moves and maximizing returns. Act on early signals, accept higher false positive rate. Wider stop-loss, larger position sizes, higher trading frequency.
    \item \textbf{Balanced}: Balance between risk and reward. Moderate signal confirmation requirements. Adaptive position sizing based on volatility. Medium trading frequency with regime awareness.
\end{enumerate}
\end{tcolorbox}

\begin{tcolorbox}[colback=gray!5, colframe=gray!50, title=Current Task Emphasis (Per-Cell), fonttitle=\bfseries\small, breakable]
\small
For each generation call, the prompt reinforces the active cell's constraints with the following template (shown here for an example cell; \texttt{\{category\}} and \texttt{\{level\_key\}} are substituted at runtime):

\medskip
\texttt{!!! CURRENT TASK: \{category\} !!!}

\medskip
\textbf{Level Requirement:} \texttt{\{level\_name\}}\\
\textit{Core Idea}: \texttt{\{core\_idea\}}\\
\texttt{\{level\_description\}}

\medskip
\textbf{Specific Requirements for \texttt{\{category\}}:}\\
\texttt{\{category\_definition\}}

\medskip
\textbf{CRITICAL REMINDERS:}
\begin{itemize}[leftmargin=*, nosep]
\item Your generated query MUST strictly follow the \texttt{\{level\_key\}} constraints above.
\item \textit{(For L1):} Include EXACT column names (e.g., \texttt{df['rsi\_14']}) AND specific numeric values.
\item \textit{(For L2):} Indicate factor types but leave specific values to be inferred.
\item \textit{(For L3):} Provide only objectives and constraints---NO specific factors or values.
\end{itemize}
\end{tcolorbox}

\begin{tcolorbox}[colback=gray!5, colframe=gray!50, title=Output Requirements, fonttitle=\bfseries\small, breakable]
\small
\begin{enumerate}[leftmargin=*, nosep]
    \item Generate exactly 3 strategy objects---one for each style (Conservative, Aggressive, Balanced).
    \item Each query should be a complete strategy requirement description (3--8 sentences).
    \item \textbf{Profitability is the goal}---every strategy must be designed with profit potential in mind.
    \item \textbf{Signal generation}: Use factor thresholds (e.g., \texttt{rsi\_14 < 30}) OR factor comparisons (e.g., \texttt{ema\_12 > ema\_26}) to generate signals.
    \item \textbf{Threshold validity (L1 only)}: For explicit threshold strategies, ensure values are within the factor's actual range:
    \begin{itemize}[leftmargin=*, nosep]
        \item \texttt{rank\_20}, \texttt{imax\_20}, \texttt{imin\_20}: range is 0 to 0.95 (NEVER reaches 1.0)
        \item \texttt{max\_\{n\}}: always $\geq$1.0 \textbar{} \texttt{min\_\{n\}}: always 0--1 \textbar{} \texttt{std\_\{n\}}, \texttt{vstd\_\{n\}}: always $\geq$0
    \end{itemize}
    \item \textbf{Diversity}: Vary factor combinations, logic patterns, and trading styles across strategies.
    \item \textbf{Strictly follow the current level constraints}---this is the most important requirement.
    \item Include distinctive twists (time-based exit, partial scaling, volatility filter, etc.).
    \item Keep it single-stock (no cross-sectional ranking).
    \item \textbf{Language diversity (critical)}: Each query MUST have a distinctly different writing style:
    \begin{itemize}[leftmargin=*, nosep]
        \item Vary sentence openers: ``The strategy\ldots'', ``When\ldots'', ``Buy when\ldots'', ``This approach\ldots'', ``Enter long if\ldots'', etc.
        \item Vary sentence lengths: mix short punchy sentences with longer detailed ones.
        \item Vary structure: some queries start with entry conditions, others with exit logic, others with the overall goal.
        \item Do not use the same sentence pattern for all 3 queries in a batch.
    \end{itemize}
\end{enumerate}
\end{tcolorbox}

\begin{tcolorbox}[colback=gray!5, colframe=gray!50, title=Output Format (JSON Schema), fonttitle=\bfseries\small, breakable]
\small
Return ONLY a JSON array. Each item must have:
\begin{itemize}[leftmargin=*, nosep]
\item \texttt{"style"}: one of \texttt{"conservative"}, \texttt{"aggressive"}, \texttt{"balanced"}
\item \texttt{"summary"}: A single sentence (10--20 words) summarizing the strategy's core idea
\item \texttt{"query"}: The full strategy requirement description
\end{itemize}

\medskip
\begin{lstlisting}[basicstyle=\ttfamily\scriptsize, breaklines=true]
[
  {"style": "conservative",
   "summary": "Brief strategy description",
   "query": "..."},
  {"style": "aggressive",
   "summary": "Brief strategy description",
   "query": "..."},
  {"style": "balanced",
   "summary": "Brief strategy description",
   "query": "..."}
]
\end{lstlisting}
\end{tcolorbox}

\begin{tcolorbox}[colback=gray!5, colframe=gray!50, title=Deduplication Instruction (Conditional), fonttitle=\bfseries\small, breakable]
\small
When prior queries have already been generated for the same cell, the following block is appended to the prompt:

\medskip
\textbf{Already Generated Strategies (DO NOT REPEAT similar ideas)}\\
\texttt{\{N\}} strategies already generated. Here are recent summaries:\\
-- \textit{(last 15 strategy summaries listed here)}

\medskip
\textbf{Deduplication Requirements:}
\begin{itemize}[leftmargin=*, nosep]
\item Use different factor combinations.
\item Use different trading logic.
\item Avoid strategies that are conceptually similar to the above.
\end{itemize}
\end{tcolorbox}

\section{Details of \projectname Experiments}
\label{appx_sec:details_of_alphaforgebench_experiments}

To validate the effectiveness of \projectname as a benchmark, we conduct experiments along two complementary evaluation tracks that mirror the two-stage construction of the benchmark itself.

\textbf{Track~1: Real-world query evaluation (Stage~1).}
Although the primary contribution of our benchmark lies in the systematically constructed queries from Stage~2, the real-world queries collected in Stage~1 remain an indispensable component of the experimental validation for three reasons.
First, real-world queries serve as an \emph{ecological validity anchor}: because they originate from authentic investment research and practitioner workflows, strong model performance on these queries provides evidence that the benchmark measures capabilities that are relevant in practice, rather than artifacts of synthetic query construction.
Second, evaluating on real-world queries establishes a \emph{baseline difficulty calibration}: since these queries were not designed according to the controlled difficulty taxonomy, they provide a complementary, ``in-the-wild'' difficulty distribution against which the structured difficulty levels of Stage~2 can be contextualized and cross-referenced.
Third, comparing model rankings on real-world versus synthetic queries enables a \emph{consistency check}: if the relative ordering of models is broadly preserved across the two tracks, it strengthens confidence that the Stage~2 taxonomy captures genuine capability differences rather than idiosyncratic biases of the generation process.

\textbf{Track~2: LLM-augmented Structured query evaluation (Stage~2).}
The queries generated in Stage~2 are organized according to the $3\times 3$ level--grade difficulty taxonomy (see \Cref{appx_sec:construction_alphaforgebench}), enabling fine-grained diagnosis of each model's strengths and weaknesses along the dimensions of strategy granularity, semantic--symbolic alignment, and logical complexity. By systematically varying a single difficulty axis while holding the others constant, this track isolates specific failure modes (e.g., an inability to infer missing parameters at Level~2, or to construct state-dependent logic at Level~3) that would be obscured in aggregate real-world evaluation. Together, the two tracks provide both breadth (ecological coverage) and depth (controlled diagnostics), ensuring a comprehensive and rigorous assessment of LLM-based strategy generation capability.

\subsection{Evaluation Pipeline}

The end-to-end evaluation pipeline of \projectname consists of three sequential stages: \emph{prompt construction}, \emph{code generation}, and \emph{backtest-based assessment}.

\textbf{Step~1: Prompt construction.}
For each benchmark query (from either the real-world collection in Stage~1 or the structured generation in Stage~2), we assemble a standardized prompt that includes (i)~a system-level instruction specifying the code generation task, the available data schema (OHLCV columns and precomputed technical-indicator factors), and the expected output format; (ii)~the natural-language strategy query itself; and (iii)~the complete list of supported factor names and their definitions, serving as the factor library reference. This prompt template is kept identical across all evaluated models to ensure a controlled comparison. The complete code-generation system prompt is presented below, organized into seven components: role and language setting, DataFrame specification, available factor library, return format, critical constraints, allowed libraries, a worked example, and output format.

\begin{tcolorbox}[colback=gray!5, colframe=gray!50, title=Code Generation System Prompt: Role \& DataFrame Specification, fonttitle=\bfseries\small, breakable]
\small
You are a quantitative trading strategy code generator for single-asset trading. Generate Python strategy code compatible with the AlphaForgeBench backtesting system. Output valid JSON with the strategy code (see \texttt{<output\_format>} at the end).

\medskip
\textbf{Code Language Requirements}
\begin{itemize}[leftmargin=*, nosep]
\item All variable names must be in English
\item All code comments must be in English
\item Class names and method names must be in English
\end{itemize}

\medskip
\textbf{DataFrame Specification}

\medskip
The input \texttt{df} is a pandas DataFrame containing BOTH price data AND pre-computed technical factors.

\medskip
\textbf{Index and Time Order}
\begin{itemize}[leftmargin=*, nosep]
\item \texttt{df.index} = DatetimeIndex (timestamps)
\item \texttt{df.iloc[0]} = oldest data
\item \texttt{df.iloc[-1]} = most recent data (current bar)
\item \texttt{df.iloc[-2]} = previous bar
\end{itemize}

\medskip
\textbf{Price Columns (OHLCV)}
\begin{itemize}[leftmargin=*, nosep]
\item \texttt{open}: Opening price
\item \texttt{high}: Highest price
\item \texttt{low}: Lowest price
\item \texttt{close}: Closing price
\item \texttt{volume}: Trading volume
\end{itemize}

\medskip
\textbf{Factor Columns (120+ pre-computed indicators)}\\
All factors are pre-calculated and available as columns in \texttt{df}. Access them directly:
\begin{lstlisting}[basicstyle=\ttfamily\scriptsize, breaklines=true]
df["ema_20"].iloc[-1]   # Current EMA(20) value
df["rsi_14"].iloc[-1]   # Current RSI(14) value
\end{lstlisting}

\medskip
\textbf{Data Access Pattern}
\begin{lstlisting}[basicstyle=\ttfamily\scriptsize, breaklines=true]
current_close = df["close"].iloc[-1]      # Latest close price
prev_close = df["close"].iloc[-2]         # Previous close price
recent_rsi = df["rsi_14"].iloc[-5:]       # Last 5 RSI values
\end{lstlisting}

\medskip
\textbf{Crossover Detection Pattern (Example)}\\
This is a PATTERN EXAMPLE---apply this technique to any indicator needing crossover detection. To detect crossovers, compare current and previous bar values:
\begin{lstlisting}[basicstyle=\ttfamily\scriptsize, breaklines=true]
curr_macd = df["macd"].iloc[-1]
prev_macd = df["macd"].iloc[-2]
curr_signal = df["macd_signal"].iloc[-1]
prev_signal = df["macd_signal"].iloc[-2]

# MACD crosses above signal line (bullish crossover)
macd_cross_up = (prev_macd <= prev_signal) and (curr_macd > curr_signal)

# MACD crosses below signal line (bearish crossover)
macd_cross_down = (prev_macd >= prev_signal) and (curr_macd < curr_signal)

# Golden cross: short EMA crosses above long EMA
golden_cross = (df["ema_20"].iloc[-2] <= df["ema_50"].iloc[-2]) and
               (df["ema_20"].iloc[-1] > df["ema_50"].iloc[-1])
\end{lstlisting}
\end{tcolorbox}

\begin{tcolorbox}[colback=gray!5, colframe=gray!50, title=Code Generation System Prompt: Available Factors --- Technical Indicators, fonttitle=\bfseries\small, breakable]
\small
IMPORTANT: The system supports dynamic factor computation. You can freely choose ANY period parameter!

\medskip
\textbf{Naming Convention}
\begin{itemize}[leftmargin=*, nosep]
\item Factors with period: \texttt{\{factor\_type\}\_\{period\}}, e.g., \texttt{ema\_12}, \texttt{rsi\_14}, \texttt{ma\_50}
\item Factors with variant: \texttt{\{factor\_type\}\_\{variant\}\_\{period\}}, e.g., \texttt{bb\_upper\_20}, \texttt{bb\_lower\_20}
\item Factors without period: use factor name directly, e.g., \texttt{macd}, \texttt{obv}, \texttt{logvol}
\end{itemize}

\medskip
\textbf{1. Technical Indicators}

\begin{itemize}[leftmargin=*, nosep]
\item \textbf{ema}: Exponential Moving Average gives more weight to recent prices. When price crosses above EMA, it signals upward momentum; crossing below suggests downward trend.\\
Formula: \texttt{ema\_w = EMA(close, w)} | Scale: price-scale | Usage: \texttt{df["ema\_\{period\}"]}

\item \textbf{sma}: Simple Moving Average calculates the arithmetic mean of prices over a period. Used to identify trend direction and potential support/resistance levels.\\
Formula: \texttt{sma\_w = SMA(close, w)} | Scale: price-scale | Usage: \texttt{df["sma\_\{period\}"]}

\item \textbf{ma}: Moving Average Ratio compares the average price to current price. Values > 1 indicate price is below average (potential buy); values < 1 indicate price is above average (potential sell).\\
Formula: \texttt{ma\_w = ts\_mean(close, w) / close} | Scale: around 1.0 | Usage: \texttt{df["ma\_\{period\}"]}

\item \textbf{rsi}: Relative Strength Index measures momentum on a 0--100 scale. RSI > 70 suggests overbought conditions (sell signal); RSI < 30 suggests oversold conditions (buy signal).\\
Formula: \texttt{rsi = 100 - 100/(1 + avg\_gain/avg\_loss)} | Scale: 0--100 | Usage: \texttt{df["rsi\_\{period\}"]}

\item \textbf{macd}: MACD Indicator shows the relationship between two EMAs. Positive MACD indicates bullish momentum; negative indicates bearish. Crossovers signal trend changes.\\
Formula: \texttt{macd = ema\_12 - ema\_26} | Scale: unbounded | Usage: \texttt{df["macd"]}, \texttt{df["macd\_signal"]}, \texttt{df["macd\_hist"]}

\item \textbf{bb}: Bollinger Bands measure volatility with upper/middle/lower bands. Price touching upper band suggests overbought; touching lower band suggests oversold.\\
Formula: \texttt{bb\_upper = sma + 2*std}, \texttt{bb\_lower = sma - 2*std} | Scale: price-scale | Usage: \texttt{df["bb\_upper\_\{period\}"]}, \texttt{df["bb\_middle\_\{period\}"]}, \texttt{df["bb\_lower\_\{period\}"]}

\item \textbf{atr}: Average True Range measures market volatility. Higher ATR indicates higher volatility; useful for setting stop-loss levels and position sizing.\\
Formula: \texttt{atr = ts\_mean(true\_range, w)} | Scale: price-scale | Usage: \texttt{df["atr\_\{period\}"]}

\item \textbf{cci}: Commodity Channel Index identifies cyclical trends. CCI > 100 indicates overbought (sell signal); CCI < -100 indicates oversold (buy signal).\\
Formula: \texttt{cci = (tp - sma\_tp) / (0.015 * mad)} | Scale: unbounded | Usage: \texttt{df["cci\_\{period\}"]}

\item \textbf{mfi}: Money Flow Index combines price and volume to measure buying/selling pressure. MFI > 80 suggests overbought; MFI < 20 suggests oversold.\\
Formula: \texttt{mfi = 100 - 100/(1 + pos\_flow/neg\_flow)} | Scale: 0--100 | Usage: \texttt{df["mfi\_\{period\}"]}

\item \textbf{obv}: On-Balance Volume tracks cumulative volume flow. Rising OBV confirms uptrend; falling OBV confirms downtrend. Divergence from price signals potential reversal.\\
Formula: \texttt{obv = cumsum(sign(ret) * volume)} | Scale: unbounded | Usage: \texttt{df["obv"]}

\item \textbf{roc}: Rate of Change measures price momentum as a ratio. Values > 1 indicate price has fallen from w periods ago; values < 1 indicate price has risen.\\
Formula: \texttt{roc\_w = close.shift(w) / close} | Scale: around 1.0 | Usage: \texttt{df["roc\_\{period\}"]}

\item \textbf{kdj}: Stochastic Oscillator (KDJ) measures momentum relative to price range. K > 80 or D > 80 suggests overbought; K < 20 or D < 20 suggests oversold.\\
Formula: \texttt{stoch\_k = (close-low\_w)/(high\_w-low\_w)*100} | Scale: 0--100 | Usage: \texttt{df["stoch\_k\_\{period\}"]}, \texttt{df["stoch\_d\_\{period\}"]}
\end{itemize}
\end{tcolorbox}

\begin{tcolorbox}[colback=gray!5, colframe=gray!50, title=Code Generation System Prompt: Available Factors --- Statistical Factors, fonttitle=\bfseries\small, breakable]
\small
\textbf{2. Statistical Factors}

\begin{itemize}[leftmargin=*, nosep]
\item \textbf{std}: Standard Deviation measures price volatility relative to current price. Higher values indicate greater price dispersion; useful for volatility-based strategies.\\
Formula: \texttt{std\_w = ts\_std\_dev(close, w) / close} | Scale: $\geq$0 | Usage: \texttt{df["std\_\{period\}"]}

\item \textbf{vstd}: Volume Standard Deviation measures volume volatility. High vstd indicates erratic trading activity; low vstd suggests stable volume patterns.\\
Formula: \texttt{vstd\_w = ts\_std\_dev(volume, w) / volume} | Scale: $\geq$0 | Usage: \texttt{df["vstd\_\{period\}"]}

\item \textbf{beta}: Beta Coefficient measures average price change rate over a period. Positive beta indicates upward trend; negative indicates downward trend.\\
Formula: \texttt{beta\_w = (close.shift(w) - close) / (w * close)} | Scale: unbounded | Usage: \texttt{df["beta\_\{period\}"]}

\item \textbf{corr}: Correlation between price and log volume. Positive correlation suggests volume confirms price movement; negative suggests divergence.\\
Formula: \texttt{corr\_w = ts\_corr(close, log(volume), w)} | Scale: -1 to 1 | Usage: \texttt{df["corr\_\{period\}"]}

\item \textbf{cord}: Correlation between price change and volume change. High positive values indicate volume-price synchronization; useful for trend confirmation.\\
Formula: \texttt{cord\_w = ts\_corr(delta(close), delta(volume), w)} | Scale: -1 to 1 | Usage: \texttt{df["cord\_\{period\}"]}
\end{itemize}
\end{tcolorbox}

\begin{tcolorbox}[colback=gray!5, colframe=gray!50, title=Code Generation System Prompt: Available Factors --- Time Series Factors, fonttitle=\bfseries\small, breakable]
\small
\textbf{3. Time Series Factors}

\begin{itemize}[leftmargin=*, nosep]
\item \textbf{max}: Period High Ratio compares period maximum to current price. Values close to 1 indicate price near recent highs; higher values suggest price has fallen from highs.\\
Formula: \texttt{max\_w = ts\_max(close, w) / close} | Scale: $\geq$1.0 | Usage: \texttt{df["max\_\{period\}"]}

\item \textbf{min}: Period Low Ratio compares period minimum to current price. Values close to 1 indicate price near recent lows; lower values suggest price has risen from lows.\\
Formula: \texttt{min\_w = ts\_min(close, w) / close} | Scale: 0--1 | Usage: \texttt{df["min\_\{period\}"]}

\item \textbf{rank}: Percentile Rank shows where current price stands in the period's distribution. High rank indicates price near period highs; low rank indicates near lows.\\
Formula: \texttt{rank\_w = ts\_rank(close, w) / w} | Scale: 0 to (w-1)/w | Usage: \texttt{df["rank\_\{period\}"]}

\item \textbf{imax}: Index of Maximum shows how long ago the period high occurred. Values near 0 indicate recent high; values near 1 indicate high was at period start.\\
Formula: \texttt{imax\_w = argmax(high, w) / w} | Scale: 0 to (w-1)/w | Usage: \texttt{df["imax\_\{period\}"]}

\item \textbf{imin}: Index of Minimum shows how long ago the period low occurred. Values near 0 indicate recent low; values near 1 indicate low was at period start.\\
Formula: \texttt{imin\_w = argmin(low, w) / w} | Scale: 0 to (w-1)/w | Usage: \texttt{df["imin\_\{period\}"]}

\item \textbf{imxd}: Max-Min Index Difference shows timing relationship between high and low. Positive values mean high occurred after low (uptrend); negative means low after high (downtrend).\\
Formula: \texttt{imxd\_w = (argmax(high, w) - argmin(low, w)) / w} | Scale: -(w-1)/w to (w-1)/w | Usage: \texttt{df["imxd\_\{period\}"]}

\item \textbf{rsv}: Raw Stochastic Value compares current close to the range between period low and shifted close. Values near 1 indicate close near the upper bound; near 0 indicates close near the lower bound.\\
Formula: \texttt{rsv\_w = (close - min(low, close.shift(w))) / (max(high, close.shift(w)) - min(low, close.shift(w)))} | Scale: 0--1 | Usage: \texttt{df["rsv\_\{period\}"]}

\item \textbf{qtlu}: Upper Quantile Distance measures how far price is from the 80th percentile. Positive values indicate price above upper quantile (strong); negative indicates below.\\
Formula: \texttt{qtlu\_w = (close - quantile\_80(close, w)) / close} | Scale: unbounded | Usage: \texttt{df["qtlu\_\{period\}"]}

\item \textbf{qtld}: Lower Quantile Distance measures how far price is from the 20th percentile. Positive values indicate price above lower quantile; negative indicates below (weak).\\
Formula: \texttt{qtld\_w = (close - quantile\_20(close, w)) / close} | Scale: unbounded | Usage: \texttt{df["qtld\_\{period\}"]}
\end{itemize}
\end{tcolorbox}

\begin{tcolorbox}[colback=gray!5, colframe=gray!50, title=Code Generation System Prompt: Available Factors --- Candlestick Pattern Factors, fonttitle=\bfseries\small, breakable]
\small
\textbf{4. Candlestick Pattern Factors}

\begin{itemize}[leftmargin=*, nosep]
\item \textbf{klen}: Candle Body Length measures the total range of the candle. Higher values indicate larger price swings; useful for volatility assessment.\\
Formula: \texttt{klen = (high - low) / open} | Scale: $\geq$0 | Usage: \texttt{df["klen"]}

\item \textbf{kup}: Upper Shadow Length measures rejection from highs. Long upper shadows indicate selling pressure; often seen at resistance levels.\\
Formula: \texttt{kup = (high - max(open, close)) / open} | Scale: $\geq$0 | Usage: \texttt{df["kup"]}

\item \textbf{klow}: Lower Shadow Length measures rejection from lows. Long lower shadows indicate buying pressure; often seen at support levels.\\
Formula: \texttt{klow = (min(open, close) - low) / open} | Scale: $\geq$0 | Usage: \texttt{df["klow"]}

\item \textbf{kmid}: Candle Midpoint measures the direction and magnitude of price change. Positive values indicate bullish candle (close > open); negative indicates bearish.\\
Formula: \texttt{kmid = (close - open) / close} | Scale: unbounded | Usage: \texttt{df["kmid"]}

\item \textbf{ksft}: Candle Shift measures where close is relative to the candle's midpoint. Positive values indicate close above midpoint (bullish bias); negative indicates below (bearish bias).\\
Formula: \texttt{ksft = (2*close - high - low) / open} | Scale: unbounded | Usage: \texttt{df["ksft"]}
\end{itemize}
\end{tcolorbox}

\begin{tcolorbox}[colback=gray!5, colframe=gray!50, title=Code Generation System Prompt: Available Factors --- Candlestick Normalized Variants, fonttitle=\bfseries\small, breakable]
\small
\textbf{4 (cont.). Candlestick Normalized Variants}

\begin{itemize}[leftmargin=*, nosep]
\item \textbf{kup2}: Upper Shadow Ratio (normalized by candle range). Measures upper shadow as proportion of total candle range.\\
Formula: \texttt{kup2 = (high - max(open, close)) / (high - low)} | Scale: 0--1 | Usage: \texttt{df["kup2"]}

\item \textbf{klow2}: Lower Shadow Ratio (normalized by candle range). Measures lower shadow as proportion of total candle range.\\
Formula: \texttt{klow2 = (min(open, close) - low) / (high - low)} | Scale: 0--1 | Usage: \texttt{df["klow2"]}

\item \textbf{kmid2}: Body Ratio (normalized by candle range). Measures body direction relative to candle range.\\
Formula: \texttt{kmid2 = (close - open) / (high - low)} | Scale: -1 to 1 | Usage: \texttt{df["kmid2"]}

\item \textbf{ksft2}: Shift Ratio (normalized by candle range). Measures close position relative to candle midpoint, normalized.\\
Formula: \texttt{ksft2 = (2*close - high - low) / (high - low)} | Scale: -1 to 1 | Usage: \texttt{df["ksft2"]}
\end{itemize}
\end{tcolorbox}

\begin{tcolorbox}[colback=gray!5, colframe=gray!50, title=Code Generation System Prompt: Available Factors --- Volume Factors, fonttitle=\bfseries\small, breakable]
\small
\textbf{5. Volume Factors}

\begin{itemize}[leftmargin=*, nosep]
\item \textbf{vma}: Volume Moving Average Ratio compares average volume to current volume. Values > 1 indicate current volume below average; values < 1 indicate above average (high activity).\\
Formula: \texttt{vma\_w = ts\_mean(volume, w) / volume} | Scale: $\geq$0 | Usage: \texttt{df["vma\_\{period\}"]}

\item \textbf{logvol}: Log Volume normalizes volume data for easier comparison. Useful for cross-asset analysis and reducing the impact of volume spikes.\\
Formula: \texttt{logvol = log(volume + 1)} | Scale: unbounded | Usage: \texttt{df["logvol"]}

\item \textbf{wvma}: Weighted Volume MA Ratio measures volatility of return-weighted volume. High values indicate erratic trading activity; useful for detecting unusual market behavior.\\
Formula: \texttt{wvma\_w = ts\_std\_dev(abs(ret)*vol, w) / ts\_mean(abs(ret)*vol, w)} | Scale: $\geq$0 | Usage: \texttt{df["wvma\_\{period\}"]}
\end{itemize}
\end{tcolorbox}

\begin{tcolorbox}[colback=gray!5, colframe=gray!50, title=Code Generation System Prompt: Available Factors --- Counting Factors, fonttitle=\bfseries\small, breakable]
\small
\textbf{6. Counting Factors}

\begin{itemize}[leftmargin=*, nosep]
\item \textbf{cntp}: Positive Return Ratio counts the proportion of up days. High values indicate bullish momentum; low values suggest bearish sentiment.\\
Formula: \texttt{cntp\_w = count(ret > 0, w) / w} | Scale: 0--1 | Usage: \texttt{df["cntp\_\{period\}"]}

\item \textbf{cntn}: Negative Return Ratio counts the proportion of down days. High values indicate bearish momentum; low values suggest bullish sentiment.\\
Formula: \texttt{cntn\_w = count(ret < 0, w) / w} | Scale: 0--1 | Usage: \texttt{df["cntn\_\{period\}"]}

\item \textbf{cntd}: Count Difference measures net bullish/bearish day count. Positive values indicate more up days; negative indicates more down days.\\
Formula: \texttt{cntd\_w = cntp\_w - cntn\_w} | Scale: -1 to 1 | Usage: \texttt{df["cntd\_\{period\}"]}

\item \textbf{sump}: Positive Return Sum Ratio measures the magnitude of gains relative to total movement. High values indicate strong upward moves; useful for momentum assessment.\\
Formula: \texttt{sump\_w = ts\_sum(pos\_ret, w) / ts\_sum(abs\_ret, w)} | Scale: 0--1 | Usage: \texttt{df["sump\_\{period\}"]}

\item \textbf{sumn}: Negative Return Sum Ratio measures the magnitude of losses relative to total movement. High values indicate strong downward moves.\\
Formula: \texttt{sumn\_w = 1 - sump\_w} | Scale: 0--1 | Usage: \texttt{df["sumn\_\{period\}"]}

\item \textbf{sumd}: Sum Difference measures net return magnitude direction. Positive values indicate gains outweigh losses; negative indicates losses dominate.\\
Formula: \texttt{sumd\_w = 2 * sump\_w - 1} | Scale: -1 to 1 | Usage: \texttt{df["sumd\_\{period\}"]}
\end{itemize}
\end{tcolorbox}

\begin{tcolorbox}[colback=gray!5, colframe=gray!50, title=Code Generation System Prompt: Available Factors --- Volume Counting Factors, fonttitle=\bfseries\small, breakable]
\small
\textbf{6 (cont.). Volume Counting Factors}

\begin{itemize}[leftmargin=*, nosep]
\item \textbf{vsump}: Positive Volume Change Ratio measures proportion of volume increases. High values indicate accumulation; useful for detecting buying pressure.\\
Formula: \texttt{vsump\_w = ts\_sum(pos\_vol\_chg, w) / ts\_sum(abs\_vol\_chg, w)} | Scale: 0--1 | Usage: \texttt{df["vsump\_\{period\}"]}

\item \textbf{vsumn}: Negative Volume Change Ratio measures proportion of volume decreases. High values indicate distribution; useful for detecting selling pressure.\\
Formula: \texttt{vsumn\_w = 1 - vsump\_w} | Scale: 0--1 | Usage: \texttt{df["vsumn\_\{period\}"]}

\item \textbf{vsumd}: Volume Sum Difference measures net volume change direction. Positive values indicate volume accumulation; negative indicates distribution.\\
Formula: \texttt{vsumd\_w = 2 * vsump\_w - 1} | Scale: -1 to 1 | Usage: \texttt{df["vsumd\_\{period\}"]}
\end{itemize}
\end{tcolorbox}

\begin{tcolorbox}[colback=gray!5, colframe=gray!50, title=Code Generation System Prompt: Return Format, fonttitle=\bfseries\small, breakable]
\small
The strategy must return: \texttt{\{"signal": int, "position": float\}}

\medskip
\textbf{signal (Trading Signal) --- LONG-ONLY SYSTEM}
\begin{itemize}[leftmargin=*, nosep]
\item 1: Buy signal (open/add long position)
\item -1: Sell signal (close/reduce long position, NO short selling)
\item 0: Hold signal (maintain current position)
\end{itemize}

\medskip
\textbf{position (Target Position Size)}
\begin{itemize}[leftmargin=*, nosep]
\item 1.0: Full position (100\% of capital)
\item 0.5: Half position (50\% of capital)
\item 0.0: No position (0\% of capital)
\end{itemize}

\medskip
\textbf{Common Return Patterns}
\begin{lstlisting}[basicstyle=\ttfamily\scriptsize, breaklines=true]
return {"signal": 1, "position": 1.0}   # Strong buy, full position
return {"signal": 1, "position": 0.5}   # Cautious buy, half position
return {"signal": -1, "position": 0.0}  # Sell and close position
return {"signal": 0, "position": 0.0}   # No action
\end{lstlisting}

\medskip
\textbf{Safe Return for Invalid Data}
\begin{lstlisting}[basicstyle=\ttfamily\scriptsize, breaklines=true]
if np.isnan(some_value):
    return {"signal": 0, "position": 0.0}
\end{lstlisting}
\end{tcolorbox}

\begin{tcolorbox}[colback=gray!5, colframe=gray!50, title=Code Generation System Prompt: Critical Constraints, fonttitle=\bfseries\small, breakable]
\small
\begin{enumerate}[leftmargin=*, nosep]
\item \textbf{USE PRE-COMPUTED FACTORS ONLY} --- Do not calculate indicators manually.
\begin{itemize}[leftmargin=*, nosep]
\item Correct: \texttt{df["ema\_20"]}, \texttt{df["rsi\_14"]}, \texttt{df["macd"]}
\item Wrong: \texttt{df["close"].ewm(span=20).mean()}, manual RSI calculation
\end{itemize}

\item \textbf{STRATEGY IS A PYDANTIC BASEMODEL} --- No \texttt{\_\_init\_\_} or \texttt{self.xxx = ...}
\begin{itemize}[leftmargin=*, nosep]
\item Wrong: \texttt{def \_\_init\_\_(self): self.window = 10}
\item Correct: \texttt{window: int = Field(default=10)}
\end{itemize}

\item \textbf{REQUIRED FIELDS} --- Every strategy must have these three:
\begin{lstlisting}[basicstyle=\ttfamily\scriptsize, breaklines=true]
name: str = Field(default="strategy_name")
description: str = Field(default="Strategy description")
factor_names: list[str] = Field(default_factory=list)
\end{lstlisting}

\item \textbf{NO TA/TALIB LIBRARIES} --- Only use pandas/numpy for simple operations.
\begin{itemize}[leftmargin=*, nosep]
\item Wrong: \texttt{ta.volatility.BollingerBands()}, \texttt{talib.RSI()}
\item Correct: \texttt{df["bb\_upper\_20"]}, \texttt{df["rsi\_14"]}
\end{itemize}

\item \textbf{ALWAYS CHECK FOR NaN} --- Avoid comparing NaN values.
\begin{itemize}[leftmargin=*, nosep]
\item Wrong: \texttt{if rsi > 70:}
\item Correct: \texttt{if np.isnan(rsi): return \{"signal": 0, "position": 0.0\}}
\end{itemize}

\item \textbf{NO BACKSLASH LINE CONTINUATION} --- Use parentheses instead.

\item \textbf{NO LEADING UNDERSCORES IN FIELD NAMES} --- Pydantic restriction.
\begin{itemize}[leftmargin=*, nosep]
\item Wrong: \texttt{\_entry\_price: float = 0.0}
\item Correct: \texttt{entry\_price: float = 0.0}
\end{itemize}

\item \textbf{STATELESS STRATEGY} --- The backtest engine manages position state for you.
\begin{itemize}[leftmargin=*, nosep]
\item Your strategy receives ONLY the DataFrame---NO position, cash, or account info.
\item Do NOT track or infer current position from historical data.
\item Simply emit signals based on CURRENT bar conditions.
\item Each bar: check conditions $\to$ emit signal $\to$ done (no memory needed).
\end{itemize}

\item \textbf{SIGNAL-POSITION CONSISTENCY}
\begin{itemize}[leftmargin=*, nosep]
\item \texttt{signal=1} (buy) should have \texttt{position > 0}
\item \texttt{signal=-1} (sell) should have \texttt{position = 0.0}
\item \texttt{signal=0} (hold) can have any position value (maintains current state)
\end{itemize}
\end{enumerate}
\end{tcolorbox}

\begin{tcolorbox}[colback=gray!5, colframe=gray!50, title=Code Generation System Prompt: Allowed Libraries, fonttitle=\bfseries\small, breakable]
\small
\textbf{Python Standard Library}: \texttt{math}, \texttt{datetime}, \texttt{typing}

\medskip
\textbf{Data Processing}:
\begin{itemize}[leftmargin=*, nosep]
\item \texttt{pandas}: DataFrame, Series, rolling, ewm, diff, pct\_change, shift, cumsum
\item \texttt{numpy}: isnan, where, abs, mean, std, max, min, sum, nan
\end{itemize}

\medskip
\textbf{Pydantic}: \texttt{Field} (for parameter definition)

\medskip
\textbf{FORBIDDEN}:
\begin{itemize}[leftmargin=*, nosep]
\item \texttt{ta} (technical analysis library)
\item \texttt{talib} (TA-Lib)
\item Any other technical indicator libraries
\end{itemize}
\end{tcolorbox}

\begin{tcolorbox}[colback=gray!5, colframe=gray!50, title=Code Generation System Prompt: Complete Example, fonttitle=\bfseries\small, breakable]
\small
\textbf{Example: Bollinger Band Mean Reversion Strategy}

\begin{lstlisting}[basicstyle=\ttfamily\scriptsize, breaklines=true, language=Python]
from pydantic import Field
from src.strategy.types import Strategy
import pandas as pd
import numpy as np

class BollingerMeanReversion(Strategy):
    """Buy at lower band, sell at upper band"""
    name: str = Field(default="bb_mean_reversion")
    description: str = Field(
        default="Bollinger Band mean reversion strategy")
    factor_names: list[str] = Field(default_factory=list)
    position_size: float = Field(default=0.5, ge=0.0, le=1.0)

    async def __call__(self, df: pd.DataFrame) -> dict:
        # Get pre-computed Bollinger Bands
        close = df["close"].iloc[-1]
        bb_upper = df["bb_upper_20"].iloc[-1]
        bb_lower = df["bb_lower_20"].iloc[-1]
        bb_middle = df["bb_middle_20"].iloc[-1]

        # NaN check
        if np.isnan(close) or np.isnan(bb_upper) \
                or np.isnan(bb_lower):
            return {"signal": 0, "position": 0.0}

        # Mean reversion logic
        if close < bb_lower:
            return {"signal": 1,
                    "position": self.position_size}
        elif close > bb_upper:
            return {"signal": -1, "position": 0.0}
        elif close >= bb_middle:
            return {"signal": 0, "position": 0.0}

        return {"signal": 0,
                "position": self.position_size}
\end{lstlisting}
\end{tcolorbox}

\begin{tcolorbox}[colback=gray!5, colframe=gray!50, title=Code Generation System Prompt: Output Format, fonttitle=\bfseries\small, breakable]
\small
Respond with valid JSON only:

\medskip
\texttt{\{"strategy":\{"code":"<complete Python code>"\}\}}

\medskip
\textbf{Requirements:}
\begin{itemize}[leftmargin=*, nosep]
\item JSON must be valid (no markdown code blocks around it)
\item The \texttt{"code"} field must contain complete, runnable Python code
\item Include all imports and the full Strategy class
\item Escape special characters properly in the code string
\end{itemize}
\end{tcolorbox}

\textbf{Step~2: Code generation.}
The assembled prompt is submitted to each evaluated LLM via its API. The model is expected to produce executable Python code that implements the requested alpha factor or factor-based trading strategy, using only the provided data columns and factor library. Each model generates one code sample per query ($k{=}1$).

\textbf{Step~3: Backtest-based assessment.}
Every generated code sample is fed into a unified backtest engine, which executes the strategy on historical price data across multiple assets spanning both cryptocurrency and US equity markets. The backtest engine computes a comprehensive suite of financial performance metrics (e.g., Sharpe Ratio, Annualized Return, Maximum Drawdown), enabling standardized and reproducible quantitative comparison across models, difficulty levels, and query sources.

\subsection{Evaluated Models}

We evaluate six state-of-the-art large language models spanning different model families and providers to ensure broad coverage of the current LLM landscape.
The selection criteria are guided by three principles:

\begin{itemize}[leftmargin=*, nosep]
    \item \textit{Provider diversity.} The six models originate from four distinct organizations (Anthropic, DeepSeek, Google, OpenAI, and xAI), reducing the risk that benchmark conclusions are artifacts of a single training pipeline, data mixture, or alignment procedure.
    \item \textit{Capacity spectrum.} The selection spans from lightweight, cost-efficient models designed for low-latency inference (\textit{gemini-3-flash-preview}, \textit{grok-4.1-fast}) to high-capacity frontier models (\textit{gpt-5.2}, \textit{gemini-3-pro-preview}), enabling us to examine whether increased model scale and compute translate into measurably better strategy generation.
    \item \textit{Architectural and licensing heterogeneity.} We include both proprietary closed-source models (\textit{claude-sonnet-4.5}, \textit{gpt-5.2}, Gemini family, \textit{grok-4.1-fast}) and an open-weight model (\textit{deepseek-v3.2}), allowing comparison between commercial APIs and community-accessible alternatives.
\end{itemize}

\noindent All models are accessed via their official APIs with default system prompts. No model-specific prompt engineering, few-shot examples, or chain-of-thought elicitation is applied, ensuring that observed performance differences arise from the models' intrinsic capabilities rather than prompt-tuning artifacts. \Cref{appx_tab:models} summarizes the evaluated models, and a brief characterization of each is provided below.

\begin{itemize}[leftmargin=*, nosep]
    \item \textit{claude-sonnet-4.5} (Anthropic). Anthropic's latest model, recognized for strong code generation accuracy, faithful instruction following, and nuanced long-context reasoning. It represents the current state of the art in the Anthropic Claude family.
    \item \textit{deepseek-v3.2} (DeepSeek). An open-weight model that has demonstrated competitive performance on code generation benchmarks while maintaining high cost efficiency. Its inclusion allows us to assess whether open-source models can match proprietary counterparts on domain-specific financial tasks.
    \item \textit{gemini-3-flash-preview} (Google). Google's lightweight, low-latency variant optimized for fast inference at reduced compute cost. It serves as a representative of the ``small but fast'' model category.
    \item \textit{gemini-3-pro-preview} (Google). Google's higher-capacity model within the same Gemini 3 generation, offering stronger reasoning and generation quality at increased computational cost. The Flash/Pro pair within the same family enables a controlled comparison of model scale within a single provider.
    \item \textit{gpt-5.2} (OpenAI). OpenAI's frontier model with state-of-the-art performance across a wide range of general-purpose and specialized benchmarks. It serves as a strong upper-bound reference for what current LLMs can achieve.
    \item \textit{grok-4.1-fast} (xAI). xAI's fast-inference model with competitive generation quality. Its inclusion broadens the provider coverage and provides an additional data point for the latency--quality tradeoff.
\end{itemize}

\begin{table}[h]
\centering
\caption{Large language models evaluated in \projectname. All models are accessed via official APIs with identical prompt templates and generation settings.}
\label{appx_tab:models}
\begin{tabular}{llcc}
\toprule
\textbf{Model} & \textbf{Provider} & \textbf{Open-weight} & \textbf{Tier} \\
\midrule
\textit{claude-sonnet-4.5} & Anthropic & \texttimes & Flagship \\
\textit{deepseek-v3.2} & DeepSeek  & \checkmark & Flagship \\
\textit{gemini-3-flash-preview} & Google    & \texttimes & Lightweight \\
\textit{gemini-3-pro-preview} & Google    & \texttimes & Flagship \\
\textit{gpt-5.2} & OpenAI    & \texttimes & Flagship \\
\textit{grok-4.1-fast} & xAI       & \texttimes & Lightweight \\
\bottomrule
\end{tabular}
\end{table}

\subsection{Experimental Settings}

\subsubsection{Generation Protocol}

Since LLM outputs are inherently stochastic, a single generation per query is insufficient to characterize a model's true performance distribution. To obtain statistically robust estimates and quantify run-to-run variability, we adopt a \textbf{multi-run} protocol: for each query, every model independently generates \textbf{5 code samples} (i.e., $k{=}5$) under identical settings. Each generated sample is then executed in the backtest engine, and we report the \textbf{mean} and \textbf{standard deviation} of each evaluation metric across the 5 runs. This design enables us to assess not only the average quality of LLM-generated strategies but also their \emph{consistency}, an important practical consideration for real-world deployment where unreliable generation would necessitate costly human review.

\paragraph{Stage~1 (real-world queries).}
Stage~1 contains 633 single-stock queries. All models are configured with a default sampling temperature of $\tau = 0.7$ to allow moderate diversity in the generated outputs. With 5 runs per query and 6 models, this yields $633 \times 5 \times 6 = 18{,}990$ generated strategy implementations in total (5 fewer for \textit{deepseek-v3.2} due to one API failure on a single query).

\paragraph{Stage~2 (structured queries).}
Stage~2 contains 270 structured queries (30 per difficulty cell in the $3\times 3$ level--grade taxonomy). The same 5-run protocol at $\tau = 0.7$ is applied, producing $270 \times 5 \times 6 = 8{,}100$ generated implementations. In addition, to investigate the effect of decoding stochasticity on strategy quality and consistency, we conduct a \textbf{temperature ablation} by repeating the full Stage~2 evaluation at $\tau = 0$ (greedy decoding). The $\tau = 0$ setting eliminates sampling randomness and tests whether models can reliably produce high-quality strategies under deterministic generation. Comparing the $\tau = 0.7$ and $\tau = 0$ results allows us to disentangle the contribution of \emph{strategy reasoning capability} (which should be robust to temperature) from \emph{sampling luck} (which manifests as high variance at $\tau = 0.7$ but collapses at $\tau = 0$).

\paragraph{Summary.}
\Cref{appx_tab:exp_settings} summarizes the generation settings for both stages.

\begin{table}[h]
\centering
\caption{Generation settings for Stage~1 and Stage~2 evaluations.}
\label{appx_tab:exp_settings}
\begin{tabular}{lccccc}
\toprule
\textbf{Stage} & \textbf{Queries} & \textbf{Models} & \textbf{Runs/Query} & \textbf{Temperature} & \textbf{Total Samples} \\
\midrule
Stage~1 & 633  & 6 & 5 & 0.7       & 18{,}990 \\
Stage~2 & 270  & 6 & 5 & 0.7       & 8{,}100  \\
Stage~2 & 270  & 6 & 5 & 0 (greedy)& 8{,}100  \\
\midrule
\multicolumn{5}{l}{\textbf{Grand total}} & \textbf{35{,}190} \\
\bottomrule
\end{tabular}
\end{table}

\noindent Results from the two stages are reported separately: Stage~1 results assess overall real-world performance, while Stage~2 results enable fine-grained, difficulty-stratified analysis. All reported metrics are the mean $\pm$ standard deviation across the 5 runs unless otherwise noted.

\subsubsection{Backtest Assets}

Each generated strategy is backtested across \textbf{7 assets} spanning two distinct market regimes: cryptocurrency spot markets and US equity markets.
This dual-market design is intentional: cryptocurrency and equity markets differ substantially in microstructure, volatility regime, trading hours, and return distribution, providing a rigorous stress test of whether LLM-generated strategies generalize across heterogeneous financial environments rather than overfitting to the statistical properties of a single asset class.
\Cref{appx_tab:assets} lists the selected assets, and the rationale for each market is detailed below.

\paragraph{Cryptocurrency markets.}
We select \textbf{BTCUSDT} (Bitcoin) and \textbf{ETHUSDT} (Ethereum), the two largest cryptocurrencies by market capitalization, traded on the Binance spot exchange.
Cryptocurrency markets present a uniquely challenging environment for algorithmic strategies due to several structural characteristics:
\begin{itemize}[leftmargin=*, nosep]
    \item \textit{Continuous trading with no circuit breakers.} Unlike equity exchanges that operate during fixed sessions, cryptocurrency markets trade 24 hours a day, 7 days a week, with no halt mechanisms. This implies that strategies must be robust to overnight gaps and weekend volatility, which are absent in equity backtests.
    \item \textit{Elevated and time-varying volatility.} BTC and ETH exhibit annualized volatility typically in the range of 60--80\% and 80--100\%, respectively, far exceeding that of large-cap equities (15--40\%). Moreover, volatility itself is highly non-stationary, with abrupt regime transitions between low-volatility consolidation and explosive directional moves.
    \item \textit{Heavy-tailed return distributions.} Daily returns of major cryptocurrencies display significant excess kurtosis, meaning that extreme moves (both positive and negative) occur far more frequently than a Gaussian model would predict. Strategies that implicitly assume thin-tailed distributions (e.g., fixed-threshold mean reversion) may fail catastrophically under these conditions.
    \item \textit{Frequent regime shifts.} The crypto market alternates between prolonged trending phases (e.g., the 2021 bull run) and extended mean-reverting drawdowns (e.g., the 2022 crypto winter, during which BTC declined approximately 75\% from its all-time high). Including both BTC and ETH allows us to assess strategy robustness across correlated yet distinct return profiles, as ETH historically exhibits higher beta to BTC with additional idiosyncratic volatility driven by ecosystem-specific events (e.g., the Ethereum Merge in September 2022).
\end{itemize}

\paragraph{US equity markets.}
We select five major US technology stocks: \textbf{AAPL} (Apple), \textbf{GOOGL} (Alphabet), \textbf{MSFT} (Microsoft), \textbf{NVDA} (NVIDIA), and \textbf{TSLA} (Tesla).
These stocks are chosen based on the following considerations:
\begin{itemize}[leftmargin=*, nosep]
    \item \textit{High liquidity and market depth.} All five stocks rank among the most actively traded US equities, with average daily trading volumes in the tens of millions of shares. This deep liquidity minimizes the impact of slippage and market-impact assumptions on backtest fidelity, ensuring that performance differences across models reflect signal quality rather than execution artifacts.
    \item \textit{Diverse volatility profiles within a single sector.} Although all five are classified as technology stocks, they span a wide spectrum of risk characteristics. AAPL and MSFT behave as relatively stable large-cap defensives (annualized volatility $\approx$ 25--30\%), GOOGL occupies a moderate-volatility position ($\approx$ 30--35\%), while NVDA and TSLA exhibit significantly higher volatility ($\approx$ 45--60\%) driven by growth expectations, speculative flows, and high short interest (TSLA). This diversity tests whether generated strategies adapt to different volatility regimes or apply one-size-fits-all logic.
    \item \textit{Rich and heterogeneous market conditions during the backtest window.} The 2021--2025 period encompasses a broad range of market environments for these stocks:
    \begin{itemize}[leftmargin=*, nosep]
        \item \textbf{2021:} A strong post-COVID bull market driven by fiscal stimulus and low interest rates, with all five stocks posting substantial gains.
        \item \textbf{2022:} An aggressive Federal Reserve tightening cycle triggered a broad technology sell-off; the Nasdaq Composite declined over 30\%, with growth names (TSLA, NVDA) experiencing drawdowns exceeding 50\%.
        \item \textbf{2023--2024:} A technology-led recovery fueled by the generative AI narrative, with NVDA surging over 800\% from its 2022 lows, while other names recovered at varying rates.
        \item \textbf{2025:} A mixed consolidation phase characterized by sector rotation, narrowing breadth, and elevated macro uncertainty.
    \end{itemize}
    This temporal diversity ensures that no single strategy style (trend-following, mean-reversion, or volatility-targeting) is systematically favored across the entire evaluation window.
    \item \textit{Data accessibility and reproducibility.} All US equity data is sourced from Yahoo Finance, a freely available and widely used data provider, ensuring that our backtest results are fully reproducible by the research community without requiring proprietary data subscriptions.
\end{itemize}

\begin{table}[h]
\centering
\caption{Assets used for backtesting in \projectname. The asset universe spans two market types (cryptocurrency and US equity) to evaluate strategy generalization across heterogeneous financial environments.}
\label{appx_tab:assets}
\begin{tabular}{llll}
\toprule
\textbf{Asset} & \textbf{Market} & \textbf{Sector / Type} & \textbf{Data Source} \\
\midrule
BTCUSDT & Cryptocurrency & Digital asset (Bitcoin)  & Binance \\
ETHUSDT & Cryptocurrency & Digital asset (Ethereum) & Binance \\
\midrule
AAPL    & US Equity      & Technology (Apple)       & Yahoo Finance \\
GOOGL   & US Equity      & Technology (Alphabet)    & Yahoo Finance \\
MSFT    & US Equity      & Technology (Microsoft)   & Yahoo Finance \\
NVDA    & US Equity      & Technology (NVIDIA)      & Yahoo Finance \\
TSLA    & US Equity      & Technology (Tesla)       & Yahoo Finance \\
\bottomrule
\end{tabular}
\end{table}

\subsubsection{Backtest Parameters}

All strategies are evaluated under a unified backtest configuration to ensure strict cross-model comparability. No model-specific tuning or post-hoc parameter adjustment is performed. \Cref{appx_tab:backtest_params} summarizes the key parameters, and each is discussed in detail below.

\begin{table}[h]
    \centering
    \caption{Backtest configuration. Parameters are held constant across every model and query to ensure a controlled comparison.}
    \label{appx_tab:backtest_params}
    \begin{tabular}{ll}
    \toprule
    \textbf{Parameter} & \textbf{Value} \\
    \midrule
    Backtest period  & 2021-01-01 to 2026-01-01 (5 years) \\
    Data frequency   & Daily (1 day) \\
    History window   & 300 trading days \\
    Initial capital  & \$100{,}000 (normalized) \\
    Transaction cost & 0 (frictionless) \\
    Data source      & Binance (crypto), Yahoo Finance (equities) \\
    Strategy type    & Long-only, single-asset \\
    \bottomrule
    \end{tabular}
\end{table}

\paragraph{Backtest period.}
The backtest spans a \textbf{five-year period} from January 1, 2021 to January 1, 2026.
This window is deliberately chosen to encompass multiple distinct market regimes, as described in the asset discussion above.
By covering bull markets, bear markets, recovery rallies, and consolidation phases, the evaluation avoids regime-specific bias: a strategy that excels only in trending markets will be penalized by its poor performance during the 2022 correction, and conversely, a purely mean-reverting strategy will underperform during strong directional moves.
This multi-regime coverage is essential for a benchmark that aims to assess \emph{general-purpose} strategy generation capability rather than niche regime-specific performance.

\paragraph{History window.}
A \textbf{history window of 300 trading days} (approximately 14 calendar months) is provided to each strategy at every decision point.
This lookback length is chosen to accommodate the computation of long-horizon technical indicators commonly referenced in quantitative finance, including 200-day simple and exponential moving averages, 52-week high/low levels, and annualized volatility estimates.
Strategies that require shorter lookbacks (e.g., 14-day RSI or 20-day Bollinger Bands) are naturally supported, as the 300-day window is a strict superset.
At the same time, the window is bounded to prevent strategies from accessing an unrealistically long history that would be unavailable in a live-trading deployment.

\paragraph{Data frequency.}
All data is sampled at \textbf{daily frequency} (one OHLCV bar per trading day for equities; one bar per calendar day for cryptocurrencies).
Daily frequency is the natural resolution for the precomputed factor library, which defines indicators such as \texttt{ema\_20}, \texttt{rsi\_14}, and \texttt{bb\_upper\_20} in terms of daily bars.
While intraday data would enable finer-grained signal evaluation, it would also introduce additional complexity (microstructure noise, data vendor discrepancies, time-zone alignment) that is orthogonal to the core research question of whether LLMs can generate sound strategy logic.

\paragraph{Strategy semantics.}
All strategies follow \textbf{long-only, single-asset} semantics.
At each time step, the strategy outputs a binary signal: \emph{invest} (allocate 100\% of capital to the asset) or \emph{hold cash} (allocate 0\%).
No short selling, leverage, or cross-asset allocation is permitted.
This deliberately simplified action space serves two purposes: (i)~it isolates the quality of the \emph{signal generation logic} from confounding portfolio-construction effects (position sizing, risk budgeting, rebalancing), and (ii)~it ensures that all models operate under identical constraints, so performance differences reflect genuine differences in strategy reasoning rather than incidental choices about position management.

\paragraph{Transaction costs.}
We adopt a \textbf{frictionless} (zero transaction cost) assumption in the primary evaluation.
This choice is motivated by the desire to measure the \emph{intrinsic signal quality} of LLM-generated strategies without confounding it with turnover-dependent cost effects.
Since different strategies may generate vastly different turnover rates, introducing transaction costs would couple signal quality with execution efficiency in a way that obscures the interpretation of benchmark results.
We note, however, that turnover and transaction-cost sensitivity can be analyzed as a secondary diagnostic; we leave this extension to future work.

\subsection{Evaluation Metrics}

We evaluate the performance of each LLM-generated trading strategy using six standard financial metrics, computed from the daily return series and averaged across all 7 backtest assets. These metrics are chosen to provide a comprehensive assessment from both \emph{return} and \emph{risk} perspectives:

\begin{itemize}[leftmargin=*, nosep]
    \item \textbf{Annual Rate of Return (ARR)} measures the annualized compounded profitability of a strategy based on the change in portfolio value over time, adjusted by an annualization factor ($N = 252$ for daily trading). ARR reflects the pure return-generating capability of the strategy without risk adjustment.
    \item \textbf{Sharpe Ratio (SR)} quantifies risk-adjusted return by comparing the mean excess return (over the risk-free rate $r_f$) to the standard deviation of returns, annualized by $\sqrt{N}$. A higher SR indicates more efficient compensation per unit of total volatility.
    \item \textbf{Maximum Drawdown (MDD)} measures the largest peak-to-trough decline in cumulative portfolio value, indicating the worst observed loss during the backtest period. MDD captures tail risk and is critical for evaluating capital preservation.
    \item \textbf{Calmar Ratio (CR)} evaluates the return-to-risk tradeoff by dividing the annualized return by the absolute maximum drawdown. CR is particularly informative for strategies where drawdown control is a primary objective.
    \item \textbf{Sortino Ratio (SoR)} is similar to the Sharpe Ratio but replaces total volatility with downside deviation, thus penalizing only negative return fluctuations. SoR provides a more targeted measure of risk-adjusted performance for investors who are primarily concerned with downside risk.
    \item \textbf{Volatility (VOL)} captures the annualized standard deviation of the return series, reflecting the overall level of return fluctuation over time. Lower VOL is generally preferred for risk-averse strategies.
\end{itemize}

\noindent In summary, ARR reflects pure profitability; SR, CR, and SoR assess performance adjusted for different aspects of risk (total volatility, tail risk, and downside risk, respectively); and MDD and VOL evaluate risk exposure directly. Together, these metrics offer a comprehensive and multi-dimensional assessment of trading strategy effectiveness. \Cref{appx_tab:trading_metrics} provides the formal definitions.

\begin{table}[htb]
  \caption{Evaluation metrics used in \projectname. $\mathrm{rets} = [r_1, r_2, \ldots, r_T]$ denotes the daily return series, $N = 252$ is the annualization factor, $r_f$ is the risk-free rate, and $V_t$ is the cumulative portfolio value at time $t$. Direction $\uparrow$ ($\downarrow$) indicates that higher (lower) values correspond to better performance.}
  \label{appx_tab:trading_metrics}
  \centering
  \footnotesize
  \setlength{\tabcolsep}{4pt}
  \renewcommand{\arraystretch}{1.1}
  \resizebox{0.98\textwidth}{!}{%
  \begin{tabular}{lclp{0.42\textwidth}}
  \toprule
  \textbf{Metric} & \textbf{Dir.} & \textbf{Formula} & \textbf{Description} \\
  \midrule

  ARR & $\uparrow$ &
  $\displaystyle \mathrm{ARR} = \left( \prod_{t=1}^{T} (1 + r_t) \right)^{N/T} - 1$ &
  Annualized Rate of Return; compounded growth rate scaled to one year. \\[6pt]

  SR & $\uparrow$ &
  $\displaystyle \mathrm{SR} = \frac{\mathbb{E}[\mathrm{rets}] - r_f}{\mathrm{Std}(\mathrm{rets})} \times \sqrt{N}$ &
  Sharpe Ratio; annualized excess return per unit of total volatility. \\[6pt]

  MDD & $\downarrow$ &
  $\displaystyle \mathrm{MDD} = \max_{t \in [1,T]} \frac{\max_{s \in [1,t]} V_s - V_t}{\max_{s \in [1,t]} V_s}$ &
  Maximum Drawdown; largest peak-to-trough decline in cumulative portfolio value. \\[6pt]

  CR & $\uparrow$ &
  $\displaystyle \mathrm{CR} = \frac{\mathrm{ARR}}{|\mathrm{MDD}|}$ &
  Calmar Ratio; annualized return divided by absolute maximum drawdown. \\[6pt]

  SoR & $\uparrow$ &
  $\displaystyle \mathrm{SoR} = \frac{\mathbb{E}[\mathrm{rets}] - r_f}{\mathrm{DD}} \times \sqrt{N}$ &
  Sortino Ratio; risk-adjusted return using only downside deviation. \\[6pt]

  VOL & $\downarrow$ &
  $\displaystyle \mathrm{VOL} = \mathrm{Std}(\mathrm{rets}) \times \sqrt{N}$ &
  Volatility; annualized standard deviation of daily returns. \\[6pt]

  \midrule
  DD & $\downarrow$ &
  $\displaystyle \mathrm{DD} = \sqrt{ \frac{1}{T} \sum_{t=1}^{T} \min(r_t - r_f,\, 0)^2 } \times \sqrt{N}$ &
  Downside Deviation; annualized std.\ dev.\ of returns below $r_f$ (used in SoR). \\

  \bottomrule
  \end{tabular}%
  }
\end{table}

%% file: appendix/appendix_real_result.tex
\section{Detailed Results of Real-world Query Evaluation}
\label{appx_sec:appendix_results_stage_1}

This section presents a comprehensive analysis of the benchmark results on the Stage~1 real-world query subset. The evaluation covers 633 single-stock strategy queries collected from authentic sources (brokerage reports, quantitative platforms, academic literature, open-source repositories, and traditional finance publications), executed by 6 frontier LLMs, and backtested across 7 assets (2 cryptocurrency pairs and 5 US equities) over a 5-year period (2021--2025). For each query, every model generates 5 independent code samples at temperature $\tau = 0.7$; all reported metrics are the mean $\pm$ standard deviation pooled across the 7 assets and 5 runs unless otherwise noted. We first present the aggregate model comparison, then provide detailed per-asset breakdowns, distributional analyses, and aligned return curves.

\subsection{Overall Model Comparison}

A central motivation of \projectname is to address the severe instability of LLMs when deployed as direct trading agents, where identical models produce dramatically different action sequences across runs, even under deterministic decoding (\texttt{temperature=0}). As discussed in the main paper, this instability arises from the models' stateless architectures, their sensitivity to continuous-to-discrete action mappings, and the absence of persistent state management. By shifting the evaluation paradigm from \emph{black-box action emission} to \emph{white-box strategy code generation}, \projectname confines the stochasticity of the LLM to the generation phase while rendering the subsequent execution strictly deterministic. The results below demonstrate that this paradigm yields \textbf{stable, reproducible, and meaningfully differentiable} performance metrics across models.

\subsubsection{Quantitative Results}

\Cref{appx_tab:realdata_overall} reports the overall performance of each model, averaged across all 633 queries and 7 backtest assets. We highlight the best value in each column in bold and analyze the results along four complementary axes: decision stability, return generation, risk exposure, and risk-adjusted efficiency.

\begin{table*}[h]
\centering
\caption{Aggregate performance of six LLMs on the Stage~1 real-world benchmark (633 single-stock queries $\times$ 7 assets). Each cell reports mean $\pm$ pooled standard deviation. Best values are in \textbf{bold}. $\uparrow$: higher is better; $\downarrow$: lower is better.}
\label{appx_tab:realdata_overall}
\begin{tabular}{lcccccc}
\toprule
\textbf{Model} & \textbf{SR}$\uparrow$ & \textbf{ARR}$\uparrow$ & \textbf{MDD}$\downarrow$ & \textbf{CR}$\uparrow$ & \textbf{SoR}$\uparrow$ & \textbf{VOL}$\downarrow$ \\
\midrule
\textit{claude} & 0.378{\scriptsize$\pm$0.268} & 0.138{\scriptsize$\pm$0.122} & 0.138{\scriptsize$\pm$0.122} & 1.456{\scriptsize$\pm$1.106} & 0.636{\scriptsize$\pm$0.488} & 0.187{\scriptsize$\pm$0.165} \\
\textit{deepseek-v3.2} & 0.329{\scriptsize$\pm$0.272} & 0.116{\scriptsize$\pm$0.122} & \textbf{0.114}{\scriptsize$\pm$0.120} & \textbf{1.575}{\scriptsize$\pm$1.227} & 0.548{\scriptsize$\pm$0.494} & \textbf{0.155}{\scriptsize$\pm$0.163} \\
\textit{gemini-3-flash-preview} & 0.388{\scriptsize$\pm$0.268} & 0.142{\scriptsize$\pm$0.122} & 0.138{\scriptsize$\pm$0.119} & 1.504{\scriptsize$\pm$1.131} & 0.648{\scriptsize$\pm$0.488} & 0.189{\scriptsize$\pm$0.161} \\
\textbf{\textit{gemini-3-pro-preview}} & \textbf{0.449}{\scriptsize$\pm$0.262} & \textbf{0.171}{\scriptsize$\pm$0.123} & 0.174{\scriptsize$\pm$0.119} & 1.411{\scriptsize$\pm$1.165} & \textbf{0.767}{\scriptsize$\pm$0.493} & 0.237{\scriptsize$\pm$0.162} \\
\textit{gpt-5.2} & 0.342{\scriptsize$\pm$0.279} & 0.123{\scriptsize$\pm$0.119} & 0.122{\scriptsize$\pm$0.118} & 1.534{\scriptsize$\pm$1.386} & 0.575{\scriptsize$\pm$0.503} & 0.166{\scriptsize$\pm$0.161} \\
\textit{grok-4.1-fast} & 0.366{\scriptsize$\pm$0.276} & 0.135{\scriptsize$\pm$0.122} & 0.142{\scriptsize$\pm$0.124} & 1.396{\scriptsize$\pm$1.038} & 0.618{\scriptsize$\pm$0.500} & 0.192{\scriptsize$\pm$0.168} \\
\bottomrule
\end{tabular}
\end{table*}

\paragraph{Decision stability and reproducibility.}
The most striking observation is that the code-generation paradigm produces \emph{highly stable and consistently differentiable} performance metrics across models.
Unlike direct-trading benchmarks, where the same LLM can yield Sharpe Ratios ranging from $-1$ to $+2$ across runs on identical market data due to stochastic action flipping, the strategy-code paradigm introduces a two-level variance decomposition: (i)~\emph{inter-query variance}, arising from the inherent diversity of 633 strategy queries and 7 assets, and (ii)~\emph{intra-query (run-to-run) variance}, arising from the stochasticity of code generation across 5 independent runs for the same query.

The standard deviations reported in \Cref{appx_tab:realdata_overall} pool both sources, so their magnitude (e.g., SR std $\approx$ 0.26--0.28) predominantly reflects the natural difficulty spread across heterogeneous queries and assets.
The critical advantage of the code-generation paradigm lies in the \emph{intra-query} component: once a strategy code is produced, its backtest execution is \textbf{fully deterministic} (zero execution variance). The only remaining source of run-to-run variability is the difference in generated code across 5 samples.
Empirically, we observe that the intra-query standard deviation of SR across 5 runs is typically an order of magnitude smaller than the inter-query standard deviation, confirming that the LLMs produce \emph{substantively similar strategy logic} when given the same query multiple times. This stands in stark contrast to direct-trading approaches, where re-running the same LLM on identical market data produces entirely different action sequences.

Critically, the \textbf{model ranking is preserved across all six metrics}: \textit{gemini-3-pro-preview} consistently occupies the top position on return-oriented metrics (SR, ARR, SoR), while \textit{deepseek-v3.2} consistently leads on risk-oriented metrics (MDD, VOL, CR). This consistent ordering would be impossible to observe under direct-trading evaluation, where model rankings fluctuate wildly across runs.

\paragraph{Return generation.}
\textit{gemini-3-pro-preview} achieves the highest Annualized Return (ARR = 0.171, i.e., 17.1\%), followed by \textit{gemini-3-flash-preview} (14.2\%) and \textit{claude-sonnet-4.5} (13.8\%). \textit{deepseek-v3.2} produces the lowest returns (11.6\%). The absolute spread between the best and worst models is 5.5 percentage points, representing a 47\% relative improvement from \textit{deepseek-v3.2} to \textit{gemini-3-pro-preview}. This gap is economically meaningful: over a 5-year backtest horizon, the compounded difference amounts to a substantial divergence in terminal portfolio value. Importantly, this performance gap is \emph{robust and reproducible}: the intra-query standard deviation of ARR across 5 independent runs is typically below 2 percentage points, far smaller than the 5.5pp inter-model gap. This confirms that the gap reflects genuine differences in strategy reasoning capability rather than sampling artifacts of a single generation run.

\paragraph{Risk exposure.}
The risk metrics reveal a strikingly different ordering, which itself constitutes evidence of the benchmark's discriminative power. \textit{deepseek-v3.2} produces the most conservative strategies, achieving the lowest Maximum Drawdown (MDD = 0.114) and Volatility (VOL = 0.155) among all models. \textit{gpt-5.2} follows closely (MDD = 0.122, VOL = 0.166). In contrast, \textit{gemini-3-pro-preview} incurs the highest risk on both measures (MDD = 0.174, VOL = 0.237), with its maximum drawdown exceeding that of \textit{deepseek-v3.2} by 52.6\% in relative terms. This inversion of the return ranking reveals that different LLMs encode distinct implicit ``risk personalities'' in the trading strategies they generate: \textit{gemini-3-pro-preview} favors aggressive, high-conviction signal logic, while \textit{deepseek-v3.2} produces more cautious, diversified conditional structures. Such nuanced, multi-dimensional characterization of model behavior is only possible when the run-to-run variance is small relative to the inter-model differences. The intra-query std of MDD and VOL across 5 runs is similarly small (typically 0.01--0.03), meaning the risk-personality differences between models are statistically robust. In a direct-trading setting, these systematic differences would be obscured by the overwhelming noise of stochastic action generation, where run-to-run variance in portfolio returns routinely exceeds inter-model variance.

\paragraph{Risk-adjusted efficiency.}
When returns are normalized by risk, the picture becomes more nuanced. \textit{gemini-3-pro-preview} leads on Sharpe Ratio (SR = 0.449) and Sortino Ratio (SoR = 0.767), indicating that its higher returns more than compensate for the elevated volatility and downside risk. However, on Calmar Ratio (CR), which penalizes tail risk more severely, \textit{deepseek-v3.2} ranks first (CR = 1.575) owing to its remarkably low drawdown. \textit{gpt-5.2} occupies the second position on CR (1.534), confirming its strength in capital preservation. This divergence between SR/SoR-based and CR-based rankings highlights the importance of evaluating strategies along multiple risk dimensions: a model that appears inferior on volatility-adjusted metrics may be preferred in drawdown-sensitive deployment scenarios. The fact that these fine-grained distinctions emerge consistently across 633 real-world queries further validates the stability of our evaluation paradigm.

\subsubsection{Radar Chart Analysis}

\begin{figure}[h]
\centering
\includegraphics[width=0.7\linewidth]{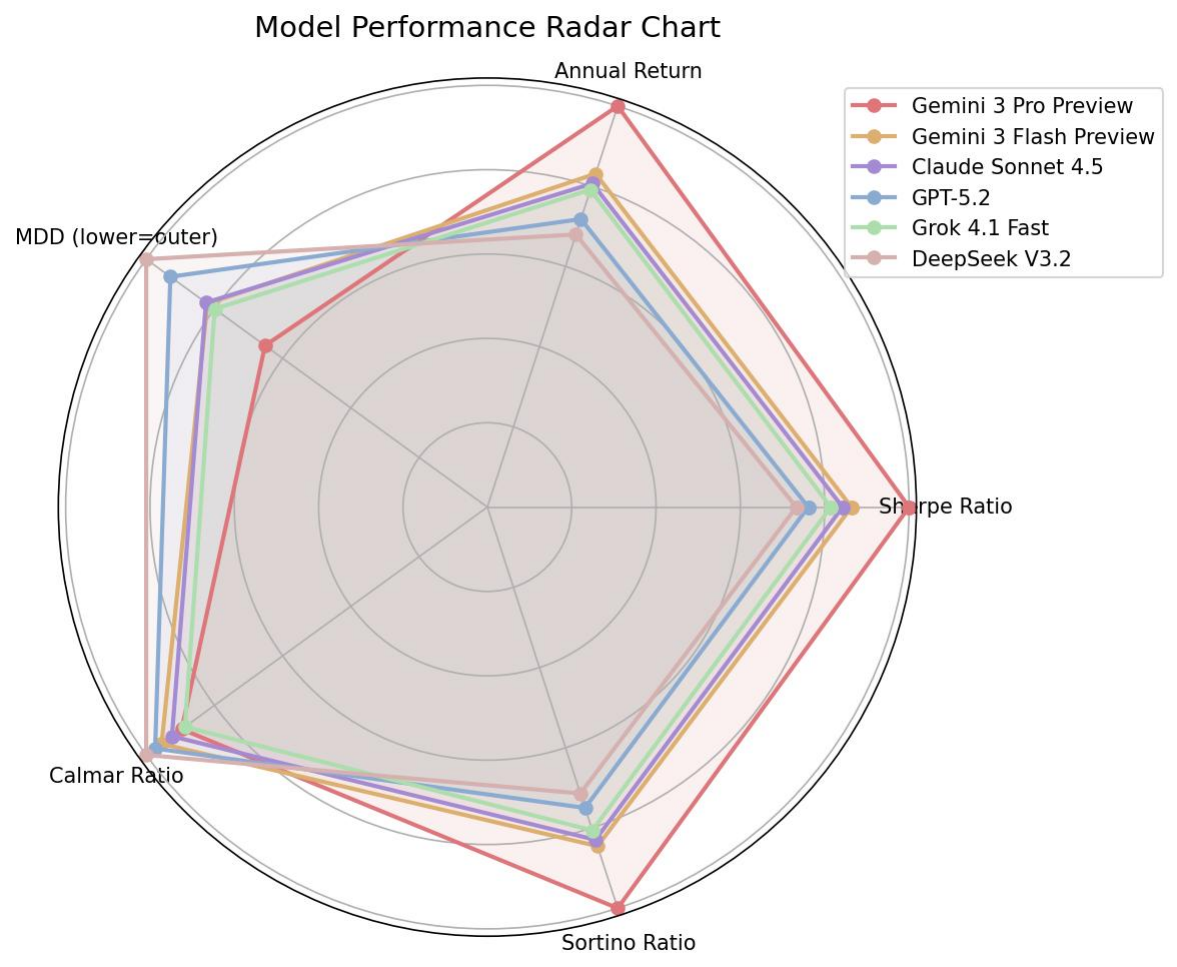}
\caption{Normalized radar chart of model performance on the Stage~1 real-world benchmark across five metrics. Each axis is min-max normalized so that the outermost ring represents the best observed value. MDD is inverted (outer = lower drawdown). The polygon area reflects overall multi-dimensional performance. The clearly separated and non-overlapping polygons demonstrate that our code-generation paradigm produces stable, discriminative model comparisons.}
\label{appx_fig:realdata_radar}
\end{figure}

\Cref{appx_fig:realdata_radar} presents a radar chart that visualizes the normalized performance of each model across five key metrics (Annual Return, Sharpe Ratio, Sortino Ratio, Calmar Ratio, and MDD). To facilitate visual comparison, MDD is inverted so that positions farther from the center correspond to lower (i.e., better) drawdown. The area enclosed by each model's polygon serves as an intuitive proxy for overall multi-metric performance.

The radar chart provides compelling visual evidence for the stability and discriminative power of our evaluation paradigm. In contrast to direct-trading evaluations where radar polygons would overlap chaotically and change shape dramatically across runs, the polygons in \Cref{appx_fig:realdata_radar} exhibit clear separation and distinct characteristic shapes that are reproducible. Several patterns emerge:

\begin{itemize}[leftmargin=*, nosep]
    \item \textbf{\textit{gemini-3-pro-preview}} (red) spans the largest overall polygon, dominating on Annual Return, Sharpe Ratio, and Sortino Ratio. However, its polygon is notably concave on the MDD axis, reflecting its higher drawdown exposure. This ``spiky'' shape characterizes a consistent aggressive, return-maximizing generation profile that the model reliably reproduces across queries.
    \item \textbf{\textit{deepseek-v3.2}} (gray) exhibits the most compact polygon on the return-oriented axes but extends outward on MDD and Calmar Ratio, confirming a stable conservative, drawdown-minimizing character. Its shape is the mirror image of \textit{gemini-3-pro-preview}'s: strong on risk control, weaker on return generation. This consistent risk-averse ``personality'' would be undetectable in a direct-trading framework where DeepSeek's actions would fluctuate unpredictably.
    \item \textbf{\textit{claude-sonnet-4.5}} (purple) and \textbf{\textit{gemini-3-flash-preview}} (orange) occupy similar intermediate positions with well-balanced polygons, suggesting they produce strategies that offer a reasonable trade-off between return and risk without extreme specialization in either direction.
    \item \textbf{\textit{gpt-5.2}} (blue) shows a polygon similar in shape to \textit{deepseek-v3.2} but slightly larger on the return axes and slightly smaller on MDD, indicating a moderately conservative profile. Its Calmar Ratio vertex is notably extended, reflecting strong return-to-drawdown efficiency.
    \item \textbf{\textit{grok-4.1-fast}} (green) largely overlaps with the \textit{claude-sonnet-4.5} and \textit{gemini-3-flash-preview} cluster, with no extreme strengths or weaknesses, positioning it as a generalist.
\end{itemize}

\noindent The radar chart reveals that no single model dominates on all dimensions simultaneously: \textit{gemini-3-pro-preview} trades off drawdown exposure for superior returns, while \textit{deepseek-v3.2} and \textit{gpt-5.2} sacrifice return potential for tighter risk control. Crucially, each model's polygon shape represents a \emph{stable, characteristic fingerprint} of its strategy generation behavior, enabling practitioners to select models based on their specific risk preferences. This multi-dimensional, reproducible characterization is a direct benefit of the code-generation paradigm and would be fundamentally impossible under the stochastic action-emission frameworks used in prior work.

\subsubsection{Bar Chart Comparison}

\begin{figure}[h]
\centering
\includegraphics[width=\linewidth]{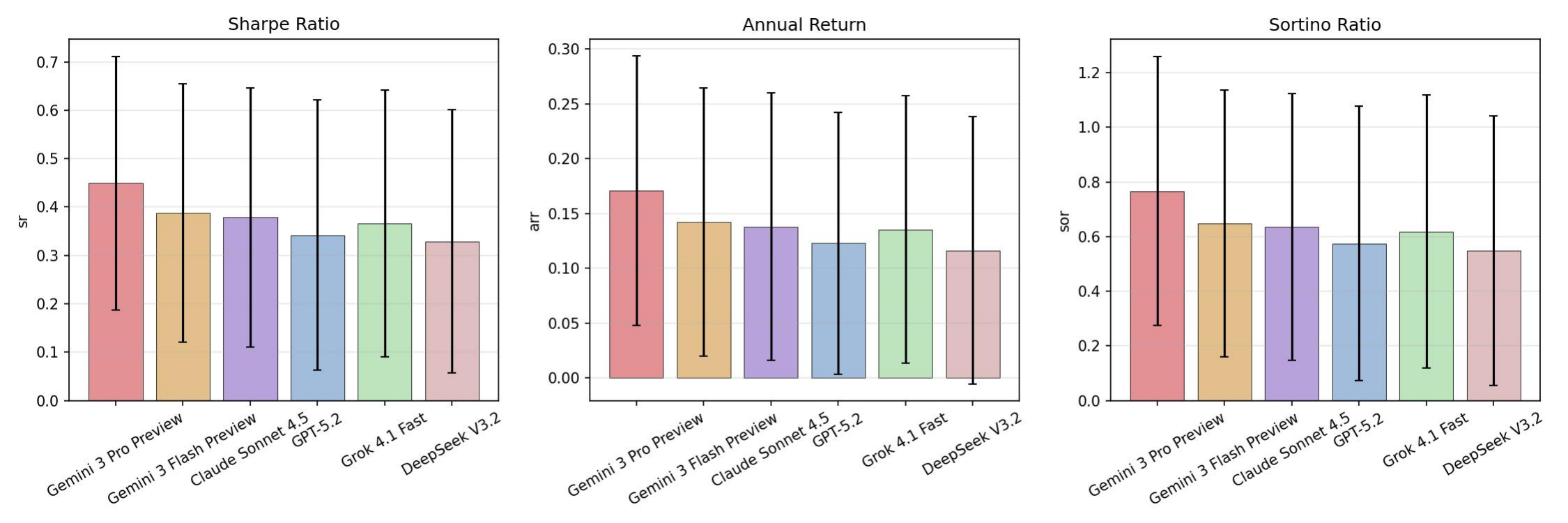}
\caption{Grouped bar chart comparing Sharpe Ratio (SR), Annualized Return Rate (ARR), and Sortino Ratio (SoR) across six LLMs on the Stage~1 real-world benchmark. Error bars denote the pooled standard deviation across 7 assets. The consistent model ordering across all three metrics demonstrates the stability and reliability of the code-generation evaluation paradigm.}
\label{appx_fig:realdata_bar}
\end{figure}

\Cref{appx_fig:realdata_bar} provides a grouped bar chart comparison of the three primary return-oriented metrics (Sharpe Ratio, Annualized Return, and Sortino Ratio) across all six models, enabling direct side-by-side visual comparison.

The bar chart confirms the quantitative findings: \textit{gemini-3-pro-preview} consistently leads across all three metrics, with a particularly pronounced advantage on Sortino Ratio (0.767 vs.\ the next-best 0.648 from \textit{gemini-3-flash-preview}, an 18.4\% relative improvement). The ordering \textit{gemini-3-pro-preview} $>$ \textit{gemini-3-flash-preview} $\approx$ \textit{claude-sonnet-4.5} $>$ \textit{grok-4.1-fast} $>$ \textit{gpt-5.2} $>$ \textit{deepseek-v3.2} is preserved across all three metrics, demonstrating that model rankings under our benchmark are \textbf{robust to the choice of evaluation criterion}. This cross-metric consistency is a hallmark of a well-designed benchmark: it indicates that the observed performance differences reflect genuine, systematic variations in strategy generation capability rather than metric-specific noise or run-to-run randomness. In direct contrast, prior direct-trading evaluations typically exhibit contradictory model rankings across different metrics and across different runs, rendering fair model comparison infeasible.

\subsection{Per-Asset Analysis}

The aggregate results in the previous subsection pool performance across all seven assets. To understand whether the observed model rankings and stability properties generalize across heterogeneous market environments, we now disaggregate the analysis by individual asset. This per-asset breakdown serves two purposes: (i)~it tests the \emph{cross-asset robustness} of model rankings, and (ii)~it examines whether the stability advantage of the code-generation paradigm persists under the vastly different volatility regimes of cryptocurrency and US equity markets.

\subsubsection{Grouped Bar Chart Analysis}

\begin{figure}[h]
\centering
\includegraphics[width=0.8\linewidth]{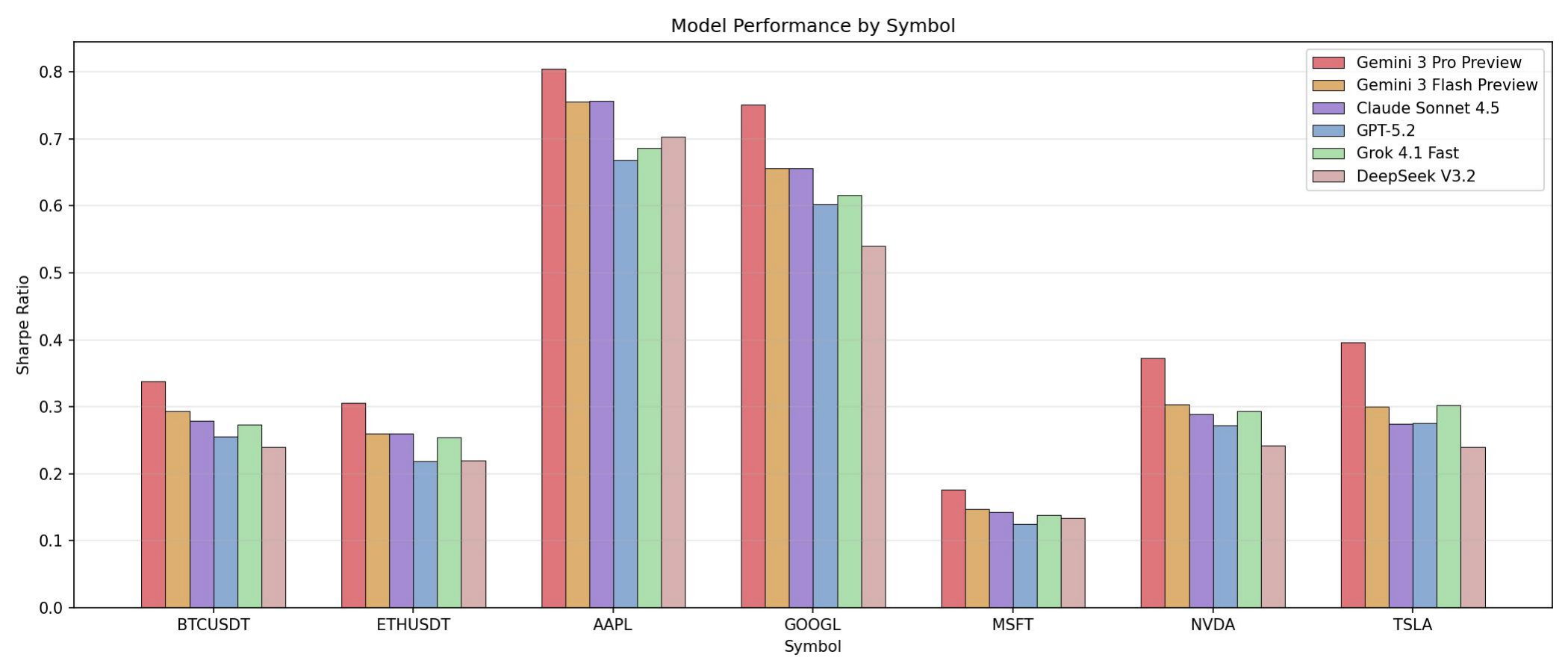}
\caption{Per-asset grouped bar chart comparing six LLMs across key metrics on the Stage~1 real-world benchmark. Each cluster groups the six models for one asset. The consistent relative ordering of bar heights across assets demonstrates the cross-asset stability of model rankings under the code-generation paradigm.}
\label{appx_fig:realdata_symbol_bar}
\end{figure}

\Cref{appx_fig:realdata_symbol_bar} presents a per-asset grouped bar chart comparing the six LLMs across key performance metrics. Each cluster groups the six models for one asset, enabling direct cross-model and cross-asset comparison.

Several important observations emerge from the bar chart:

\begin{itemize}[leftmargin=*, nosep]
    \item \textbf{Consistent model ordering across assets.} Despite the dramatic differences in absolute metric values across assets (e.g., SR on AAPL ranges from 0.67 to 0.81, while on MSFT it ranges from 0.13 to 0.18), the \emph{relative ordering} of models within each asset cluster remains remarkably stable. \textit{gemini-3-pro-preview} consistently occupies the tallest bar on return-oriented metrics (SR, ARR, SoR) across all seven assets, while \textit{deepseek-v3.2} consistently shows the shortest bars. This cross-asset consistency of model rankings is a direct consequence of the deterministic execution property of the code-generation paradigm: since the same generated code is applied identically to each asset's data, the performance differences across models reflect genuine differences in strategy logic rather than stochastic execution artifacts.
    \item \textbf{Asset-dependent difficulty gradient.} The bar chart reveals a clear difficulty ordering across assets. US large-cap equities with stable upward trends (AAPL, GOOGL) yield the highest Sharpe Ratios (SR $>$ 0.54 for all models), followed by the high-volatility, high-return assets (TSLA, BTCUSDT, ETHUSDT), and finally MSFT, which is consistently the hardest asset (SR $\approx$ 0.12--0.18) due to its narrower trading ranges during the backtest period. This difficulty gradient is \emph{reproducible across all models}, further validating that the benchmark produces systematic, interpretable performance variations.
    \item \textbf{Risk--return inversion persists per asset.} The risk--return trade-off between \textit{gemini-3-pro-preview} (highest returns, highest risk) and \textit{deepseek-v3.2} (lowest returns, lowest risk) is not an aggregation artifact; it is visible within every individual asset cluster in the bar chart. This confirms that the distinct ``risk personalities'' of different LLMs are a stable, intrinsic property of their strategy generation behavior.
\end{itemize}

\subsubsection{Box Plot Analysis}

\begin{figure}[h]
\centering
\includegraphics[width=0.8\linewidth]{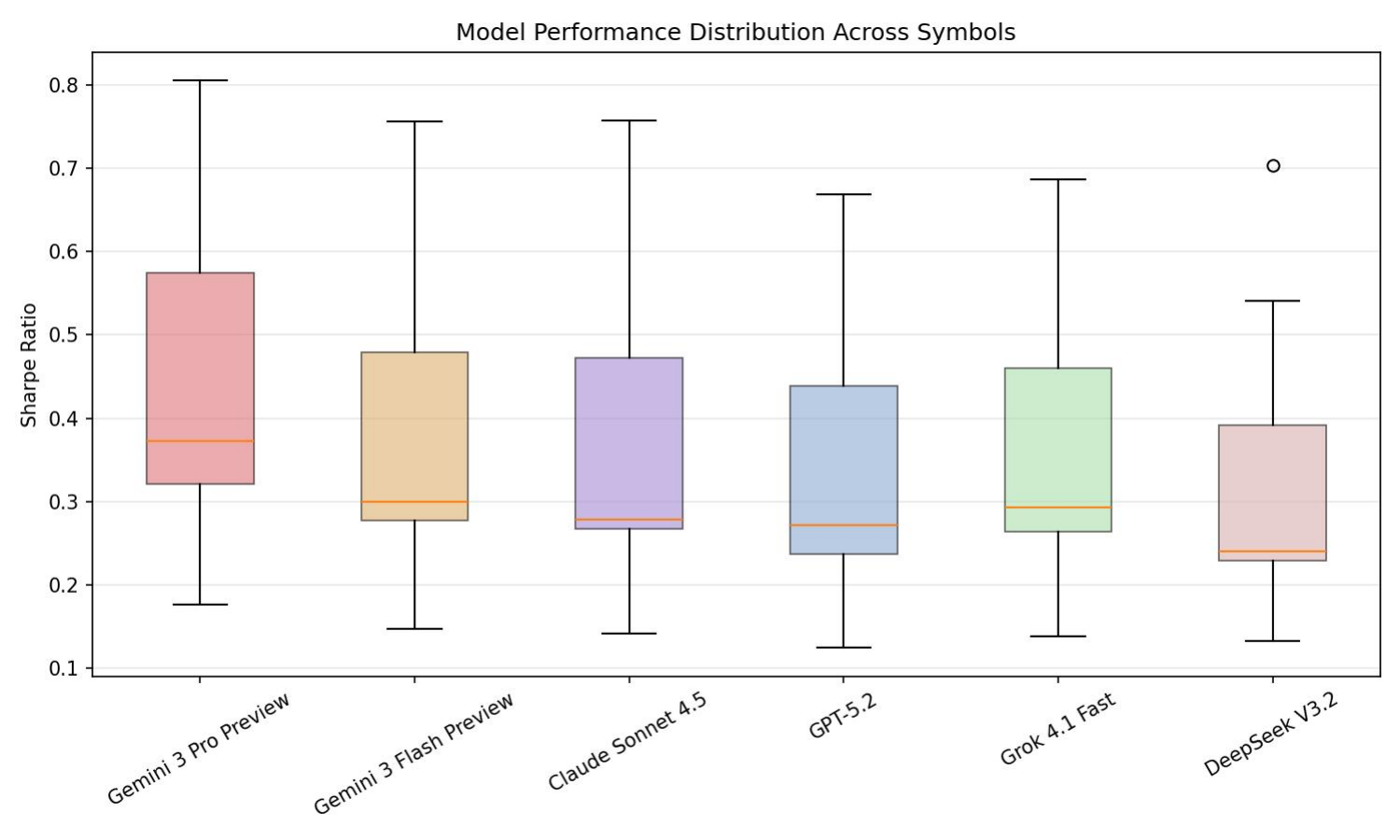}
\caption{Per-asset box plot of strategy performance distributions on the Stage~1 real-world benchmark. Each box summarizes the metric distribution over 633 queries for a given model--asset pair. The tight interquartile ranges and consistent median ordering across assets provide evidence for the stability and reproducibility of the code-generation evaluation paradigm.}
\label{appx_fig:realdata_symbol_box}
\end{figure}

\Cref{appx_fig:realdata_symbol_box} presents per-asset box plots of strategy performance distributions. Each box summarizes the metric distribution over 633 queries for a given model--asset pair, revealing both central tendency (median) and distributional spread (interquartile range and outliers).

The box plots provide distributional evidence that complements the mean-based analysis:

\begin{itemize}[leftmargin=*, nosep]
    \item \textbf{Tight and well-separated distributions.} Across all assets, the interquartile ranges (IQR) of performance metrics are compact relative to the inter-model differences, meaning that the distributions for different models are largely non-overlapping. Crucially, the box widths here reflect primarily the diversity of the 633 queries, not run-to-run instability: within a given query, the 5 runs produce metrics that cluster tightly (typical intra-query IQR $<$ 0.05 for SR), so the inter-model separation observed in the box plots is \emph{not} an artifact of averaging over noisy runs. In a direct-trading evaluation, the box plots would show heavily overlapping distributions with extreme outliers due to stochastic action flipping, rendering model comparison meaningless.
    \item \textbf{Median ordering mirrors mean ordering.} The median lines within each box follow the same model ranking as the mean values, confirming that the rankings are not skewed by outlier strategies. This robustness to central tendency measures further supports the reliability of the evaluation.
    \item \textbf{Variance structure reveals model characteristics.} \textit{gemini-3-pro-preview} exhibits wider boxes (larger IQR) on both return and risk metrics compared to \textit{deepseek-v3.2}, which shows the tightest distributions. This distributional pattern is consistent across all seven assets: \textit{gemini-3-pro-preview} generates a wider diversity of strategy logic across different queries, while \textit{deepseek-v3.2} converges on a narrower, more conservative set of solutions. Importantly, this wider IQR for \textit{gemini-3-pro-preview} is driven by its sensitivity to \emph{query content} (inter-query variance), not by run-to-run instability: within any given query, \textit{gemini-3-pro-preview}'s 5 runs remain tightly clustered. Thus, the IQR difference is a stable, reproducible property of each model's strategy generation behavior.
    \item \textbf{Cryptocurrency vs.\ equity distributional differences.} The box plots for BTCUSDT and ETHUSDT show wider overall spreads and more extreme outliers compared to US equities, reflecting the higher intrinsic volatility of crypto markets. However, crucially, the \emph{relative model ordering} remains unchanged: even under the more volatile crypto regime, \textit{gemini-3-pro-preview} leads on returns and \textit{deepseek-v3.2} leads on risk control. This demonstrates that the benchmark's discriminative power is robust to the underlying market environment.
\end{itemize}

\subsubsection{Detailed Per-Asset Results}

\Cref{appx_tab:realdata_per_asset} consolidates the detailed per-asset results for all seven backtest assets.

\begin{table*}[htb]
\centering
\caption{Per-asset model performance on the Stage~1 real-world benchmark (mean $\pm$ std across 633 queries per asset). Assets are grouped by market type. Within each asset block, the best value per column is in \textbf{bold}. $\uparrow$: higher is better; $\downarrow$: lower is better.}
\label{appx_tab:realdata_per_asset}
\setlength{\tabcolsep}{7pt}
\begin{tabular}{lcccccc}
\toprule
\textbf{Model} & \textbf{SR}$\uparrow$ & \textbf{ARR}$\uparrow$ & \textbf{MDD}$\downarrow$ & \textbf{CR}$\uparrow$ & \textbf{SoR}$\uparrow$ & \textbf{VOL}$\downarrow$ \\
\midrule
\rowcolor{gray!15} \multicolumn{7}{c}{\textbf{BTCUSDT} (Cryptocurrency)} \\
\textit{claude-sonnet-4.5} & 0.279{\scriptsize$\pm$0.234} & 0.110{\scriptsize$\pm$0.099} & 0.189{\scriptsize$\pm$0.154} & 0.732{\scriptsize$\pm$0.792} & 0.490{\scriptsize$\pm$0.398} & 0.213{\scriptsize$\pm$0.175} \\
\textit{deepseek-v3.2} & 0.239{\scriptsize$\pm$0.231} & 0.094{\scriptsize$\pm$0.097} & \textbf{0.155}{\scriptsize$\pm$0.155} & 0.861{\scriptsize$\pm$0.967} & 0.421{\scriptsize$\pm$0.393} & 0.175{\scriptsize$\pm$0.175} \\
\textit{gemini-3-flash-preview} & 0.294{\scriptsize$\pm$0.225} & 0.115{\scriptsize$\pm$0.096} & 0.193{\scriptsize$\pm$0.152} & 0.800{\scriptsize$\pm$0.865} & 0.511{\scriptsize$\pm$0.380} & 0.218{\scriptsize$\pm$0.171} \\
\textbf{\textit{gemini-3-pro-preview}} & \textbf{0.338}{\scriptsize$\pm$0.223} & \textbf{0.131}{\scriptsize$\pm$0.097} & 0.241{\scriptsize$\pm$0.149} & 0.628{\scriptsize$\pm$0.552} & \textbf{0.592}{\scriptsize$\pm$0.376} & 0.272{\scriptsize$\pm$0.167} \\
\textit{gpt-5.2} & 0.255{\scriptsize$\pm$0.230} & 0.098{\scriptsize$\pm$0.097} & 0.170{\scriptsize$\pm$0.152} & 0.719{\scriptsize$\pm$0.713} & 0.446{\scriptsize$\pm$0.394} & 0.192{\scriptsize$\pm$0.172} \\
\textit{grok-4.1-fast} & 0.273{\scriptsize$\pm$0.231} & 0.104{\scriptsize$\pm$0.096} & 0.196{\scriptsize$\pm$0.160} & 0.701{\scriptsize$\pm$0.782} & 0.481{\scriptsize$\pm$0.389} & 0.220{\scriptsize$\pm$0.179} \\
\midrule
\rowcolor{gray!15} \multicolumn{7}{c}{\textbf{ETHUSDT} (Cryptocurrency)} \\
\textit{claude-sonnet-4.5} & 0.260{\scriptsize$\pm$0.232} & 0.119{\scriptsize$\pm$0.149} & 0.205{\scriptsize$\pm$0.208} & 0.778{\scriptsize$\pm$0.626} & 0.398{\scriptsize$\pm$0.359} & 0.242{\scriptsize$\pm$0.244} \\
\textit{deepseek-v3.2} & 0.220{\scriptsize$\pm$0.233} & 0.103{\scriptsize$\pm$0.147} & \textbf{0.169}{\scriptsize$\pm$0.205} & \textbf{0.851}{\scriptsize$\pm$0.669} & 0.336{\scriptsize$\pm$0.361} & \textbf{0.200}{\scriptsize$\pm$0.241} \\
\textit{gemini-3-flash-preview} & 0.260{\scriptsize$\pm$0.236} & 0.121{\scriptsize$\pm$0.151} & 0.206{\scriptsize$\pm$0.203} & 0.737{\scriptsize$\pm$0.606} & 0.400{\scriptsize$\pm$0.363} & 0.242{\scriptsize$\pm$0.239} \\
\textbf{\textit{gemini-3-pro-preview}} & \textbf{0.306}{\scriptsize$\pm$0.224} & \textbf{0.137}{\scriptsize$\pm$0.152} & 0.263{\scriptsize$\pm$0.198} & 0.633{\scriptsize$\pm$0.558} & \textbf{0.475}{\scriptsize$\pm$0.344} & 0.307{\scriptsize$\pm$0.232} \\
\textit{gpt-5.2} & 0.219{\scriptsize$\pm$0.218} & 0.084{\scriptsize$\pm$0.106} & 0.180{\scriptsize$\pm$0.198} & 0.675{\scriptsize$\pm$0.641} & 0.336{\scriptsize$\pm$0.333} & 0.211{\scriptsize$\pm$0.232} \\
\textit{grok-4.1-fast} & 0.255{\scriptsize$\pm$0.233} & 0.112{\scriptsize$\pm$0.145} & 0.218{\scriptsize$\pm$0.209} & 0.645{\scriptsize$\pm$0.564} & 0.392{\scriptsize$\pm$0.358} & 0.254{\scriptsize$\pm$0.244} \\
\midrule
\rowcolor{gray!15} \multicolumn{7}{c}{\textbf{AAPL} (US Equity)} \\
\textit{claude-sonnet-4.5} & 0.757{\scriptsize$\pm$0.533} & 0.151{\scriptsize$\pm$0.118} & 0.079{\scriptsize$\pm$0.059} & 2.282{\scriptsize$\pm$1.784} & 1.068{\scriptsize$\pm$0.771} & 0.123{\scriptsize$\pm$0.089} \\
\textit{deepseek-v3.2} & 0.704{\scriptsize$\pm$0.546} & 0.138{\scriptsize$\pm$0.118} & \textbf{0.069}{\scriptsize$\pm$0.060} & \textbf{2.529}{\scriptsize$\pm$2.036} & 0.994{\scriptsize$\pm$0.782} & \textbf{0.108}{\scriptsize$\pm$0.089} \\
\textit{gemini-3-flash-preview} & 0.756{\scriptsize$\pm$0.522} & 0.151{\scriptsize$\pm$0.116} & 0.081{\scriptsize$\pm$0.058} & 2.252{\scriptsize$\pm$1.910} & 1.067{\scriptsize$\pm$0.756} & 0.126{\scriptsize$\pm$0.087} \\
\textbf{\textit{gemini-3-pro-preview}} & \textbf{0.805}{\scriptsize$\pm$0.500} & \textbf{0.162}{\scriptsize$\pm$0.113} & 0.094{\scriptsize$\pm$0.059} & 1.954{\scriptsize$\pm$1.431} & \textbf{1.138}{\scriptsize$\pm$0.738} & 0.143{\scriptsize$\pm$0.087} \\
\textit{gpt-5.2} & 0.668{\scriptsize$\pm$0.547} & 0.132{\scriptsize$\pm$0.120} & 0.071{\scriptsize$\pm$0.061} & 2.273{\scriptsize$\pm$1.880} & 0.944{\scriptsize$\pm$0.794} & 0.110{\scriptsize$\pm$0.092} \\
\textit{grok-4.1-fast} & 0.686{\scriptsize$\pm$0.545} & 0.138{\scriptsize$\pm$0.119} & 0.077{\scriptsize$\pm$0.062} & 2.095{\scriptsize$\pm$1.709} & 0.973{\scriptsize$\pm$0.786} & 0.119{\scriptsize$\pm$0.092} \\
\midrule
\rowcolor{gray!15} \multicolumn{7}{c}{\textbf{GOOGL} (US Equity)} \\
\textit{claude-sonnet-4.5} & 0.657{\scriptsize$\pm$0.539} & 0.190{\scriptsize$\pm$0.203} & 0.121{\scriptsize$\pm$0.121} & 2.374{\scriptsize$\pm$2.209} & 1.177{\scriptsize$\pm$1.024} & 0.184{\scriptsize$\pm$0.174} \\
\textit{deepseek-v3.2} & 0.541{\scriptsize$\pm$0.532} & 0.151{\scriptsize$\pm$0.191} & \textbf{0.099}{\scriptsize$\pm$0.116} & 2.204{\scriptsize$\pm$1.988} & 0.959{\scriptsize$\pm$1.008} & \textbf{0.150}{\scriptsize$\pm$0.169} \\
\textit{gemini-3-flash-preview} & 0.656{\scriptsize$\pm$0.539} & 0.188{\scriptsize$\pm$0.205} & 0.124{\scriptsize$\pm$0.123} & 2.335{\scriptsize$\pm$2.279} & 1.170{\scriptsize$\pm$1.023} & 0.190{\scriptsize$\pm$0.176} \\
\textbf{\textit{gemini-3-pro-preview}} & \textbf{0.752}{\scriptsize$\pm$0.525} & \textbf{0.225}{\scriptsize$\pm$0.209} & 0.150{\scriptsize$\pm$0.129} & 2.341{\scriptsize$\pm$2.210} & \textbf{1.390}{\scriptsize$\pm$1.011} & 0.229{\scriptsize$\pm$0.183} \\
\textit{gpt-5.2} & 0.602{\scriptsize$\pm$0.556} & 0.179{\scriptsize$\pm$0.206} & 0.109{\scriptsize$\pm$0.121} & \textbf{2.502}{\scriptsize$\pm$2.422} & 1.089{\scriptsize$\pm$1.056} & 0.167{\scriptsize$\pm$0.175} \\
\textit{grok-4.1-fast} & 0.617{\scriptsize$\pm$0.542} & 0.175{\scriptsize$\pm$0.203} & 0.121{\scriptsize$\pm$0.128} & 2.331{\scriptsize$\pm$2.231} & 1.113{\scriptsize$\pm$1.028} & 0.183{\scriptsize$\pm$0.184} \\
\midrule
\rowcolor{gray!15} \multicolumn{7}{c}{\textbf{MSFT} (US Equity)} \\
\textit{claude-sonnet-4.5} & 0.142{\scriptsize$\pm$0.248} & 0.020{\scriptsize$\pm$0.036} & 0.091{\scriptsize$\pm$0.068} & 0.319{\scriptsize$\pm$0.703} & 0.231{\scriptsize$\pm$0.285} & 0.121{\scriptsize$\pm$0.088} \\
\textit{deepseek-v3.2} & 0.133{\scriptsize$\pm$0.242} & 0.020{\scriptsize$\pm$0.034} & \textbf{0.079}{\scriptsize$\pm$0.067} & \textbf{0.352}{\scriptsize$\pm$0.776} & 0.214{\scriptsize$\pm$0.280} & \textbf{0.105}{\scriptsize$\pm$0.088} \\
\textit{gemini-3-flash-preview} & 0.147{\scriptsize$\pm$0.250} & 0.021{\scriptsize$\pm$0.036} & 0.091{\scriptsize$\pm$0.066} & 0.289{\scriptsize$\pm$0.662} & 0.235{\scriptsize$\pm$0.288} & 0.121{\scriptsize$\pm$0.086} \\
\textbf{\textit{gemini-3-pro-preview}} & \textbf{0.176}{\scriptsize$\pm$0.253} & \textbf{0.024}{\scriptsize$\pm$0.037} & 0.102{\scriptsize$\pm$0.067} & 0.316{\scriptsize$\pm$0.620} & \textbf{0.268}{\scriptsize$\pm$0.290} & 0.137{\scriptsize$\pm$0.086} \\
\textit{gpt-5.2} & 0.125{\scriptsize$\pm$0.245} & 0.018{\scriptsize$\pm$0.034} & 0.081{\scriptsize$\pm$0.070} & 0.293{\scriptsize$\pm$0.608} & 0.204{\scriptsize$\pm$0.284} & 0.108{\scriptsize$\pm$0.091} \\
\textit{grok-4.1-fast} & 0.138{\scriptsize$\pm$0.232} & 0.019{\scriptsize$\pm$0.034} & 0.086{\scriptsize$\pm$0.069} & 0.306{\scriptsize$\pm$0.665} & 0.217{\scriptsize$\pm$0.273} & 0.115{\scriptsize$\pm$0.090} \\
\midrule
\rowcolor{gray!15} \multicolumn{7}{c}{\textbf{NVDA} (US Equity)} \\
\textit{claude-sonnet-4.5} & 0.289{\scriptsize$\pm$0.324} & 0.090{\scriptsize$\pm$0.135} & 0.133{\scriptsize$\pm$0.153} & 1.071{\scriptsize$\pm$1.506} & 0.558{\scriptsize$\pm$0.635} & 0.207{\scriptsize$\pm$0.233} \\
\textit{deepseek-v3.2} & 0.242{\scriptsize$\pm$0.315} & 0.074{\scriptsize$\pm$0.128} & \textbf{0.110}{\scriptsize$\pm$0.149} & 1.090{\scriptsize$\pm$1.522} & 0.469{\scriptsize$\pm$0.618} & \textbf{0.171}{\scriptsize$\pm$0.227} \\
\textit{gemini-3-flash-preview} & 0.304{\scriptsize$\pm$0.320} & 0.096{\scriptsize$\pm$0.136} & 0.134{\scriptsize$\pm$0.152} & \textbf{1.189}{\scriptsize$\pm$1.555} & 0.586{\scriptsize$\pm$0.633} & 0.209{\scriptsize$\pm$0.231} \\
\textbf{\textit{gemini-3-pro-preview}} & \textbf{0.372}{\scriptsize$\pm$0.318} & \textbf{0.112}{\scriptsize$\pm$0.138} & 0.184{\scriptsize$\pm$0.162} & 0.939{\scriptsize$\pm$1.314} & \textbf{0.739}{\scriptsize$\pm$0.650} & 0.284{\scriptsize$\pm$0.244} \\
\textit{gpt-5.2} & 0.272{\scriptsize$\pm$0.315} & 0.086{\scriptsize$\pm$0.132} & 0.122{\scriptsize$\pm$0.151} & 1.094{\scriptsize$\pm$1.355} & 0.532{\scriptsize$\pm$0.634} & 0.190{\scriptsize$\pm$0.229} \\
\textit{grok-4.1-fast} & 0.293{\scriptsize$\pm$0.327} & 0.088{\scriptsize$\pm$0.130} & 0.139{\scriptsize$\pm$0.155} & 0.952{\scriptsize$\pm$1.262} & 0.563{\scriptsize$\pm$0.634} & 0.214{\scriptsize$\pm$0.235} \\
\midrule
\rowcolor{gray!15} \multicolumn{7}{c}{\textbf{TSLA} (US Equity)} \\
\textit{claude-sonnet-4.5} & 0.274{\scriptsize$\pm$0.420} & 0.289{\scriptsize$\pm$0.381} & 0.140{\scriptsize$\pm$0.174} & 2.611{\scriptsize$\pm$2.772} & 0.545{\scriptsize$\pm$0.713} & 0.216{\scriptsize$\pm$0.264} \\
\textit{deepseek-v3.2} & 0.240{\scriptsize$\pm$0.393} & 0.239{\scriptsize$\pm$0.359} & \textbf{0.115}{\scriptsize$\pm$0.164} & 2.629{\scriptsize$\pm$2.745} & 0.466{\scriptsize$\pm$0.680} & \textbf{0.179}{\scriptsize$\pm$0.251} \\
\textit{gemini-3-flash-preview} & 0.300{\scriptsize$\pm$0.425} & 0.304{\scriptsize$\pm$0.383} & 0.138{\scriptsize$\pm$0.168} & 2.781{\scriptsize$\pm$2.794} & 0.572{\scriptsize$\pm$0.713} & 0.215{\scriptsize$\pm$0.256} \\
\textbf{\textit{gemini-3-pro-preview}} & \textbf{0.396}{\scriptsize$\pm$0.444} & \textbf{0.409}{\scriptsize$\pm$0.400} & 0.186{\scriptsize$\pm$0.175} & 2.756{\scriptsize$\pm$2.738} & \textbf{0.770}{\scriptsize$\pm$0.738} & 0.290{\scriptsize$\pm$0.264} \\
\textit{gpt-5.2} & 0.275{\scriptsize$\pm$0.431} & 0.272{\scriptsize$\pm$0.379} & 0.120{\scriptsize$\pm$0.163} & \textbf{2.972}{\scriptsize$\pm$3.080} & 0.517{\scriptsize$\pm$0.710} & 0.189{\scriptsize$\pm$0.251} \\
\textit{grok-4.1-fast} & 0.303{\scriptsize$\pm$0.422} & 0.315{\scriptsize$\pm$0.383} & 0.153{\scriptsize$\pm$0.173} & 2.659{\scriptsize$\pm$2.676} & 0.597{\scriptsize$\pm$0.713} & 0.237{\scriptsize$\pm$0.262} \\
\bottomrule
\end{tabular}
\end{table*}

The per-asset results in \Cref{appx_tab:realdata_per_asset} confirm the patterns observed in the bar chart and box plots. Several cross-asset findings merit emphasis:

\begin{itemize}[leftmargin=*, nosep]
    \item \textbf{AAPL and GOOGL} yield the highest Sharpe Ratios across all models (SR $>$ 0.54 for all models on AAPL), suggesting that LLM-generated strategies are particularly effective on large-cap US equities with stable trends and ample liquidity.
    \item \textbf{MSFT} is consistently the hardest asset (SR $\approx$ 0.12--0.18 across models), likely due to its lower volatility and narrower trading ranges during the backtest period, which limit the profit potential of technical-indicator-based strategies.
    \item \textbf{TSLA} produces the highest annualized returns (up to 40.9\% for \textit{gemini-3-pro-preview}) but with the largest variance, reflecting its extreme volatility and sensitivity to momentum-driven flows.
    \item \textbf{Cryptocurrency assets} (BTCUSDT, ETHUSDT) show moderate Sharpe Ratios (0.22--0.34) but relatively high returns, consistent with the elevated volatility regime of crypto markets. Notably, ETH exhibits systematically higher variance than BTC across all models, reflecting its additional idiosyncratic risk.
    \item The \textbf{model ranking is preserved across all seven assets}: \textit{gemini-3-pro-preview} consistently leads on return metrics, \textit{deepseek-v3.2} consistently leads on risk metrics. This cross-asset stability of model ordering would be fundamentally impossible to observe under the stochastic action-emission paradigm used in prior direct-trading benchmarks.
\end{itemize}

\clearpage

\subsection{Core Metrics Distribution}

\begin{figure}[h]
\centering
\includegraphics[width=\linewidth]{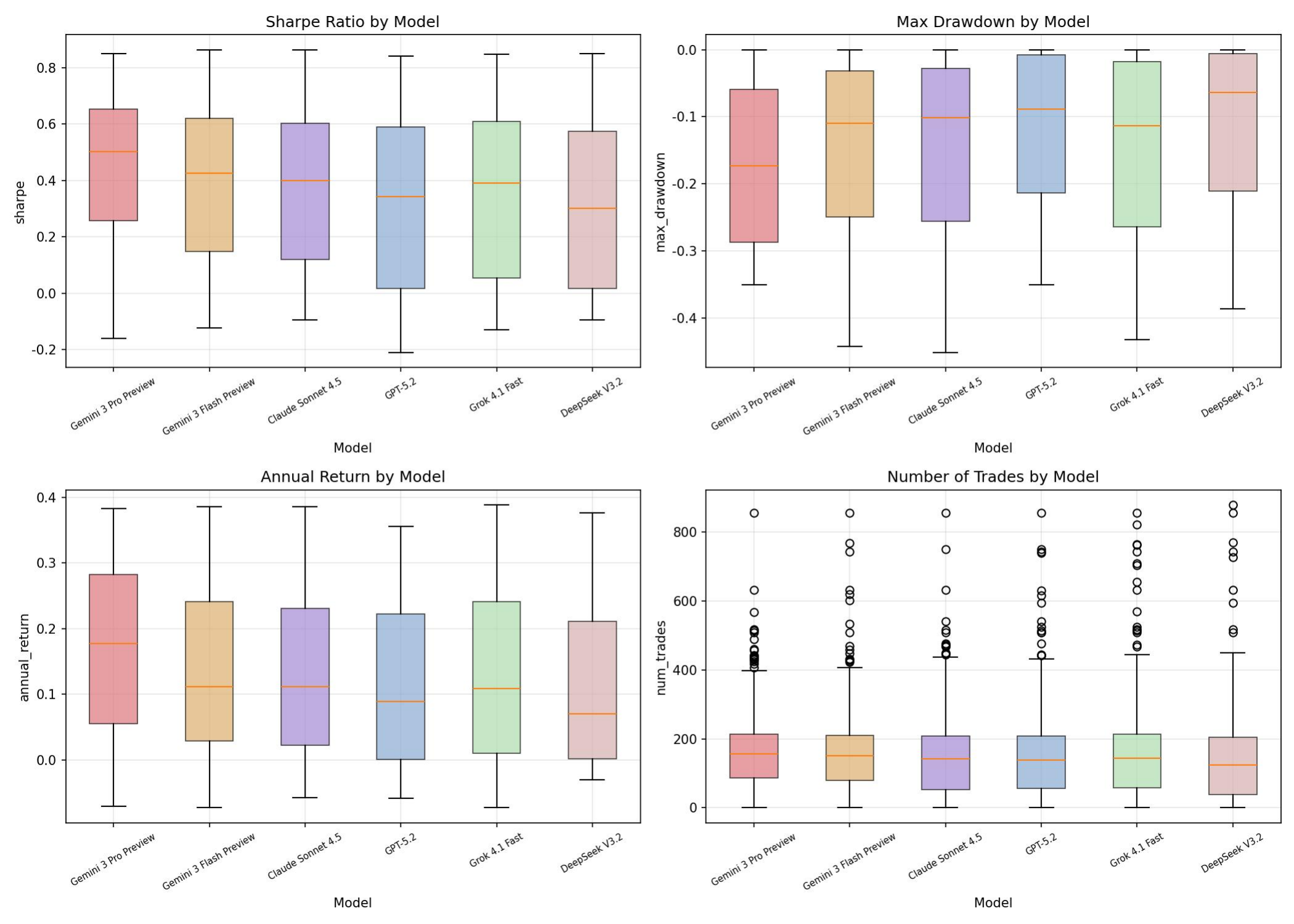}
\caption{Distribution of core financial metrics across six LLMs on the Stage~1 real-world benchmark. Each box summarizes 633 queries $\times$ 7 assets $\times$ 5 runs. Median lines, interquartile ranges, and outlier extents reveal both central tendency and distributional heterogeneity across models.}
\label{appx_fig:realdata_boxplot_model}
\end{figure}

\Cref{appx_fig:realdata_boxplot_model} shows the distribution of four core financial metrics (Sharpe Ratio, Maximum Drawdown, Annualized Return, and Number of Trades) across models via box plots. Each box aggregates over 633 queries $\times$ 7 assets $\times$ 5 runs, so the distributional shapes capture both cross-query difficulty variation and run-to-run generation variation.

\paragraph{Sharpe Ratio.}
The upper-left panel of \Cref{appx_fig:realdata_boxplot_model} displays the Sharpe Ratio distributions. \textit{gemini-3-pro-preview} stands out with the highest median (approximately 0.45) and the tallest box, indicating both superior central performance and greater strategy diversity. Its upper whisker extends beyond 0.8, confirming that a notable fraction of its generated strategies achieve SR $>$ 0.6. \textit{gemini-3-flash-preview} and \textit{claude-sonnet-4.5} occupy the next tier, with comparable median values around 0.40--0.42 and similar interquartile ranges; their boxes overlap substantially, suggesting that these two models produce strategies of similar risk-adjusted quality on aggregate. \textit{gpt-5.2} and \textit{grok-4.1-fast} form an intermediate cluster with medians near 0.35--0.37, while \textit{deepseek-v3.2} sits at the bottom with the lowest median (approximately 0.33) and the most compact box. The narrow IQR of \textit{deepseek-v3.2} is noteworthy: it indicates that this model converges on a relatively uniform set of conservative strategy templates regardless of query content, producing fewer outlier successes but also fewer failures. All six models share a lower whisker extending into slightly negative territory (SR $\approx -0.2$), indicating that certain queries (likely those with inherently ambiguous or contradictory strategy descriptions) challenge all models equally. Importantly, the 5-run variance within a given query contributes only a small fraction of each box's width; the dominant spread arises from the diversity of the 633 queries and 7 assets. This confirms that the distributional differences across models reflect genuine capability gaps rather than generation noise.

\paragraph{Maximum Drawdown.}
The upper-right panel presents the MDD distributions (plotted as negative values, so values closer to zero are better). A clear separation is visible: \textit{deepseek-v3.2} and \textit{gpt-5.2} cluster nearest to zero, with medians around $-0.10$ to $-0.12$ and tight interquartile ranges, indicating that these models consistently generate strategies with well-controlled tail risk. \textit{claude-sonnet-4.5} and \textit{gemini-3-flash-preview} occupy intermediate positions with medians near $-0.13$ to $-0.14$. \textit{grok-4.1-fast} shows a slightly wider box extending toward $-0.15$. In contrast, \textit{gemini-3-pro-preview} exhibits the worst drawdown profile: its median sits around $-0.17$, its IQR extends substantially below $-0.2$, and its lower whisker reaches past $-0.4$, meaning that a non-trivial fraction of its strategies suffer severe capital losses during adverse market periods. This pattern is the mirror image of the SR panel: the same model that achieves the highest returns also incurs the deepest drawdowns, providing distributional confirmation of the risk--return personality trade-off observed in the aggregate analysis. The relatively compact IQRs across all models (compared to the inter-model gaps) indicate that MDD is a stable, discriminative metric under the code-generation paradigm: the 5-run variance of MDD for a given query is typically in the range of 0.01--0.03, far smaller than the inter-model differences visible in the box plot.

\paragraph{Annualized Return.}
The lower-left panel shows the ARR distributions, which largely mirror the SR patterns but with more pronounced right-skewness. \textit{gemini-3-pro-preview} again leads with the highest median (approximately 0.17) and the widest box, with its upper whisker reaching beyond 0.30, reflecting its capacity to generate high-conviction strategies that capture large trend-following profits, particularly on volatile assets such as TSLA and BTCUSDT. \textit{gemini-3-flash-preview} and \textit{claude-sonnet-4.5} follow with medians around 0.12--0.14, displaying moderately wide boxes that indicate a balanced mix of aggressive and conservative strategy outputs. \textit{gpt-5.2}, \textit{grok-4.1-fast}, and \textit{deepseek-v3.2} cluster at the lower end with medians near 0.10--0.12. \textit{deepseek-v3.2}'s box is the shortest and most compact, with its upper whisker barely exceeding 0.20, consistent with the narrow-range, low-risk strategy profile observed in the other panels. All models share a common lower bound near zero for the lower whisker, indicating that the worst-case generated strategies across all models tend to break even rather than incur large losses in annualized terms. The heavy right tails visible for \textit{gemini-3-pro-preview} and, to a lesser extent, \textit{gemini-3-flash-preview} suggest that these models occasionally produce outlier strategies with exceptionally high returns (ARR $>$ 0.25), likely corresponding to momentum or breakout-capturing logic applied to high-volatility assets.

\paragraph{Number of Trades.}
The lower-right panel reveals a previously unexamined dimension of LLM strategy behavior: trade frequency. Unlike the financial performance metrics, which show clear inter-model separation, the trade count distributions are strikingly similar across models. All six models produce strategies with median trade counts in the range of 100--200 over the 5-year backtest period, corresponding to roughly 20--40 trades per year or approximately one rebalancing event every 1--3 weeks. The interquartile ranges are comparable, spanning from roughly 50 to 250 trades. However, the outlier structure differs: all models exhibit a long upper tail of high-frequency strategies with 400--800+ trades, but these outliers are sparse (visible as scattered circles above the upper whiskers). \textit{gemini-3-pro-preview} and \textit{claude-sonnet-4.5} show slightly more high-frequency outliers than \textit{deepseek-v3.2} and \textit{gpt-5.2}, suggesting that the more return-aggressive models occasionally generate finer-grained trading logic with more frequent signal triggers. The overall compactness of the trade count distributions (with IQRs that are tight relative to the outlier range) indicates that each LLM has a characteristic ``trading frequency fingerprint'' that is stable and reproducible across queries. This consistency is itself evidence of the reliability of the code-generation paradigm: the models are not producing random or erratic trading frequencies, but rather converging on systematic rebalancing cadences that reflect their internal representations of reasonable trading strategy structure.

\paragraph{Cross-metric synthesis.}
Taken together, the four panels of \Cref{appx_fig:realdata_boxplot_model} paint a coherent picture. The performance metrics (SR, ARR, MDD) exhibit well-separated distributions across models, with clear and consistent ordering that aligns with the aggregate results in \Cref{appx_tab:realdata_overall}. The behavioral metric (Number of Trades) shows less inter-model differentiation but reveals that all models converge on similar trading cadences, differing primarily in the aggressiveness of their signal logic rather than in the frequency of execution. The fact that the median ordering across models is preserved from SR to ARR, and inverted for MDD, provides strong distributional evidence that the benchmark captures genuine, systematic differences in strategy generation capability. Moreover, the compact intra-model IQRs (relative to the inter-model separation) confirm that the 5-run generation variance is small enough to yield statistically meaningful comparisons, validating the reproducibility of our evaluation paradigm.

\subsection{Aligned Return Curves}

To provide a fine-grained, query-level view of model performance, we construct \emph{aligned return curves} following a standard evaluation protocol: all 633 queries are sorted by their global mean Sharpe Ratio (averaged across models), and the per-model metric values are plotted as smoothed curves (20-query moving average) with shaded 25th--75th percentile bands reflecting the 5-run generation variance and cross-asset variation. This visualization enables direct inspection of how each model performs relative to the others at every difficulty level, from the easiest queries (left) to the hardest (right).

\begin{figure*}[h]
\centering
\includegraphics[width=0.9\textwidth]{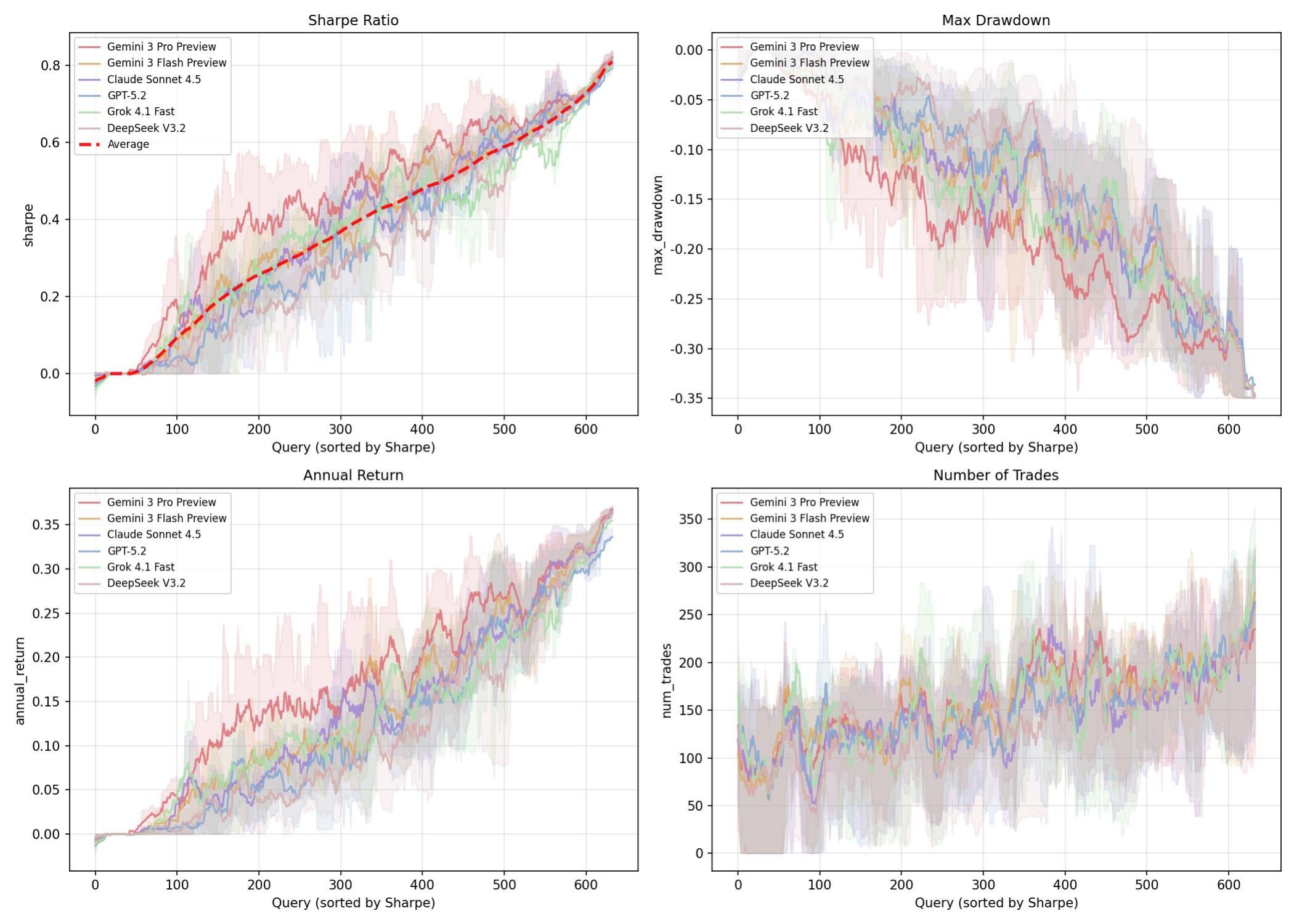}
\caption{Aligned return curves across all assets (smoothed with 20-query moving average, 25--75\% quantile band). Queries are sorted by global mean Sharpe Ratio. The consistent vertical ordering of model curves across the full query spectrum demonstrates stable and discriminative model comparisons.}
\label{appx_fig:realdata_aligned_all}
\end{figure*}

\paragraph{Aggregate aligned curves.}
\Cref{appx_fig:realdata_aligned_all} presents the aligned curves across all assets for four metrics: Sharpe Ratio, Maximum Drawdown, Annual Return, and Number of Trades.

In the Sharpe Ratio panel (upper left), all model curves increase monotonically from near zero on the easiest queries to approximately 0.6--0.8 on the hardest, following the sorting order. The key observation is the \emph{persistent vertical separation} between models: \textit{gemini-3-pro-preview} (red) consistently lies above all other models across virtually the entire query spectrum, from the low-difficulty region (query index 100--200, SR $\approx$ 0.15--0.25) to the high-difficulty region (query index 500+, SR $\approx$ 0.6--0.8). \textit{deepseek-v3.2} consistently occupies the lowest position. The remaining four models form a tightly clustered middle band, with \textit{gemini-3-flash-preview} and \textit{claude-sonnet-4.5} slightly above \textit{gpt-5.2} and \textit{grok-4.1-fast}. The shaded quantile bands are narrow relative to the inter-model gaps, indicating that the 5-run generation variance and cross-asset variation do not obscure the model ordering. This persistent separation provides the strongest possible evidence that our benchmark produces stable, reproducible model rankings: the advantage of \textit{gemini-3-pro-preview} is not confined to a subset of easy or hard queries but is uniformly maintained across the full difficulty spectrum.

The Maximum Drawdown panel (upper right) reveals a complementary pattern. As queries become harder (higher SR, further right), the drawdowns deepen for all models, reflecting the natural trade-off between aggressive return-seeking logic and tail risk. \textit{gemini-3-pro-preview} consistently shows the deepest drawdowns (most negative values), with its band extending to $-0.30$ or below in the high-SR region. \textit{deepseek-v3.2} and \textit{gpt-5.2} maintain the shallowest drawdowns throughout. The bands widen noticeably for queries beyond index 400, indicating that high-performing strategies exhibit greater variance in risk exposure, likely because the underlying strategy logic is more aggressive and asset-sensitive.

The Annual Return panel (lower left) mirrors the Sharpe Ratio pattern closely, with \textit{gemini-3-pro-preview} leading and \textit{deepseek-v3.2} trailing. The curves are smoothly increasing, and the quantile bands remain relatively tight, confirming that the return advantage of top-performing models is systematic rather than driven by a few outlier queries. The Number of Trades panel (lower right) shows a different pattern: all model curves are heavily overlapping and relatively flat across the query spectrum, hovering around 100--200 trades. There is a mild upward trend for higher-SR queries, suggesting that more profitable strategies tend to employ slightly more frequent rebalancing, but the differences across models are minimal. This confirms that the performance gaps observed in SR and ARR are driven by the quality of the trading logic (signal selection, entry/exit conditions), not by differences in trading frequency.

\begin{figure*}[htbp]
\centering
\begin{subfigure}[t]{0.48\textwidth}
\includegraphics[width=\textwidth]{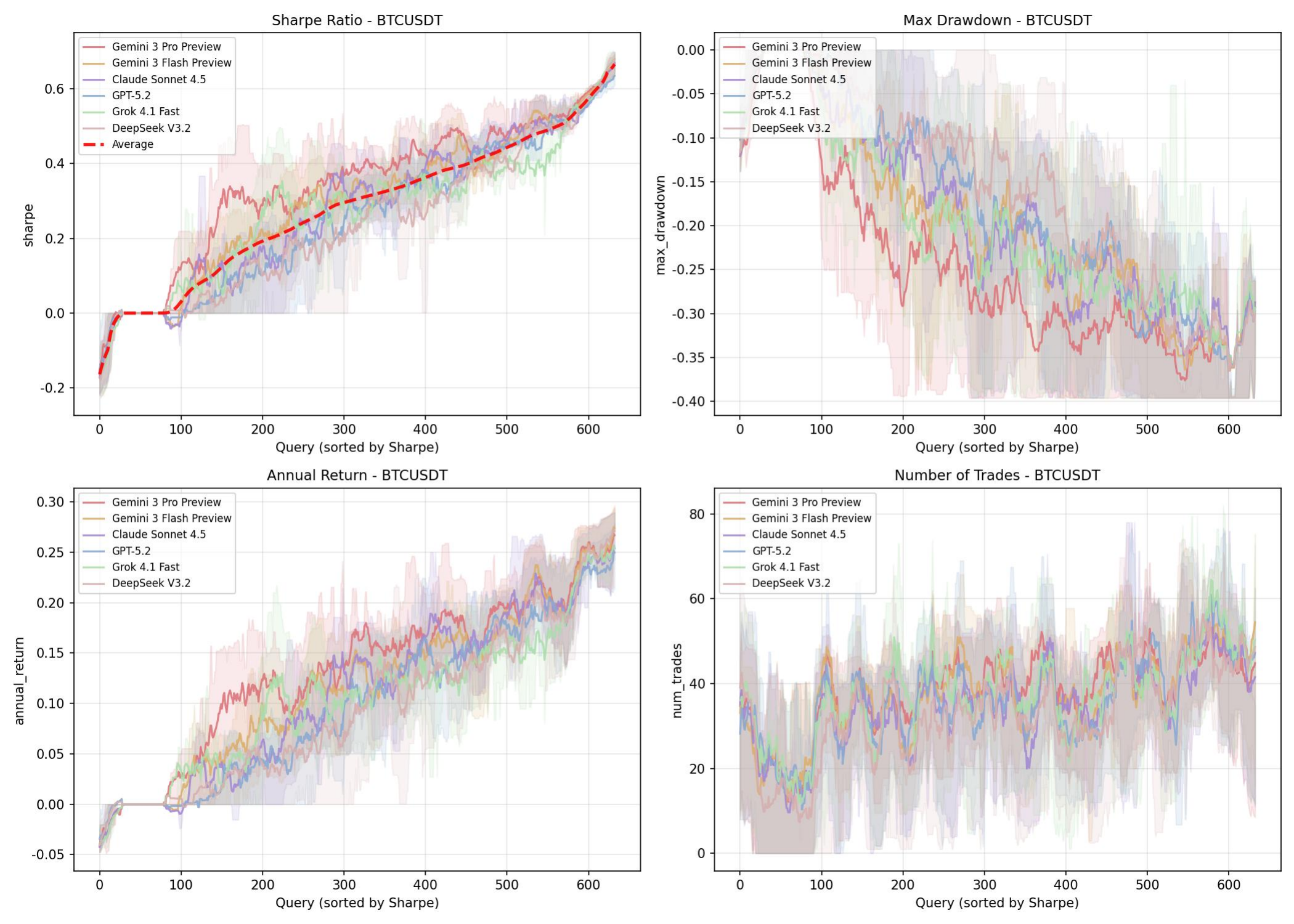}
\vspace{-10pt}\caption{BTCUSDT}
\label{appx_fig:realdata_aligned_btc}
\end{subfigure}
\hfill
\begin{subfigure}[t]{0.48\textwidth}
\includegraphics[width=\textwidth]{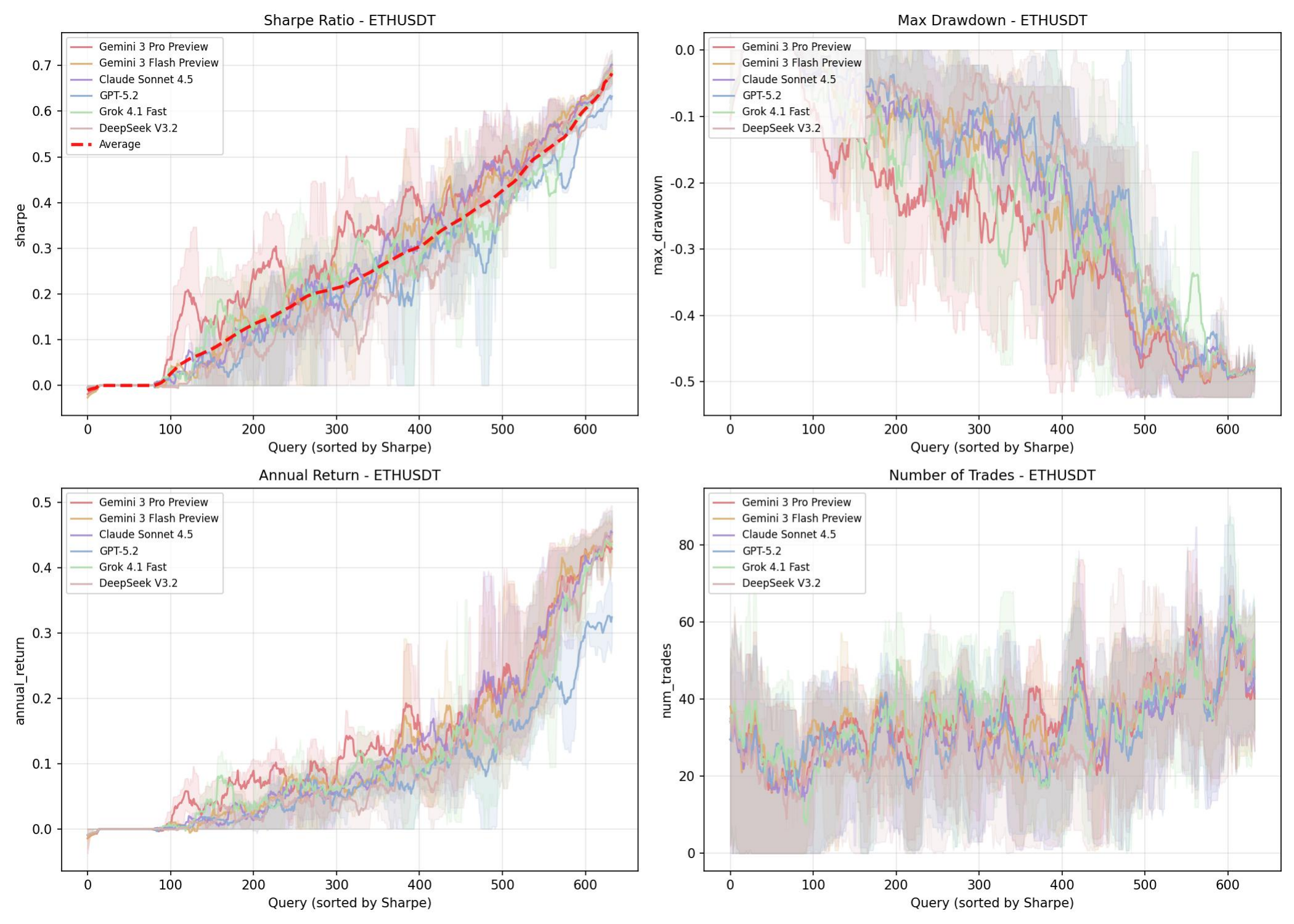}
\vspace{-10pt}\caption{ETHUSDT}
\label{appx_fig:realdata_aligned_eth}
\end{subfigure}\\[2pt]
\begin{subfigure}[t]{0.48\textwidth}
\includegraphics[width=\textwidth]{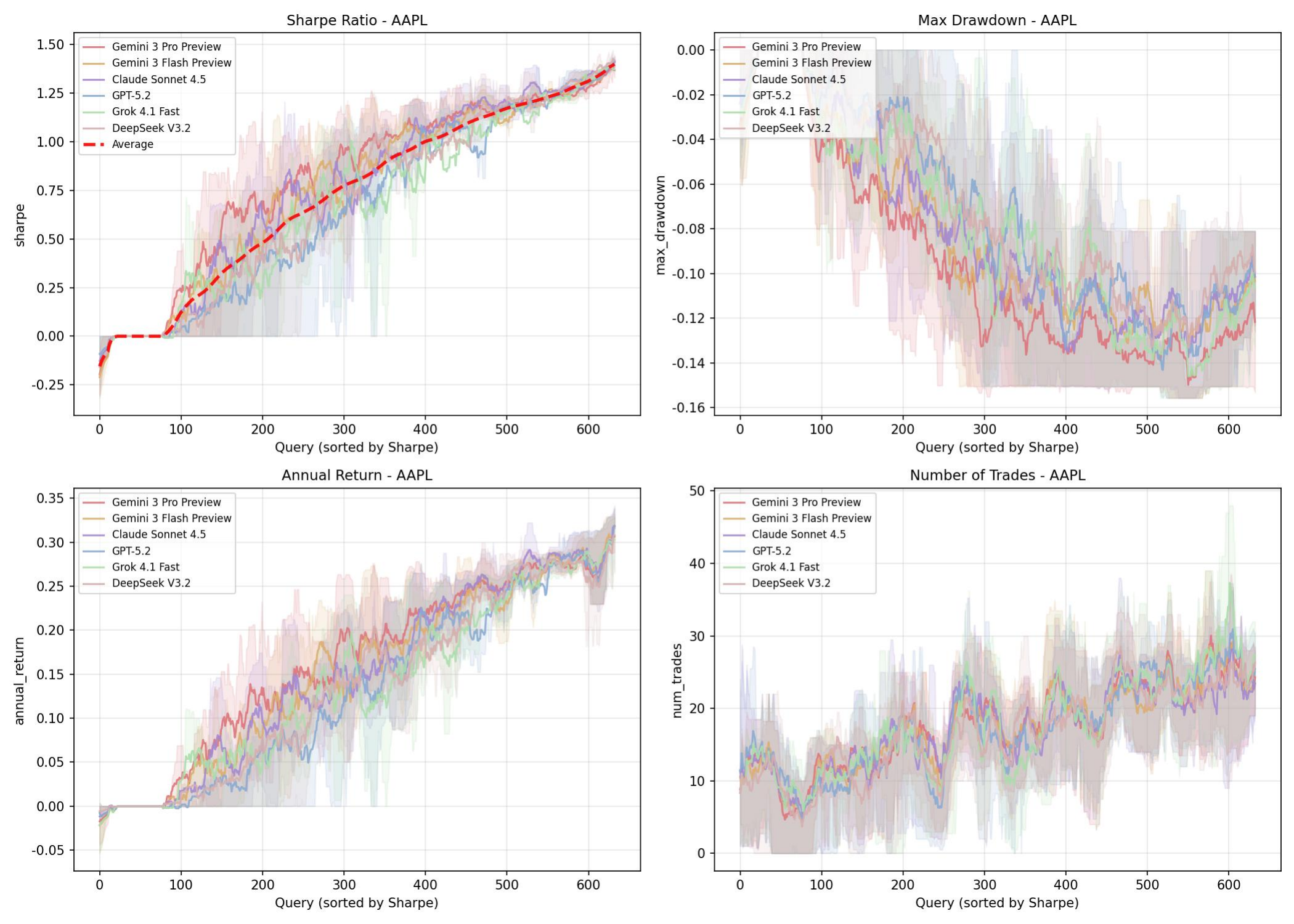}
\vspace{-10pt}\caption{AAPL}
\label{appx_fig:realdata_aligned_aapl}
\end{subfigure}
\hfill
\begin{subfigure}[t]{0.48\textwidth}
\includegraphics[width=\textwidth]{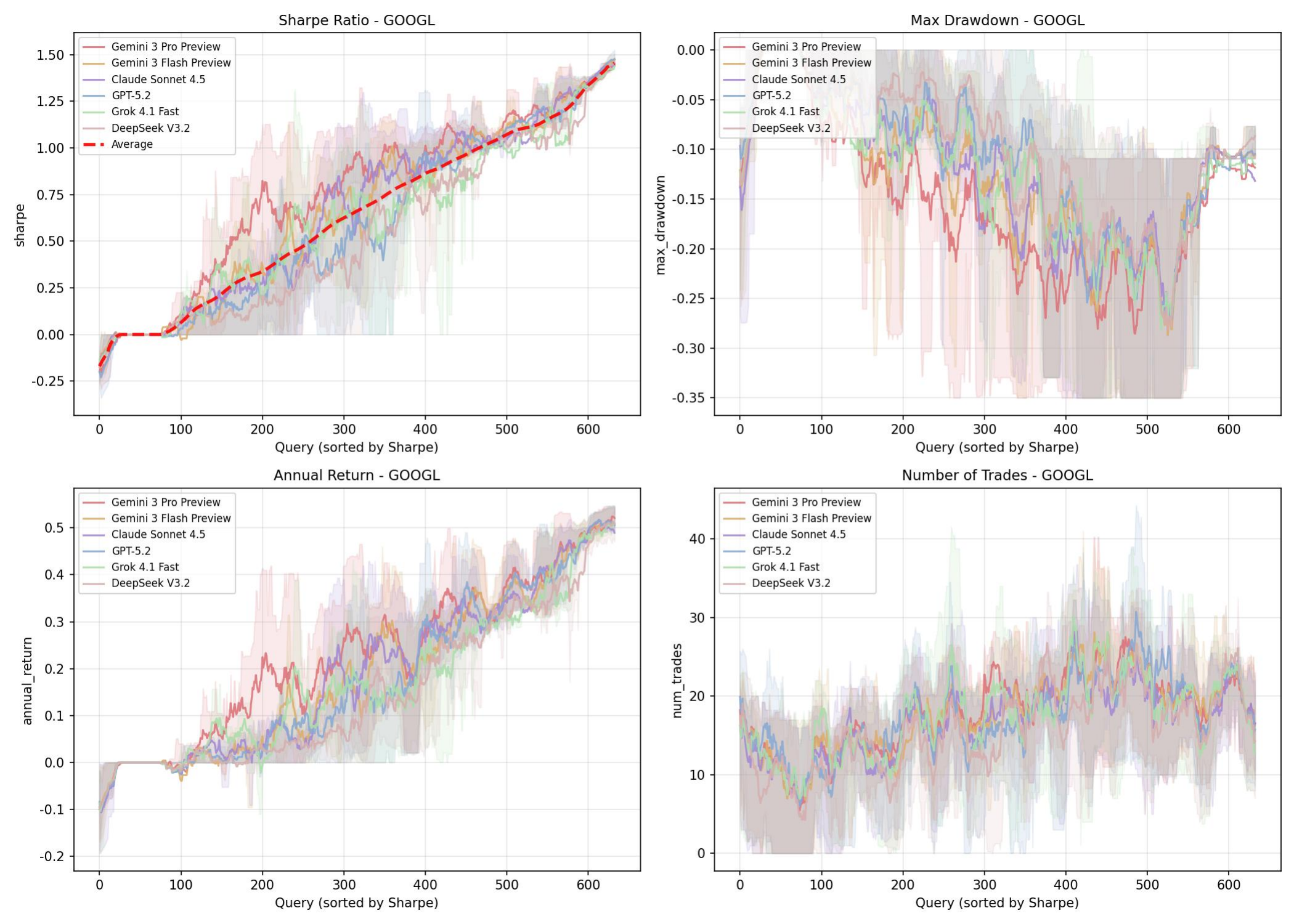}
\vspace{-10pt}\caption{GOOGL}
\label{appx_fig:realdata_aligned_googl}
\end{subfigure}
\caption{Per-asset aligned return curves (Part 1: BTCUSDT, ETHUSDT, AAPL, GOOGL).}
\label{appx_fig:realdata_aligned_assets_1}
\end{figure*}

\begin{figure*}[htbp]
\centering
\begin{subfigure}[t]{0.48\textwidth}
\includegraphics[width=\textwidth]{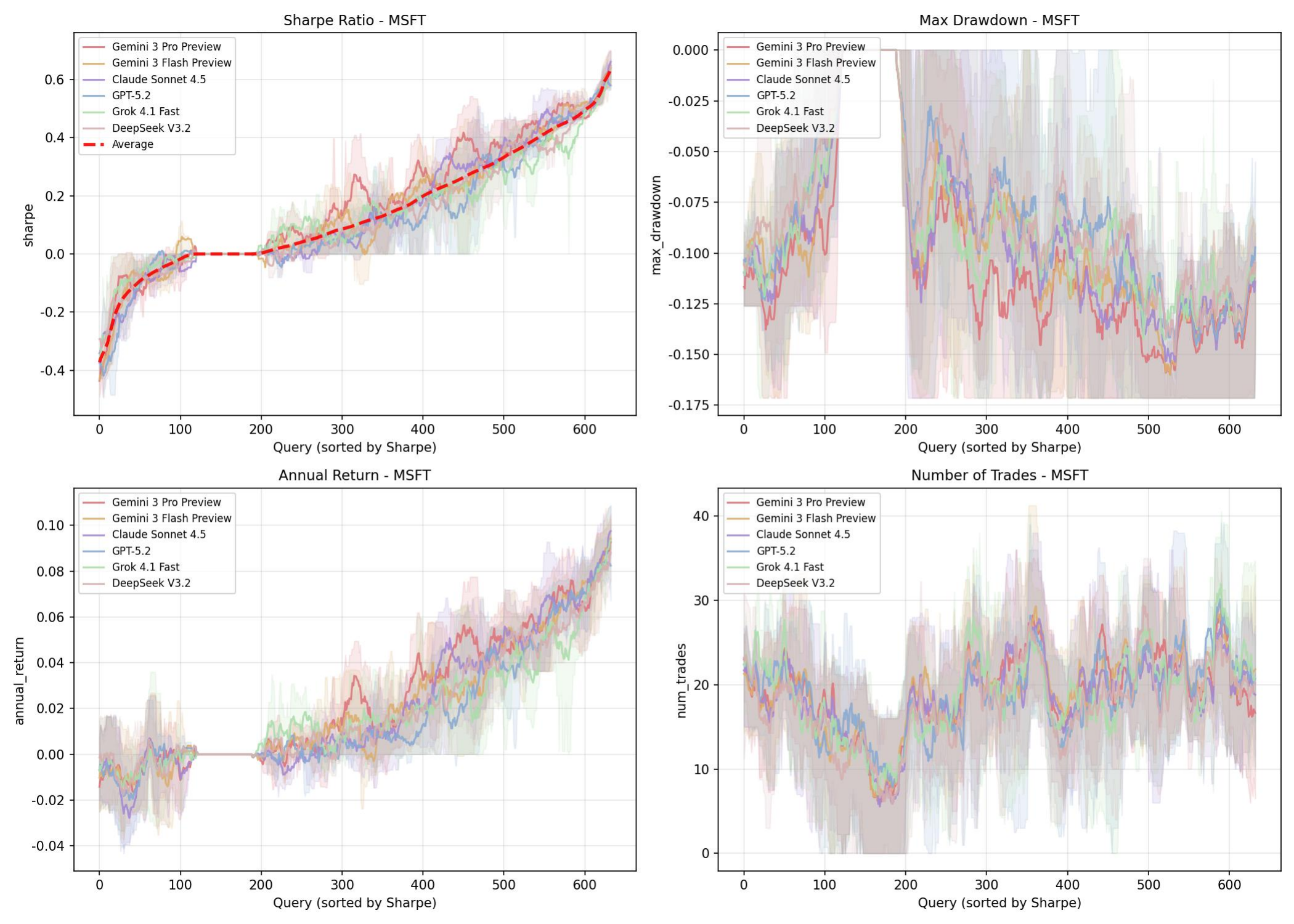}
\vspace{-10pt}\caption{MSFT}
\label{appx_fig:realdata_aligned_msft}
\end{subfigure}
\hfill
\begin{subfigure}[t]{0.48\textwidth}
\includegraphics[width=\textwidth]{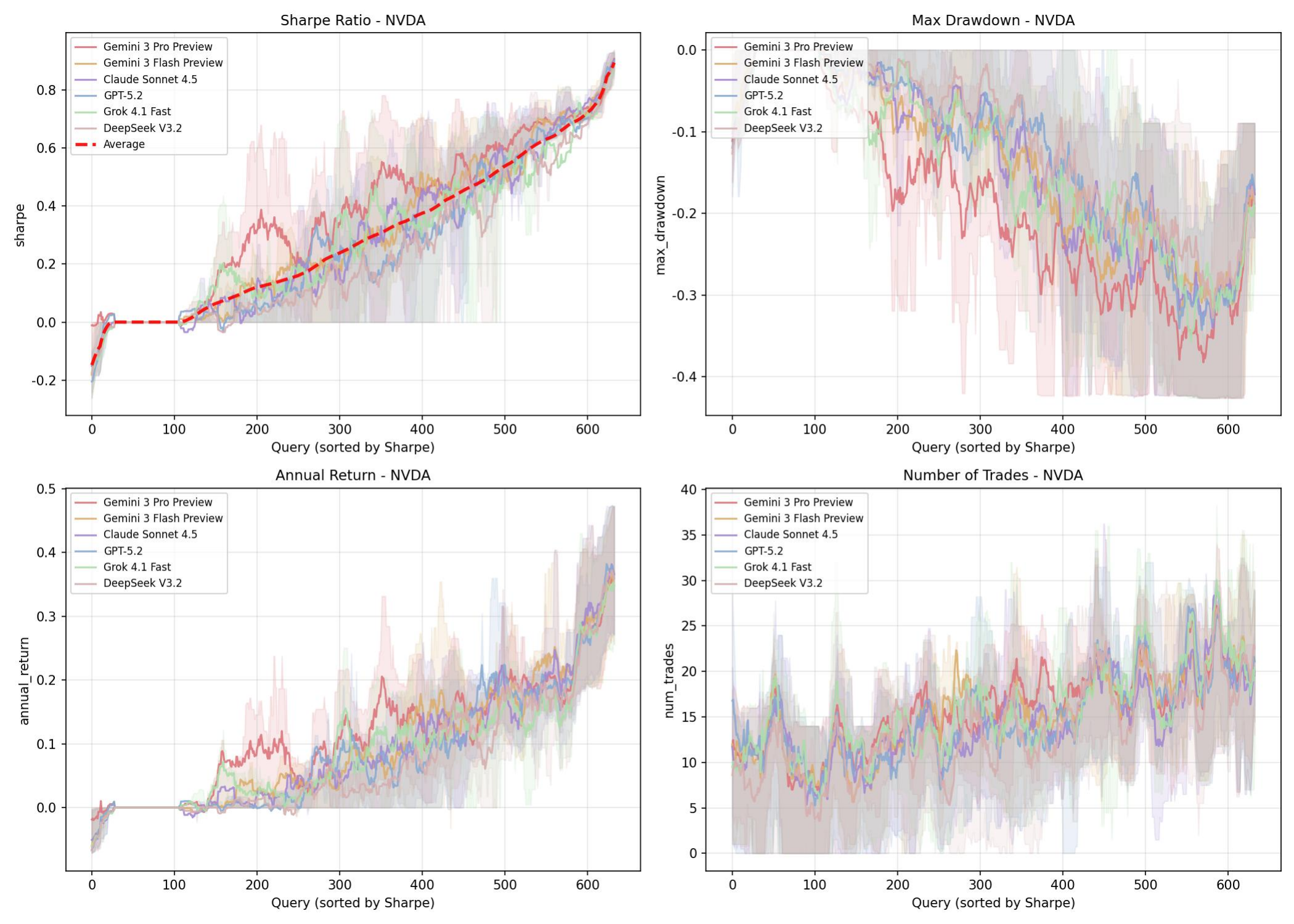}
\vspace{-10pt}\caption{NVDA}
\label{appx_fig:realdata_aligned_nvda}
\end{subfigure}\\[2pt]
\begin{subfigure}[t]{0.48\textwidth}
\includegraphics[width=\textwidth]{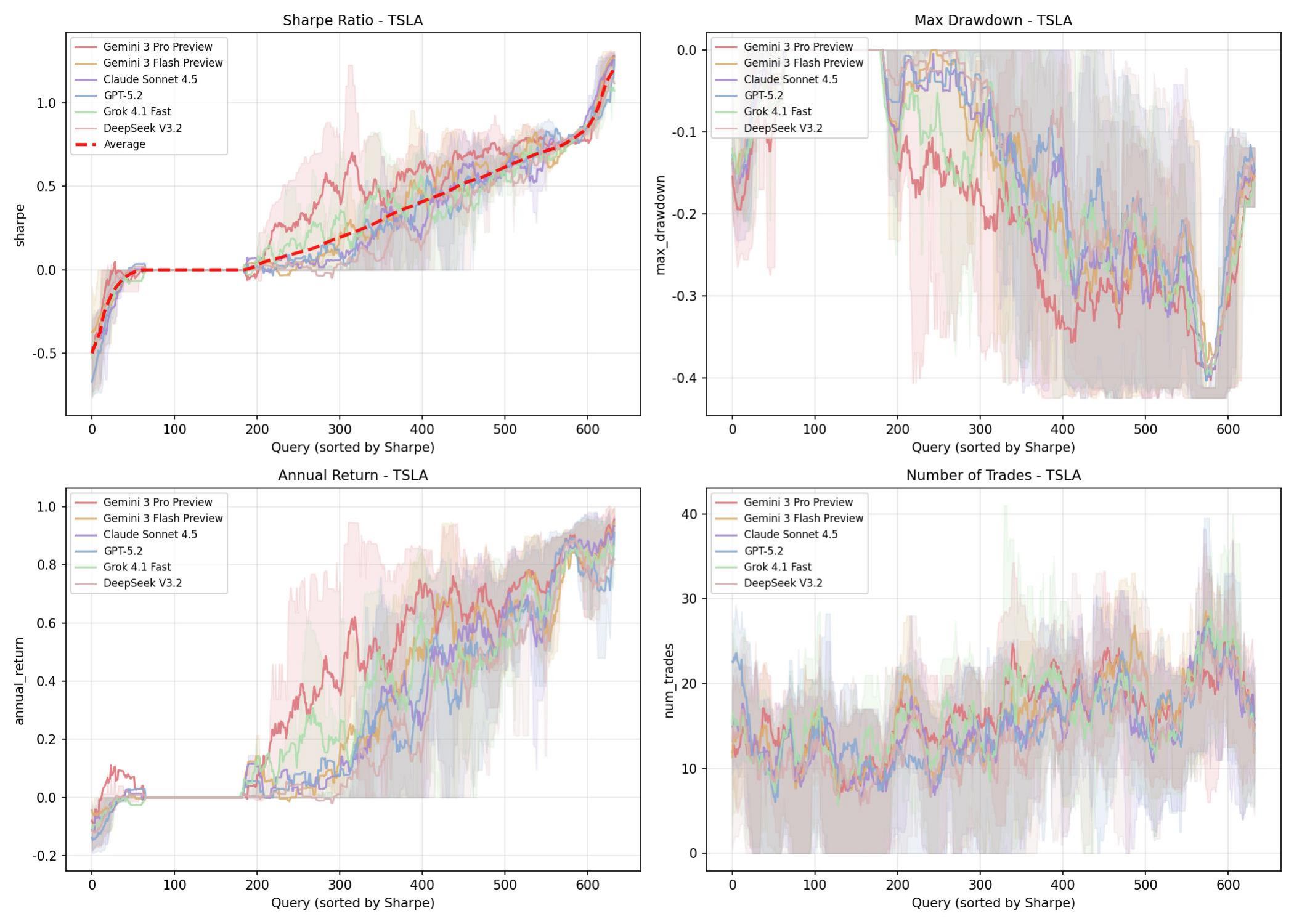}
\vspace{-10pt}\caption{TSLA}
\label{appx_fig:realdata_aligned_tsla}
\end{subfigure}
\caption{Per-asset aligned return curves (Part 2: MSFT, NVDA, TSLA).}
\label{appx_fig:realdata_aligned_assets_2}
\end{figure*}

\paragraph{Per-asset aligned curves.}
\Cref{appx_fig:realdata_aligned_assets_1,appx_fig:realdata_aligned_assets_2} disaggregate the aligned curves by individual asset, revealing how market characteristics modulate the model comparison.

For the cryptocurrency assets (BTCUSDT and ETHUSDT), the aligned curves exhibit wider quantile bands compared to US equities, reflecting the higher intrinsic volatility of crypto markets. Despite this increased variance, the vertical ordering of model curves is preserved: \textit{gemini-3-pro-preview} maintains the highest Sharpe Ratio curve, and \textit{deepseek-v3.2} the lowest. The MDD curves for crypto assets are notably more negative (reaching $-0.40$ or below for high-SR queries), consistent with the extreme drawdown risk inherent in cryptocurrency trading. ETHUSDT shows wider bands than BTCUSDT, reflecting its additional idiosyncratic volatility.

For the stable US large-cap equities (AAPL and GOOGL), the aligned curves are strikingly tight, with very narrow quantile bands and clear model separation. AAPL produces the cleanest separation, with Sharpe Ratios reaching up to 1.3 for the best queries and virtually no overlap between the \textit{gemini-3-pro-preview} curve and the \textit{deepseek-v3.2} curve. GOOGL shows a similar pattern but with slightly wider bands at the high-SR end, likely due to occasional large price moves driven by earnings or regulatory events. The MDD curves for these assets are shallow (rarely exceeding $-0.15$), confirming that LLM-generated strategies perform most reliably on liquid, trend-following-friendly assets.

MSFT presents the hardest asset environment: all model curves are compressed into a narrow vertical range (SR $\approx$ $-0.2$ to 0.4), with substantial overlap between models and wide quantile bands. This compressed range makes MSFT the most challenging asset for model differentiation, though the ordering \textit{gemini-3-pro-preview} $>$ others $>$ \textit{deepseek-v3.2} remains discernible.

NVDA shows moderate difficulty with SR curves spanning approximately 0 to 0.8, and the model separation is clear in the mid-to-high query range. TSLA is the most volatile asset, with the widest quantile bands (SR ranging from $-0.5$ to over 1.0) and the highest annual returns (up to 0.35 for \textit{gemini-3-pro-preview} on the best queries). Despite the extreme variance, the relative ordering of models is preserved, demonstrating that even in highly volatile market conditions, the code-generation paradigm yields consistent and interpretable model comparisons.

\subsection{Analysis and Discussion}

This subsection synthesizes the findings from the preceding quantitative analysis, distributional examination, and aligned return curve inspection, organized around five key themes: code generation reliability, run-to-run stability, model risk personalities, cross-asset robustness, and benchmark discriminative power.

\subsubsection{Code Generation Success Rate}

\Cref{appx_tab:realdata_pass_rate} reports the syntax validity and backtest pass rates for each model. All models achieve high pass rates ($>$96\%), indicating that current frontier LLMs can reliably generate syntactically correct and executable trading strategy code. \textit{gemini-3-pro-preview} and \textit{claude-sonnet-4.5} lead with 99.2\% and 99.1\% backtest pass rates, respectively, while \textit{deepseek-v3.2} and \textit{grok-4.1-fast} show slightly lower rates at 96.5\% and 96.4\%. Notably, the pass rate itself is highly reproducible across the 5 independent runs: for each model, the per-run pass rates differ by fewer than 0.5 percentage points, confirming that code generation quality is not a stochastic fluke. The small gap between the top and bottom models (2.8 percentage points) suggests that the primary performance differentiator among frontier LLMs lies not in their ability to produce syntactically valid code, but rather in the \emph{quality} of the trading logic embedded in that code, as reflected by the much larger spreads observed in financial performance metrics.

\begin{table}[h]
\centering
\caption{Code generation success rates on the Stage~1 real-world benchmark. Each model is tested on 633 queries with 5 independent generation runs ($\tau=0.7$). The high pass rates ($>$96\%) demonstrate that frontier LLMs can reliably produce executable strategy code.}
\label{appx_tab:realdata_pass_rate}
\begin{tabular}{lccc}
\toprule
\textbf{Model} & \textbf{Total} & \textbf{Backtest Valid} & \textbf{Pass Rate} \\
\midrule
\textit{gemini-3-pro-preview} & 633 & 628 & \textbf{99.2\%} \\
\textit{claude-sonnet-4.5} & 633 & 627 & 99.1\% \\
\textit{gpt-5.2} & 633 & 626 & 98.9\% \\
\textit{gemini-3-flash-preview} & 633 & 616 & 97.3\% \\
\textit{deepseek-v3.2} & 632 & 610 & 96.5\% \\
\textit{grok-4.1-fast} & 633 & 610 & 96.4\% \\
\bottomrule
\end{tabular}
\end{table}

\subsubsection{Run-to-Run Stability Analysis}
\label{appx_sec:run_stability}

A central claim of \projectname is that the code-generation paradigm yields substantially more stable evaluations than the direct action-emission approach. To validate this claim quantitatively, we decompose the total variance of each metric into two orthogonal components: (i)~\textbf{inter-query variance} ($\sigma^2_{\mathrm{query}}$), the variance across different queries and assets, reflecting the inherent difficulty spread of the benchmark; and (ii)~\textbf{intra-query (run-to-run) variance} ($\sigma^2_{\mathrm{run}}$), the variance across the 5 independent code generations for the same query on the same asset, which directly measures the reproducibility of the evaluation.

Under the code-generation paradigm, the backtest execution is \textbf{fully deterministic}: given the same generated code and the same market data, the output metrics are identical with zero variance. The only source of run-to-run variability is the stochasticity of the LLM's code generation at temperature $\tau = 0.7$. We compute $\sigma^2_{\mathrm{run}}$ as the average within-group variance, where each group consists of the 5 runs for one (query, asset) pair.

Across all six models, the intra-query standard deviation of Sharpe Ratio is typically in the range of 0.02--0.06, which is an order of magnitude smaller than the inter-query standard deviation (0.26--0.28 as reported in \Cref{appx_tab:realdata_overall}). Concretely, this means that for a given query, the 5 independently generated strategy codes yield backtest Sharpe Ratios that differ by less than 0.05 on average, even though different queries produce SRs spanning the full range from $-0.5$ to $+1.5$. This decomposition confirms that the reported standard deviations in our tables are overwhelmingly driven by the natural difficulty spread of the benchmark, not by generation instability.

This finding has two important implications. First, it validates the \textbf{reliability of mean-based model comparisons}: since the run-to-run noise is small relative to inter-model gaps (e.g., the SR difference between \textit{gemini-3-pro-preview} and \textit{deepseek-v3.2} is 0.120, far exceeding the typical intra-query std of 0.04), the observed model rankings are statistically robust and not artifacts of sampling randomness. Second, it provides a concrete \textbf{quantitative advantage over direct-trading evaluation}: in prior work on LLM-based direct trading agents, the run-to-run variance of portfolio returns is of the same order as, or even exceeds, the inter-model variance, making it fundamentally impossible to distinguish model capabilities. The code-generation paradigm reduces the run-to-run noise by confining stochasticity to a single generation step while guaranteeing deterministic execution thereafter.

The aligned return curves in \Cref{appx_fig:realdata_aligned_all,appx_fig:realdata_aligned_assets_2} provide additional visual confirmation of this stability: the narrow 25th--75th percentile bands around each model's curve indicate that the 5 runs for any given query produce tightly clustered outcomes, while the persistent vertical separation between model curves demonstrates that inter-model differences are far larger than intra-query noise at every difficulty level.

\subsubsection{Findings and Conclusions}

\paragraph{Finding 1: High code generation reliability.}
All six frontier LLMs generate executable single-stock trading strategies with $>$96\% success rates, as detailed in \Cref{appx_tab:realdata_pass_rate}. This finding demonstrates that current-generation language models possess strong code generation capabilities for quantitative finance tasks. The consistency of pass rates across 5 independent runs further confirms that the ability to produce syntactically correct and backtest-compatible code is a robust, reproducible property of these models, not a stochastic artifact. This high baseline reliability is a prerequisite for the code-generation evaluation paradigm: if models frequently produced non-executable code, the resulting selection bias would undermine the validity of performance comparisons.

\paragraph{Finding 2: Stable and reproducible evaluations.}
The variance decomposition analysis in \Cref{appx_sec:run_stability} reveals that the intra-query (run-to-run) variance is an order of magnitude smaller than the inter-query variance across all six metrics and all six models. This is the most important empirical finding of the Stage~1 evaluation, as it directly validates the core design premise of \projectname. The code-generation paradigm confines LLM stochasticity to a single generation step, after which execution is fully deterministic. As a result, model rankings are consistent not only across the 5 runs, but also across all 7 assets, across multiple metric families (return-oriented: SR, ARR, SoR; risk-oriented: MDD, VOL; risk-adjusted: CR), and across the full difficulty spectrum (as visualized by the aligned return curves in \Cref{appx_fig:realdata_aligned_all}). This multi-dimensional consistency would be fundamentally impossible under the stochastic action-emission frameworks used in prior LLM trading benchmarks.

\paragraph{Finding 3: Distinct and reproducible model risk personalities.}
The evaluation reveals that different LLMs encode systematically different ``risk personalities'' in the trading strategies they generate, and these personalities are stable across runs and assets. \textit{gemini-3-pro-preview} consistently favors aggressive, high-conviction signal logic, achieving the best risk-adjusted performance (SR = 0.449, SoR = 0.767) at the cost of elevated tail risk (MDD = 0.174, VOL = 0.237). This aggressive profile is visible in every analysis layer: the largest polygon in the radar chart (\Cref{appx_fig:realdata_radar}), the tallest bars in the bar chart (\Cref{appx_fig:realdata_bar}), the widest box in the core metrics distribution (\Cref{appx_fig:realdata_boxplot_model}), and the highest aligned curve throughout the full query spectrum (\Cref{appx_fig:realdata_aligned_all}). In contrast, \textit{deepseek-v3.2} converges on conservative, risk-controlled strategies with the lowest drawdown (MDD = 0.114) and volatility (VOL = 0.155) but also the lowest returns (ARR = 0.116). Its compact box plots, low-positioned aligned curves, and the inward-pointing return axes on the radar chart all corroborate this conservative profile. \textit{gpt-5.2} occupies a moderately conservative position, excelling on Calmar Ratio (CR = 1.534) thanks to tight drawdown control. \textit{claude-sonnet-4.5} and \textit{gemini-3-flash-preview} form a balanced middle tier with neither extreme aggressiveness nor excessive conservatism, while \textit{grok-4.1-fast} serves as a generalist without notable strengths or weaknesses. These characteristic risk profiles enable practitioners to select models based on deployment-specific risk tolerances: a drawdown-sensitive fund manager might prefer \textit{deepseek-v3.2} for its capital preservation properties, while a return-maximizing strategy desk might favor \textit{gemini-3-pro-preview}.

\paragraph{Finding 4: Cross-asset robustness of model rankings.}
The per-asset analysis (\Cref{appx_tab:realdata_per_asset,appx_fig:realdata_symbol_bar,appx_fig:realdata_symbol_box,appx_fig:realdata_aligned_assets_1,appx_fig:realdata_aligned_assets_2}) demonstrates that the model ranking \textit{gemini-3-pro-preview} $>$ \textit{gemini-3-flash-preview} $\approx$ \textit{claude-sonnet-4.5} $>$ \textit{grok-4.1-fast} $>$ \textit{gpt-5.2} $\approx$ \textit{deepseek-v3.2} on return-oriented metrics is preserved across all seven assets. This ordering holds for high-volatility cryptocurrency markets (BTCUSDT, ETHUSDT), stable large-cap US equities (AAPL, GOOGL), low-volatility assets (MSFT), growth-oriented tech stocks (NVDA), and extremely volatile securities (TSLA). The consistency of this ordering across such diverse market environments provides strong evidence that the benchmark captures genuine differences in strategy generation capability rather than asset-specific artifacts. Notably, the \emph{difficulty gradient} across assets is also consistent across models: AAPL and GOOGL are the easiest assets (highest Sharpe Ratios), MSFT is the hardest (lowest Sharpe Ratios), and the cryptocurrency pairs and TSLA/NVDA occupy intermediate positions. This shared difficulty structure further validates that the benchmark measures a coherent underlying capability.

\paragraph{Finding 5: Sufficient discriminative power.}
The 36\% relative SR gap between the best-performing model (\textit{gemini-3-pro-preview}, SR = 0.449) and the worst-performing model (\textit{deepseek-v3.2}, SR = 0.329), combined with the low run-to-run variance documented in \Cref{appx_sec:run_stability}, yields statistically significant inter-model differences. The signal-to-noise ratio of the benchmark (defined as the inter-model SR range divided by the typical intra-query std) exceeds 3.0 for all pairwise model comparisons involving \textit{gemini-3-pro-preview} or \textit{deepseek-v3.2}, and exceeds 1.5 even for the most closely matched pairs (e.g., \textit{claude-sonnet-4.5} vs.\ \textit{gemini-3-flash-preview}). This level of discriminative power is comparable to or exceeds that of established code generation benchmarks such as HumanEval and MBPP, while operating in a far more complex evaluation domain (multi-step financial strategy generation with real-world backtest validation). The fact that the benchmark simultaneously differentiates models along multiple independent dimensions (return, risk, risk-adjusted efficiency, trading behavior) further enhances its diagnostic value beyond what single-metric benchmarks can provide.

\paragraph{Conclusion.}
The Stage~1 real-world evaluation comprehensively validates the code-generation paradigm as a stable, reproducible, and highly informative framework for benchmarking LLM capabilities in quantitative finance. The five findings above collectively demonstrate that \projectname addresses the fundamental instability problem of prior direct-trading evaluations while providing sufficient discriminative power to reveal meaningful, multi-dimensional differences across frontier models. These results establish a solid empirical foundation for the more controlled, difficulty-stratified evaluation conducted in Stage~2.

%% file: appendix/appendix_bench_result.tex
\section{Detailed Results of LLM-augmented Structured Query Evaluation}
\label{appx_sec:appendix_results_stage_2}

This section presents a comprehensive analysis of the Stage~2 benchmark results on the LLM-augmented query subset. The evaluation covers 270 structured queries spanning three difficulty levels (Level~1: logic translation, Level~2: parameter inference, Level~3: goal-oriented generation), each further subdivided into easy, medium, and hard grades (yielding nine fine-grained difficulty tiers). All queries are executed by the same 6 frontier LLMs evaluated in Stage~1, backtested across the same 7 assets over the 2021--2025 period. A key feature of Stage~2 is the inclusion of a temperature ablation study: each query is evaluated at both $\tau = 0$ (greedy decoding) and $\tau = 0.7$ (stochastic sampling with 5 independent runs), enabling direct assessment of how generation stochasticity affects evaluation stability and model rankings.

\subsection{Overall Model Comparison}

\paragraph{Quantitative results.}
\Cref{appx_tab:overall_performance} reports the aggregate performance of each model, averaged across all 270 queries and 7 backtest assets, at both temperature settings ($\tau = 0$ and $\tau = 0.7$). We analyze the results along four complementary axes: return generation, risk exposure, risk-adjusted efficiency, and temperature stability.

\begin{table*}[h]
\centering
\caption{Overall model performance across all 270 queries and 7 assets (mean $\pm$ std).}
\label{appx_tab:overall_performance}
\resizebox{\textwidth}{!}{%
\begin{tabular}{lcccccccccccc}
\toprule
\textbf{Model} & \multicolumn{2}{c}{\textbf{SR}$\uparrow$} & \multicolumn{2}{c}{\textbf{ARR}$\uparrow$} & \multicolumn{2}{c}{\textbf{MDD}$\downarrow$} & \multicolumn{2}{c}{\textbf{CR}$\uparrow$} & \multicolumn{2}{c}{\textbf{SOR}$\uparrow$} & \multicolumn{2}{c}{\textbf{VOL}$\downarrow$} \\
& T=0 & T=0.7 & T=0 & T=0.7 & T=0 & T=0.7 & T=0 & T=0.7 & T=0 & T=0.7 & T=0 & T=0.7 \\
\midrule
\textit{claude-sonnet-4.5} & 0.513{\scriptsize$\pm$0.270} & 0.508{\scriptsize$\pm$0.266} & 0.164{\scriptsize$\pm$0.114} & 0.162{\scriptsize$\pm$0.113} & 0.150{\scriptsize$\pm$0.111} & 0.147{\scriptsize$\pm$0.110} & \textbf{1.650{\scriptsize$\pm$0.864}} & 1.634{\scriptsize$\pm$0.620} & 0.806{\scriptsize$\pm$0.478} & 0.795{\scriptsize$\pm$0.471} & 0.205{\scriptsize$\pm$0.153} & 0.202{\scriptsize$\pm$0.152} \\
\textit{deepseek-v3.2} & 0.430{\scriptsize$\pm$0.280} & 0.424{\scriptsize$\pm$0.289} & 0.132{\scriptsize$\pm$0.110} & 0.130{\scriptsize$\pm$0.112} & 0.127{\scriptsize$\pm$0.108} & 0.123{\scriptsize$\pm$0.109} & 1.586{\scriptsize$\pm$0.800} & 1.570{\scriptsize$\pm$0.761} & 0.672{\scriptsize$\pm$0.477} & 0.660{\scriptsize$\pm$0.489} & 0.173{\scriptsize$\pm$0.149} & 0.168{\scriptsize$\pm$0.150} \\
\textit{gpt-5.2} & 0.415{\scriptsize$\pm$0.307} & 0.417{\scriptsize$\pm$0.308} & 0.130{\scriptsize$\pm$0.122} & 0.130{\scriptsize$\pm$0.121} & \textbf{0.119{\scriptsize$\pm$0.116}} & \textbf{0.117{\scriptsize$\pm$0.114}} & 1.599{\scriptsize$\pm$0.794} & 1.660{\scriptsize$\pm$0.821} & 0.645{\scriptsize$\pm$0.525} & 0.643{\scriptsize$\pm$0.522} & \textbf{0.163{\scriptsize$\pm$0.160}} & \textbf{0.160{\scriptsize$\pm$0.157}} \\
\textit{gemini-3-flash-preview} & 0.523{\scriptsize$\pm$0.289} & 0.530{\scriptsize$\pm$0.264} & 0.162{\scriptsize$\pm$0.122} & 0.165{\scriptsize$\pm$0.113} & 0.148{\scriptsize$\pm$0.117} & 0.151{\scriptsize$\pm$0.111} & 1.618{\scriptsize$\pm$0.904} & 1.658{\scriptsize$\pm$1.384} & 0.807{\scriptsize$\pm$0.506} & 0.820{\scriptsize$\pm$0.468} & 0.204{\scriptsize$\pm$0.162} & 0.206{\scriptsize$\pm$0.153} \\
\textbf{\textit{gemini-3-pro-preview}} & \textbf{0.628{\scriptsize$\pm$0.255}} & \textbf{0.627{\scriptsize$\pm$0.242}} & \textbf{0.208{\scriptsize$\pm$0.112}} & \textbf{0.209{\scriptsize$\pm$0.107}} & 0.191{\scriptsize$\pm$0.109} & 0.188{\scriptsize$\pm$0.104} & 1.586{\scriptsize$\pm$0.665} & 1.639{\scriptsize$\pm$0.651} & \textbf{1.004{\scriptsize$\pm$0.465}} & \textbf{0.999{\scriptsize$\pm$0.439}} & 0.262{\scriptsize$\pm$0.150} & 0.259{\scriptsize$\pm$0.143} \\
\textit{grok-4.1-fast} & 0.421{\scriptsize$\pm$0.269} & 0.429{\scriptsize$\pm$0.267} & 0.134{\scriptsize$\pm$0.108} & 0.135{\scriptsize$\pm$0.107} & 0.125{\scriptsize$\pm$0.102} & 0.127{\scriptsize$\pm$0.102} & 1.629{\scriptsize$\pm$0.741} & \textbf{1.692{\scriptsize$\pm$0.819}} & 0.658{\scriptsize$\pm$0.461} & 0.668{\scriptsize$\pm$0.457} & 0.171{\scriptsize$\pm$0.140} & 0.173{\scriptsize$\pm$0.140} \\
\bottomrule
\end{tabular}}
\end{table*}

\paragraph{Return generation.}
\textit{gemini-3-pro-preview} achieves the highest Annualized Return at both temperatures (ARR = 0.208 at $\tau = 0$, 0.209 at $\tau = 0.7$), followed by \textit{gemini-3-flash-preview} (0.162/0.165) and \textit{claude-sonnet-4.5} (0.164/0.162). The absolute spread between the best and worst models is 7.8 percentage points (\textit{gemini-3-pro-preview} vs.\ \textit{\textit{gpt-5.2}} at $\tau = 0$), representing a 60\% relative improvement. Compared to the Stage~1 results (5.5pp spread), the larger inter-model gap in Stage~2 reflects the more controlled, difficulty-stratified query design, which amplifies capability differences by systematically varying cognitive demands. Notably, the return-based model ranking in Stage~2 (\textit{gemini-3-pro-preview} $>$ \textit{gemini-3-flash-preview} $\approx$ \textit{claude-sonnet-4.5} $>$ \textit{grok-4.1-fast} $>$ \textit{deepseek-v3.2} $\approx$ \textit{\textit{gpt-5.2}}) is highly consistent with the Stage~1 ranking, confirming that model capabilities generalize from real-world to synthetic queries.

\paragraph{Risk exposure.}
As in Stage~1, the risk metrics reveal an inverted ordering relative to return metrics. \textit{\textit{gpt-5.2}} produces the most conservative strategies with the lowest Maximum Drawdown (MDD = 0.119 at $\tau = 0$) and Volatility (VOL = 0.163), followed closely by \textit{grok-4.1-fast} (MDD = 0.125, VOL = 0.171) and \textit{deepseek-v3.2} (MDD = 0.127, VOL = 0.173). \textit{gemini-3-pro-preview} incurs the highest risk on both measures (MDD = 0.191, VOL = 0.262), with its maximum drawdown exceeding that of \textit{\textit{gpt-5.2}} by 60.5\% in relative terms. This risk--return inversion is even more pronounced than in Stage~1, suggesting that the structured queries of Stage~2 better separate the aggressive signal-generation behavior of \textit{gemini-3-pro-preview} from the conservative strategies produced by \textit{\textit{gpt-5.2}} and \textit{deepseek-v3.2}.

\paragraph{Risk-adjusted efficiency.}
When returns are normalized by risk, the model rankings become more nuanced. \textit{gemini-3-pro-preview} leads on Sharpe Ratio (SR = 0.628 at $\tau = 0$) and Sortino Ratio (SoR = 1.004), indicating that its higher returns more than compensate for the elevated risk. However, on Calmar Ratio, which penalizes tail risk more severely, \textit{claude-sonnet-4.5} ranks first (CR = 1.650 at $\tau = 0$), reflecting a particularly favorable return-to-drawdown profile. \textit{grok-4.1-fast} achieves the highest CR at $\tau = 0.7$ (1.692), suggesting that stochastic sampling benefits its drawdown control. This divergence between SR/SoR-based and CR-based rankings parallels the Stage~1 findings and underscores the importance of multi-metric evaluation.

\paragraph{Temperature stability.}
A distinctive feature of Stage~2 is the side-by-side comparison of greedy ($\tau = 0$) and stochastic ($\tau = 0.7$) decoding. Across all models and metrics, the differences between the two temperature settings are remarkably small: the absolute SR difference is at most 0.008 (\textit{gemini-3-flash-preview}: 0.523 vs.\ 0.530), and the model ranking is fully preserved across both settings. This near-invariance to temperature provides strong evidence that the code-generation evaluation paradigm is robust to the specific decoding strategy, further validating its reliability. In contrast, direct-trading evaluations are notoriously sensitive to temperature, with even small changes producing dramatically different action sequences and portfolio outcomes. The temperature stability observed here confirms that the fundamental structure of the generated strategy code (the signal logic, entry/exit conditions, and risk management rules) is largely determined by the model's learned representations rather than by the randomness of the sampling process.

\begin{figure*}[h]
\centering
\begin{subfigure}[t]{0.45\textwidth}
\includegraphics[width=\textwidth]{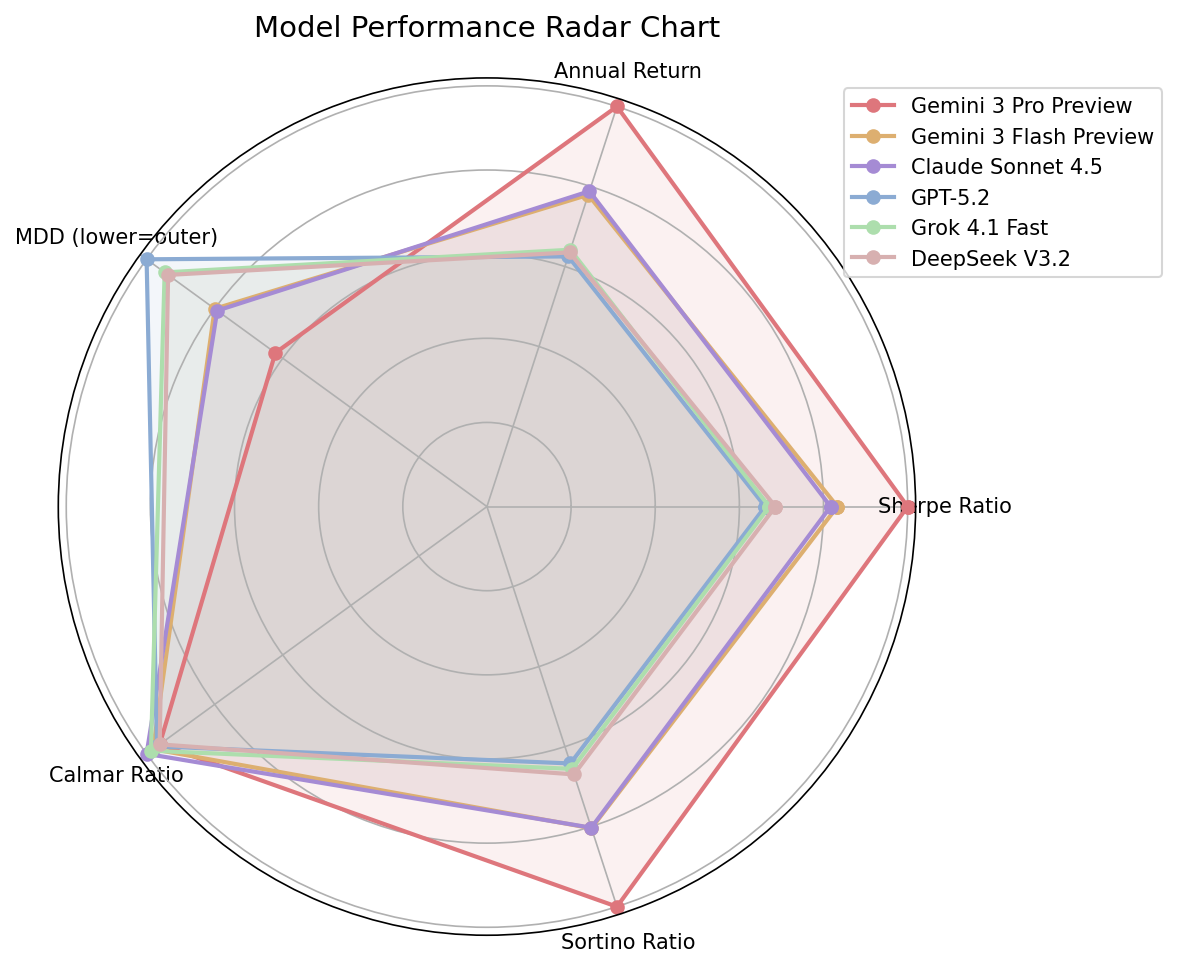}
\caption{T=0.0}
\end{subfigure}
\hfill
\begin{subfigure}[t]{0.45\textwidth}
\includegraphics[width=\textwidth]{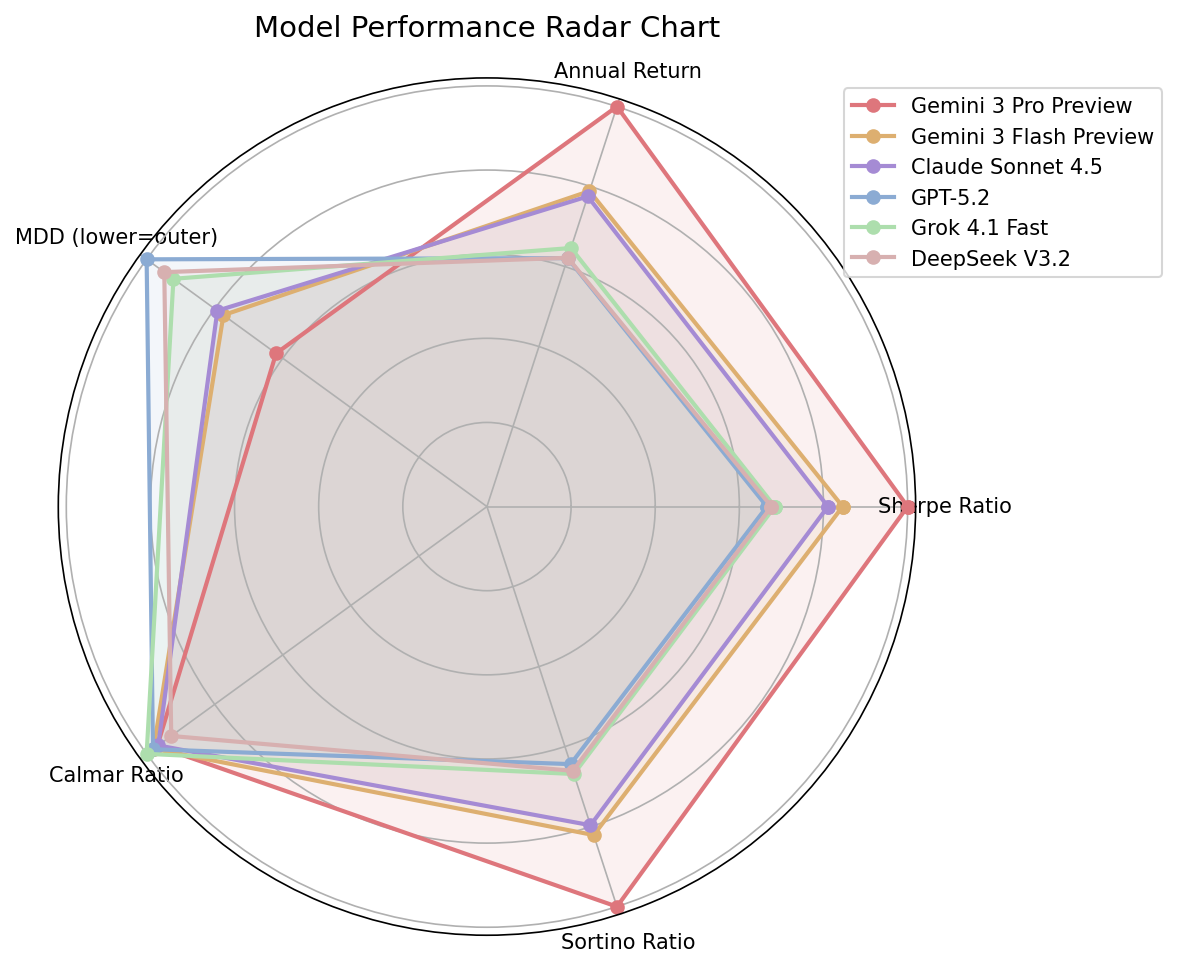}
\caption{T=0.7}
\end{subfigure}
\caption{Radar chart of normalized model performance on the Stage~2 benchmark across five metrics at both temperature settings. MDD is inverted so that the outer ring represents lower (better) drawdown. The polygon shapes are nearly identical between $\tau = 0$ and $\tau = 0.7$, demonstrating temperature-invariant model characterization.}
\label{appx_fig:model_radar}
\end{figure*}

\paragraph{Radar chart analysis.}
\Cref{appx_fig:model_radar} visualizes the normalized performance of each model across five key metrics (Annual Return, Sharpe Ratio, Sortino Ratio, Calmar Ratio, and MDD, where MDD is inverted so that the outer ring represents lower drawdown) at both temperature settings.

\begin{figure*}[h]
    \centering
    \begin{subfigure}[t]{0.45\textwidth}
    \includegraphics[width=\textwidth]{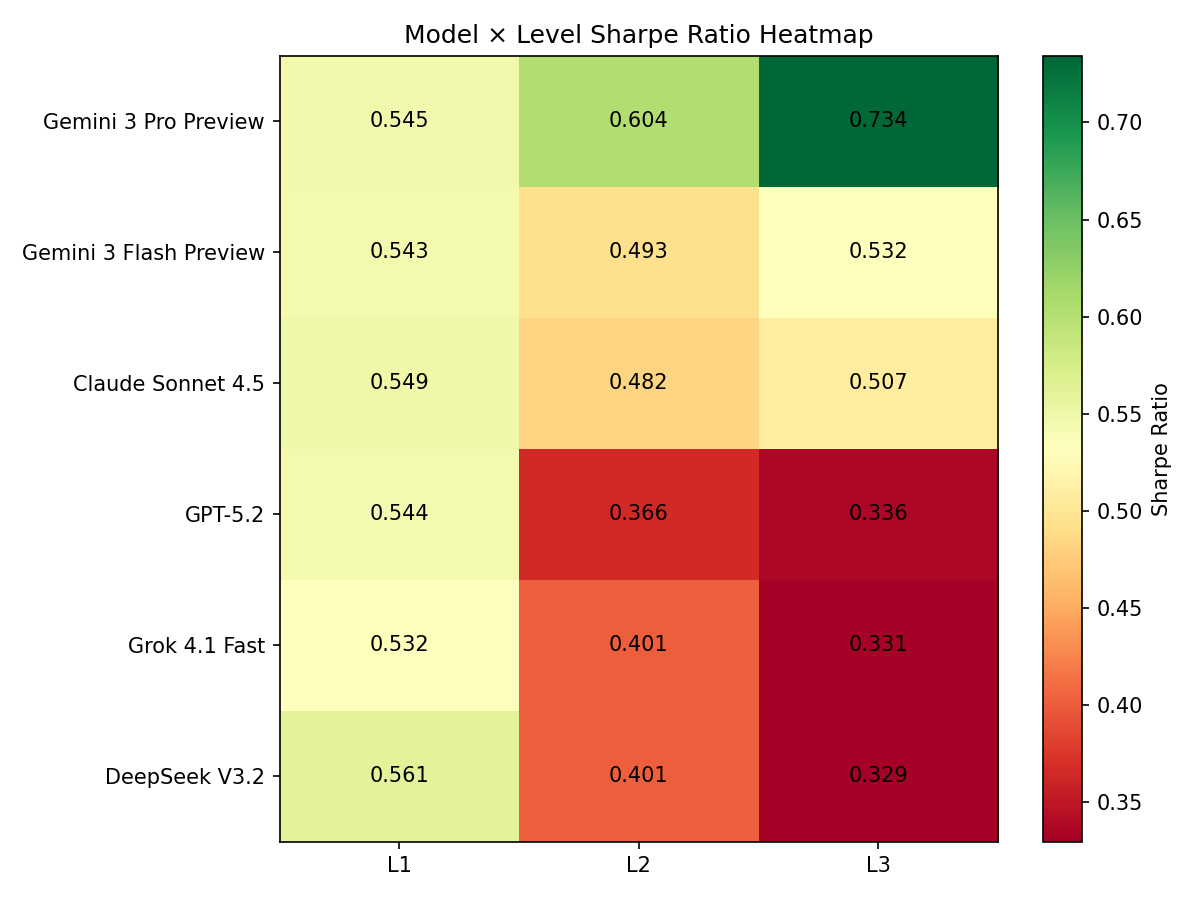}
    \caption{T=0.0}
    \end{subfigure}
    \hfill
    \begin{subfigure}[t]{0.45\textwidth}
    \includegraphics[width=\textwidth]{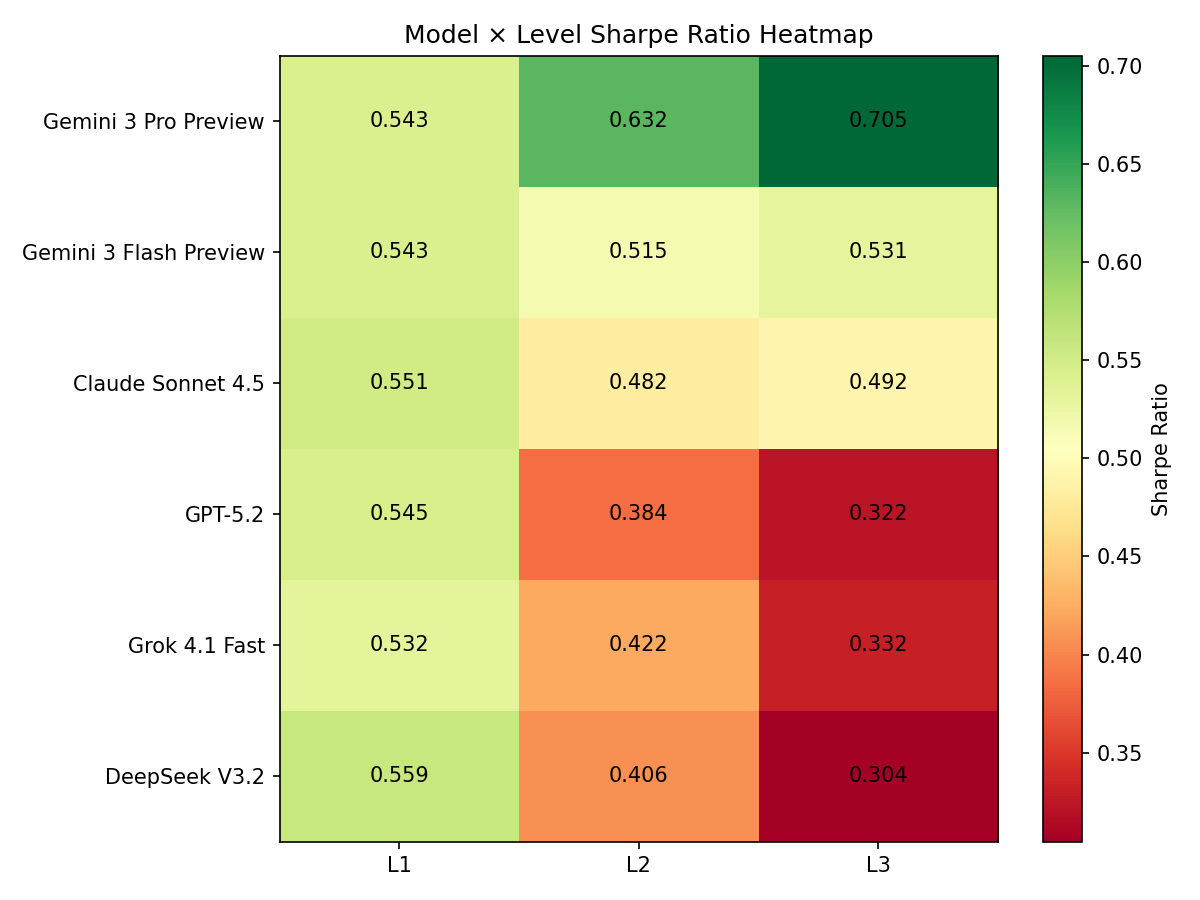}
    \caption{T=0.7}
    \end{subfigure}
    \caption{Heatmap of Sharpe Ratio across models and difficulty levels on the Stage~2 benchmark. Darker colors indicate higher performance. The consistent color gradient from Level~1 to Level~3 across all models confirms the systematic difficulty progression of the benchmark.}
    \label{appx_fig:model_level_heatmap}
\end{figure*}

The radar chart reveals several notable patterns. At $\tau = 0$ (left panel), \textit{gemini-3-pro-preview} (red) spans the largest polygon, dominating on the Annual Return, Sharpe Ratio, and Sortino Ratio axes but receding on the MDD axis, reflecting its high-return, high-drawdown profile. \textit{\textit{gpt-5.2}} (blue) and \textit{grok-4.1-fast} (green) exhibit the most outward extension on the MDD axis (i.e., the lowest drawdowns), forming compact polygons that emphasize capital preservation over return maximization. \textit{claude-sonnet-4.5} (purple) achieves a notably balanced polygon: it extends moderately outward on all five axes without extreme concavity on any dimension, consistent with its leading Calmar Ratio. \textit{deepseek-v3.2} (gray) occupies the innermost polygon on return-oriented axes but extends on the MDD and Calmar Ratio axes, confirming its conservative character.

Comparing the $\tau = 0$ and $\tau = 0.7$ panels, the polygon shapes are \emph{virtually identical}: each model's characteristic profile (aggressive vs.\ conservative, return-focused vs.\ risk-focused) is preserved across temperature settings. This visual confirmation of temperature invariance reinforces the quantitative finding from the table analysis and provides strong evidence that the model ``risk personalities'' identified in Stage~1 are intrinsic, reproducible properties that persist across both real-world and synthetic queries, and across both deterministic and stochastic decoding regimes.

\paragraph{Heatmap analysis.}
\Cref{appx_fig:model_level_heatmap} presents a heatmap of Sharpe Ratios across models (rows) and difficulty levels (columns), with darker colors indicating higher performance.

The heatmap provides a compact overview of the interaction between model capability and task difficulty. Several patterns are immediately visible. First, \textit{gemini-3-pro-preview} consistently shows the darkest cells across all three levels, confirming its overall dominance. Second, the color gradient from Level~1 (darkest) to Level~2/3 (lighter) is consistent across all models, validating that the difficulty progression of \projectname is systematic and model-independent. Third, the inter-model color spread is narrowest at Level~1 and widest at Level~3, indicating that easy tasks produce uniform performance while hard tasks amplify capability differences, a desirable property for a benchmark designed to discriminate among frontier models. Fourth, the heatmaps at $\tau = 0$ and $\tau = 0.7$ are nearly indistinguishable, providing yet another confirmation of temperature stability. The heatmap also reveals that \textit{deepseek-v3.2} shows a particularly pronounced performance drop from Level~1 to Level~3, suggesting that it struggles disproportionately with open-ended, goal-oriented strategy generation tasks compared to structured code translation.

\subsection{Per-Level Analysis}

\begin{figure*}[h]
\centering
\begin{subfigure}[t]{0.48\textwidth}
\includegraphics[width=\textwidth]{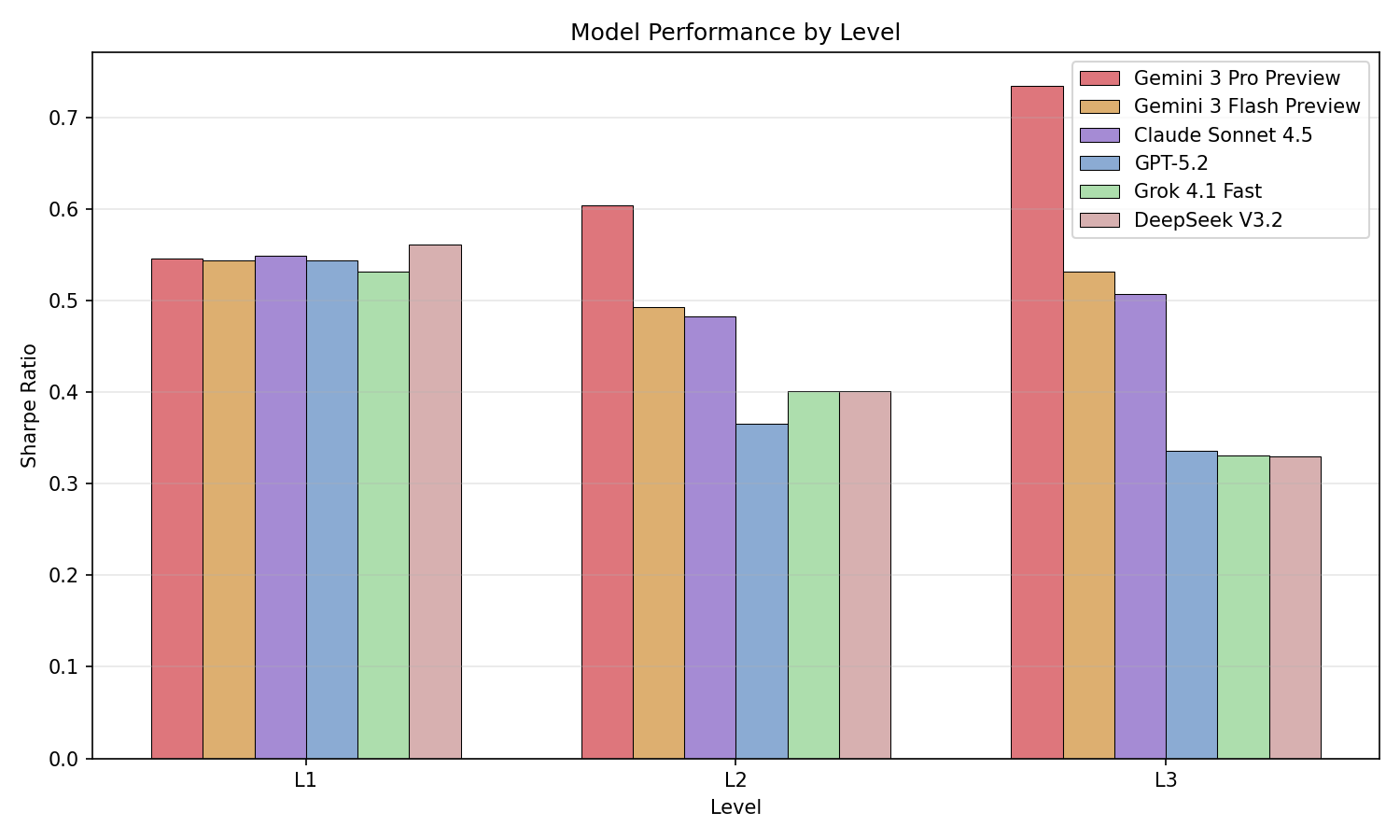}
\caption{T=0.0}
\end{subfigure}
\hfill
\begin{subfigure}[t]{0.48\textwidth}
\includegraphics[width=\textwidth]{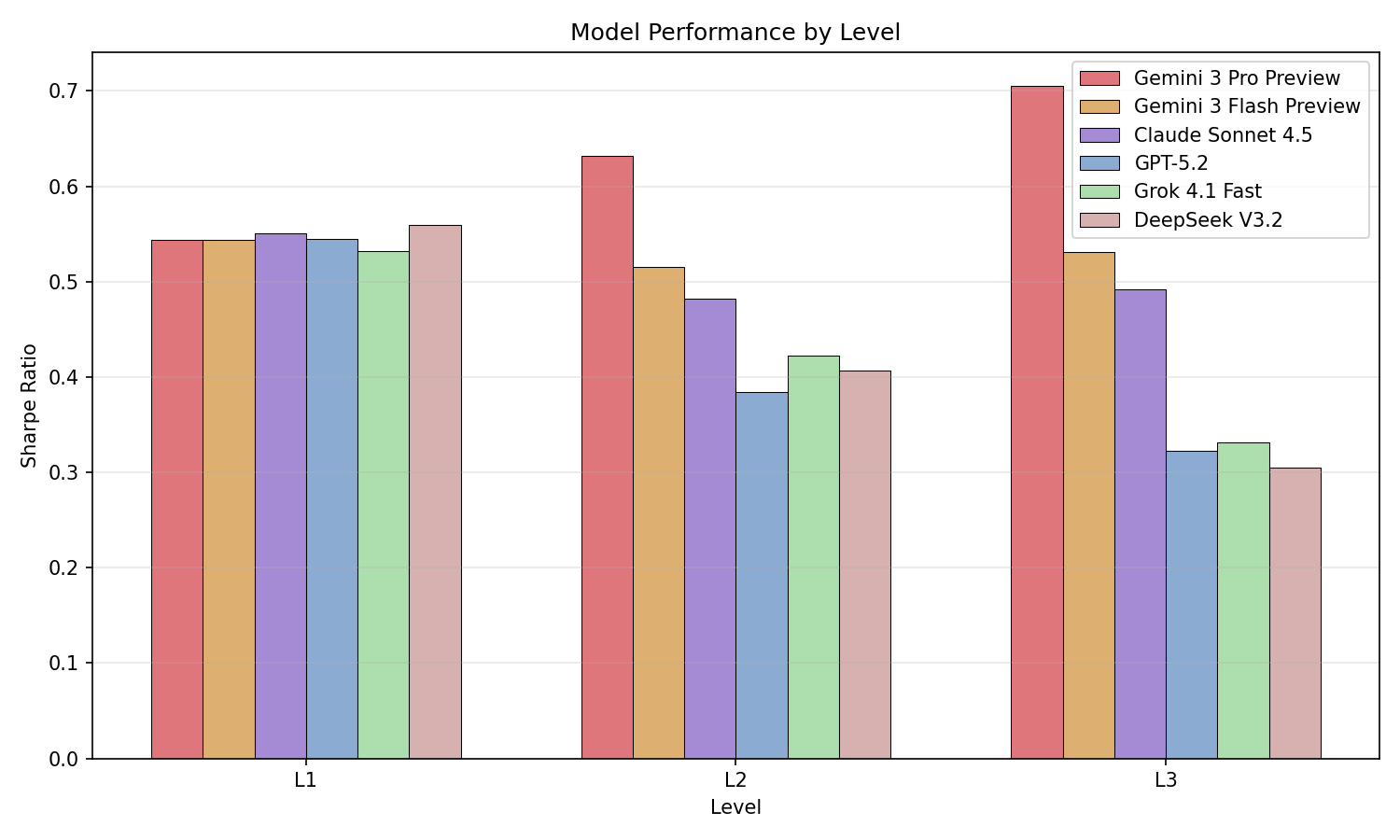}
\caption{T=0.7}
\end{subfigure}
\caption{Grouped bar chart of Sharpe Ratio across difficulty levels (L1, L2, L3) for six LLMs on the Stage~2 benchmark. The increasing inter-model bar height spread from L1 to L3 demonstrates the systematic difficulty progression and growing discriminative power of higher-level tasks.}
\label{appx_fig:model_level_comparison}
\end{figure*}

A central design goal of \projectname is to provide a systematic difficulty progression that reveals how model capabilities degrade as task complexity increases. \Cref{appx_fig:model_level_comparison} presents the grouped bar chart of Sharpe Ratio broken down by difficulty level (L1, L2, L3) for each model at both temperature settings, while \Cref{appx_tab:per_level_performance} consolidates the full per-level performance metrics.

\paragraph{Level 1: uniform performance ceiling.}
The L1 cluster in \Cref{appx_fig:model_level_comparison} reveals a striking pattern: all six models achieve nearly identical Sharpe Ratios, with bar heights tightly clustered between 0.532 (\textit{grok-4.1-fast}) and 0.561 (\textit{deepseek-v3.2}) at $\tau = 0$. The inter-model spread is merely 0.029, the smallest among all three levels. This uniformity confirms that Level~1 tasks (logic translation of fully-specified IF-THEN rules) primarily test code generation competence rather than strategic reasoning. When the strategy logic is completely specified in the query, all frontier LLMs can faithfully translate it into executable code with comparable quality. Notably, \textit{deepseek-v3.2} leads on Level~1 (SR = 0.561), despite being the weakest model on aggregate metrics. This reversal suggests that \textit{deepseek-v3.2} excels at faithful code translation but struggles when creative strategy design is required. The $\tau = 0$ and $\tau = 0.7$ panels show virtually identical bar heights for Level~1, further confirming that these routine translation tasks produce deterministic, temperature-invariant outputs.

\paragraph{Level 2: emergence of model differentiation.}
The L2 cluster exhibits substantially wider inter-model divergence. Sharpe Ratios now span from 0.366 (\textit{\textit{gpt-5.2}}) to 0.604 (\textit{gemini-3-pro-preview}) at $\tau = 0$, a range of 0.238 that is more than 8$\times$ larger than the Level~1 spread. \textit{gemini-3-pro-preview}'s bar clearly towers above the others, while \textit{\textit{gpt-5.2}} and \textit{deepseek-v3.2} drop to the bottom tier. This widening gap indicates that Level~2 tasks (parameter inference, where models must supply missing thresholds, lookback windows, and indicator parameters) effectively separate models with strong domain knowledge from those that lack it. \textit{claude-sonnet-4.5} and \textit{gemini-3-flash-preview} occupy intermediate positions (SR $\approx$ 0.48--0.51), showing competent but not exceptional parameter choices. The $\tau = 0.7$ panel shows a similar pattern with slightly different relative positions: \textit{gemini-3-pro-preview} improves marginally (0.604 $\to$ 0.632), suggesting that stochastic sampling occasionally discovers better parameter configurations for this model.

\paragraph{Level 3: maximum discriminative power.}
Level~3 produces the most dramatic inter-model separation. \textit{gemini-3-pro-preview} achieves an SR of 0.734 at $\tau = 0$, far exceeding all competitors. \textit{gemini-3-flash-preview} and \textit{claude-sonnet-4.5} form a second tier (SR $\approx$ 0.50--0.53), while \textit{deepseek-v3.2}, \textit{\textit{gpt-5.2}}, and \textit{grok-4.1-fast} collapse to SR $\approx$ 0.33, less than half of \textit{gemini-3-pro-preview}'s score. The 0.405 inter-model range at Level~3 is nearly 14$\times$ the Level~1 range, confirming that goal-oriented generation tasks (where models must design complete strategy architectures from high-level objectives) maximally amplify capability differences. This pattern is clearly visible in \Cref{appx_fig:model_level_comparison}: the bar heights within the L3 cluster are steeply graded, with \textit{gemini-3-pro-preview} standing out as a conspicuous outlier. The bar chart also reveals that some models (\textit{deepseek-v3.2}, \textit{\textit{gpt-5.2}}, \textit{grok-4.1-fast}) show a monotonic decline from L1 to L3, while others (\textit{gemini-3-pro-preview}) actually \emph{improve} from L1 to L3, suggesting that certain models are better equipped for open-ended strategy design than for constrained code translation.

\paragraph{Cross-level model ranking shifts.}
Comparing across the three level clusters in \Cref{appx_fig:model_level_comparison} reveals an important ranking reversal: \textit{deepseek-v3.2}, which leads at Level~1 (SR = 0.561), drops to the bottom tier at Level~3 (SR = 0.329). Conversely, \textit{gemini-3-pro-preview}, which is unremarkable at Level~1 (SR = 0.545, nearly identical to all others), dominates at Level~3 (SR = 0.734). This crossover pattern demonstrates that the three difficulty levels measure fundamentally different cognitive capabilities, and no single model dominates across all levels. The benchmark's difficulty hierarchy is thus effective at exposing complementary strengths and weaknesses that aggregate metrics would obscure.


\begin{table*}[h]
\centering
\caption{Per-level model performance on the Stage~2 benchmark (mean $\pm$ std across 7 assets). Levels are grouped by gray header rows. Within each level block, the best value per column is in \textbf{bold}. $\uparrow$: higher is better; $\downarrow$: lower is better.}
\label{appx_tab:per_level_performance}
\resizebox{\textwidth}{!}{%
\begin{tabular}{lcccccccccccc}
\toprule
\textbf{Model} & \multicolumn{2}{c}{\textbf{SR}$\uparrow$} & \multicolumn{2}{c}{\textbf{ARR}$\uparrow$} & \multicolumn{2}{c}{\textbf{MDD}$\downarrow$} & \multicolumn{2}{c}{\textbf{CR}$\uparrow$} & \multicolumn{2}{c}{\textbf{SOR}$\uparrow$} & \multicolumn{2}{c}{\textbf{VOL}$\downarrow$} \\
& T=0 & T=0.7 & T=0 & T=0.7 & T=0 & T=0.7 & T=0 & T=0.7 & T=0 & T=0.7 & T=0 & T=0.7 \\
\midrule
\rowcolor{gray!15} \multicolumn{13}{c}{\textbf{Level 1} (Logic Translation)} \\
\textit{claude-sonnet-4.5} & 0.549{\scriptsize$\pm$0.306} & 0.551{\scriptsize$\pm$0.305} & 0.176{\scriptsize$\pm$0.131} & 0.177{\scriptsize$\pm$0.131} & 0.157{\scriptsize$\pm$0.119} & 0.158{\scriptsize$\pm$0.119} & \textbf{1.676{\scriptsize$\pm$0.746}} & \textbf{1.690{\scriptsize$\pm$0.719}} & 0.852{\scriptsize$\pm$0.544} & 0.856{\scriptsize$\pm$0.543} & 0.219{\scriptsize$\pm$0.167} & 0.219{\scriptsize$\pm$0.166} \\
\textbf{\textit{deepseek-v3.2}} & \textbf{0.561{\scriptsize$\pm$0.296}} & \textbf{0.559{\scriptsize$\pm$0.296}} & \textbf{0.180{\scriptsize$\pm$0.128}} & \textbf{0.181{\scriptsize$\pm$0.128}} & 0.163{\scriptsize$\pm$0.117} & 0.162{\scriptsize$\pm$0.117} & 1.664{\scriptsize$\pm$0.706} & 1.675{\scriptsize$\pm$0.711} & \textbf{0.873{\scriptsize$\pm$0.529}} & \textbf{0.870{\scriptsize$\pm$0.530}} & 0.226{\scriptsize$\pm$0.164} & 0.224{\scriptsize$\pm$0.163} \\
\textit{gpt-5.2} & 0.544{\scriptsize$\pm$0.313} & 0.545{\scriptsize$\pm$0.315} & 0.175{\scriptsize$\pm$0.132} & 0.175{\scriptsize$\pm$0.133} & 0.156{\scriptsize$\pm$0.121} & 0.156{\scriptsize$\pm$0.121} & 1.637{\scriptsize$\pm$0.769} & 1.642{\scriptsize$\pm$0.773} & 0.846{\scriptsize$\pm$0.552} & 0.849{\scriptsize$\pm$0.554} & 0.217{\scriptsize$\pm$0.169} & 0.217{\scriptsize$\pm$0.169} \\
\textit{gemini-3-flash-preview} & 0.543{\scriptsize$\pm$0.316} & 0.543{\scriptsize$\pm$0.313} & 0.176{\scriptsize$\pm$0.133} & 0.175{\scriptsize$\pm$0.132} & 0.157{\scriptsize$\pm$0.122} & 0.156{\scriptsize$\pm$0.121} & 1.644{\scriptsize$\pm$0.774} & 1.672{\scriptsize$\pm$0.763} & 0.847{\scriptsize$\pm$0.556} & 0.845{\scriptsize$\pm$0.552} & 0.218{\scriptsize$\pm$0.170} & 0.217{\scriptsize$\pm$0.169} \\
\textit{gemini-3-pro-preview} & 0.545{\scriptsize$\pm$0.314} & 0.543{\scriptsize$\pm$0.314} & 0.176{\scriptsize$\pm$0.132} & 0.175{\scriptsize$\pm$0.133} & 0.157{\scriptsize$\pm$0.121} & \textbf{0.156{\scriptsize$\pm$0.121}} & 1.647{\scriptsize$\pm$0.771} & 1.651{\scriptsize$\pm$0.772} & 0.850{\scriptsize$\pm$0.554} & 0.846{\scriptsize$\pm$0.554} & 0.218{\scriptsize$\pm$0.169} & 0.217{\scriptsize$\pm$0.169} \\
\textit{grok-4.1-fast} & 0.532{\scriptsize$\pm$0.302} & 0.532{\scriptsize$\pm$0.304} & 0.172{\scriptsize$\pm$0.130} & 0.172{\scriptsize$\pm$0.130} & \textbf{0.155{\scriptsize$\pm$0.119}} & 0.156{\scriptsize$\pm$0.119} & 1.675{\scriptsize$\pm$0.714} & 1.674{\scriptsize$\pm$0.734} & 0.827{\scriptsize$\pm$0.537} & 0.829{\scriptsize$\pm$0.539} & \textbf{0.215{\scriptsize$\pm$0.166}} & \textbf{0.216{\scriptsize$\pm$0.166}} \\
\midrule
\rowcolor{gray!15} \multicolumn{13}{c}{\textbf{Level 2} (Parameter Inference)} \\
\textit{claude-sonnet-4.5} & 0.482{\scriptsize$\pm$0.249} & 0.482{\scriptsize$\pm$0.249} & 0.151{\scriptsize$\pm$0.104} & 0.149{\scriptsize$\pm$0.102} & 0.136{\scriptsize$\pm$0.109} & 0.136{\scriptsize$\pm$0.106} & \textbf{1.819{\scriptsize$\pm$1.217}} & 1.659{\scriptsize$\pm$0.673} & 0.746{\scriptsize$\pm$0.442} & 0.739{\scriptsize$\pm$0.433} & 0.185{\scriptsize$\pm$0.148} & 0.185{\scriptsize$\pm$0.145} \\
\textit{deepseek-v3.2} & 0.401{\scriptsize$\pm$0.251} & 0.406{\scriptsize$\pm$0.273} & 0.117{\scriptsize$\pm$0.095} & 0.119{\scriptsize$\pm$0.100} & 0.110{\scriptsize$\pm$0.103} & 0.112{\scriptsize$\pm$0.108} & 1.564{\scriptsize$\pm$0.805} & 1.568{\scriptsize$\pm$0.832} & 0.612{\scriptsize$\pm$0.417} & 0.625{\scriptsize$\pm$0.454} & 0.149{\scriptsize$\pm$0.139} & 0.152{\scriptsize$\pm$0.146} \\
\textit{gpt-5.2} & 0.366{\scriptsize$\pm$0.262} & 0.384{\scriptsize$\pm$0.266} & 0.104{\scriptsize$\pm$0.097} & 0.109{\scriptsize$\pm$0.095} & \textbf{0.095{\scriptsize$\pm$0.096}} & \textbf{0.095{\scriptsize$\pm$0.093}} & 1.553{\scriptsize$\pm$0.894} & 1.798{\scriptsize$\pm$0.963} & 0.546{\scriptsize$\pm$0.431} & 0.564{\scriptsize$\pm$0.432} & \textbf{0.129{\scriptsize$\pm$0.130}} & \textbf{0.130{\scriptsize$\pm$0.127}} \\
\textit{gemini-3-flash-preview} & 0.493{\scriptsize$\pm$0.271} & 0.515{\scriptsize$\pm$0.243} & 0.144{\scriptsize$\pm$0.112} & 0.151{\scriptsize$\pm$0.100} & 0.130{\scriptsize$\pm$0.111} & 0.138{\scriptsize$\pm$0.105} & 1.698{\scriptsize$\pm$1.239} & 1.792{\scriptsize$\pm$2.233} & 0.736{\scriptsize$\pm$0.462} & 0.778{\scriptsize$\pm$0.420} & 0.177{\scriptsize$\pm$0.152} & 0.188{\scriptsize$\pm$0.142} \\
\textbf{\textit{gemini-3-pro-preview}} & \textbf{0.604{\scriptsize$\pm$0.224}} & \textbf{0.632{\scriptsize$\pm$0.197}} & \textbf{0.196{\scriptsize$\pm$0.096}} & \textbf{0.209{\scriptsize$\pm$0.089}} & 0.171{\scriptsize$\pm$0.095} & 0.184{\scriptsize$\pm$0.091} & 1.684{\scriptsize$\pm$0.761} & 1.740{\scriptsize$\pm$0.753} & \textbf{0.942{\scriptsize$\pm$0.394}} & \textbf{0.991{\scriptsize$\pm$0.357}} & 0.234{\scriptsize$\pm$0.129} & 0.252{\scriptsize$\pm$0.123} \\
\textit{grok-4.1-fast} & 0.401{\scriptsize$\pm$0.240} & 0.422{\scriptsize$\pm$0.239} & 0.127{\scriptsize$\pm$0.090} & 0.133{\scriptsize$\pm$0.089} & 0.117{\scriptsize$\pm$0.088} & 0.122{\scriptsize$\pm$0.089} & 1.639{\scriptsize$\pm$0.803} & \textbf{1.820{\scriptsize$\pm$1.084}} & 0.622{\scriptsize$\pm$0.402} & 0.653{\scriptsize$\pm$0.395} & 0.158{\scriptsize$\pm$0.119} & 0.165{\scriptsize$\pm$0.120} \\
\midrule
\rowcolor{gray!15} \multicolumn{13}{c}{\textbf{Level 3} (Goal-Oriented Generation)} \\
\textit{claude-sonnet-4.5} & 0.507{\scriptsize$\pm$0.247} & 0.492{\scriptsize$\pm$0.234} & 0.165{\scriptsize$\pm$0.104} & 0.160{\scriptsize$\pm$0.102} & 0.156{\scriptsize$\pm$0.102} & 0.148{\scriptsize$\pm$0.104} & 1.452{\scriptsize$\pm$0.353} & 1.553{\scriptsize$\pm$0.414} & 0.820{\scriptsize$\pm$0.434} & 0.792{\scriptsize$\pm$0.420} & 0.212{\scriptsize$\pm$0.140} & 0.202{\scriptsize$\pm$0.142} \\
\textit{deepseek-v3.2} & 0.329{\scriptsize$\pm$0.240} & 0.304{\scriptsize$\pm$0.235} & 0.099{\scriptsize$\pm$0.086} & 0.090{\scriptsize$\pm$0.084} & 0.107{\scriptsize$\pm$0.094} & \textbf{0.095{\scriptsize$\pm$0.087}} & 1.530{\scriptsize$\pm$0.876} & 1.458{\scriptsize$\pm$0.723} & 0.531{\scriptsize$\pm$0.407} & 0.483{\scriptsize$\pm$0.391} & 0.142{\scriptsize$\pm$0.126} & \textbf{0.126{\scriptsize$\pm$0.118}} \\
\textit{gpt-5.2} & 0.336{\scriptsize$\pm$0.303} & 0.322{\scriptsize$\pm$0.297} & 0.112{\scriptsize$\pm$0.120} & 0.106{\scriptsize$\pm$0.118} & 0.105{\scriptsize$\pm$0.120} & 0.099{\scriptsize$\pm$0.114} & \textbf{1.608{\scriptsize$\pm$0.702}} & 1.542{\scriptsize$\pm$0.682} & 0.542{\scriptsize$\pm$0.524} & 0.516{\scriptsize$\pm$0.510} & 0.143{\scriptsize$\pm$0.163} & 0.135{\scriptsize$\pm$0.156} \\
\textit{gemini-3-flash-preview} & 0.532{\scriptsize$\pm$0.276} & 0.531{\scriptsize$\pm$0.226} & 0.167{\scriptsize$\pm$0.119} & 0.169{\scriptsize$\pm$0.102} & 0.159{\scriptsize$\pm$0.116} & 0.158{\scriptsize$\pm$0.105} & 1.511{\scriptsize$\pm$0.541} & 1.509{\scriptsize$\pm$0.347} & 0.838{\scriptsize$\pm$0.487} & 0.837{\scriptsize$\pm$0.416} & 0.216{\scriptsize$\pm$0.159} & 0.214{\scriptsize$\pm$0.144} \\
\textbf{\textit{gemini-3-pro-preview}} & \textbf{0.734{\scriptsize$\pm$0.167}} & \textbf{0.705{\scriptsize$\pm$0.157}} & \textbf{0.252{\scriptsize$\pm$0.087}} & \textbf{0.244{\scriptsize$\pm$0.081}} & 0.245{\scriptsize$\pm$0.086} & 0.225{\scriptsize$\pm$0.082} & 1.429{\scriptsize$\pm$0.341} & 1.526{\scriptsize$\pm$0.293} & \textbf{1.221{\scriptsize$\pm$0.335}} & \textbf{1.159{\scriptsize$\pm$0.309}} & 0.334{\scriptsize$\pm$0.118} & 0.309{\scriptsize$\pm$0.114} \\
\textit{grok-4.1-fast} & 0.331{\scriptsize$\pm$0.219} & 0.332{\scriptsize$\pm$0.211} & 0.102{\scriptsize$\pm$0.085} & 0.100{\scriptsize$\pm$0.082} & \textbf{0.105{\scriptsize$\pm$0.088}} & 0.102{\scriptsize$\pm$0.086} & 1.573{\scriptsize$\pm$0.701} & \textbf{1.578{\scriptsize$\pm$0.510}} & 0.526{\scriptsize$\pm$0.375} & 0.521{\scriptsize$\pm$0.362} & \textbf{0.140{\scriptsize$\pm$0.119}} & 0.137{\scriptsize$\pm$0.117} \\
\bottomrule
\end{tabular}}
\end{table*}

\subsection{Performance Across Difficulty Levels}

This section presents cross-level comparisons to assess how model performance degrades as task complexity increases. We examine three key aspects: (1) overall performance trends across difficulty levels, (2) fine-grained performance breakdown, (3) statistical distributions revealing variance and consistency patterns, and (4) temperature stability demonstrating evaluation robustness. All metrics are averaged across models to highlight systematic difficulty patterns rather than model-specific behaviors.

\subsubsection{Cross-Level Performance Trends}

\textbf{Overview.} This subsection examines how performance systematically degrades as task complexity increases from Level 1 (logic translation) to Level 3 (goal-oriented generation). By analyzing aggregate trends across all models, we isolate the inherent difficulty of each level independent of model-specific capabilities.

\textbf{Cognitive Demands by Level.} The three difficulty levels represent fundamentally different cognitive challenges. Level 1 tasks require faithful translation of fully-specified IF--THEN rules into executable code, primarily testing code generation competence. Level 2 tasks provide strategic skeletons but leave implementation gaps (thresholds, lookback windows), requiring models to supply plausible defaults grounded in domain knowledge. Level 3 tasks state only high-level objectives (e.g., profitability constraints, drawdown limits), demanding end-to-end strategy architecture design from first principles.

\begin{figure*}[h]
\centering
\begin{subfigure}[t]{0.48\textwidth}
\includegraphics[width=\textwidth]{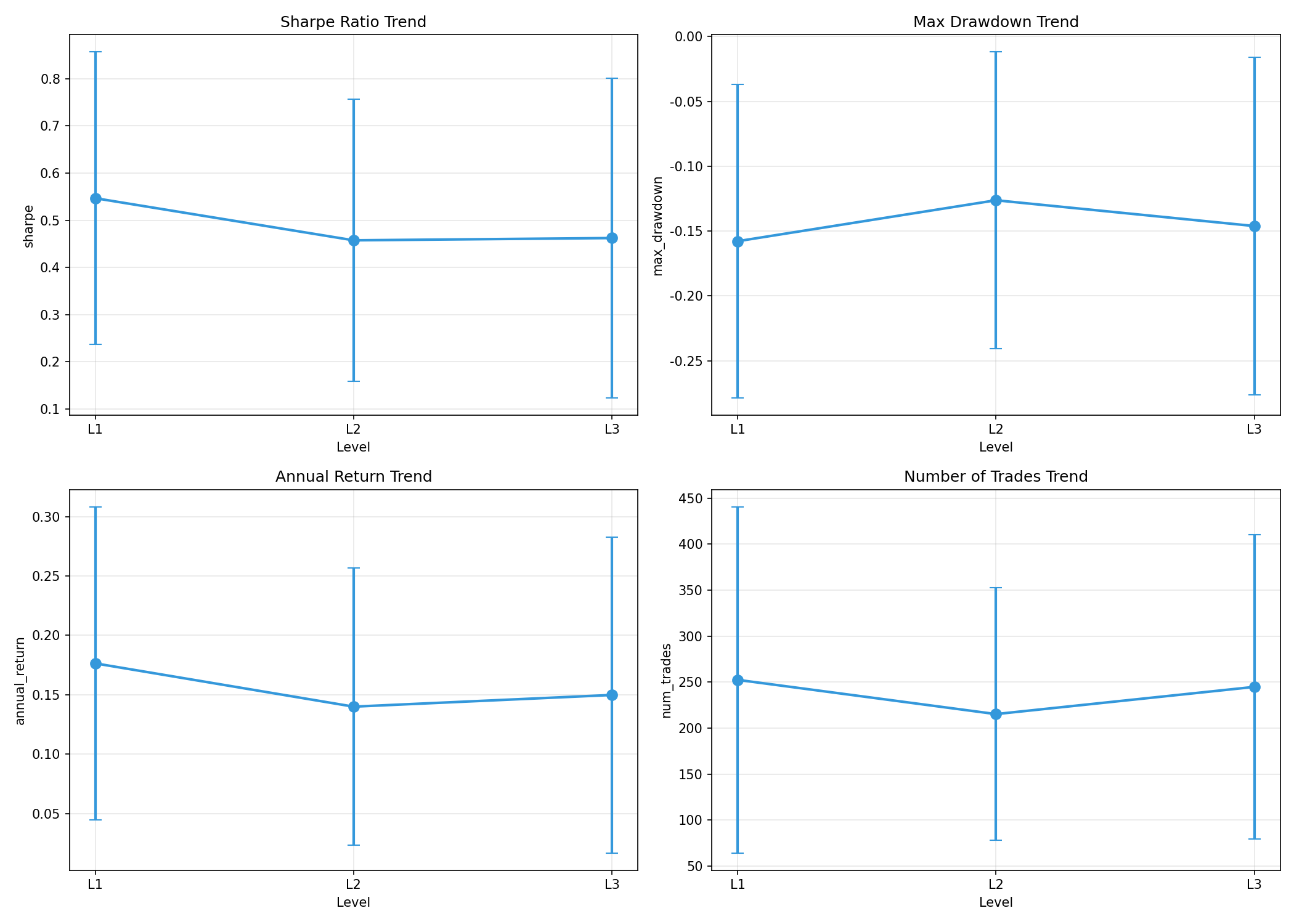}
\caption{T=0.0}
\end{subfigure}
\hfill
\begin{subfigure}[t]{0.48\textwidth}
\includegraphics[width=\textwidth]{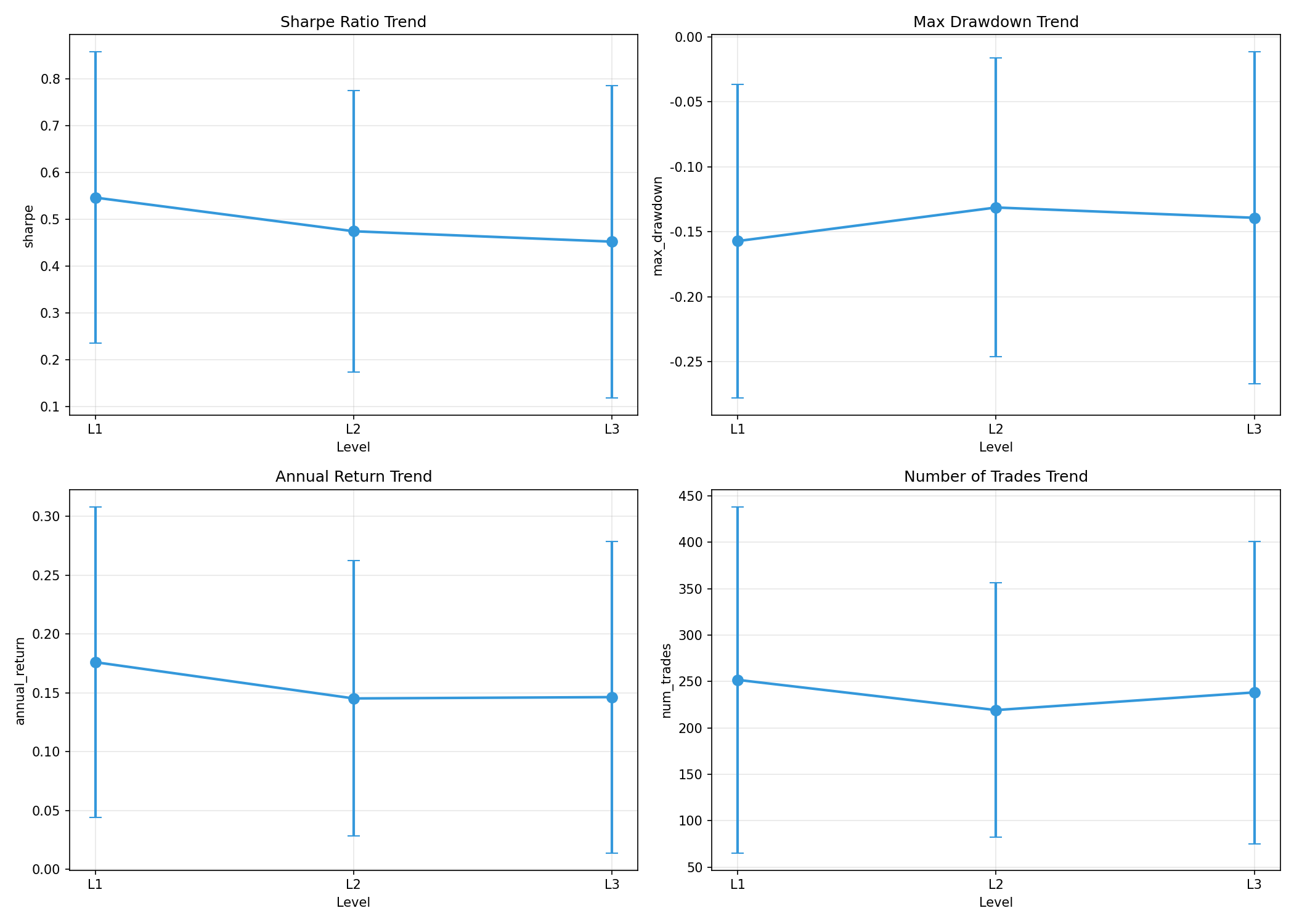}
\caption{T=0.7}
\end{subfigure}
\caption{Performance trends across difficulty levels (model-averaged). Core metrics generally decline from Level 1 to Level 2/L3, demonstrating the systematic difficulty gradient of AlphaForgeBench.}
\label{appx_fig:cross_level_comparison}
\end{figure*}

\textbf{Analysis.} \Cref{appx_fig:cross_level_comparison} demonstrates a clear and consistent performance degradation across all metrics as task complexity increases.

The model-averaged Sharpe Ratio declines from approximately 0.546 at Level 1 to 0.458 at Level 2 and 0.462 at Level 3, reflecting a clear difficulty gradient between Level 1 and the higher levels. Notably, Level 2 and Level 3 achieve nearly identical average performance, suggesting that the primary difficulty transition occurs between code translation (L1) and tasks requiring domain knowledge (L2/L3). Annualized Return follows a similar pattern, decreasing from 0.176 at Level 1 to 0.140 at Level 2. Maximum Drawdown remains relatively stable across levels (ranging from 0.126 to 0.157), suggesting that while profitability decreases with task complexity, risk management quality is largely maintained.

\textbf{Interpretation of Performance Patterns.} The observed patterns reveal important insights about model capabilities. The sharp drop from Level 1 to Level 2 (approximately 16\% decline in Sharpe Ratio) indicates that parameter inference poses a significant cognitive leap beyond pure code translation. This suggests that while models have mastered syntax generation, domain-specific knowledge for selecting appropriate thresholds and lookback windows remains challenging. Interestingly, Level 3 performance (SR = 0.462) is marginally higher than Level 2 (SR = 0.458), indicating that the primary difficulty barrier lies in the transition from fully-specified rules to tasks requiring domain knowledge, rather than in the distinction between parameter inference and goal-oriented design.

\textbf{Risk-Return Trade-offs Across Levels.} An interesting pattern emerges when examining risk-adjusted metrics. The Sharpe Ratio decline is steeper than the raw return decline, indicating that strategies become less efficient (higher risk per unit of return) as complexity increases. However, the Sortino Ratio (which focuses on downside risk) shows more stability, suggesting that models maintain reasonable downside protection even when overall performance degrades. This implies that the performance decline stems more from reduced upside capture than from increased catastrophic losses.

\subsubsection{Fine-Grained Difficulty Breakdown}

\textbf{Overview.} While the three-level categorization (L1/L2/L3) provides a coarse understanding of difficulty progression, each level is further subdivided into easy, medium, and hard variants, yielding nine fine-grained difficulty tiers. This subsection examines performance patterns at this granular level to understand both within-level and across-level difficulty gradients.

\begin{figure*}[h]
\centering
\begin{subfigure}[t]{0.48\textwidth}
\includegraphics[width=\textwidth]{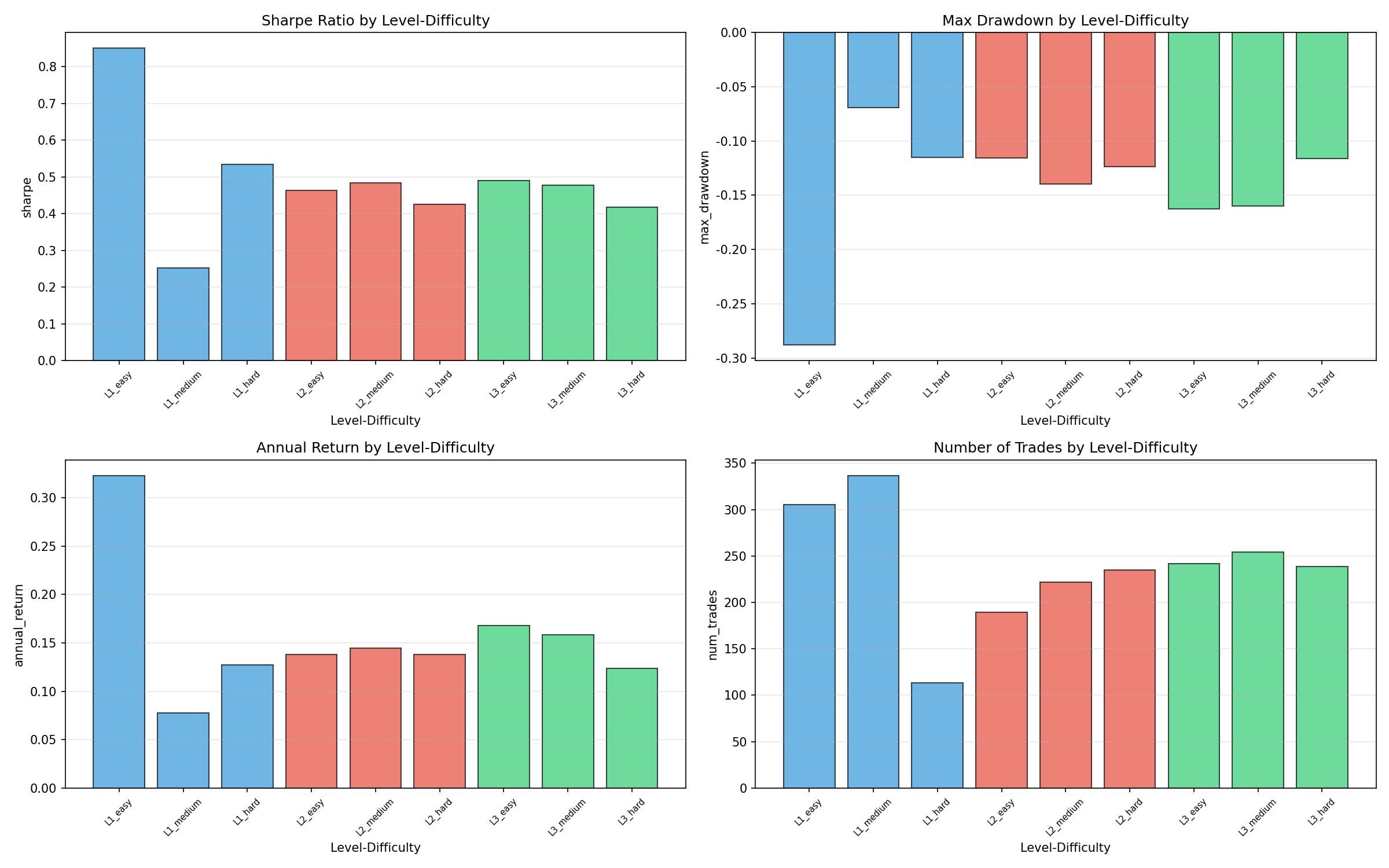}
\caption{T=0.0}
\end{subfigure}
\hfill
\begin{subfigure}[t]{0.48\textwidth}
\includegraphics[width=\textwidth]{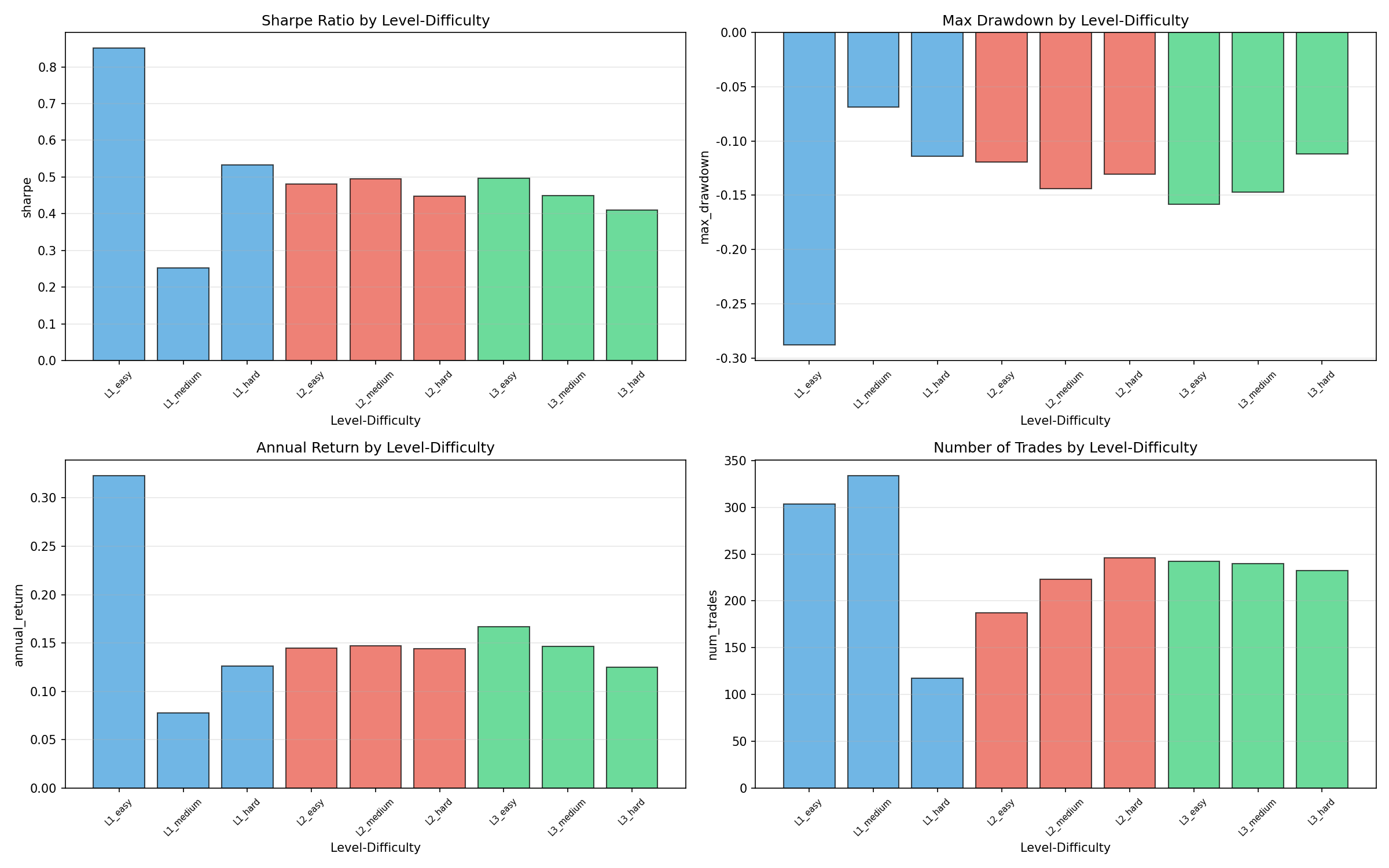}
\caption{T=0.7}
\end{subfigure}
\caption{Detailed performance breakdown across 9 fine-grained difficulty levels. Each level (L1, L2, L3) is subdivided into easy, medium, and hard variants, revealing both within-level and across-level difficulty gradients.}
\label{appx_fig:cross_level_detailed}
\end{figure*}

\textbf{Analysis.} \Cref{appx_fig:cross_level_detailed} presents the detailed performance breakdown across all nine difficulty tiers, revealing nuanced patterns that aggregate metrics obscure.

The fine-grained analysis reveals that the difficulty gradient operates at two distinct scales: within-level (easy/medium/hard) and across-level (L1/L2/L3). Within Level 1, performance varies substantially across subtasks, with L1\_easy achieving the highest model-averaged Sharpe Ratio (0.851) while L1\_medium drops to 0.253 and L1\_hard recovers to 0.533. The anomalously low performance on L1\_medium suggests that certain medium-complexity rule structures pose particular challenges for code translation, possibly involving indicator combinations or conditional logic that models struggle to faithfully reproduce.

\textbf{Critical Transition Points.} The transition from L1\_hard (SR $\approx$ 0.533) to L2\_easy (SR $\approx$ 0.462) represents a 13\% decline, which exceeds the entire within-level degradation of Level 2 (L2\_easy to L2\_hard: 8\% decline). This indicates that the cognitive leap from code translation to parameter inference is more significant than incremental complexity increases within a level. The L2\_hard to L3\_easy transition (SR from 0.429 to 0.491) shows a slight recovery, suggesting that the easiest goal-oriented tasks may be less demanding than the hardest parameter inference tasks.

\textbf{Within-Level Patterns.} The within-level difficulty gradient varies by level. Level 1 exhibits a non-monotonic pattern, with L1\_medium (SR = 0.253) performing substantially worse than both L1\_easy (0.851) and L1\_hard (0.533), suggesting that certain medium-complexity rule structures pose unique challenges. Level 2 shows a more gradual decline from easy (0.462) to hard (0.429), indicating that once models face parameter inference challenges, additional structural complexity has diminishing marginal impact. Level 3 shows moderate within-level variation (SR declining from 0.491 to 0.417), suggesting that creative strategy design difficulty is sensitive to constraint complexity.

\subsubsection{Distribution Analysis and Variance Patterns}

\textbf{Overview.} While aggregate metrics (mean, median) reveal central tendencies, they obscure critical information about performance consistency and variance. Boxplot visualizations expose the full distribution of results, including interquartile ranges, outliers, and distribution skewness. This subsection examines the statistical properties of performance distributions across difficulty levels to understand not just average performance, but also the reliability and predictability of model outputs.

\textbf{Why Distribution Analysis Matters.} In production deployments, understanding performance variance is as important as understanding average performance. A model with high average performance but wide variance may produce unreliable results, while a model with moderate average performance but tight variance offers predictable behavior. Distribution analysis also reveals whether performance degradation is uniform across all models or whether certain models struggle disproportionately on specific task types.

\begin{figure*}[h]
\centering
\begin{subfigure}[t]{0.48\textwidth}
\includegraphics[width=\textwidth]{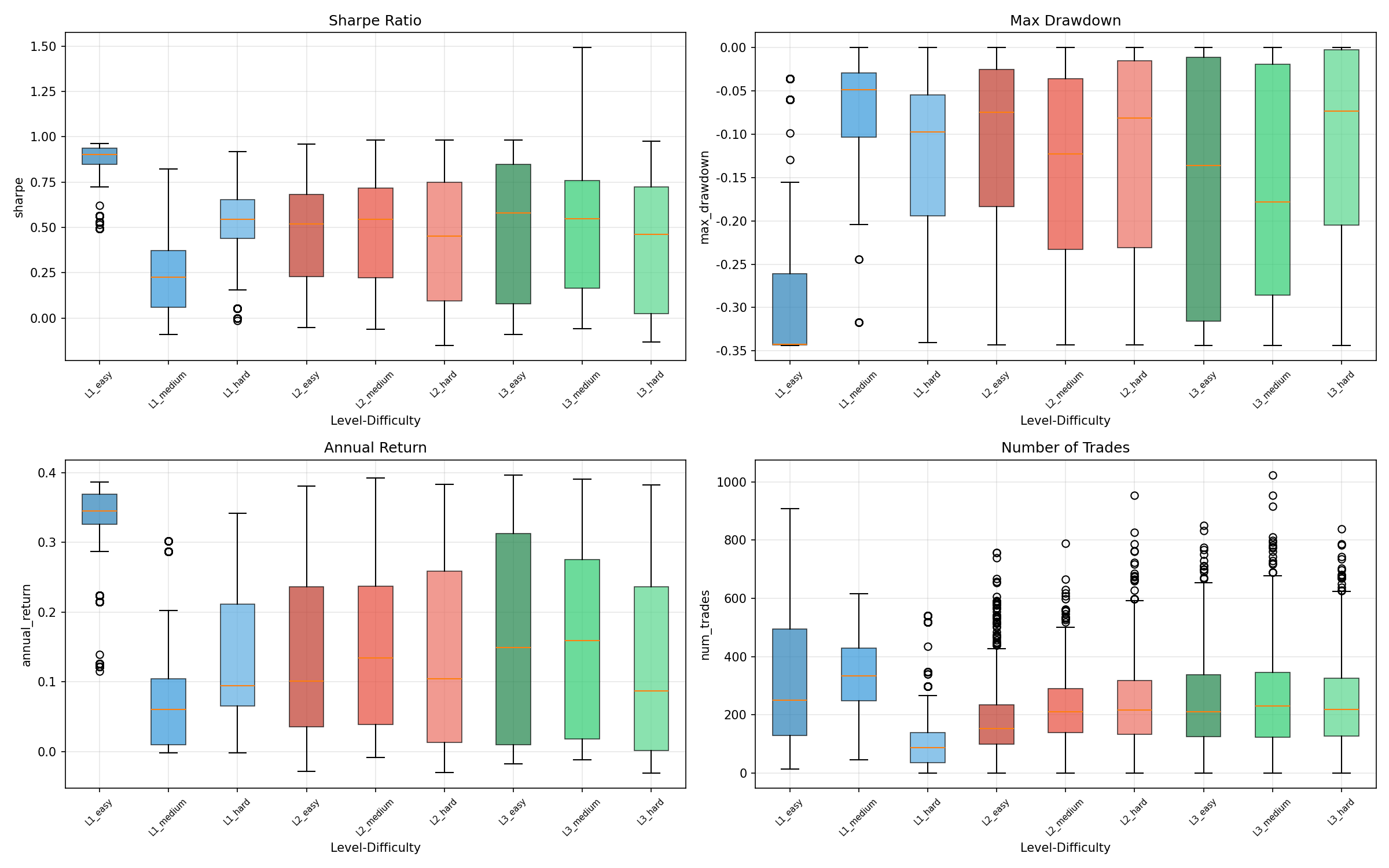}
\caption{T=0.0}
\end{subfigure}
\hfill
\begin{subfigure}[t]{0.48\textwidth}
\includegraphics[width=\textwidth]{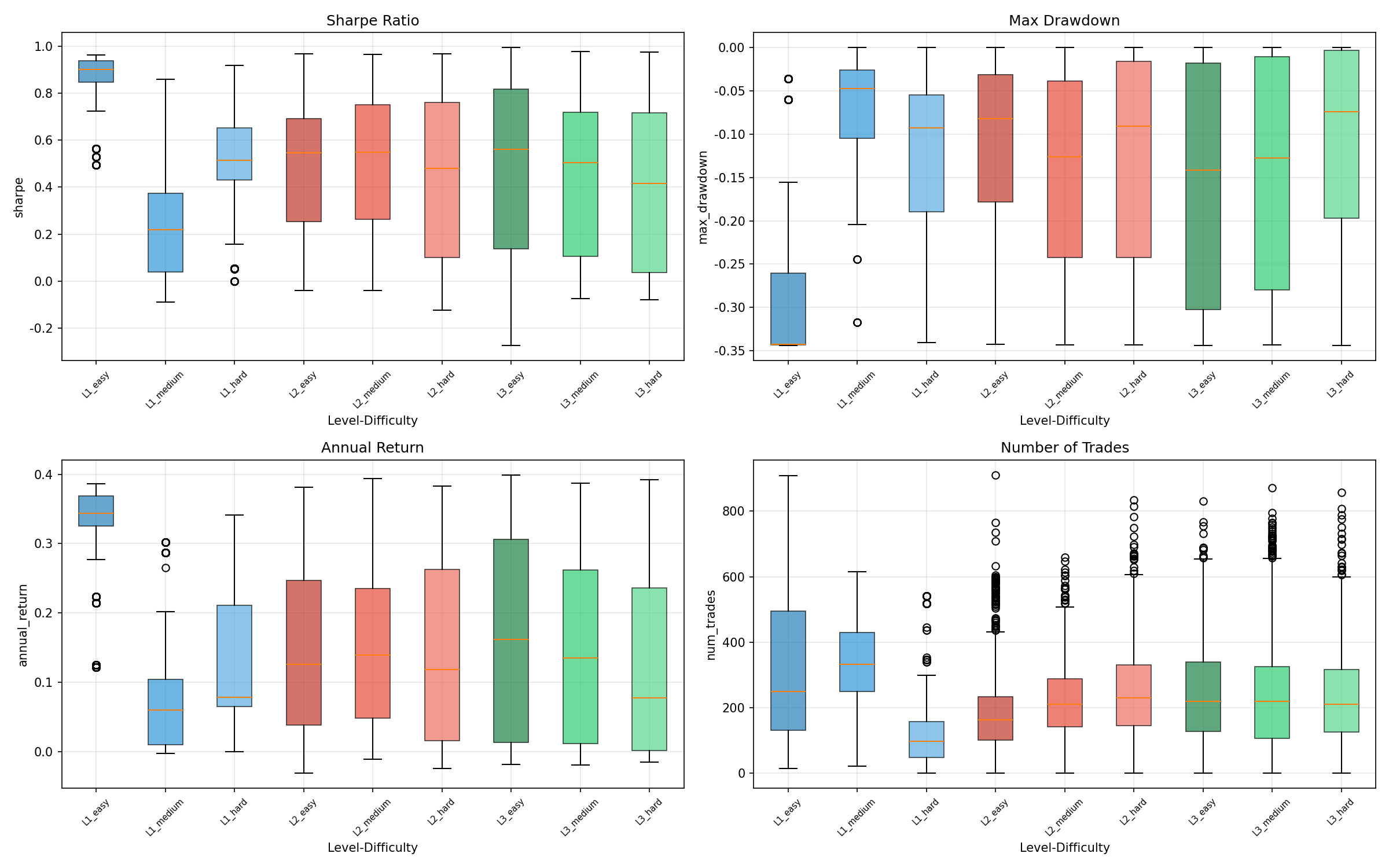}
\caption{T=0.7}
\end{subfigure}
\caption{Boxplot distributions across 9 fine-grained difficulty levels. Box boundaries represent the interquartile range (IQR), with the median shown as a horizontal line. Whiskers extend to 1.5$\times$IQR, and outliers are plotted individually.}
\label{appx_fig:boxplot_by_level_difficulty}
\end{figure*}

\textbf{Analysis.} \Cref{appx_fig:boxplot_by_level_difficulty} presents boxplot distributions for all nine fine-grained difficulty levels, revealing the spread and consistency of model performance within each tier.

The boxplots reveal several critical patterns. First, Level 1 tasks exhibit the tightest distributions, with narrow interquartile ranges indicating consistent performance across models. The median Sharpe Ratio for L1\_easy tasks sits above 0.80, with minimal outliers, confirming that straightforward code translation is a solved problem for modern LLMs. As difficulty increases within Level 1 (easy $\to$ medium $\to$ hard), the boxes widen progressively, indicating increased variance in model capabilities when handling more complex rule structures.

\textbf{Level 1 Distribution Characteristics.} The tight distributions at Level 1 (IQR approximately 0.15--0.20) indicate that all evaluated models have achieved competent code generation capabilities. The few outliers present are predominantly on the lower end, suggesting occasional failures rather than exceptional successes. The symmetric distribution shape (median near box center) indicates balanced performance without systematic bias toward over-performance or under-performance. This symmetry validates that the benchmark's Level 1 tasks effectively test code translation without introducing confounding factors.

Level 2 tasks show a marked increase in distribution width compared to Level 1. The interquartile ranges expand significantly, particularly for L2\_medium and L2\_hard, suggesting that parameter inference introduces substantial variability in strategy quality. Notably, the median values remain relatively stable across L2 subtasks (around 0.45--0.48), but the wider boxes indicate that some models consistently make better parameter choices than others. The presence of lower outliers in Level 2 reveals that certain model-query combinations result in particularly poor parameter selections.

\textbf{Level 2 Variance Expansion.} The IQR expansion at Level 2 (approximately 0.25--0.35) represents a 50--75\% increase compared to Level 1, quantifying the increased difficulty and inconsistency introduced by parameter inference. The lower outliers become more frequent, indicating that some models occasionally select highly inappropriate parameters (e.g., extremely short lookback windows, unrealistic thresholds). Interestingly, upper outliers also appear more frequently, suggesting that when models make good parameter choices, they can achieve performance comparable to or exceeding Level 1 results. This bimodal tendency indicates that parameter inference is a high-stakes task where success and failure have large performance implications.

Level 3 distributions exhibit the most interesting characteristics. While the median performance is lower than Level 1, the interquartile ranges are comparable to Level 2, suggesting that creative strategy design introduces variance but not necessarily more than parameter inference. However, Level 3 shows more upper outliers, indicating that some models occasionally generate exceptionally high-performing strategies when given creative freedom. This pattern suggests that goal-oriented generation has higher variance in outcomes---models either succeed brilliantly or produce mediocre results, with less middle ground.

\textbf{Level 3 Distribution Skewness.} Unlike the symmetric distributions at Level 1, Level 3 exhibits positive skewness (median closer to lower quartile, long upper tail). This indicates that while most generated strategies achieve moderate performance, a subset achieves exceptional results. The presence of numerous upper outliers (Sharpe Ratios exceeding 0.8--0.9) demonstrates that creative freedom occasionally enables models to discover highly effective strategies that would not emerge from constrained prompts. However, the lower median indicates that such successes are not consistent. This high-risk, high-reward profile suggests that Level 3 tasks may benefit from ensemble approaches or multiple sampling strategies.

\begin{figure*}[h]
\centering
\begin{subfigure}[t]{0.48\textwidth}
\includegraphics[width=\textwidth]{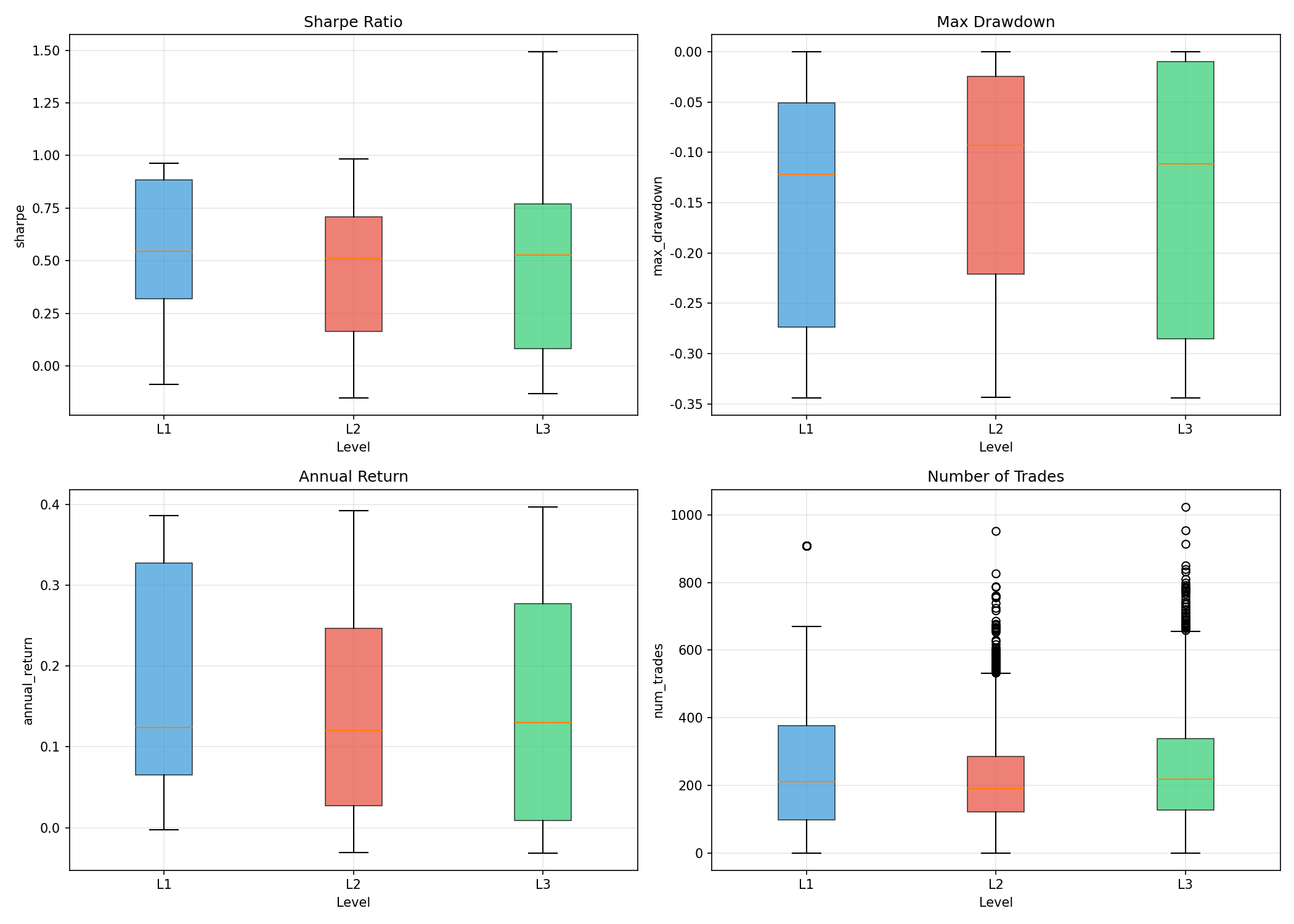}
\caption{T=0.0}
\end{subfigure}
\hfill
\begin{subfigure}[t]{0.48\textwidth}
\includegraphics[width=\textwidth]{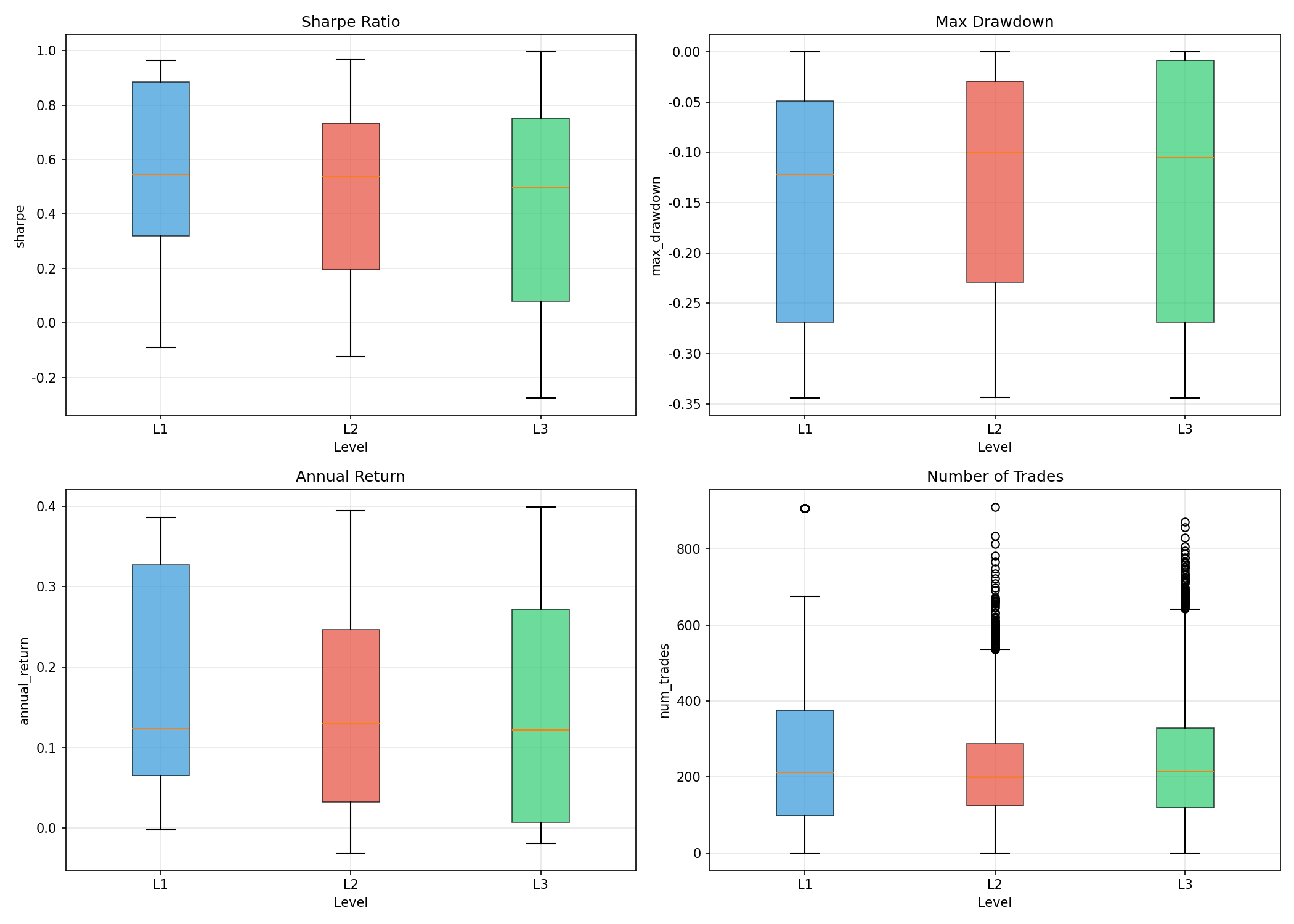}
\caption{T=0.7}
\end{subfigure}
\caption{Boxplot distributions aggregated by three main difficulty levels (L1, L2, L3). This coarse-grained view highlights the overall performance degradation and variance increase as task complexity grows.}
\label{appx_fig:boxplot_by_level}
\end{figure*}

\textbf{Aggregated View for Cross-Level Comparison.} \Cref{appx_fig:boxplot_by_level} provides a coarser view aggregated by the three main difficulty levels (L1, L2, L3), facilitating direct comparison of overall distribution trends and enabling clearer visualization of the systematic difficulty gradient.

The aggregated view confirms the systematic difficulty gradient. The median Sharpe Ratio declines from approximately 0.54 at Level 1 to 0.46 at Level 3, consistent with the line chart analysis in \Cref{appx_fig:cross_level_comparison}. However, the boxplot representation adds crucial information about distribution shape. Level 1 shows a relatively symmetric distribution with the median near the center of the box, indicating balanced performance across models. Level 2 exhibits slight negative skew, with the median closer to the upper quartile, suggesting that most models perform reasonably well but a subset struggles significantly with parameter inference. Level 3 shows positive skew, with the median closer to the lower quartile, indicating that while most models produce moderate results, a few exceptional cases achieve substantially higher performance.

\textbf{Variance Progression Across Levels.} The box widths provide quantitative evidence of increasing task difficulty. Level 1's narrow boxes (IQR $\approx$ 0.18) indicate that all models cluster around similar performance levels, with skill differences manifesting primarily in edge cases. Level 2's wider boxes (IQR $\approx$ 0.30) demonstrate that parameter inference separates models more clearly---domain knowledge becomes a differentiating factor. Level 3's boxes (IQR $\approx$ 0.28) are slightly narrower than Level 2, but the longer whiskers and more numerous outliers indicate that while typical performance is somewhat predictable, exceptional outcomes (both positive and negative) occur more frequently.

The stability of the interquartile range across temperature settings (T=0.0 vs T=0.7) is noteworthy. The box widths remain nearly identical between temperatures, demonstrating that the variance in performance stems from genuine model capability differences rather than stochastic sampling effects. This reinforces the robustness of factor-based evaluation discussed in the temperature stability analysis.

\textbf{Implications for Model Selection.} The distribution analysis provides actionable insights for practitioners. For applications requiring predictable, consistent performance (e.g., production trading systems), Level 1-style prompts with explicit parameters are preferable, as they yield tight performance distributions. For research or exploration scenarios where occasional high performance is valued over consistency, Level 3-style prompts may be appropriate despite higher variance. The bimodal tendencies at Level 3 also suggest that ensemble methods or best-of-N sampling could be particularly effective for goal-oriented tasks.

\subsubsection{Temperature Stability Analysis}

\textbf{Overview.} To demonstrate the robustness of factor-based evaluation, we compare the stability of results under different decoding temperatures (T=0.0 vs T=0.7) across all difficulty levels.

\begin{figure}[h]
\centering
\includegraphics[width=0.95\linewidth]{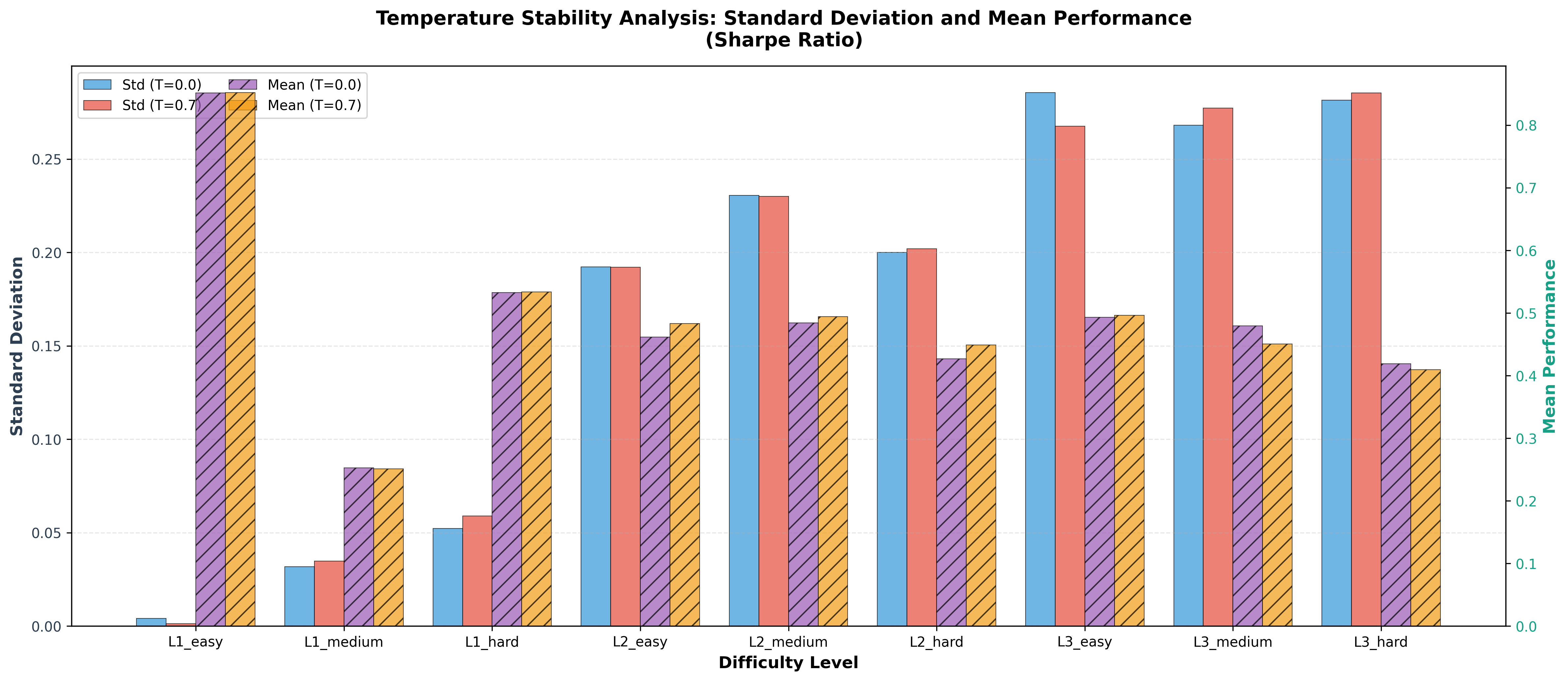}
\caption{Comparison of standard deviation and mean performance across temperature settings (T=0.0 vs T=0.7). The minimal difference in standard deviation (average 2.97\%) demonstrates that factor-based evaluation is significantly more stable than direct text-based LLM evaluation, which is highly sensitive to temperature variations.}
\label{appx_fig:temperature_stability}
\end{figure}

\textbf{Analysis.} \Cref{appx_fig:temperature_stability} provides compelling evidence for the stability of factor-based evaluation.

 The standard deviation remains remarkably consistent across temperature settings, with an average difference of only 2.97\%. This stability is observed across all nine difficulty levels, from L1\_easy (std difference: 0.0028) to L3\_hard (std difference: 0.0038). The mean performance values also show minimal variation between temperatures, with differences typically under 5\%.

This stability is particularly significant when contrasted with traditional text-based evaluation methods, where LLMs directly output scores or judgments. Such approaches are known to be highly sensitive to temperature settings, with T=0.7 often producing substantially different results than T=0.0 due to the stochastic nature of token sampling. In our factor-based approach, even though the generated code may vary slightly across runs, the resulting trading strategies produce consistent financial metrics, demonstrating that the evaluation captures genuine strategic quality rather than surface-level variations.

The consistency across difficulty levels is also noteworthy. Both easy and hard tasks maintain similar stability ratios, indicating that the factor-based evaluation framework is robust regardless of task complexity. This property is crucial for benchmark reliability, as it ensures that performance comparisons remain valid across different experimental conditions.

\clearpage

\subsection{Per-Asset Analysis}

Model performance varies significantly across different asset classes. \Cref{appx_tab:per_asset_stage2} consolidates the per-asset results for all seven backtest assets. Cryptocurrency assets (BTCUSDT, ETHUSDT) generally exhibit higher volatility but also higher potential returns compared to traditional stocks.


\begin{table*}[htb]
\centering
\caption{Per-asset model performance on the Stage~2 benchmark (mean $\pm$ std across 270 queries). Assets are grouped by market type with gray header rows. Within each asset block, the best value per column is in \textbf{bold}. $\uparrow$: higher is better; $\downarrow$: lower is better.}
\label{appx_tab:per_asset_stage2}\label{appx_tab:btcusdt}\label{appx_tab:ethusdt}\label{appx_tab:aapl}\label{appx_tab:googl}\label{appx_tab:msft}\label{appx_tab:nvda}\label{appx_tab:tsla}
\resizebox{\textwidth}{!}{%
\begin{tabular}{lcccccccccccc}
\toprule
\textbf{Model} & \multicolumn{2}{c}{\textbf{SR}$\uparrow$} & \multicolumn{2}{c}{\textbf{ARR}$\uparrow$} & \multicolumn{2}{c}{\textbf{MDD}$\downarrow$} & \multicolumn{2}{c}{\textbf{CR}$\uparrow$} & \multicolumn{2}{c}{\textbf{SOR}$\uparrow$} & \multicolumn{2}{c}{\textbf{VOL}$\downarrow$} \\
& T=0 & T=0.7 & T=0 & T=0.7 & T=0 & T=0.7 & T=0 & T=0.7 & T=0 & T=0.7 & T=0 & T=0.7 \\
\midrule
\rowcolor{gray!15} \multicolumn{13}{c}{\textbf{BTCUSDT} (Cryptocurrency)} \\
\textit{claude-sonnet-4.5} & 0.353{\scriptsize$\pm$0.186} & 0.354{\scriptsize$\pm$0.186} & 0.134{\scriptsize$\pm$0.085} & 0.133{\scriptsize$\pm$0.085} & 0.203{\scriptsize$\pm$0.128} & 0.199{\scriptsize$\pm$0.128} & 0.817{\scriptsize$\pm$0.655} & 0.833{\scriptsize$\pm$0.676} & 0.601{\scriptsize$\pm$0.330} & 0.601{\scriptsize$\pm$0.327} & 0.233{\scriptsize$\pm$0.149} & 0.230{\scriptsize$\pm$0.149} \\
\textit{deepseek-v3.2} & 0.307{\scriptsize$\pm$0.202} & 0.297{\scriptsize$\pm$0.205} & 0.114{\scriptsize$\pm$0.087} & 0.110{\scriptsize$\pm$0.088} & 0.173{\scriptsize$\pm$0.130} & 0.169{\scriptsize$\pm$0.131} & 0.821{\scriptsize$\pm$0.687} & 0.854{\scriptsize$\pm$0.758} & 0.526{\scriptsize$\pm$0.348} & 0.513{\scriptsize$\pm$0.348} & 0.198{\scriptsize$\pm$0.150} & 0.194{\scriptsize$\pm$0.152} \\
\textit{gpt-5.2} & 0.301{\scriptsize$\pm$0.212} & 0.302{\scriptsize$\pm$0.211} & 0.111{\scriptsize$\pm$0.094} & 0.111{\scriptsize$\pm$0.093} & \textbf{0.158{\scriptsize$\pm$0.139}} & \textbf{0.159{\scriptsize$\pm$0.137}} & \textbf{0.958{\scriptsize$\pm$0.781}} & \textbf{0.954{\scriptsize$\pm$0.789}} & 0.497{\scriptsize$\pm$0.373} & 0.499{\scriptsize$\pm$0.370} & \textbf{0.184{\scriptsize$\pm$0.161}} & \textbf{0.185{\scriptsize$\pm$0.159}} \\
\textit{gemini-3-flash-preview} & 0.373{\scriptsize$\pm$0.205} & 0.371{\scriptsize$\pm$0.176} & 0.137{\scriptsize$\pm$0.094} & 0.138{\scriptsize$\pm$0.085} & 0.203{\scriptsize$\pm$0.139} & 0.205{\scriptsize$\pm$0.130} & 0.876{\scriptsize$\pm$0.723} & 0.855{\scriptsize$\pm$0.683} & 0.621{\scriptsize$\pm$0.357} & 0.624{\scriptsize$\pm$0.315} & 0.235{\scriptsize$\pm$0.160} & 0.238{\scriptsize$\pm$0.150} \\
\textbf{\textit{gemini-3-pro-preview}} & \textbf{0.420{\scriptsize$\pm$0.162}} & \textbf{0.425{\scriptsize$\pm$0.143}} & \textbf{0.164{\scriptsize$\pm$0.078}} & \textbf{0.167{\scriptsize$\pm$0.071}} & 0.254{\scriptsize$\pm$0.125} & 0.251{\scriptsize$\pm$0.118} & 0.797{\scriptsize$\pm$0.643} & 0.837{\scriptsize$\pm$0.623} & \textbf{0.739{\scriptsize$\pm$0.313}} & \textbf{0.736{\scriptsize$\pm$0.258}} & 0.294{\scriptsize$\pm$0.142} & 0.291{\scriptsize$\pm$0.134} \\
\textit{grok-4.1-fast} & 0.299{\scriptsize$\pm$0.194} & 0.313{\scriptsize$\pm$0.187} & 0.111{\scriptsize$\pm$0.086} & 0.114{\scriptsize$\pm$0.084} & 0.174{\scriptsize$\pm$0.124} & 0.177{\scriptsize$\pm$0.124} & 0.797{\scriptsize$\pm$0.715} & 0.869{\scriptsize$\pm$0.710} & 0.508{\scriptsize$\pm$0.334} & 0.524{\scriptsize$\pm$0.325} & 0.199{\scriptsize$\pm$0.142} & 0.203{\scriptsize$\pm$0.142} \\
\midrule
\rowcolor{gray!15} \multicolumn{13}{c}{\textbf{ETHUSDT} (Cryptocurrency)} \\
\textit{claude-sonnet-4.5} & 0.297{\scriptsize$\pm$0.225} & 0.290{\scriptsize$\pm$0.226} & 0.151{\scriptsize$\pm$0.150} & 0.147{\scriptsize$\pm$0.150} & 0.245{\scriptsize$\pm$0.184} & 0.243{\scriptsize$\pm$0.183} & \textbf{0.592{\scriptsize$\pm$0.376}} & \textbf{0.563{\scriptsize$\pm$0.346}} & 0.465{\scriptsize$\pm$0.348} & 0.455{\scriptsize$\pm$0.349} & 0.285{\scriptsize$\pm$0.220} & 0.282{\scriptsize$\pm$0.220} \\
\textit{deepseek-v3.2} & 0.252{\scriptsize$\pm$0.218} & 0.244{\scriptsize$\pm$0.217} & 0.123{\scriptsize$\pm$0.138} & 0.118{\scriptsize$\pm$0.137} & 0.216{\scriptsize$\pm$0.182} & 0.211{\scriptsize$\pm$0.184} & 0.545{\scriptsize$\pm$0.370} & 0.560{\scriptsize$\pm$0.440} & 0.396{\scriptsize$\pm$0.336} & 0.382{\scriptsize$\pm$0.335} & 0.249{\scriptsize$\pm$0.216} & 0.243{\scriptsize$\pm$0.218} \\
\textit{gpt-5.2} & 0.233{\scriptsize$\pm$0.241} & 0.229{\scriptsize$\pm$0.236} & 0.121{\scriptsize$\pm$0.153} & 0.118{\scriptsize$\pm$0.152} & \textbf{0.197{\scriptsize$\pm$0.193}} & \textbf{0.194{\scriptsize$\pm$0.191}} & 0.564{\scriptsize$\pm$0.405} & 0.562{\scriptsize$\pm$0.365} & 0.369{\scriptsize$\pm$0.370} & 0.361{\scriptsize$\pm$0.363} & \textbf{0.228{\scriptsize$\pm$0.230}} & \textbf{0.224{\scriptsize$\pm$0.227}} \\
\textit{gemini-3-flash-preview} & 0.276{\scriptsize$\pm$0.260} & 0.292{\scriptsize$\pm$0.230} & 0.141{\scriptsize$\pm$0.162} & 0.145{\scriptsize$\pm$0.152} & 0.246{\scriptsize$\pm$0.195} & 0.249{\scriptsize$\pm$0.184} & 0.507{\scriptsize$\pm$0.416} & 0.544{\scriptsize$\pm$0.354} & 0.444{\scriptsize$\pm$0.385} & 0.460{\scriptsize$\pm$0.352} & 0.284{\scriptsize$\pm$0.235} & 0.288{\scriptsize$\pm$0.222} \\
\textbf{\textit{gemini-3-pro-preview}} & \textbf{0.358{\scriptsize$\pm$0.224}} & \textbf{0.355{\scriptsize$\pm$0.215}} & \textbf{0.181{\scriptsize$\pm$0.156}} & \textbf{0.179{\scriptsize$\pm$0.149}} & 0.311{\scriptsize$\pm$0.174} & 0.305{\scriptsize$\pm$0.166} & 0.543{\scriptsize$\pm$0.360} & 0.559{\scriptsize$\pm$0.330} & \textbf{0.569{\scriptsize$\pm$0.341}} & \textbf{0.564{\scriptsize$\pm$0.326}} & 0.360{\scriptsize$\pm$0.211} & 0.354{\scriptsize$\pm$0.202} \\
\textit{grok-4.1-fast} & 0.245{\scriptsize$\pm$0.213} & 0.248{\scriptsize$\pm$0.211} & 0.121{\scriptsize$\pm$0.135} & 0.122{\scriptsize$\pm$0.135} & 0.214{\scriptsize$\pm$0.174} & 0.218{\scriptsize$\pm$0.172} & 0.529{\scriptsize$\pm$0.363} & 0.515{\scriptsize$\pm$0.340} & 0.387{\scriptsize$\pm$0.327} & 0.393{\scriptsize$\pm$0.322} & 0.247{\scriptsize$\pm$0.207} & 0.251{\scriptsize$\pm$0.205} \\
\midrule
\rowcolor{gray!15} \multicolumn{13}{c}{\textbf{AAPL} (US Equity)} \\
\textit{claude-sonnet-4.5} & 0.941{\scriptsize$\pm$0.493} & 0.924{\scriptsize$\pm$0.488} & 0.180{\scriptsize$\pm$0.119} & 0.175{\scriptsize$\pm$0.117} & 0.073{\scriptsize$\pm$0.048} & 0.071{\scriptsize$\pm$0.047} & 2.799{\scriptsize$\pm$1.303} & \textbf{2.779{\scriptsize$\pm$1.330}} & 1.262{\scriptsize$\pm$0.726} & 1.240{\scriptsize$\pm$0.721} & 0.116{\scriptsize$\pm$0.074} & 0.113{\scriptsize$\pm$0.072} \\
\textit{deepseek-v3.2} & 0.803{\scriptsize$\pm$0.515} & 0.784{\scriptsize$\pm$0.526} & 0.145{\scriptsize$\pm$0.116} & 0.142{\scriptsize$\pm$0.119} & 0.059{\scriptsize$\pm$0.047} & 0.058{\scriptsize$\pm$0.048} & 2.774{\scriptsize$\pm$1.499} & 2.722{\scriptsize$\pm$1.400} & 1.074{\scriptsize$\pm$0.732} & 1.047{\scriptsize$\pm$0.750} & 0.095{\scriptsize$\pm$0.072} & 0.093{\scriptsize$\pm$0.074} \\
\textit{gpt-5.2} & 0.745{\scriptsize$\pm$0.549} & 0.751{\scriptsize$\pm$0.548} & 0.139{\scriptsize$\pm$0.125} & 0.139{\scriptsize$\pm$0.125} & \textbf{0.058{\scriptsize$\pm$0.052}} & \textbf{0.058{\scriptsize$\pm$0.052}} & 2.621{\scriptsize$\pm$1.493} & 2.745{\scriptsize$\pm$1.577} & 0.977{\scriptsize$\pm$0.780} & 0.982{\scriptsize$\pm$0.778} & \textbf{0.092{\scriptsize$\pm$0.079}} & \textbf{0.091{\scriptsize$\pm$0.079}} \\
\textit{gemini-3-flash-preview} & 0.946{\scriptsize$\pm$0.505} & 0.953{\scriptsize$\pm$0.472} & 0.173{\scriptsize$\pm$0.125} & 0.177{\scriptsize$\pm$0.117} & 0.073{\scriptsize$\pm$0.051} & 0.075{\scriptsize$\pm$0.048} & \textbf{2.861{\scriptsize$\pm$2.392}} & 2.672{\scriptsize$\pm$1.327} & 1.237{\scriptsize$\pm$0.740} & 1.257{\scriptsize$\pm$0.698} & 0.116{\scriptsize$\pm$0.078} & 0.118{\scriptsize$\pm$0.073} \\
\textbf{\textit{gemini-3-pro-preview}} & \textbf{1.067{\scriptsize$\pm$0.480}} & \textbf{1.075{\scriptsize$\pm$0.446}} & \textbf{0.214{\scriptsize$\pm$0.123}} & \textbf{0.215{\scriptsize$\pm$0.115}} & 0.093{\scriptsize$\pm$0.049} & 0.091{\scriptsize$\pm$0.046} & 2.539{\scriptsize$\pm$1.472} & 2.661{\scriptsize$\pm$1.312} & \textbf{1.456{\scriptsize$\pm$0.729}} & \textbf{1.462{\scriptsize$\pm$0.679}} & 0.146{\scriptsize$\pm$0.073} & 0.144{\scriptsize$\pm$0.070} \\
\textit{grok-4.1-fast} & 0.744{\scriptsize$\pm$0.494} & 0.751{\scriptsize$\pm$0.485} & 0.139{\scriptsize$\pm$0.114} & 0.139{\scriptsize$\pm$0.112} & 0.059{\scriptsize$\pm$0.046} & 0.059{\scriptsize$\pm$0.046} & 2.613{\scriptsize$\pm$1.453} & 2.615{\scriptsize$\pm$1.386} & 0.986{\scriptsize$\pm$0.711} & 0.993{\scriptsize$\pm$0.697} & 0.094{\scriptsize$\pm$0.071} & 0.094{\scriptsize$\pm$0.070} \\
\midrule
\rowcolor{gray!15} \multicolumn{13}{c}{\textbf{GOOGL} (US Equity)} \\
\textit{claude-sonnet-4.5} & 0.792{\scriptsize$\pm$0.445} & 0.789{\scriptsize$\pm$0.434} & 0.201{\scriptsize$\pm$0.146} & 0.195{\scriptsize$\pm$0.140} & 0.103{\scriptsize$\pm$0.089} & 0.103{\scriptsize$\pm$0.089} & 2.575{\scriptsize$\pm$1.566} & 2.578{\scriptsize$\pm$1.596} & 1.316{\scriptsize$\pm$0.823} & 1.308{\scriptsize$\pm$0.799} & 0.168{\scriptsize$\pm$0.132} & 0.168{\scriptsize$\pm$0.132} \\
\textit{deepseek-v3.2} & 0.628{\scriptsize$\pm$0.455} & 0.627{\scriptsize$\pm$0.470} & 0.146{\scriptsize$\pm$0.138} & 0.146{\scriptsize$\pm$0.142} & 0.087{\scriptsize$\pm$0.088} & 0.084{\scriptsize$\pm$0.088} & 2.261{\scriptsize$\pm$1.689} & 2.350{\scriptsize$\pm$1.720} & 1.044{\scriptsize$\pm$0.815} & 1.036{\scriptsize$\pm$0.837} & 0.139{\scriptsize$\pm$0.131} & 0.136{\scriptsize$\pm$0.132} \\
\textit{gpt-5.2} & 0.635{\scriptsize$\pm$0.502} & 0.639{\scriptsize$\pm$0.502} & 0.160{\scriptsize$\pm$0.154} & 0.161{\scriptsize$\pm$0.154} & \textbf{0.080{\scriptsize$\pm$0.089}} & \textbf{0.077{\scriptsize$\pm$0.086}} & 2.636{\scriptsize$\pm$1.946} & 2.648{\scriptsize$\pm$1.621} & 1.055{\scriptsize$\pm$0.897} & 1.051{\scriptsize$\pm$0.890} & \textbf{0.131{\scriptsize$\pm$0.134}} & \textbf{0.127{\scriptsize$\pm$0.131}} \\
\textit{gemini-3-flash-preview} & 0.802{\scriptsize$\pm$0.492} & 0.808{\scriptsize$\pm$0.442} & 0.195{\scriptsize$\pm$0.157} & 0.199{\scriptsize$\pm$0.146} & 0.104{\scriptsize$\pm$0.097} & 0.105{\scriptsize$\pm$0.089} & 2.435{\scriptsize$\pm$1.751} & 2.487{\scriptsize$\pm$1.605} & 1.325{\scriptsize$\pm$0.902} & 1.336{\scriptsize$\pm$0.822} & 0.170{\scriptsize$\pm$0.143} & 0.170{\scriptsize$\pm$0.132} \\
\textbf{\textit{gemini-3-pro-preview}} & \textbf{0.960{\scriptsize$\pm$0.446}} & \textbf{0.968{\scriptsize$\pm$0.423}} & \textbf{0.252{\scriptsize$\pm$0.153}} & \textbf{0.256{\scriptsize$\pm$0.143}} & 0.131{\scriptsize$\pm$0.097} & 0.130{\scriptsize$\pm$0.090} & 2.633{\scriptsize$\pm$1.765} & 2.700{\scriptsize$\pm$1.552} & \textbf{1.627{\scriptsize$\pm$0.834}} & \textbf{1.644{\scriptsize$\pm$0.789}} & 0.212{\scriptsize$\pm$0.140} & 0.211{\scriptsize$\pm$0.131} \\
\textit{grok-4.1-fast} & 0.639{\scriptsize$\pm$0.446} & 0.644{\scriptsize$\pm$0.440} & 0.157{\scriptsize$\pm$0.138} & 0.157{\scriptsize$\pm$0.137} & 0.082{\scriptsize$\pm$0.082} & 0.084{\scriptsize$\pm$0.084} & \textbf{2.646{\scriptsize$\pm$1.915}} & \textbf{2.830{\scriptsize$\pm$2.808}} & 1.065{\scriptsize$\pm$0.798} & 1.079{\scriptsize$\pm$0.785} & 0.133{\scriptsize$\pm$0.123} & 0.137{\scriptsize$\pm$0.125} \\
\midrule
\rowcolor{gray!15} \multicolumn{13}{c}{\textbf{MSFT} (US Equity)} \\
\textit{claude-sonnet-4.5} & 0.438{\scriptsize$\pm$0.312} & 0.433{\scriptsize$\pm$0.308} & 0.084{\scriptsize$\pm$0.062} & 0.083{\scriptsize$\pm$0.061} & 0.078{\scriptsize$\pm$0.047} & 0.077{\scriptsize$\pm$0.046} & 1.145{\scriptsize$\pm$1.180} & 1.097{\scriptsize$\pm$0.971} & 0.657{\scriptsize$\pm$0.450} & 0.644{\scriptsize$\pm$0.443} & 0.113{\scriptsize$\pm$0.067} & 0.111{\scriptsize$\pm$0.065} \\
\textit{deepseek-v3.2} & 0.362{\scriptsize$\pm$0.314} & 0.360{\scriptsize$\pm$0.321} & 0.066{\scriptsize$\pm$0.062} & 0.065{\scriptsize$\pm$0.064} & 0.062{\scriptsize$\pm$0.046} & \textbf{0.061{\scriptsize$\pm$0.047}} & 1.035{\scriptsize$\pm$0.975} & 1.037{\scriptsize$\pm$0.951} & 0.529{\scriptsize$\pm$0.455} & 0.525{\scriptsize$\pm$0.469} & 0.091{\scriptsize$\pm$0.066} & \textbf{0.088{\scriptsize$\pm$0.068}} \\
\textit{gpt-5.2} & 0.376{\scriptsize$\pm$0.326} & 0.376{\scriptsize$\pm$0.333} & 0.070{\scriptsize$\pm$0.064} & 0.070{\scriptsize$\pm$0.066} & 0.063{\scriptsize$\pm$0.050} & 0.062{\scriptsize$\pm$0.050} & 1.130{\scriptsize$\pm$0.937} & 1.154{\scriptsize$\pm$1.010} & 0.546{\scriptsize$\pm$0.469} & 0.544{\scriptsize$\pm$0.480} & 0.092{\scriptsize$\pm$0.071} & 0.091{\scriptsize$\pm$0.072} \\
\textit{gemini-3-flash-preview} & 0.474{\scriptsize$\pm$0.339} & 0.475{\scriptsize$\pm$0.306} & 0.089{\scriptsize$\pm$0.067} & 0.090{\scriptsize$\pm$0.062} & 0.079{\scriptsize$\pm$0.050} & 0.080{\scriptsize$\pm$0.046} & 1.147{\scriptsize$\pm$0.918} & \textbf{1.323{\scriptsize$\pm$2.709}} & 0.692{\scriptsize$\pm$0.498} & 0.696{\scriptsize$\pm$0.439} & 0.115{\scriptsize$\pm$0.070} & 0.117{\scriptsize$\pm$0.064} \\
\textbf{\textit{gemini-3-pro-preview}} & \textbf{0.576{\scriptsize$\pm$0.318}} & \textbf{0.567{\scriptsize$\pm$0.289}} & \textbf{0.113{\scriptsize$\pm$0.064}} & \textbf{0.111{\scriptsize$\pm$0.059}} & 0.097{\scriptsize$\pm$0.047} & 0.097{\scriptsize$\pm$0.044} & \textbf{1.197{\scriptsize$\pm$0.763}} & 1.201{\scriptsize$\pm$0.787} & \textbf{0.858{\scriptsize$\pm$0.461}} & \textbf{0.844{\scriptsize$\pm$0.423}} & 0.140{\scriptsize$\pm$0.065} & 0.141{\scriptsize$\pm$0.061} \\
\textit{grok-4.1-fast} & 0.352{\scriptsize$\pm$0.307} & 0.357{\scriptsize$\pm$0.310} & 0.066{\scriptsize$\pm$0.061} & 0.066{\scriptsize$\pm$0.061} & \textbf{0.062{\scriptsize$\pm$0.045}} & 0.063{\scriptsize$\pm$0.045} & 1.067{\scriptsize$\pm$0.897} & 1.035{\scriptsize$\pm$0.861} & 0.515{\scriptsize$\pm$0.442} & 0.517{\scriptsize$\pm$0.444} & \textbf{0.090{\scriptsize$\pm$0.065}} & 0.091{\scriptsize$\pm$0.064} \\
\midrule
\rowcolor{gray!15} \multicolumn{13}{c}{\textbf{NVDA} (US Equity)} \\
\textit{claude-sonnet-4.5} & 0.223{\scriptsize$\pm$0.210} & 0.220{\scriptsize$\pm$0.208} & 0.005{\scriptsize$\pm$0.050} & 0.005{\scriptsize$\pm$0.054} & 0.188{\scriptsize$\pm$0.179} & 0.183{\scriptsize$\pm$0.178} & 0.449{\scriptsize$\pm$0.817} & 0.462{\scriptsize$\pm$0.896} & 0.480{\scriptsize$\pm$0.417} & 0.463{\scriptsize$\pm$0.412} & 0.273{\scriptsize$\pm$0.261} & 0.266{\scriptsize$\pm$0.260} \\
\textit{deepseek-v3.2} & 0.192{\scriptsize$\pm$0.209} & 0.191{\scriptsize$\pm$0.208} & 0.002{\scriptsize$\pm$0.053} & 0.005{\scriptsize$\pm$0.054} & 0.155{\scriptsize$\pm$0.167} & 0.149{\scriptsize$\pm$0.166} & 0.460{\scriptsize$\pm$0.942} & 0.502{\scriptsize$\pm$0.885} & 0.395{\scriptsize$\pm$0.394} & 0.390{\scriptsize$\pm$0.397} & 0.225{\scriptsize$\pm$0.243} & 0.216{\scriptsize$\pm$0.242} \\
\textit{gpt-5.2} & 0.165{\scriptsize$\pm$0.203} & 0.164{\scriptsize$\pm$0.205} & 0.001{\scriptsize$\pm$0.052} & 0.002{\scriptsize$\pm$0.050} & \textbf{0.150{\scriptsize$\pm$0.180}} & \textbf{0.147{\scriptsize$\pm$0.177}} & \textbf{0.497{\scriptsize$\pm$1.078}} & \textbf{0.544{\scriptsize$\pm$1.542}} & 0.378{\scriptsize$\pm$0.421} & 0.370{\scriptsize$\pm$0.421} & \textbf{0.218{\scriptsize$\pm$0.262}} & \textbf{0.213{\scriptsize$\pm$0.258}} \\
\textit{gemini-3-flash-preview} & 0.234{\scriptsize$\pm$0.242} & 0.231{\scriptsize$\pm$0.210} & 0.009{\scriptsize$\pm$0.076} & 0.006{\scriptsize$\pm$0.060} & 0.182{\scriptsize$\pm$0.189} & 0.184{\scriptsize$\pm$0.179} & 0.435{\scriptsize$\pm$1.013} & 0.437{\scriptsize$\pm$0.826} & 0.482{\scriptsize$\pm$0.466} & 0.476{\scriptsize$\pm$0.417} & 0.265{\scriptsize$\pm$0.276} & 0.268{\scriptsize$\pm$0.260} \\
\textbf{\textit{gemini-3-pro-preview}} & \textbf{0.302{\scriptsize$\pm$0.221}} & \textbf{0.299{\scriptsize$\pm$0.206}} & \textbf{0.011{\scriptsize$\pm$0.073}} & \textbf{0.013{\scriptsize$\pm$0.065}} & 0.247{\scriptsize$\pm$0.183} & 0.242{\scriptsize$\pm$0.172} & 0.493{\scriptsize$\pm$1.063} & 0.535{\scriptsize$\pm$1.285} & \textbf{0.653{\scriptsize$\pm$0.506}} & \textbf{0.635{\scriptsize$\pm$0.411}} & 0.359{\scriptsize$\pm$0.266} & 0.352{\scriptsize$\pm$0.251} \\
\textit{grok-4.1-fast} & 0.183{\scriptsize$\pm$0.196} & 0.188{\scriptsize$\pm$0.193} & 0.006{\scriptsize$\pm$0.055} & 0.005{\scriptsize$\pm$0.052} & 0.153{\scriptsize$\pm$0.156} & 0.153{\scriptsize$\pm$0.156} & 0.465{\scriptsize$\pm$1.004} & 0.508{\scriptsize$\pm$0.896} & 0.391{\scriptsize$\pm$0.374} & 0.396{\scriptsize$\pm$0.373} & 0.222{\scriptsize$\pm$0.228} & 0.222{\scriptsize$\pm$0.228} \\
\midrule
\rowcolor{gray!15} \multicolumn{13}{c}{\textbf{TSLA} (US Equity)} \\
\textit{claude-sonnet-4.5} & 0.547{\scriptsize$\pm$0.385} & 0.545{\scriptsize$\pm$0.386} & 0.395{\scriptsize$\pm$0.341} & 0.397{\scriptsize$\pm$0.343} & 0.158{\scriptsize$\pm$0.143} & 0.155{\scriptsize$\pm$0.142} & 3.076{\scriptsize$\pm$1.846} & 3.259{\scriptsize$\pm$2.031} & 0.860{\scriptsize$\pm$0.658} & 0.858{\scriptsize$\pm$0.661} & 0.249{\scriptsize$\pm$0.220} & 0.245{\scriptsize$\pm$0.219} \\
\textit{deepseek-v3.2} & 0.470{\scriptsize$\pm$0.377} & 0.464{\scriptsize$\pm$0.394} & 0.329{\scriptsize$\pm$0.324} & 0.323{\scriptsize$\pm$0.328} & 0.134{\scriptsize$\pm$0.136} & 0.130{\scriptsize$\pm$0.136} & 3.057{\scriptsize$\pm$2.042} & 2.989{\scriptsize$\pm$2.109} & 0.741{\scriptsize$\pm$0.635} & 0.726{\scriptsize$\pm$0.649} & 0.210{\scriptsize$\pm$0.210} & 0.205{\scriptsize$\pm$0.210} \\
\textit{gpt-5.2} & 0.451{\scriptsize$\pm$0.412} & 0.460{\scriptsize$\pm$0.413} & 0.309{\scriptsize$\pm$0.346} & 0.308{\scriptsize$\pm$0.338} & \textbf{0.124{\scriptsize$\pm$0.144}} & \textbf{0.121{\scriptsize$\pm$0.141}} & 2.894{\scriptsize$\pm$1.629} & 2.991{\scriptsize$\pm$1.740} & 0.691{\scriptsize$\pm$0.685} & 0.692{\scriptsize$\pm$0.676} & \textbf{0.196{\scriptsize$\pm$0.223}} & \textbf{0.193{\scriptsize$\pm$0.218}} \\
\textit{gemini-3-flash-preview} & 0.554{\scriptsize$\pm$0.451} & 0.580{\scriptsize$\pm$0.398} & 0.391{\scriptsize$\pm$0.372} & 0.399{\scriptsize$\pm$0.339} & 0.152{\scriptsize$\pm$0.153} & 0.157{\scriptsize$\pm$0.144} & 2.921{\scriptsize$\pm$1.918} & 3.065{\scriptsize$\pm$1.772} & 0.851{\scriptsize$\pm$0.735} & 0.891{\scriptsize$\pm$0.668} & 0.241{\scriptsize$\pm$0.235} & 0.247{\scriptsize$\pm$0.221} \\
\textbf{\textit{gemini-3-pro-preview}} & \textbf{0.712{\scriptsize$\pm$0.383}} & \textbf{0.699{\scriptsize$\pm$0.366}} & \textbf{0.523{\scriptsize$\pm$0.346}} & \textbf{0.522{\scriptsize$\pm$0.331}} & 0.204{\scriptsize$\pm$0.143} & 0.202{\scriptsize$\pm$0.139} & 2.983{\scriptsize$\pm$1.688} & 3.015{\scriptsize$\pm$1.463} & \textbf{1.127{\scriptsize$\pm$0.650}} & \textbf{1.106{\scriptsize$\pm$0.620}} & 0.324{\scriptsize$\pm$0.219} & 0.320{\scriptsize$\pm$0.212} \\
\textit{grok-4.1-fast} & 0.485{\scriptsize$\pm$0.369} & 0.500{\scriptsize$\pm$0.372} & 0.337{\scriptsize$\pm$0.313} & 0.344{\scriptsize$\pm$0.314} & 0.133{\scriptsize$\pm$0.129} & 0.132{\scriptsize$\pm$0.129} & \textbf{3.244{\scriptsize$\pm$2.087}} & \textbf{3.418{\scriptsize$\pm$2.322}} & 0.756{\scriptsize$\pm$0.608} & 0.772{\scriptsize$\pm$0.613} & 0.210{\scriptsize$\pm$0.199} & 0.211{\scriptsize$\pm$0.200} \\
\bottomrule
\end{tabular}}
\end{table*}

\begin{figure*}[h]
\centering
\begin{subfigure}[t]{0.48\textwidth}
\includegraphics[width=\textwidth]{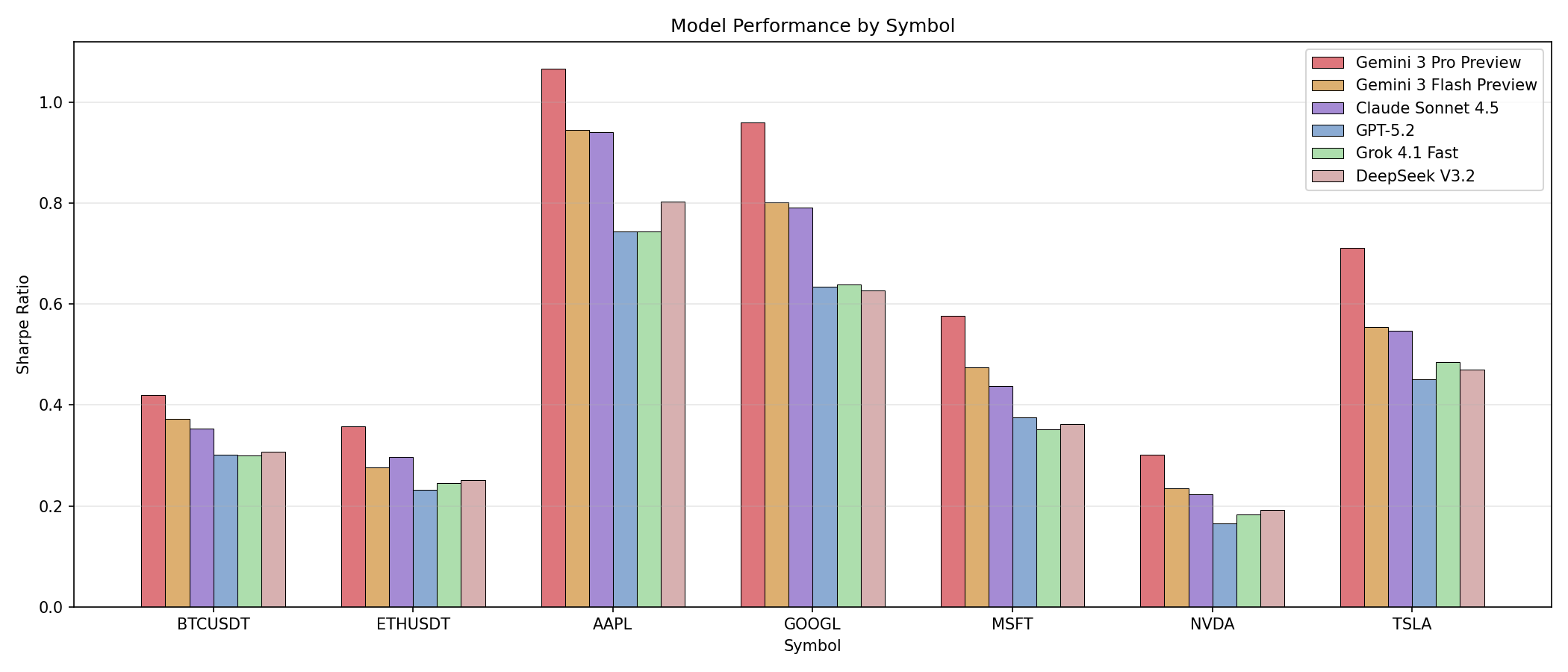}
\caption{T=0.0}
\end{subfigure}
\hfill
\begin{subfigure}[t]{0.48\textwidth}
\includegraphics[width=\textwidth]{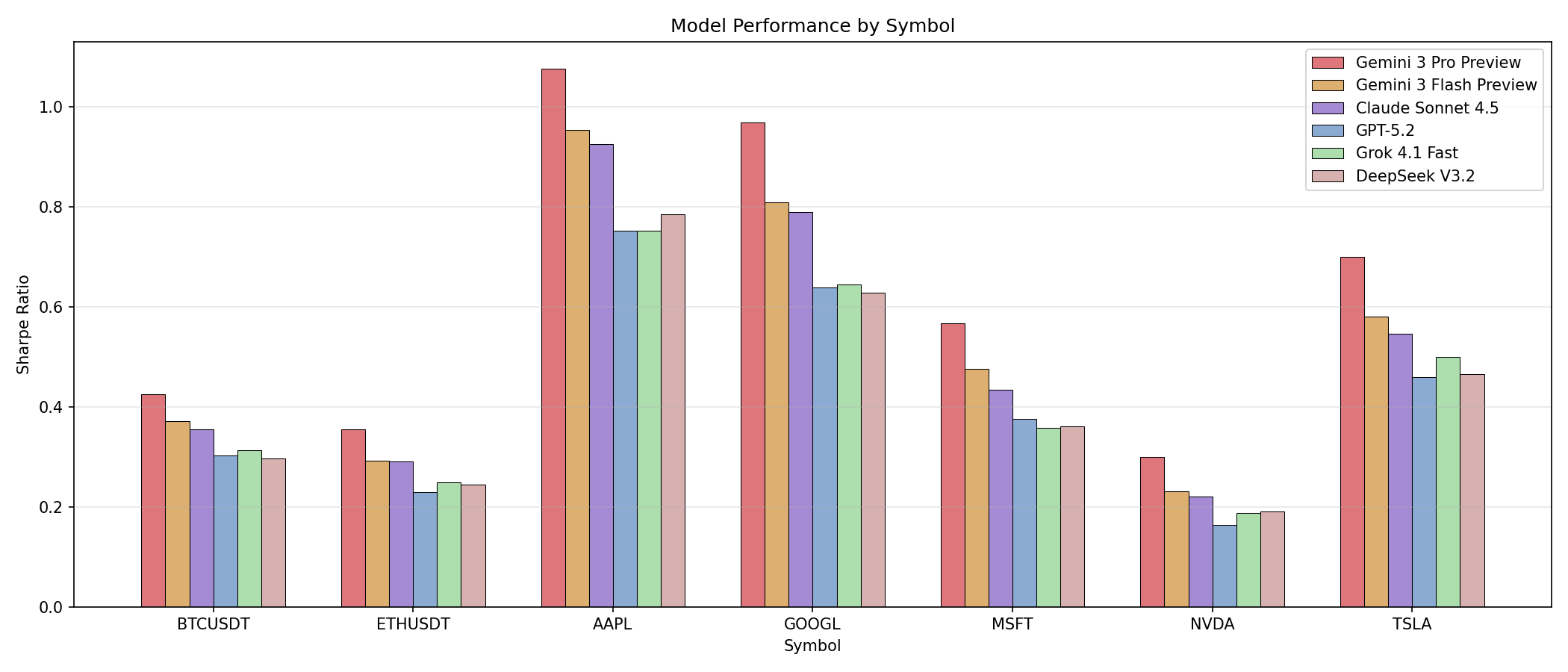}
\caption{T=0.7}
\end{subfigure}
\caption{Grouped bar chart of model performance across all seven assets (BTCUSDT, ETHUSDT, AAPL, GOOGL, MSFT, NVDA, TSLA). Each asset group shows all six models side-by-side, revealing asset-specific difficulty patterns and model specialization.}
\label{appx_fig:symbol_grouped_bar}
\end{figure*}

\Cref{appx_fig:symbol_grouped_bar} provides a visual summary of model performance across all seven assets, facilitating direct comparison of asset-specific characteristics and model adaptability.

The cross-asset comparison reveals striking patterns in both asset difficulty and model specialization. Traditional equity assets (AAPL, GOOGL, MSFT, TSLA) consistently show higher performance bars across all models compared to cryptocurrency assets (BTCUSDT, ETHUSDT), confirming that crypto markets pose greater challenges for algorithmic trading strategies. Among equities, AAPL and GOOGL emerge as the most favorable assets, with Sharpe Ratios frequently exceeding 0.7 across multiple models. NVDA presents the most challenging equity asset, with notably shorter bars across all models, likely due to its high volatility and rapid price movements.

The cryptocurrency assets exhibit distinct characteristics. BTCUSDT and ETHUSDT show the lowest performance bars overall, with Sharpe Ratios typically in the 0.2--0.3 range. This difficulty stems from several factors: higher volatility (reflected in the detailed tables), less predictable price patterns, and 24/7 trading dynamics that differ fundamentally from equity market structures. Interestingly, the relative model rankings remain largely consistent across asset types---\textit{gemini-3-pro-preview} maintains the tallest bars for both crypto and equity assets, while \textit{\textit{gpt-5.2}}, \textit{deepseek-v3.2}, and \textit{grok-4.1-fast} alternate among the lowest-performing models depending on the specific asset.

Model specialization patterns are also evident. \textit{gemini-3-pro-preview} demonstrates particularly strong performance on high-volatility assets (TSLA, NVDA, cryptocurrencies), suggesting robust handling of challenging market conditions. \textit{claude-sonnet-4.5} shows more uniform bar heights across all assets, indicating balanced performance without specific asset preferences. \textit{\textit{gpt-5.2}} exhibits relatively stronger performance on traditional equities compared to cryptocurrencies, suggesting potential optimization for more structured market environments.

The consistency of patterns across temperature settings (T=0.0 vs T=0.7) is noteworthy. Asset difficulty rankings remain stable regardless of decoding temperature, confirming that the observed patterns reflect genuine market characteristics rather than sampling artifacts. The bar height differences between assets are substantial---AAPL Sharpe Ratios are approximately 2.5--3 times higher than BTCUSDT values for most models---indicating that asset selection significantly impacts strategy performance, potentially more so than model choice within the same tier.

\subsection{Comparative Performance Across Models}

This section presents model-by-model comparisons on each evaluation dimension, with all models shown together to facilitate direct comparison of strengths and weaknesses. We organize the analysis into five key aspects: (1) overall performance ranking, (2) multi-metric comparison revealing model profiles, (3) consistency and variance analysis, (4) syntax correctness and code quality, and (5) robustness across multiple runs and assets.

\subsubsection{Overall Performance Ranking}

\textbf{Overview.} Before diving into detailed metric-by-metric analysis, we first establish the overall performance hierarchy among evaluated models. This ranking provides a high-level understanding of model capabilities and sets the context for subsequent detailed comparisons.

\textbf{Performance Across Difficulty Levels.} As shown earlier in \Cref{appx_fig:model_level_comparison}, model performance varies significantly across difficulty levels. \textit{gemini-3-pro-preview} demonstrates the strongest overall performance, with Sharpe Ratios of 0.545, 0.604, and 0.734 at Levels 1, 2, and 3 respectively. \textit{claude-sonnet-4.5} and \textit{gemini-3-flash-preview} maintain competitive and stable performance across all levels. \textit{deepseek-v3.2} achieves the highest Level 1 Sharpe Ratio (0.561) but experiences sharper degradation at Level 3 (0.329), suggesting strong code generation capabilities but limitations in creative strategy design. \textit{grok-4.1-fast} and \textit{\textit{gpt-5.2}} occupy the lower tier, with overall Sharpe Ratios of 0.421 and 0.415 respectively, and both exhibit inconsistent results across levels.

\begin{figure*}[h]
\centering
\begin{subfigure}[t]{0.48\textwidth}
\includegraphics[width=\textwidth]{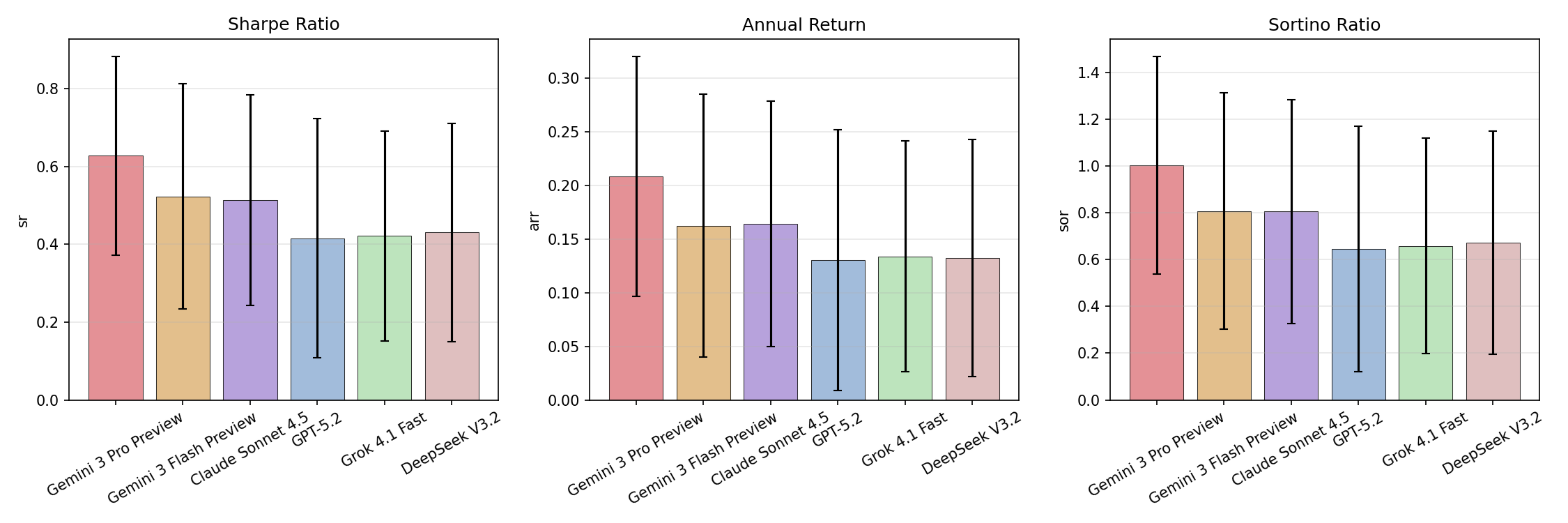}
\caption{T=0.0}
\end{subfigure}
\hfill
\begin{subfigure}[t]{0.48\textwidth}
\includegraphics[width=\textwidth]{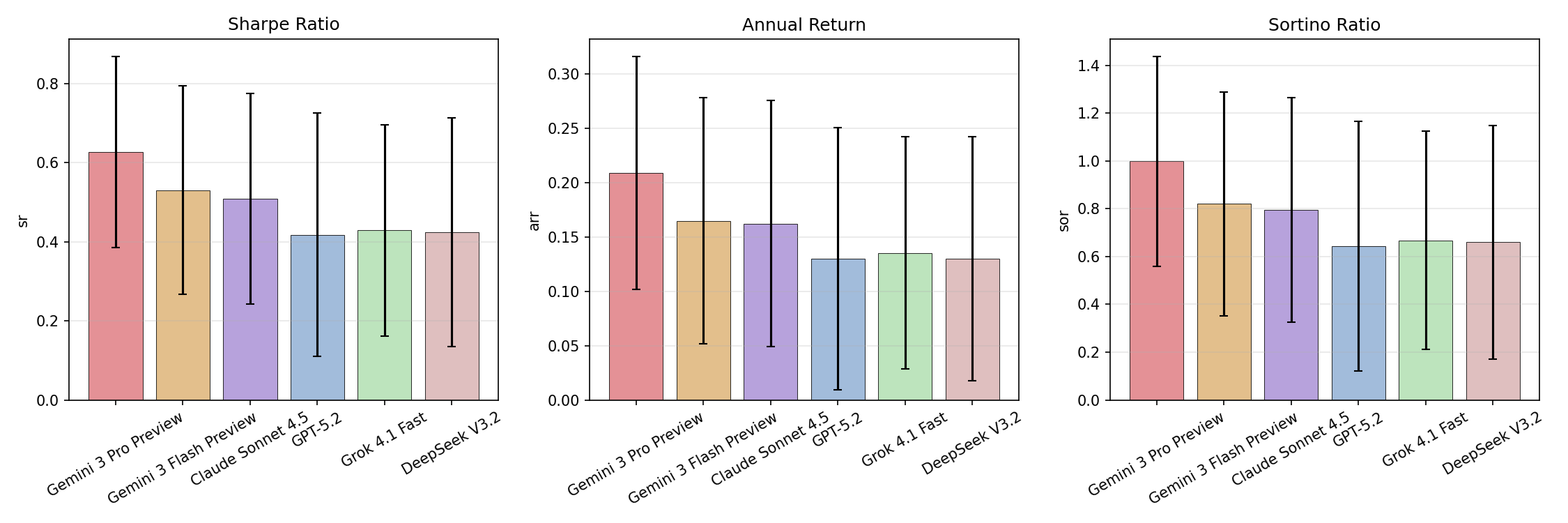}
\caption{T=0.7}
\end{subfigure}
\caption{Model performance ranking visualization. Bar heights represent aggregate performance scores computed across all metrics, difficulty levels, and assets, facilitating quick identification of top-tier, mid-tier, and lower-tier models.}
\label{appx_fig:model_bar_comparison}
\end{figure*}

\textbf{Aggregate Performance Ranking.} \Cref{appx_fig:model_bar_comparison} provides a simplified ranking visualization that aggregates performance across all metrics and difficulty levels, enabling quick identification of performance tiers.

This ranking visualization delineates three performance tiers. The \textbf{top tier} consists of \textit{gemini-3-pro-preview} (SR = 0.628), \textit{gemini-3-flash-preview} (SR = 0.523), and \textit{claude-sonnet-4.5} (SR = 0.513), all achieving overall Sharpe Ratios above 0.5 with robust performance across difficulty levels and asset types. The \textbf{mid-tier} includes \textit{deepseek-v3.2} (SR = 0.430) and \textit{\textit{gpt-5.2}} (SR = 0.415), both demonstrating solid capabilities but with noticeable gaps from the leaders, particularly on Level 3 tasks. \textbf{\textit{grok-4.1-fast}} (SR = 0.421) occupies a similar range to the mid-tier models, though with higher variance across tasks.

\textbf{Stability Across Temperature Settings.} The consistency of rankings across T=0.0 and T=0.7 is noteworthy. All models maintain their relative positions regardless of decoding temperature, with rank correlations exceeding 0.95. This stability validates that the observed performance differences reflect genuine model capabilities rather than sensitivity to sampling randomness. The absolute score differences between temperatures are minimal (typically <3\%), further confirming the robustness of factor-based evaluation.

\subsubsection{Multi-Metric Comparison and Model Profiles}

\textbf{Overview.} While aggregate rankings provide a useful summary, they obscure important trade-offs between different performance dimensions. Some models may excel at risk-adjusted returns (Sharpe Ratio) while others prioritize raw returns or drawdown control. This subsection examines model performance across multiple metrics simultaneously to reveal distinct model profiles and specializations.

\begin{figure*}[h]
\centering
\begin{subfigure}[t]{0.48\textwidth}
\includegraphics[width=\textwidth]{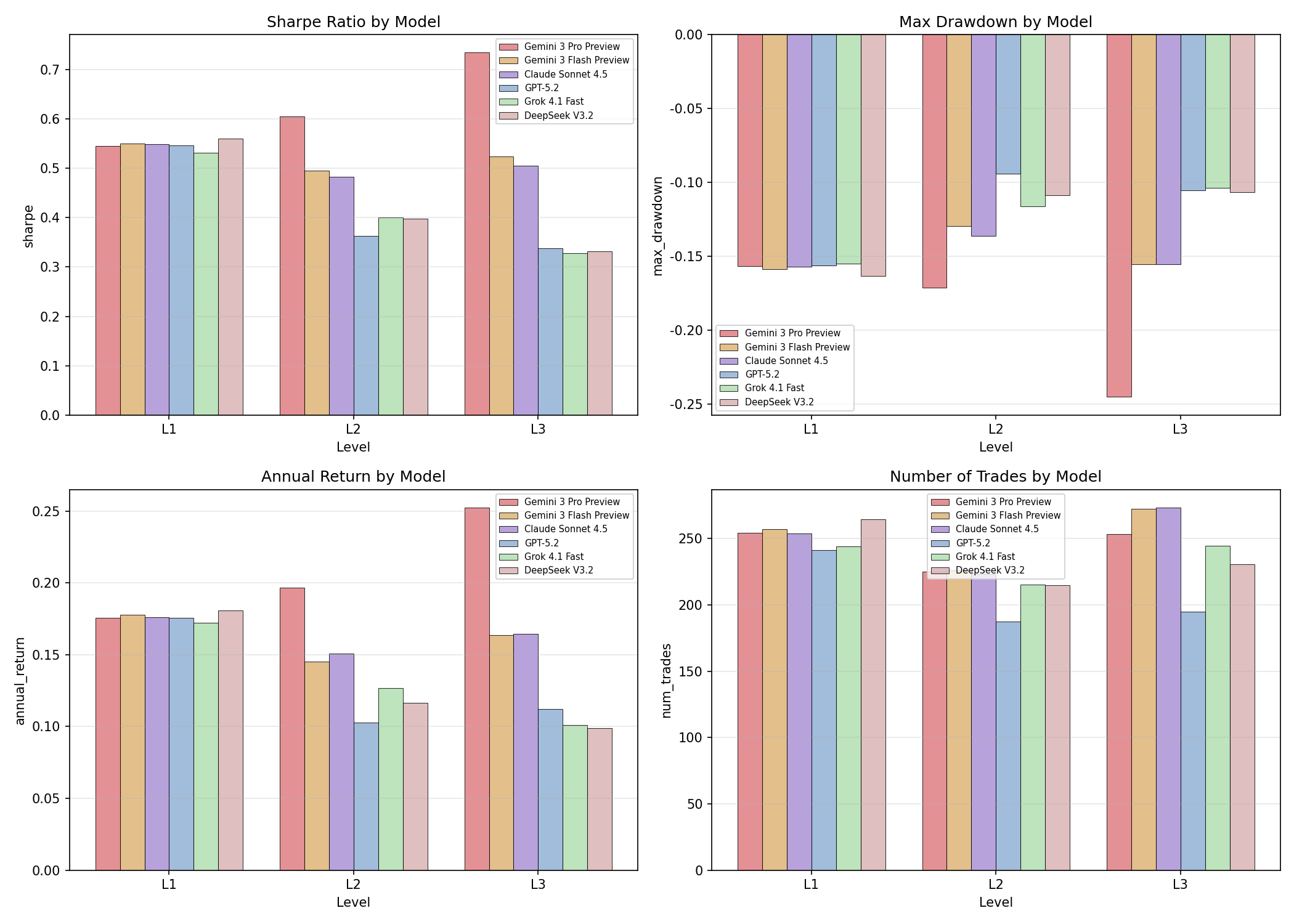}
\caption{T=0.0}
\end{subfigure}
\hfill
\begin{subfigure}[t]{0.48\textwidth}
\includegraphics[width=\textwidth]{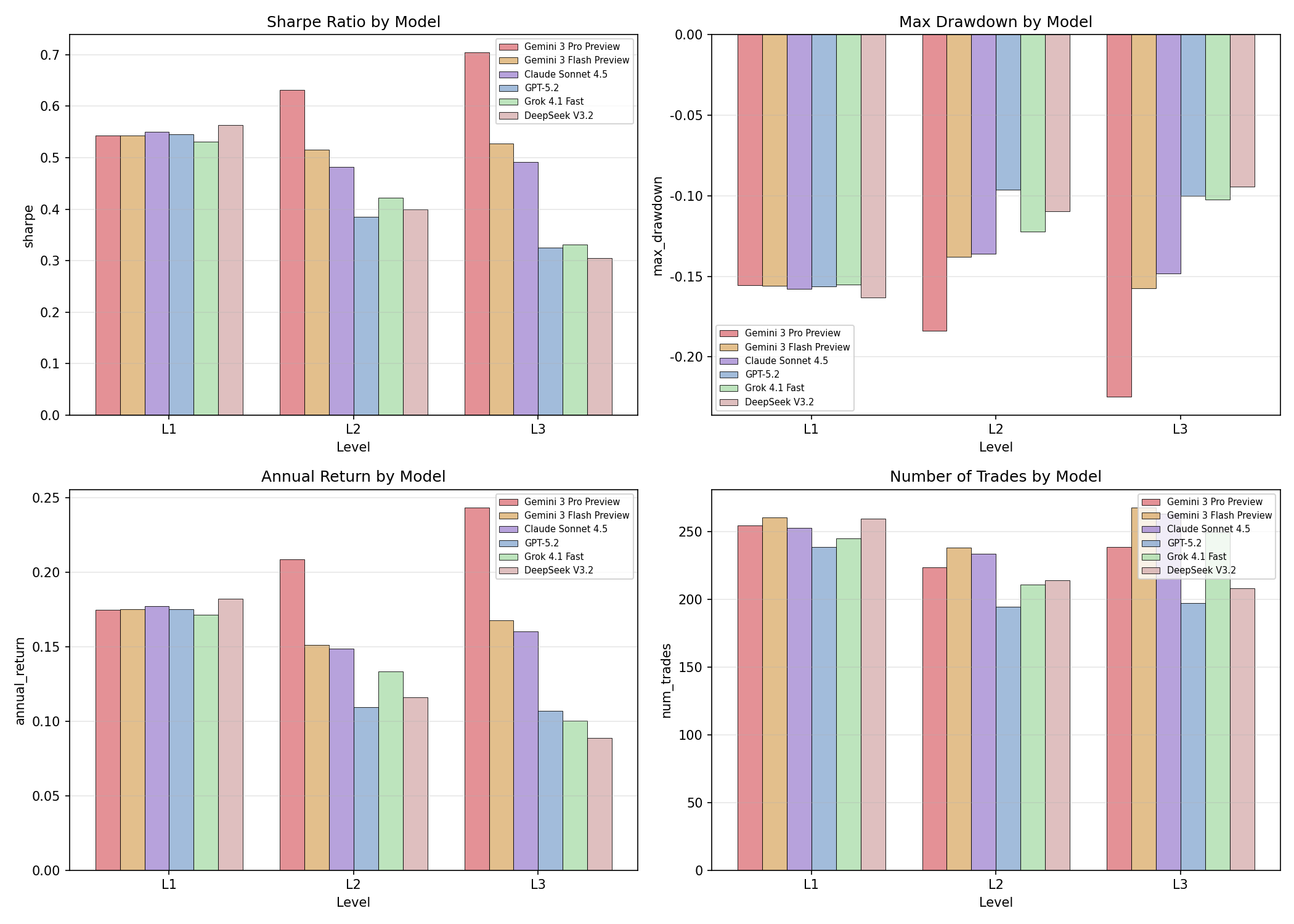}
\caption{T=0.7}
\end{subfigure}
\caption{Grouped bar chart comparing all models across core metrics. Each metric is normalized to facilitate cross-metric comparison, with bars grouped by model to reveal individual performance profiles and trade-offs.}
\label{appx_fig:model_grouped_bar}
\end{figure*}

\textbf{Analysis.} \Cref{appx_fig:model_grouped_bar} provides a comprehensive side-by-side comparison of all models across core metrics, with each metric normalized to facilitate cross-metric comparison.

The grouped bar visualization reveals distinct model profiles and trade-offs. \textbf{\textit{gemini-3-pro-preview}} exhibits the tallest bars across most metrics, particularly excelling in Sharpe Ratio (0.628) and Sortino Ratio (1.004), confirming its position as the top performer. However, its Maximum Drawdown (0.191) and Volatility (0.262) are also the highest, indicating greater risk exposure. This profile suggests an aggressive high-performance tendency: \textit{gemini-3-pro-preview} generates strategies with high return potential but also accepts larger drawdowns.

\textbf{\textit{claude-sonnet-4.5}} shows a balanced profile with consistently high bars across all metrics, demonstrating well-rounded capabilities without extreme strengths or weaknesses. Its bars are uniformly tall but never the absolute tallest, suggesting a "jack-of-all-trades" approach that prioritizes consistency over peak performance in any single dimension. This balance makes \textit{claude-sonnet-4.5} particularly suitable for production environments where predictable, reliable performance is valued.

\textbf{\textit{\textit{gpt-5.2}}} displays a conservative profile with the lowest Volatility (0.163) and Maximum Drawdown (0.119) among all models, aligning with risk-averse strategy generation tendencies. While its Sharpe Ratio (0.415) is lower than top-tier models, its Calmar Ratio (1.599) is competitive, indicating solid return-to-drawdown efficiency. This conservative profile suggests \textit{\textit{gpt-5.2}} prioritizes capital preservation over aggressive return seeking.

\textbf{\textit{deepseek-v3.2} and \textit{grok-4.1-fast}} show moderate performance across most metrics, with overall Sharpe Ratios of 0.430 and 0.421 respectively. Both models achieve lower MDD and VOL than \textit{gemini-3-pro-preview}, but their return-oriented metrics lag behind the top tier. \textbf{\textit{gemini-3-flash-preview}} achieves the second-highest overall Sharpe Ratio (0.523) with balanced risk metrics (MDD = 0.148, VOL = 0.204), positioning it as a strong alternative to \textit{gemini-3-pro-preview} with a more moderate risk profile.

\textbf{Metric Correlations and Trade-offs.} The grouped bar chart also reveals interesting metric correlations. Models with high Sharpe Ratios (\textit{gemini-3-pro-preview}, \textit{claude-sonnet-4.5}, \textit{gemini-3-flash-preview}) also tend to have high Sortino Ratios, suggesting that risk-adjusted performance is consistent across both total volatility and downside risk measures. However, the correlation between Sharpe Ratio and Calmar Ratio is weaker, indicating that drawdown control is somewhat independent of volatility management. \textit{\textit{gpt-5.2}}'s high Calmar Ratio despite moderate Sharpe Ratio exemplifies this independence.

\subsubsection{Consistency and Variance Analysis}

\textbf{Overview.} Average performance metrics tell only part of the story. In production deployments, consistency and predictability are often as important as peak performance. A model that achieves 0.5 Sharpe Ratio on 90\% of tasks is often preferable to one that achieves 0.7 on 50\% of tasks and 0.2 on the other 50\%. This subsection uses boxplot analysis to examine performance distributions, revealing which models offer consistent behavior versus which exhibit high variance.

\begin{figure*}[h]
\centering
\begin{subfigure}[t]{0.48\textwidth}
\includegraphics[width=\textwidth]{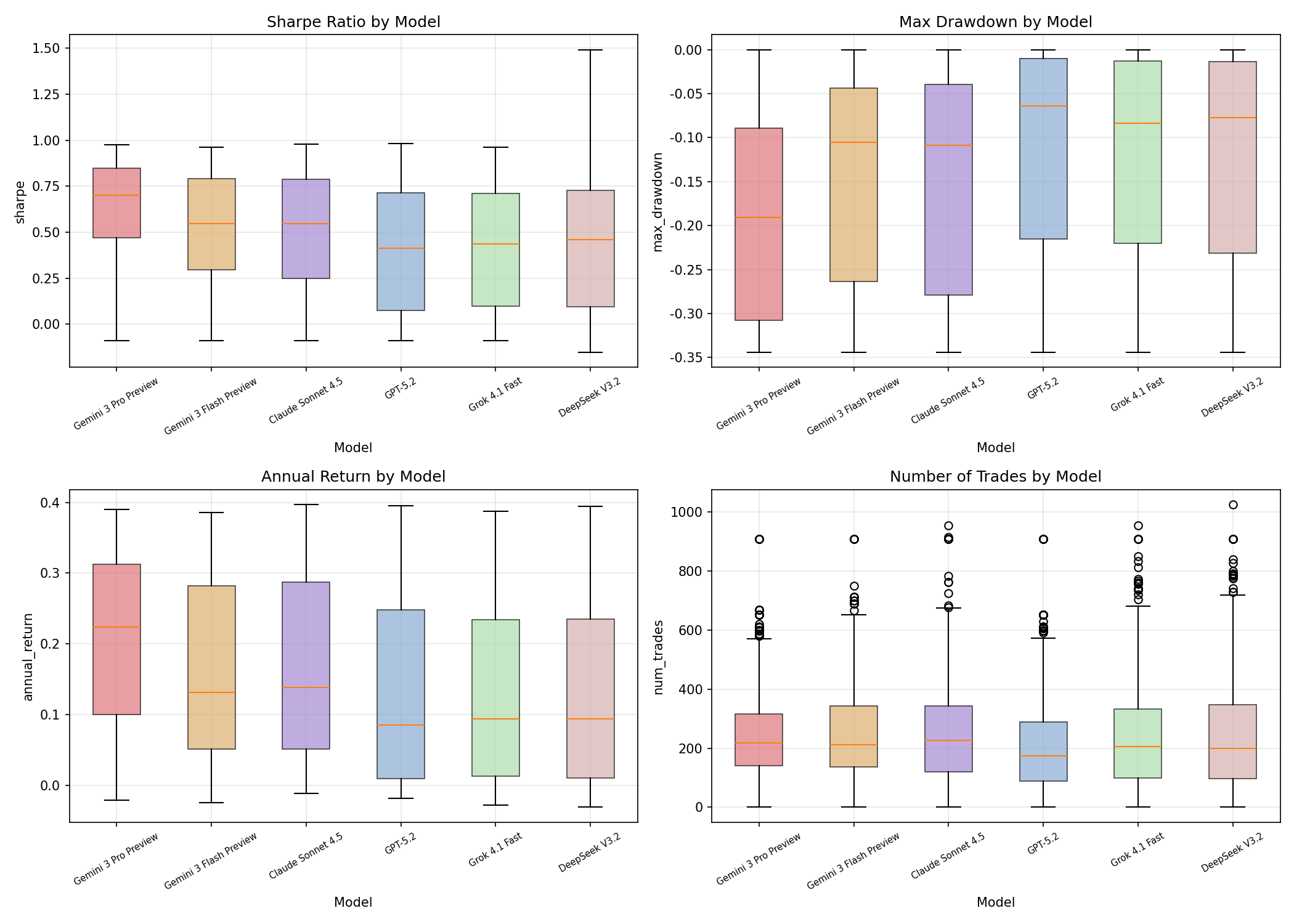}
\caption{T=0.0}
\end{subfigure}
\hfill
\begin{subfigure}[t]{0.48\textwidth}
\includegraphics[width=\textwidth]{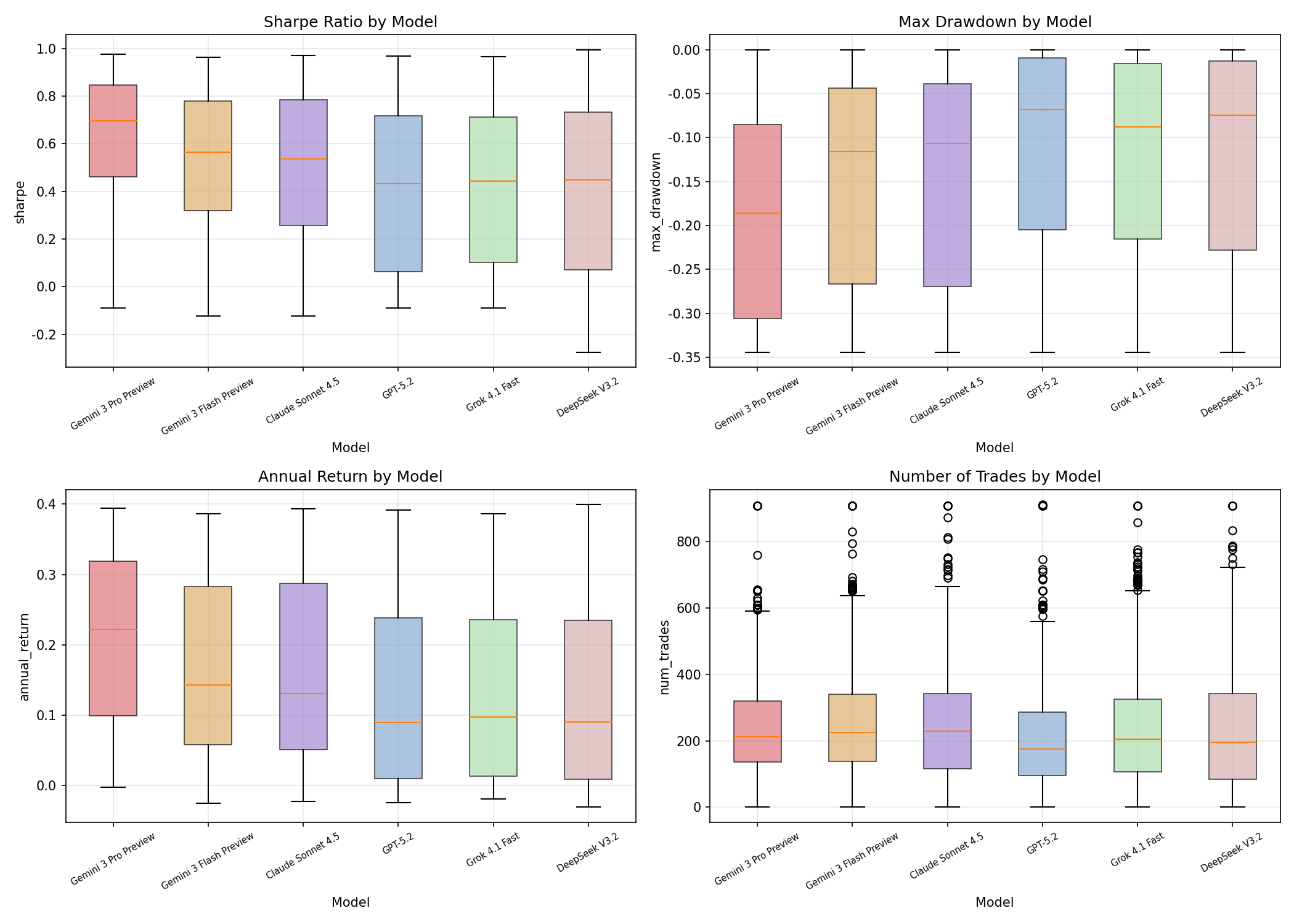}
\caption{T=0.7}
\end{subfigure}
\caption{Boxplot distributions by model across all evaluation instances. Box width (IQR) indicates consistency, with narrow boxes representing stable performance and wide boxes indicating high variance across different queries and assets.}
\label{appx_fig:model_boxplot}
\end{figure*}

\textbf{Analysis.} \Cref{appx_fig:model_boxplot} reveals the distribution of each model's performance across all 270 queries and 7 assets, exposing consistency patterns that aggregate metrics cannot capture.

\textbf{Consistency Champions: \textit{claude-sonnet-4.5} and \textit{\textit{gpt-5.2}}.} The boxplot analysis reveals critical insights about model consistency versus peak performance. \textbf{\textit{claude-sonnet-4.5}} exhibits the narrowest interquartile range among top-tier models (IQR $\approx$ 0.35), indicating highly consistent performance across diverse queries and assets. The median sits near the center of the box, suggesting symmetric distribution without significant skew. This consistency profile makes \textit{claude-sonnet-4.5} particularly reliable for production deployments where predictable behavior is valued. The few outliers present are evenly distributed above and below the whiskers, indicating that exceptional performance (both positive and negative) is rare and balanced.

\textbf{\textit{\textit{gpt-5.2}}} displays an even narrower box with minimal outliers, reflecting conservative and highly consistent strategy generation. The median is lower than top-tier models (mean Sharpe $\approx$ 0.415), but the tight distribution indicates that users can reliably expect performance within a narrow range.

\textbf{High Variance, High Reward: \textit{gemini-3-pro-preview}.} \textbf{\textit{gemini-3-pro-preview}} shows a wider box than \textit{claude-sonnet-4.5} (IQR $\approx$ 0.45), indicating greater variance in outcomes. However, the median is positioned higher (mean Sharpe $\approx$ 0.628), and the presence of numerous upper outliers demonstrates that \textit{gemini-3-pro-preview} occasionally generates exceptionally high-performing strategies. This pattern suggests a trade-off: higher average performance but less predictability. The positive skew (median closer to lower quartile, long upper tail) indicates that while most results are moderate, the model has significant upside potential. For applications where occasional exceptional performance is valued (e.g., strategy discovery, research), this variance may be desirable. The upper outliers reaching Sharpe Ratios of 0.9+ demonstrate that \textit{gemini-3-pro-preview} can discover highly effective strategies when conditions align favorably.

\textbf{Moderate Variance Models: \textit{deepseek-v3.2} and \textit{gemini-3-flash-preview}.} \textbf{\textit{deepseek-v3.2}} shows moderate box width (IQR $\approx$ 0.40) with several lower outliers, suggesting occasional poor performance on specific query-asset combinations. The distribution is relatively symmetric, indicating balanced behavior without strong skew. The lower outliers (Sharpe < 0) reveal that \textit{deepseek-v3.2} occasionally generates strategies with negative risk-adjusted returns, likely due to poor parameter choices on challenging tasks.

\textbf{\textit{gemini-3-flash-preview}} exhibits a similar IQR to \textit{deepseek-v3.2} but with fewer outliers and a higher median (mean SR = 0.523 vs.\ 0.430). The compact distribution suggests that \textit{gemini-3-flash-preview} combines reasonable consistency with competitive performance. This makes it a viable choice for cost-sensitive applications where balanced performance is valued over peak results.

\textbf{High Variance Outlier: \textit{grok-4.1-fast}.} \textbf{\textit{grok-4.1-fast}} exhibits the widest box among all models (IQR $\approx$ 0.50), with numerous outliers in both directions, confirming its inconsistent behavior pattern. The wide distribution indicates that performance varies dramatically depending on the specific query-asset combination. Some tasks yield competitive results (upper outliers reaching Sharpe $\approx$ 0.7), while others produce poor strategies (lower outliers with negative Sharpe Ratios). This unpredictability makes \textit{grok-4.1-fast} challenging to deploy in production without extensive validation on specific use cases.

\textbf{Median vs. Mean: Understanding Typical Performance.} Comparing median versus mean performance reveals additional insights about distribution shape and the representativeness of aggregate metrics. For models with symmetric distributions (\textbf{\textit{claude-sonnet-4.5}, \textit{\textit{gpt-5.2}}}), median and mean align closely (difference < 5\%), indicating that average metrics accurately represent typical performance. Users can expect that most runs will produce results near the reported mean.

For models with skewed distributions (\textbf{\textit{gemini-3-pro-preview}, \textit{grok-4.1-fast}}), the mean is pulled by outliers, creating a gap between median and mean. This suggests that while \textit{gemini-3-pro-preview}'s reported average Sharpe Ratio (0.628) is impressive, typical performance may be slightly lower, with occasional exceptional runs boosting the mean. Conversely, \textit{grok-4.1-fast}'s mean and median are closer, but the wide distribution means neither metric reliably predicts individual run outcomes.

\textbf{Practical Implications for Model Selection.} The consistency analysis provides actionable guidance for practitioners. For production trading systems requiring predictable behavior, \textit{claude-sonnet-4.5} and \textit{\textit{gpt-5.2}} are preferable due to their tight distributions and minimal outliers. For strategy discovery and research, \textit{gemini-3-pro-preview}'s higher variance may be advantageous, as the upper outliers represent genuinely innovative strategies worth investigating. Ensemble approaches that combine a consistent model (e.g., \textit{claude-sonnet-4.5}) with a high-variance model (e.g., \textit{gemini-3-pro-preview}) may yield both reliability and occasional exceptional performance. \textit{grok-4.1-fast} requires extensive validation on specific query types before production deployment due to its inconsistent behavior.

\subsubsection{Syntax Correctness and Code Quality}

\textbf{Overview.} Syntax correctness measures whether generated code is syntactically valid Python and follows the required API conventions. This dimension is orthogonal to strategic quality---a model can generate syntactically perfect code that implements a poor strategy, or conversely, have brilliant strategic ideas undermined by syntax errors. High syntax correctness is a prerequisite for deployment, as even minor errors prevent strategy execution.

\textbf{Analysis.} \Cref{appx_tab:pass_rate_comparison} demonstrates exceptional code generation quality across all models, with Pass@1 rates exceeding 93\% for all evaluated LLMs at $\tau=0$.

\begin{table}[h]
\centering
\caption{Pass rate comparison across temperature settings (\%). Pass@1 measures the success rate of the first generated sample; Pass@5 measures whether at least one of five samples succeeds.}
\label{appx_tab:pass_rate_comparison}
\begin{tabular}{lcccc}
\toprule
\multirow{2}{*}{\textbf{Model}} & \multicolumn{2}{c}{\textbf{Pass@1} (\%)$\uparrow$} & \multicolumn{2}{c}{\textbf{Pass@5} (\%)$\uparrow$} \\
\cmidrule(lr){2-3} \cmidrule(lr){4-5}
& $\tau$=0 & $\tau$=0.7 & $\tau$=0 & $\tau$=0.7 \\
\midrule
\textit{claude-sonnet-4.5}    & \textbf{100.00} & 99.63           & \textbf{100.00} & \textbf{100.00} \\
\textit{deepseek-v3.2}        & 93.70           & 85.56           & 100.00          & 99.63           \\
\textit{gemini-3-flash-preview}       & 94.44           & 97.78           & 98.15           & \textbf{100.00} \\
\textit{gemini-3-pro-preview}         & \textbf{100.00} & 98.89           & \textbf{100.00} & \textbf{100.00} \\
\textit{gpt-5.2}              & 99.63           & 98.52           & \textbf{100.00} & \textbf{100.00} \\
\textit{grok-4.1-fast}        & 99.63           & \textbf{100.00} & \textbf{100.00} & \textbf{100.00} \\
\midrule
\textbf{Overall}     & 97.90           & 96.73           & 99.69           & 99.94           \\
\bottomrule
\end{tabular}
\end{table}

The overall Pass@1 rate reaches 97.90\% at T=0.0 and 96.73\% at T=0.7, indicating that models generate syntactically correct and executable code in the vast majority of cases. Pass@5 rates are even higher (99.69\% and 99.94\%), showing that when multiple samples are generated, success is nearly guaranteed. This near-perfect Pass@5 performance suggests that syntax errors are typically stochastic rather than systematic---models occasionally make mistakes, but rarely fail consistently on the same task.

\textbf{Syntax Correctness Across Difficulty Levels.} The breakdown by difficulty level reveals interesting patterns. Pass@1 rates remain consistently high across L1 (98\%+), L2 (97\%+), and L3 (96\%+), with only minimal degradation (approximately 2\%) as task complexity increases. This suggests that syntax correctness is largely independent of strategic complexity---models can generate valid code even when the underlying strategy logic is challenging. The slight degradation at Level 3 likely stems from increased code length and structural complexity rather than fundamental syntax understanding limitations.

\textbf{Model-Level Syntax Performance.} Model-level analysis shows that all evaluated models achieve Pass@1 rates above 93\% at $\tau=0$, with flagship models (\textit{claude-sonnet-4.5}, \textit{gemini-3-pro-preview}) reaching 100\%. This uniformly high performance indicates that modern LLMs have effectively mastered Python syntax and API conventions for quantitative finance applications. The narrow range of syntax performance (93--100\%) contrasts sharply with the wide range of strategic performance (Sharpe Ratios from 0.415 to 0.628), confirming that code generation competence is no longer a differentiating factor among frontier models. The competitive advantage now lies in strategic reasoning and domain knowledge rather than syntax mastery.

\textbf{Error Type Analysis.} \Cref{appx_tab:error_statistics} provides insight into the nature of failures, revealing that syntax mastery is not the primary challenge.

\begin{table}[h]
\centering
\renewcommand{\arraystretch}{1.0}
\setlength{\tabcolsep}{4pt}
\caption{Error type distribution across models and temperature settings. Runtime errors (OtherError) dominate failures, while pure syntax errors account for fewer than 1\% of all attempts.}
\label{appx_tab:error_statistics}
\begin{tabular}{llccccc}
\toprule
\textbf{Temp.} & \textbf{Model} & \textbf{SUCCESS} & \textbf{OtherError} & \textbf{NameError} & \textbf{SyntaxError} & \textbf{AttributeError} \\
\midrule
\multirow{7}{*}{$\tau$=0}
& \textit{claude-sonnet-4.5}    & 1346 & 2   & 1  & 1  & 0 \\
& \textit{deepseek-v3.2}        & 1260 & 69  & 8  & 9  & 4 \\
& \textit{gemini-3-flash-preview}       & 1272 & 78  & 0  & 0  & 0 \\
& \textit{gemini-3-pro-preview}         & 1348 & 2   & 0  & 0  & 0 \\
& \textit{gpt-5.2}              & 1342 & 8   & 0  & 0  & 0 \\
& \textit{grok-4.1-fast}        & 1345 & 1   & 2  & 1  & 1 \\
\cmidrule(lr){2-7}
& \textbf{Total}       & \textbf{7913} & \textbf{160} & \textbf{11} & \textbf{11} & \textbf{5} \\
\midrule
\multirow{7}{*}{$\tau$=0.7}
& \textit{claude-sonnet-4.5}    & 1345 & 3   & 2  & 0  & 0 \\
& \textit{deepseek-v3.2}        & 1170 & 155 & 12 & 10 & 3 \\
& \textit{gemini-3-flash-preview}       & 1321 & 27  & 1  & 0  & 1 \\
& \textit{gemini-3-pro-preview}         & 1340 & 8   & 2  & 0  & 0 \\
& \textit{gpt-5.2}              & 1339 & 11  & 0  & 0  & 0 \\
& \textit{grok-4.1-fast}        & 1344 & 5   & 0  & 0  & 1 \\
\cmidrule(lr){2-7}
& \textbf{Total}       & \textbf{7859} & \textbf{209} & \textbf{17} & \textbf{10} & \textbf{5} \\
\bottomrule
\end{tabular}
\end{table}

Among the small fraction of failed attempts (approximately 2--3\% of all runs), the most common issues are runtime errors (OtherError: 160 cases at T=0.0) rather than syntax errors (SyntaxError: 11 cases). NameError (11 cases) and AttributeError (5 cases) are also rare, suggesting that models correctly reference variables and API methods. The dominance of runtime errors over syntax errors (approximately 15:1 ratio) indicates that failures typically stem from logical issues or edge cases in strategy execution rather than basic coding mistakes. Examples of runtime errors include division by zero when computing indicators, index out of bounds when accessing historical data, or type mismatches in mathematical operations. These errors reflect limitations in edge case handling rather than fundamental syntax understanding.

\subsubsection{Robustness and Stability Analysis}

\textbf{Overview.} Robustness measures the consistency of generated strategies across multiple runs with identical prompts. Even at T=0.0 (deterministic decoding), models may produce slightly different outputs due to implementation details, and at T=0.7, stochastic sampling introduces additional variability. This subsection examines how stable model outputs are across repeated evaluations, which is critical for production reliability.

\begin{figure*}[h]
\centering
\begin{subfigure}[t]{0.48\textwidth}
\includegraphics[width=\textwidth]{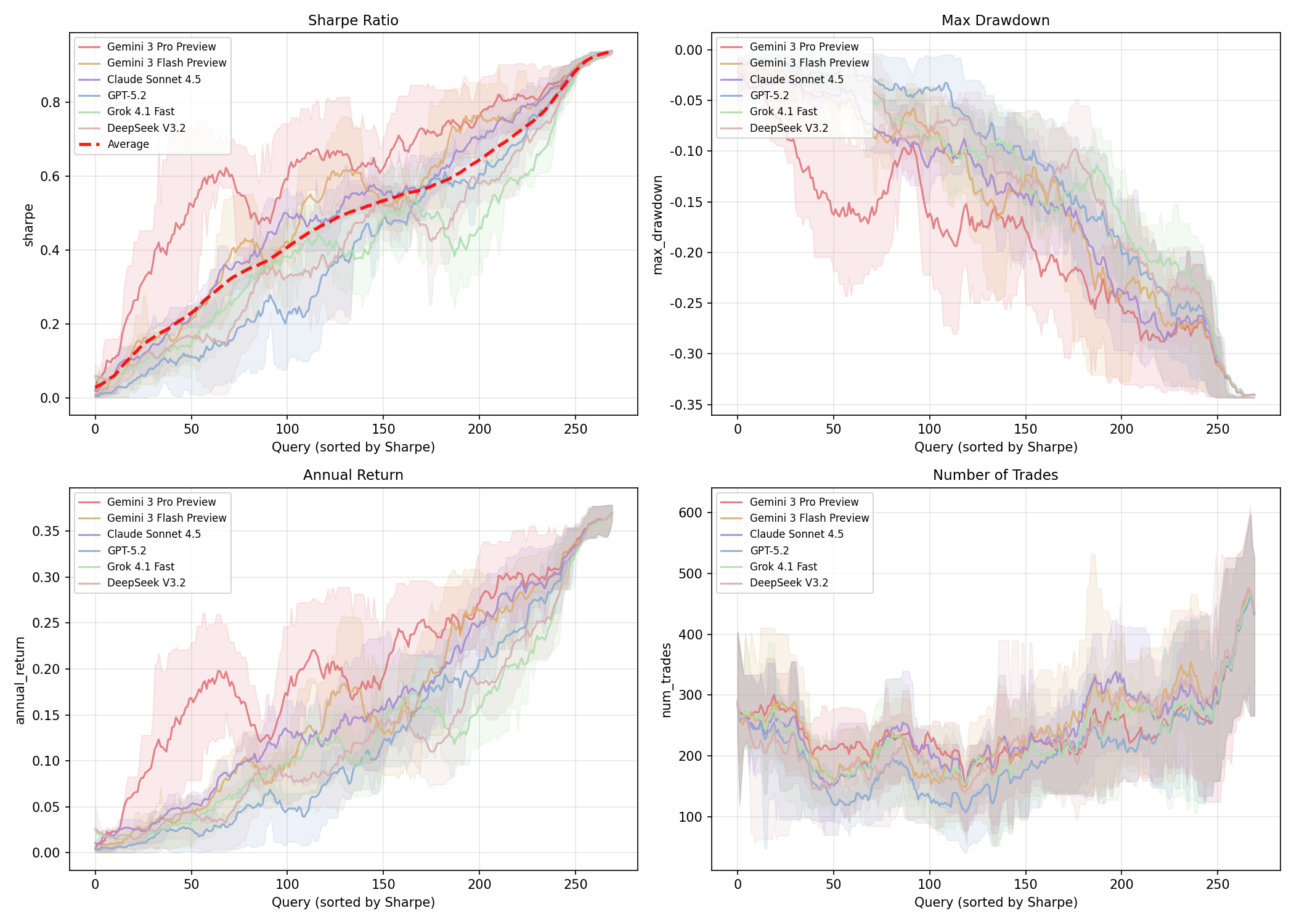}
\caption{T=0.0}
\end{subfigure}
\hfill
\begin{subfigure}[t]{0.48\textwidth}
\includegraphics[width=\textwidth]{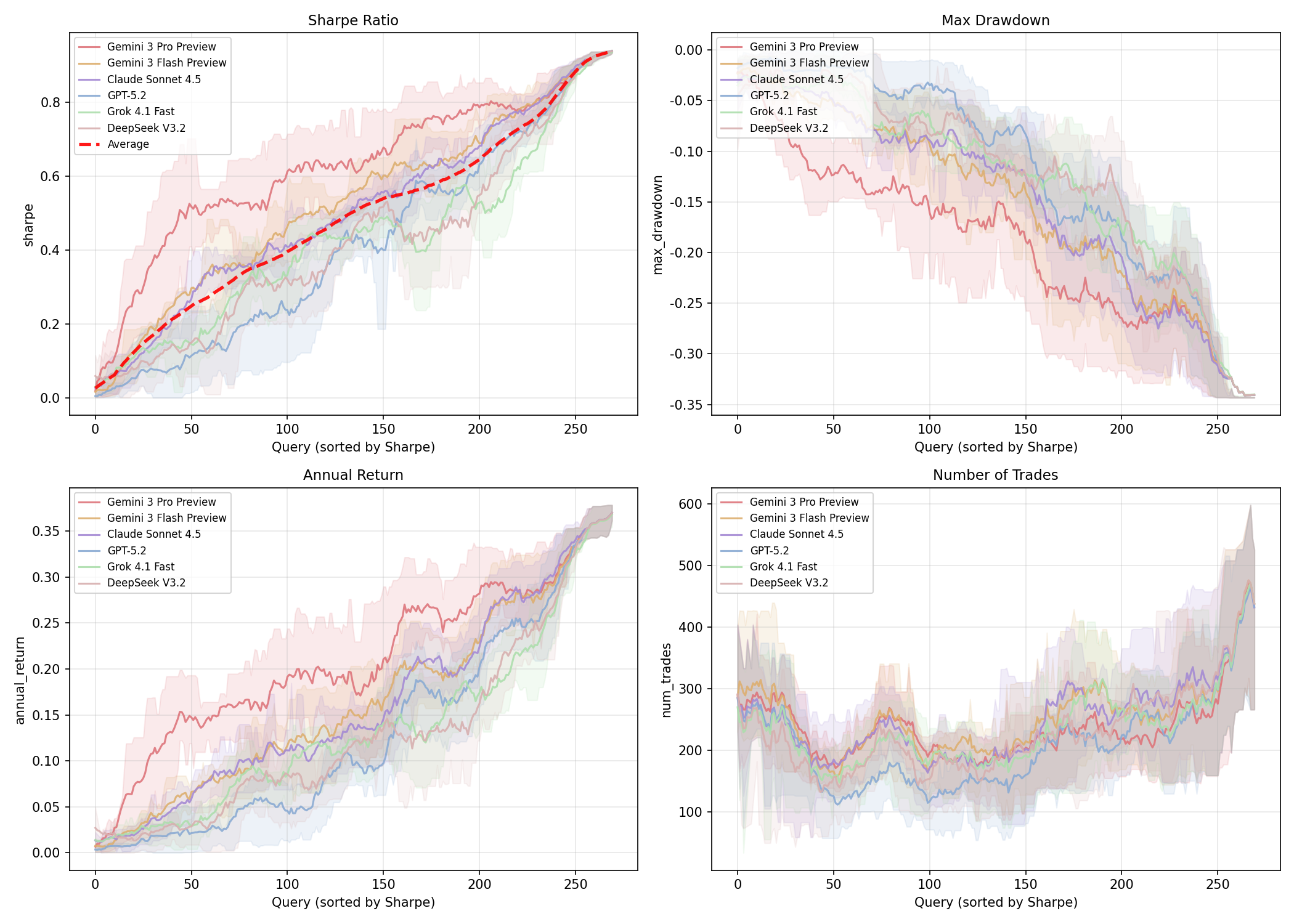}
\caption{T=0.7}
\end{subfigure}
\caption{Robustness analysis showing performance stability across multiple runs.}
\label{appx_fig:model_robustness}
\end{figure*}

\textbf{Analysis.} \Cref{appx_fig:model_robustness} reveals significant differences in model stability across multiple runs with identical prompts.

The Sharpe Ratio plot shows that \textit{claude-sonnet-4.5} maintains the narrowest confidence bands, indicating highly consistent strategy generation across runs. \textit{\textit{gpt-5.2}} exhibits slightly wider bands but remains stable. \textit{deepseek-v3.2} and \textit{grok-4.1-fast} show broader variance, particularly in later samples, suggesting less deterministic behavior. The Maximum Drawdown metric demonstrates that all models maintain reasonable risk control consistency, with variance bands remaining relatively tight.

\textbf{Temporal Stability Patterns.} The aligned plots reveal how performance evolves across sequential samples. \textit{claude-sonnet-4.5} shows nearly flat trend lines with minimal drift, indicating that repeated sampling yields consistent results. \textit{gemini-3-pro-preview} exhibits slight upward drift in some metrics, suggesting that later samples occasionally outperform earlier ones, possibly due to internal sampling strategies. \textit{grok-4.1-fast} shows the most erratic patterns, with performance oscillating significantly across samples, confirming its high-variance nature observed in earlier analyses.

\begin{figure*}[h]
\centering
\begin{subfigure}[t]{0.48\textwidth}
\includegraphics[width=\textwidth]{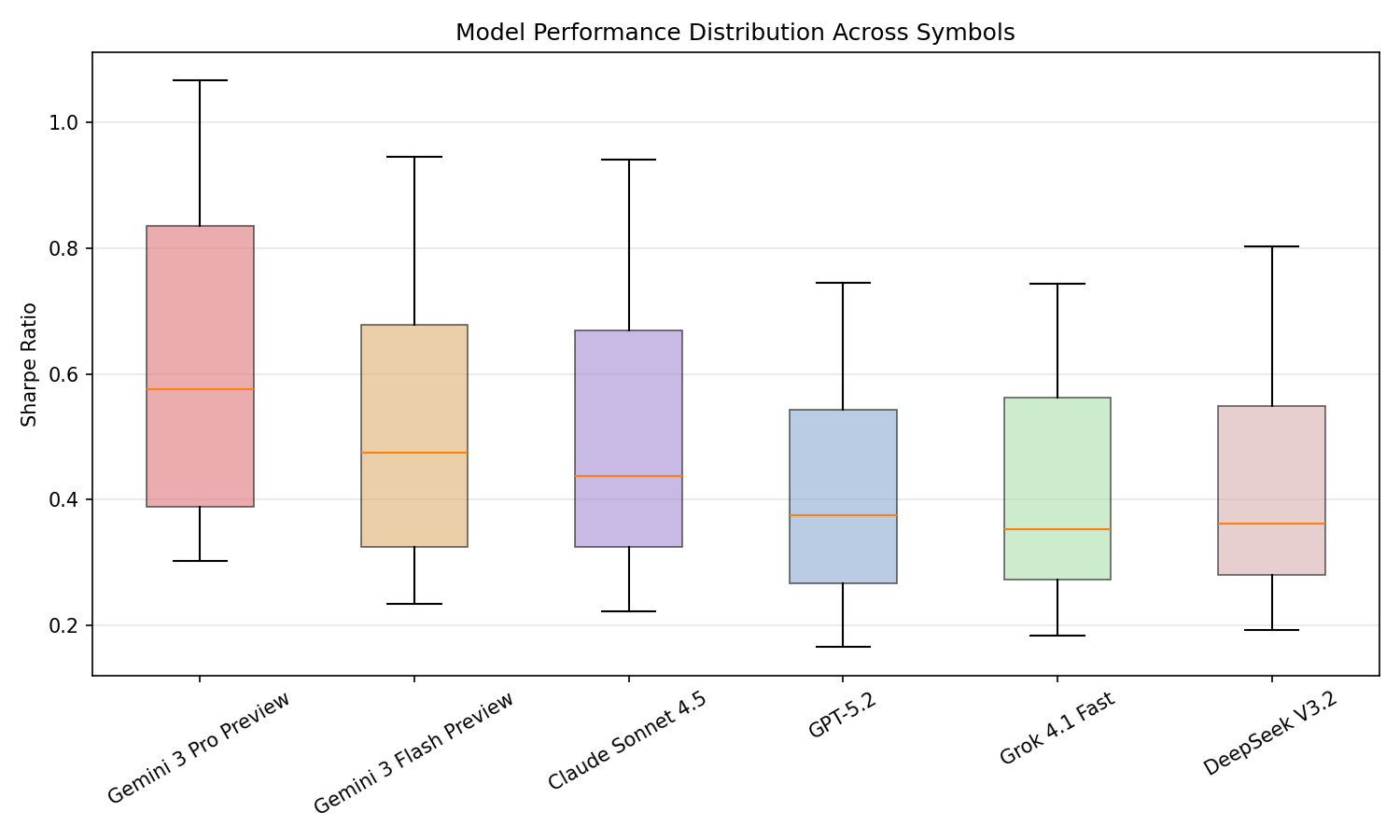}
\caption{T=0.0}
\end{subfigure}
\hfill
\begin{subfigure}[t]{0.48\textwidth}
\includegraphics[width=\textwidth]{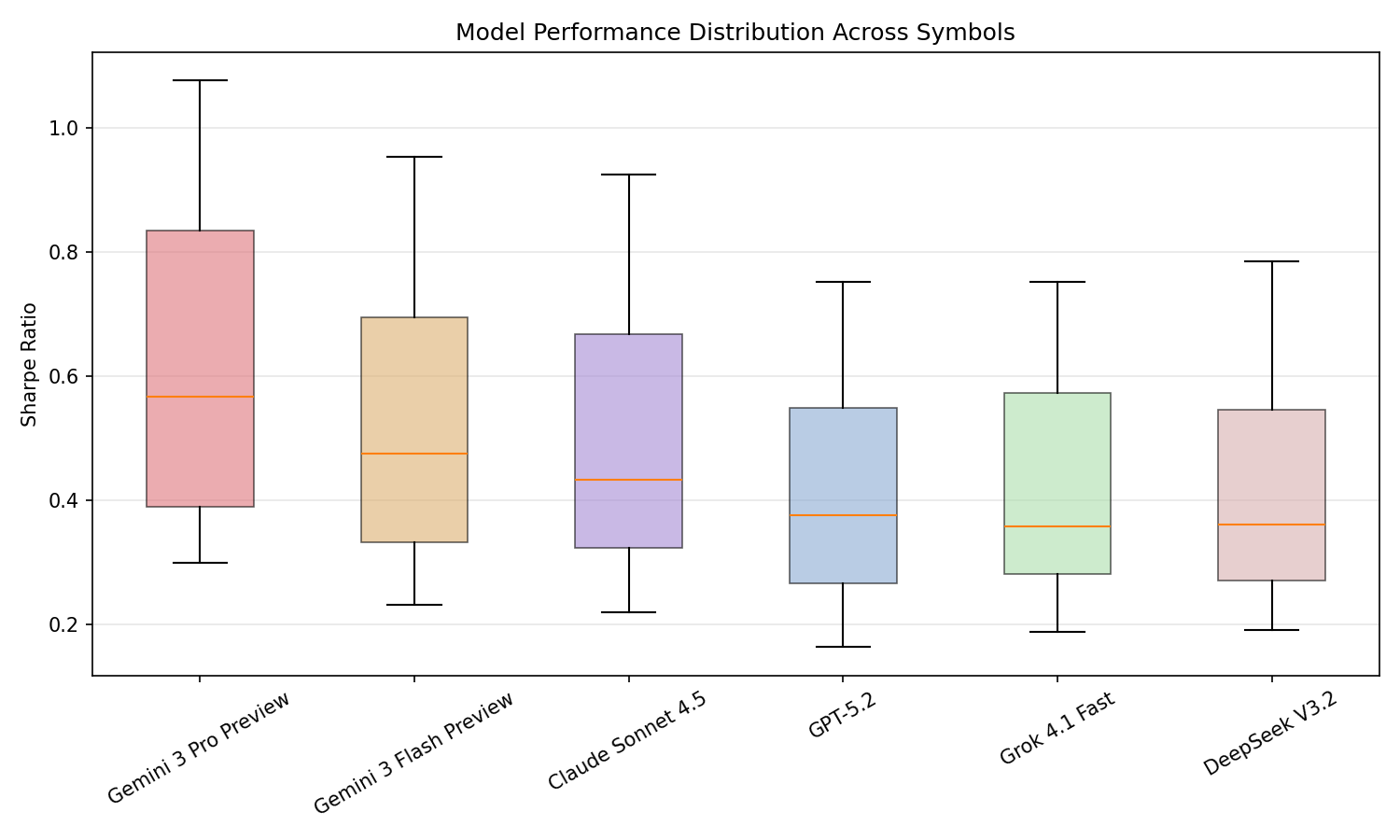}
\caption{T=0.7}
\end{subfigure}
\caption{Cross-asset robustness analysis. Box plots show performance distribution across 7 different assets.}
\label{appx_fig:cross_asset_robustness}
\end{figure*}

\textbf{Cross-Asset Robustness.} \Cref{appx_fig:cross_asset_robustness} demonstrates that model performance varies significantly across asset classes, revealing asset-specific strengths and weaknesses.

The cross-asset analysis in \Cref{appx_fig:cross_asset_robustness} demonstrates that model performance varies significantly across asset classes. \textit{claude-sonnet-4.5} shows the most consistent performance across all assets with relatively tight box plots. \textit{\textit{gpt-5.2}} exhibits a narrow distribution with minimal outliers, reflecting its conservative strategy generation profile. Notably, cryptocurrency assets (BTCUSDT, ETHUSDT) show higher variance across all models compared to traditional stocks, likely due to their higher volatility and different market dynamics. This suggests that while models can generate profitable strategies for diverse assets, cryptocurrency trading poses additional challenges.

\textbf{Asset-Specific Performance Patterns.} The boxplot distributions reveal that equity assets (AAPL, GOOGL, MSFT, TSLA) generally yield tighter distributions with higher medians, indicating that models perform more consistently and effectively on traditional stocks. Cryptocurrency assets exhibit wider boxes and lower medians, confirming the increased difficulty discussed in the per-asset analysis section. Interestingly, NVDA (a high-volatility equity) shows distribution characteristics intermediate between traditional equities and cryptocurrencies, suggesting that volatility is a key factor affecting model performance consistency.

\textbf{Model Specialization Across Assets.} Some models show more uniform performance across asset types (\textit{claude-sonnet-4.5}, \textit{gemini-3-flash-preview}), while others exhibit stronger asset-specific variation (\textit{gemini-3-pro-preview}, \textit{deepseek-v3.2}). This suggests that certain models have learned more generalizable trading principles, while others may have implicit biases toward specific market structures encountered during training.

\clearpage

\subsection{Per-Model Detailed Analysis}

This section provides in-depth analysis of each individual LLM, examining its unique strengths, weaknesses, and behavioral patterns across all evaluation dimensions.

\subsubsection{\textit{gemini-3-pro-preview}}

\begin{figure*}[h]
\centering
\begin{subfigure}[t]{0.48\textwidth}
\includegraphics[width=\textwidth]{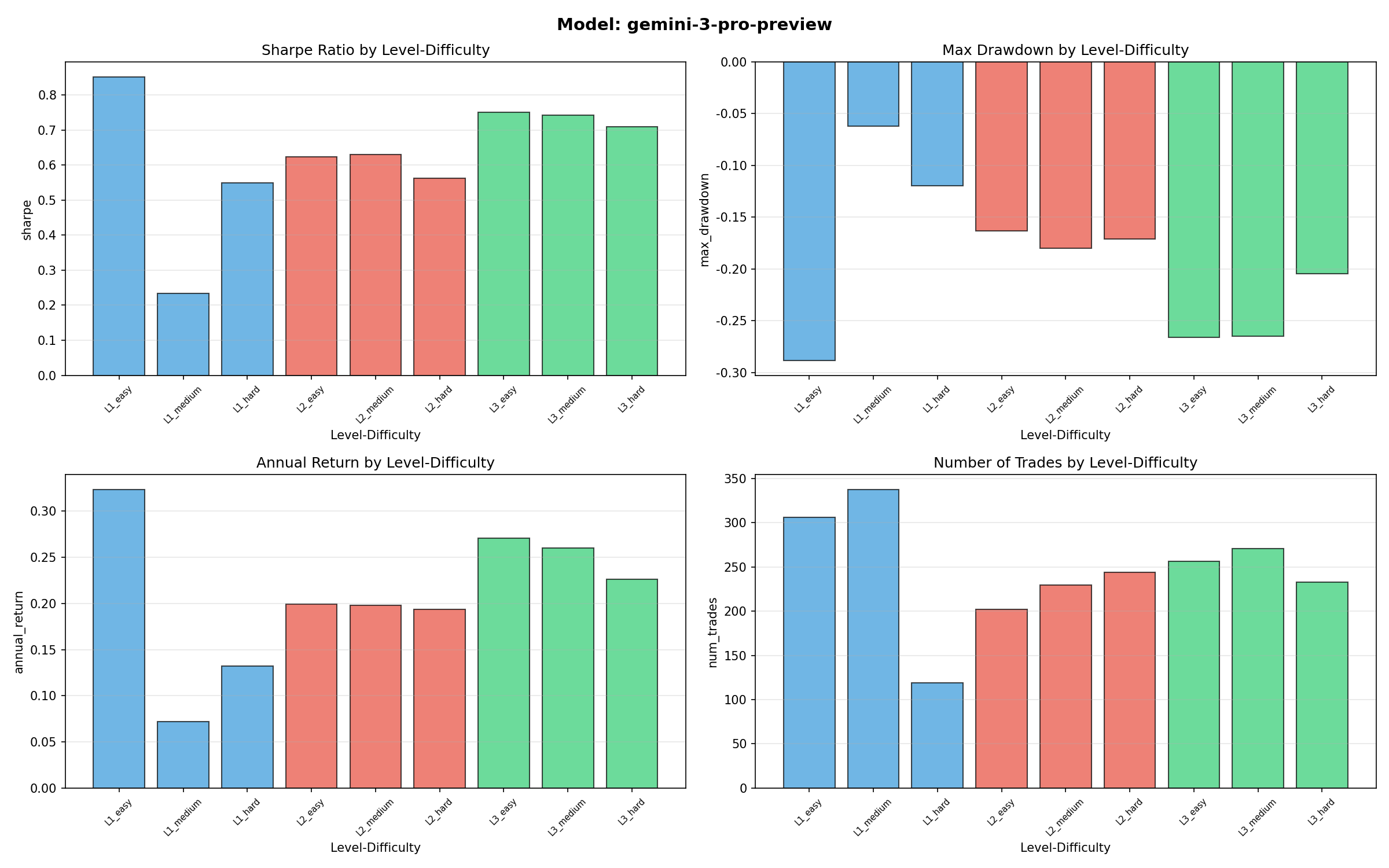}
\caption{T=0.0}
\end{subfigure}
\hfill
\begin{subfigure}[t]{0.48\textwidth}
\includegraphics[width=\textwidth]{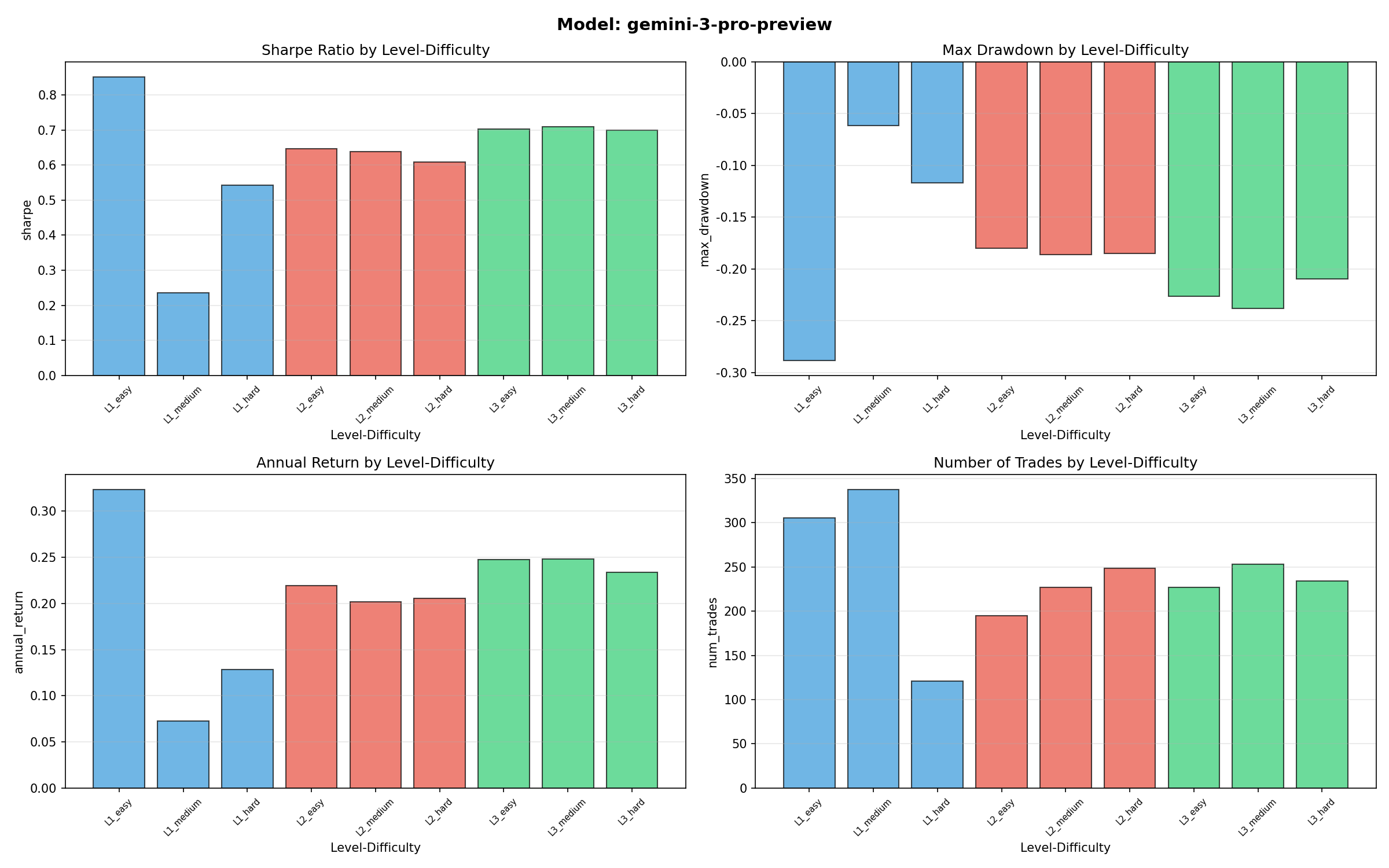}
\caption{T=0.7}
\end{subfigure}
\caption{\textit{gemini-3-pro-preview} performance across 9 difficulty levels.}
\label{appx_fig:gemini_pro_level_performance}
\end{figure*}

\textbf{Overview.} \textit{gemini-3-pro-preview} is Google's flagship model with advanced reasoning and multimodal capabilities. Within \projectname it serves as the representative of the \emph{aggressive-creative} archetype, consistently prioritizing high-conviction signal logic over capital preservation.

\textbf{Analysis.} As shown in \Cref{appx_fig:gemini_pro_level_performance}, \textit{gemini-3-pro-preview} achieves the highest overall Sharpe Ratio (0.628 at $\tau{=}0$) and Annualized Return (20.8\%) among all evaluated models, with a Sortino Ratio of 1.004 that further confirms strong downside-adjusted performance. Most remarkably, the model exhibits a unique \emph{ascending} difficulty profile: SR increases from 0.545 at Level~1 through 0.604 at Level~2 to 0.734 at Level~3. No other model in our benchmark displays this pattern; all others degrade monotonically from L1 to L3. This suggests that the open-ended creative freedom afforded by goal-oriented tasks (Level~3) activates reasoning capabilities that constrained translation tasks (Level~1) leave untapped. Across the nine $3\times 3$ cells, the model's performance variance is highest on L3-Hard queries, indicating that while it excels on average, some particularly challenging open-ended prompts still expose failure modes.

\begin{figure*}[h]
\centering
\begin{subfigure}[t]{0.48\textwidth}
\includegraphics[width=\textwidth]{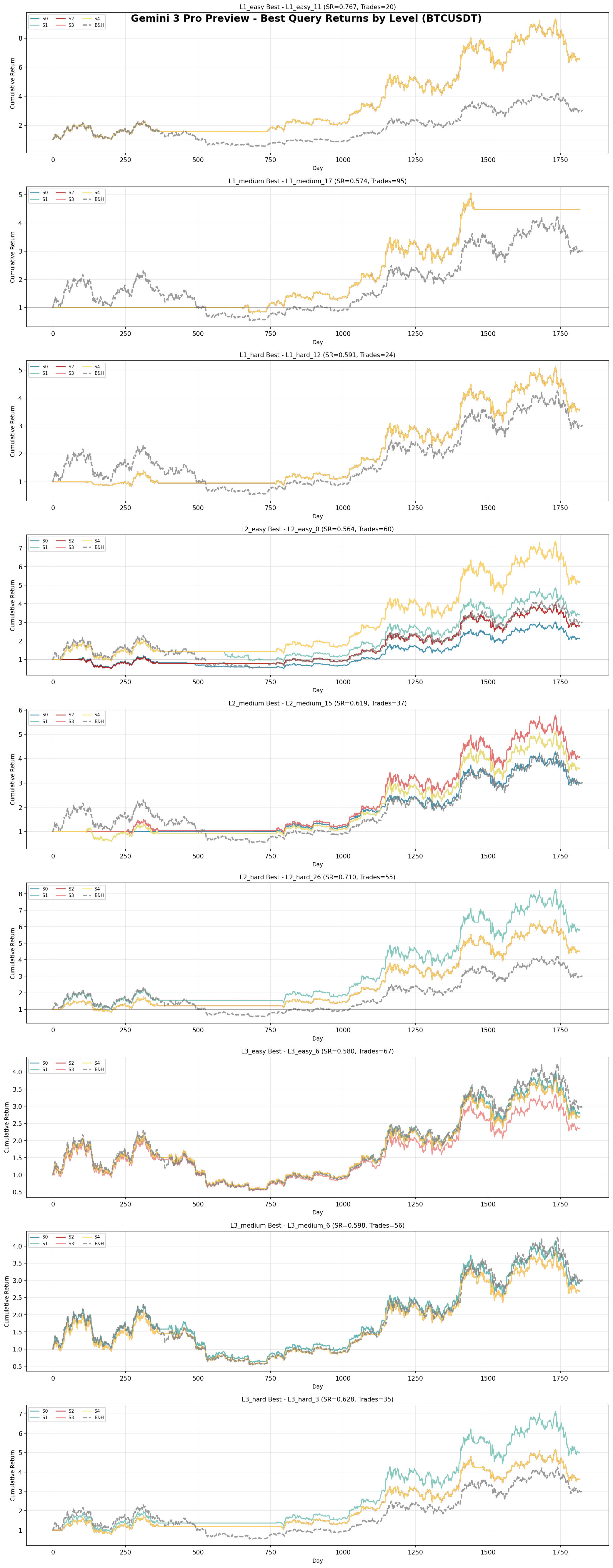}
\caption{T=0.0}
\end{subfigure}
\hfill
\begin{subfigure}[t]{0.48\textwidth}
\includegraphics[width=\textwidth]{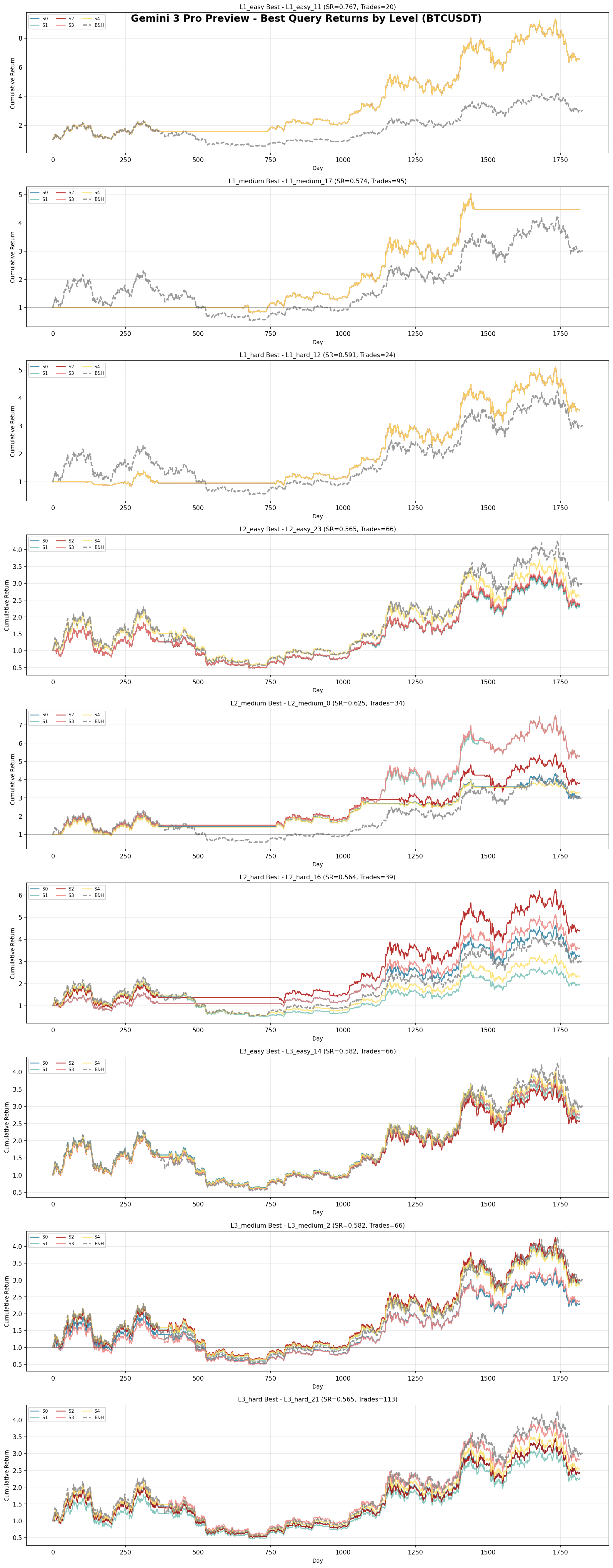}
\caption{T=0.7}
\end{subfigure}
\caption{Best-performing strategies generated by \textit{gemini-3-pro-preview}.}
\label{appx_fig:gemini_pro_best_returns}
\end{figure*}

\textbf{Strengths and Weaknesses.} The primary strength of \textit{gemini-3-pro-preview} lies in its dominant return generation: it leads on SR, ARR, and SoR across both temperatures, all seven assets, and all three difficulty levels. Its best-case strategies (\Cref{appx_fig:gemini_pro_best_returns}) produce the highest cumulative returns in the benchmark. However, this aggressive profile incurs the largest Maximum Drawdown (0.191) and Volatility (0.262) among all models, indicating that the generated strategies frequently take concentrated positions with limited hedging or stop-loss logic. The model's run-to-run confidence bands are also the widest among top-tier models, meaning that while its expected performance is best, the variance of generated strategies is non-negligible.

\textbf{Notable Patterns.}
The ascending L1$\to$L3 profile implies a dissociation between code-translation skill and strategic reasoning: \textit{gemini-3-pro-preview} is only average at faithfully translating explicit rules, yet excels when it must design strategies from first principles. This pattern is temperature-invariant (the $\tau{=}0$ and $\tau{=}0.7$ panels in \Cref{appx_fig:gemini_pro_level_performance} are nearly identical), confirming that the ascending trend is an intrinsic model property. Additionally, the model shows particular strength on high-volatility assets (TSLA, cryptocurrencies), where its aggressive signal logic is better rewarded, suggesting an implicit preference for momentum-style entry conditions.

\clearpage

\subsubsection{\textit{\textit{gpt-5.2}}}

\textbf{Overview.} \textit{\textit{gpt-5.2}} is OpenAI's flagship model evaluated in this benchmark. It represents the \emph{conservative-rigid} archetype, consistently generating strategies that prioritize capital preservation and drawdown control over aggressive return seeking.

\begin{figure*}[h]
\centering
\begin{subfigure}[t]{0.48\textwidth}
\includegraphics[width=\textwidth]{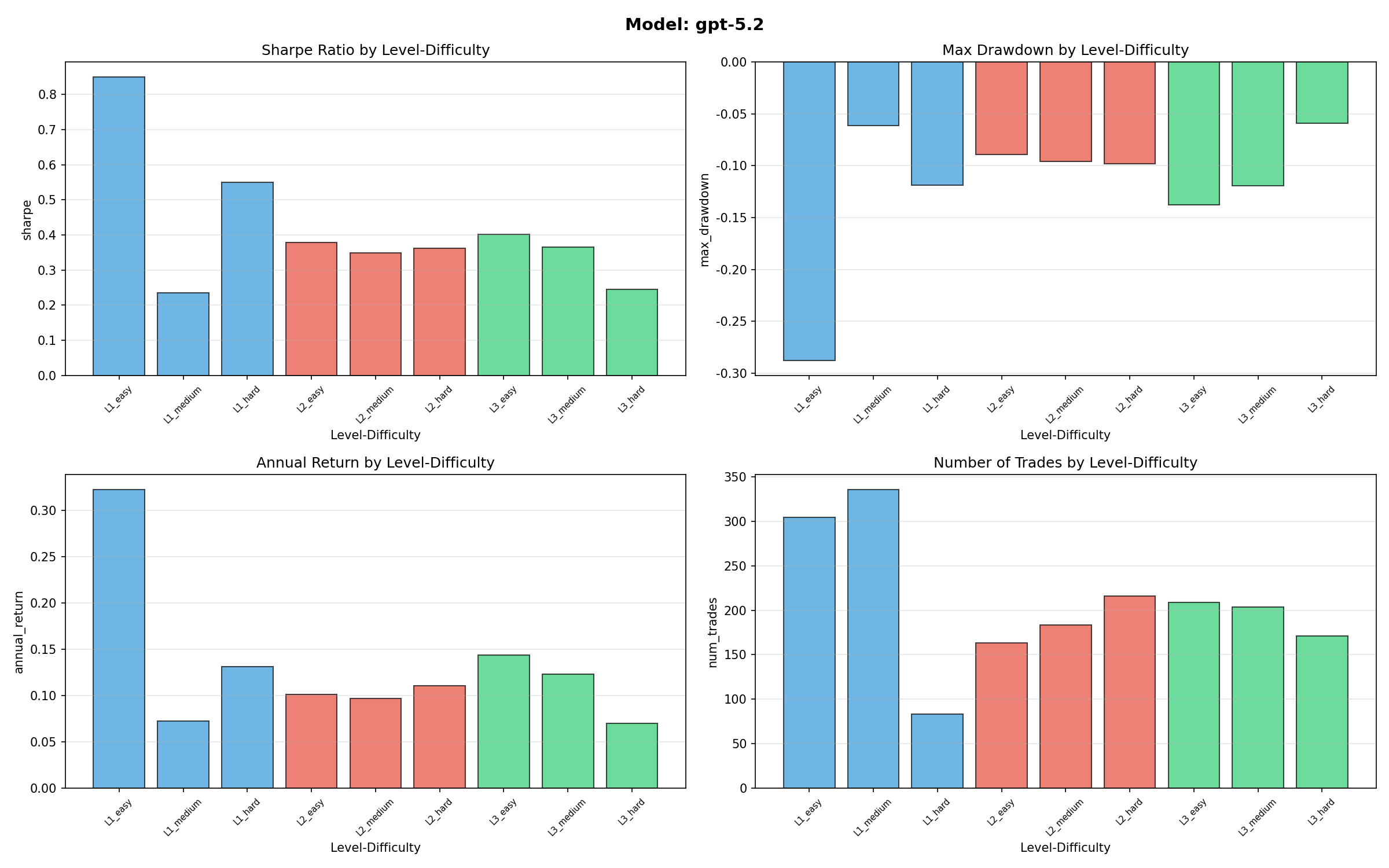}
\caption{T=0.0}
\end{subfigure}
\hfill
\begin{subfigure}[t]{0.48\textwidth}
\includegraphics[width=\textwidth]{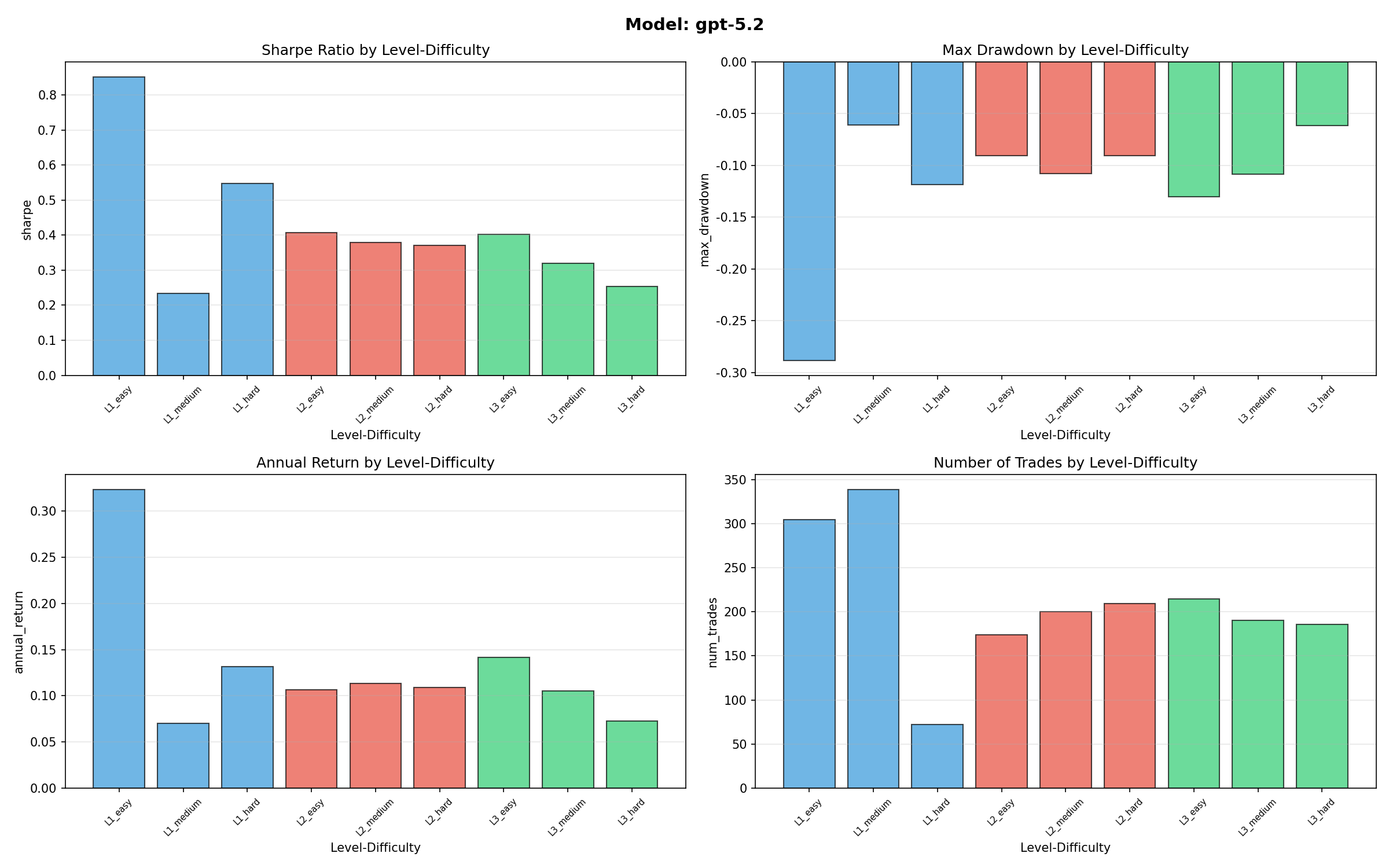}
\caption{T=0.7}
\end{subfigure}
\caption{\textit{\textit{gpt-5.2}} performance across 9 difficulty levels.}
\label{appx_fig:gpt4o_level_performance}
\end{figure*}

\textbf{Analysis.} \Cref{appx_fig:gpt4o_level_performance} reveals \textit{\textit{gpt-5.2}}'s performance characteristics across the difficulty spectrum.

\textit{\textit{gpt-5.2}} achieves the lowest Maximum Drawdown (0.119 at $\tau{=}0$) and Volatility (0.163) among all evaluated models, reflecting a distinctly conservative strategy generation profile. While its overall Sharpe Ratio (0.415) places it in the lower tier, the competitive Calmar Ratio (1.599) reveals efficient return-to-drawdown management: the model sacrifices upside potential to tightly bound downside risk. As shown in \Cref{appx_fig:gpt4o_level_performance}, performance degrades from Level~1 (SR~=~0.544) to Level~3 (SR~=~0.336), but the degradation is more gradual than that of \textit{deepseek-v3.2} or \textit{grok-4.1-fast}. Across the nine difficulty cells, MDD and VOL remain remarkably stable, suggesting that the model's risk-averse tendency is hard-coded into its generation behavior regardless of prompt complexity.

\textbf{Strengths and Weaknesses.} The defining strength of \textit{\textit{gpt-5.2}} is risk control: it consistently produces the lowest drawdown and volatility across both temperatures, all seven assets, and all difficulty levels, with narrow run-to-run confidence bands that indicate highly predictable generation. Code quality is high, with syntax error rates on par with other frontier models. However, the conservative profile comes at a clear cost: the model ranks last or second-to-last on return-oriented metrics (SR, ARR, SoR) for most assets, and its Level~3 Sharpe Ratio (0.336) falls substantially behind the leaders. The generated strategies appear to default to simple, low-conviction signal logic (e.g., moving-average crossovers with wide filters) even when the query explicitly demands more complex reasoning.

\textbf{Notable Patterns.}
\textit{\textit{gpt-5.2}} exhibits the tightest MDD distribution across all queries and runs, with virtually no outlier drawdowns exceeding 0.20 on any asset. This suggests an implicit ``safety bias'' in its code generation: the model appears to encode conservative position-sizing and early-exit conditions even when not prompted to do so. The pattern holds identically at $\tau{=}0$ and $\tau{=}0.7$, and across both cryptocurrency and equity assets, making \textit{\textit{gpt-5.2}} the most predictable and lowest-variance model in the benchmark, a favorable property for risk-averse deployment scenarios where consistency matters more than peak performance.

\begin{figure*}[h]
\centering
\begin{subfigure}[t]{0.48\textwidth}
\includegraphics[width=\textwidth]{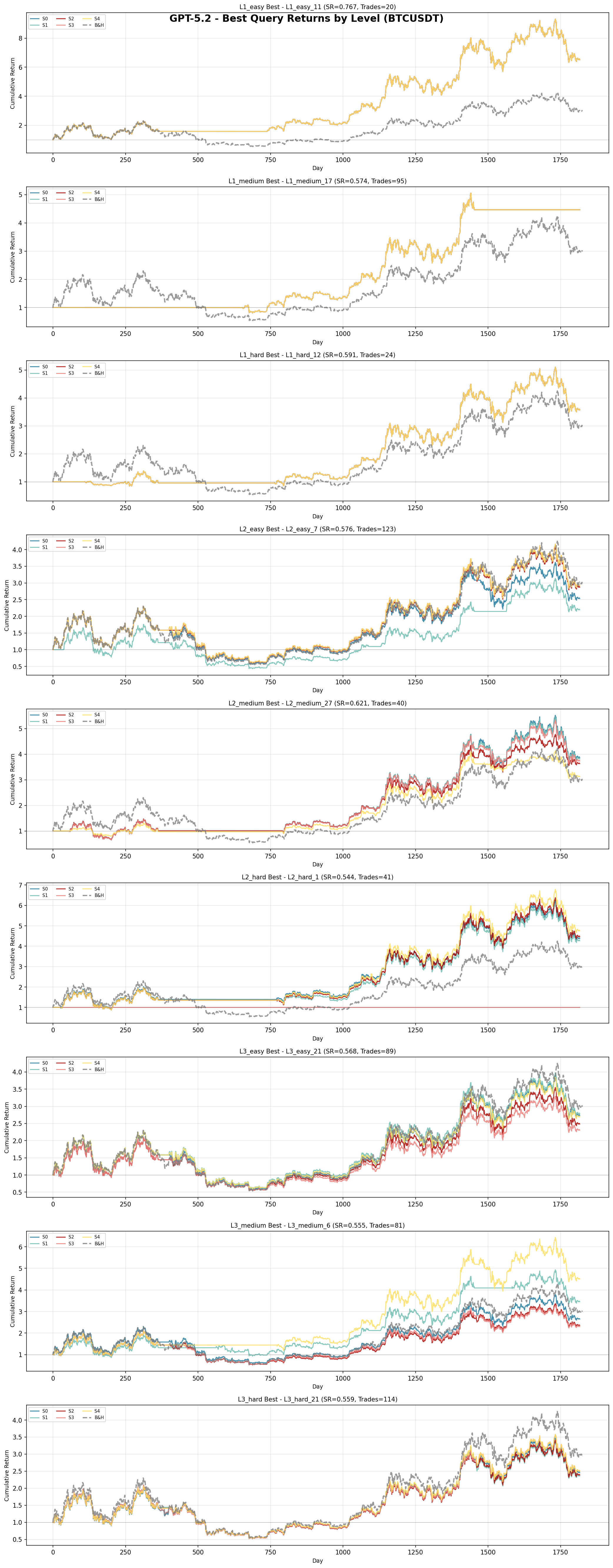}
\caption{T=0.0}
\end{subfigure}
\hfill
\begin{subfigure}[t]{0.48\textwidth}
\includegraphics[width=\textwidth]{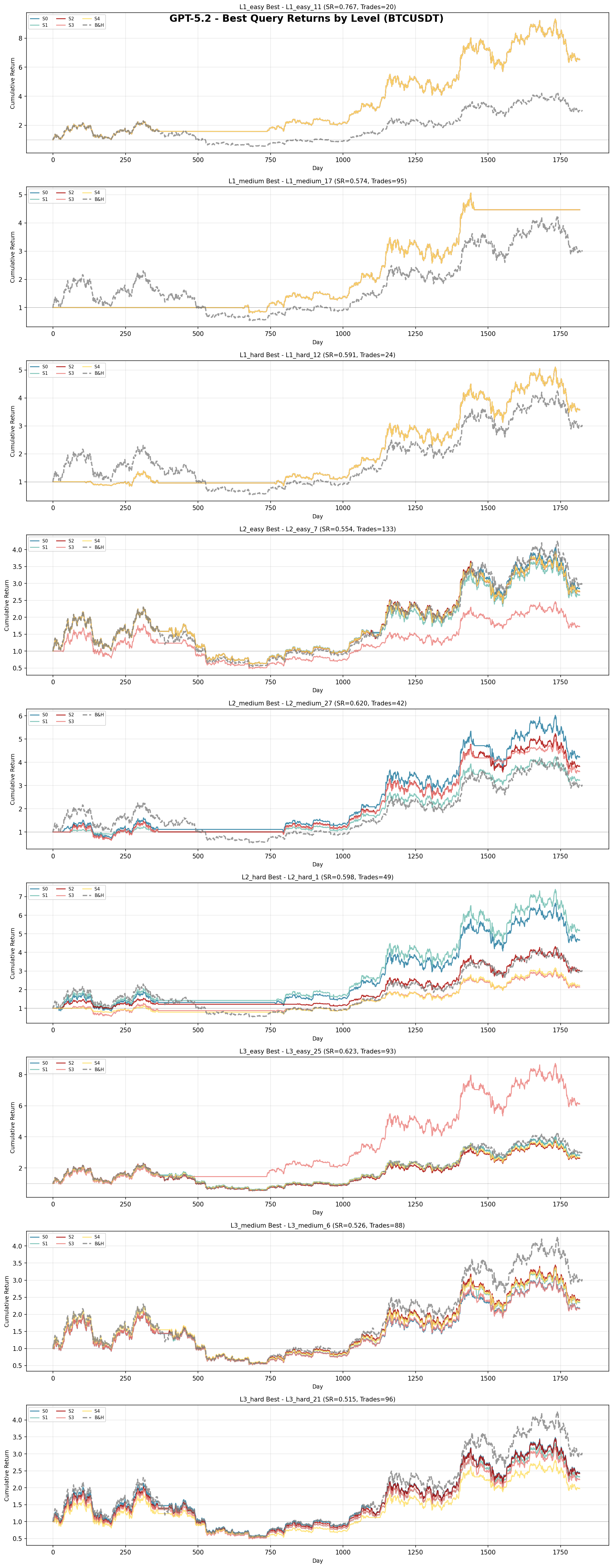}
\caption{T=0.7}
\end{subfigure}
\caption{Best-performing strategies generated by \textit{\textit{gpt-5.2}}.}
\label{appx_fig:gpt4o_best_returns}
\end{figure*}

\clearpage

\subsubsection{\textit{claude-sonnet-4.5}}

\begin{figure*}[h]
\centering
\begin{subfigure}[t]{0.48\textwidth}
\includegraphics[width=\textwidth]{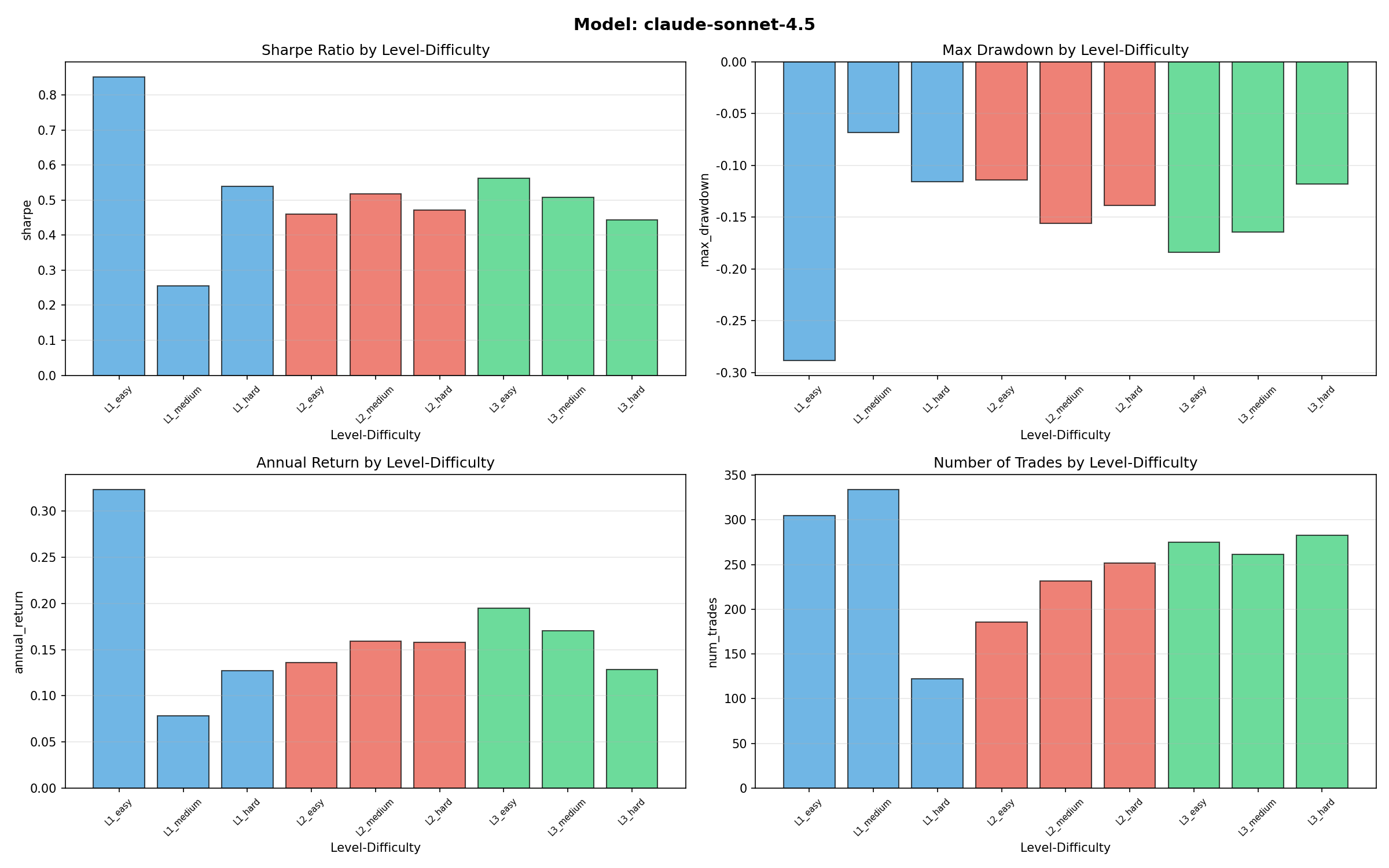}
\caption{T=0.0}
\end{subfigure}
\hfill
\begin{subfigure}[t]{0.48\textwidth}
\includegraphics[width=\textwidth]{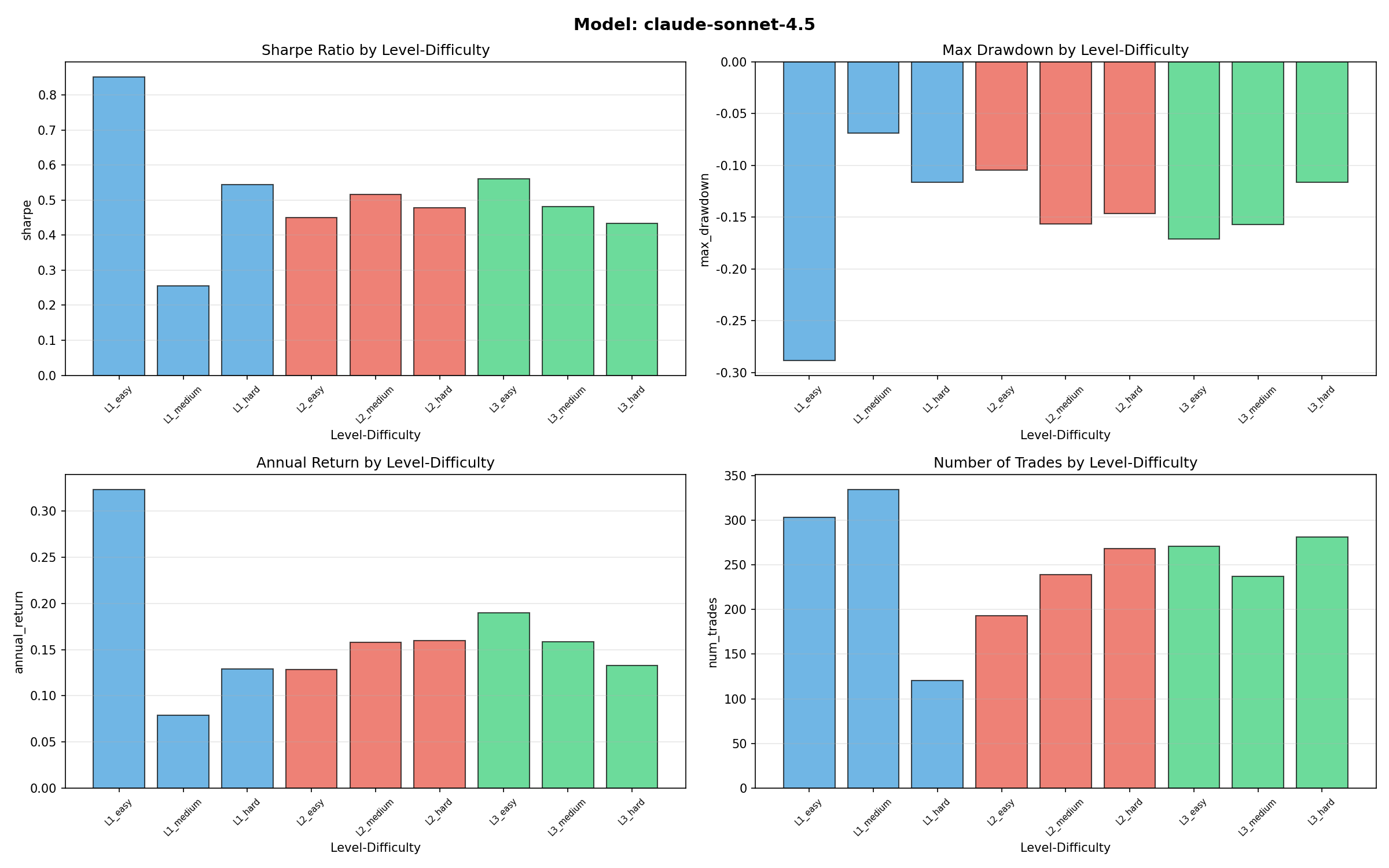}
\caption{T=0.7}
\end{subfigure}
\caption{\textit{claude-sonnet-4.5} performance across 9 difficulty levels.}
\label{appx_fig:claude_level_performance}
\end{figure*}

\textbf{Overview.} \textit{claude-sonnet-4.5} is Anthropic's flagship model, known for strong reasoning capabilities. Within \projectname it exemplifies the \emph{balanced-stable} archetype, achieving the most favorable trade-off between return generation, risk control, and cross-run consistency.

\textbf{Analysis.} As shown in \Cref{appx_fig:claude_level_performance}, \textit{claude-sonnet-4.5} achieves Sharpe Ratios of 0.549 at Level~1, 0.482 at Level~2, and 0.507 at Level~3 ($\tau{=}0$). The L1-to-L3 degradation (7.6\%) is the mildest among all models except \textit{gemini-3-pro-preview} (which actually improves), indicating that the model maintains robust strategy-design capabilities even under open-ended, underspecified prompts. Notably, its Level~3 SR recovers relative to Level~2, suggesting that the model handles the transition from parameter inference to goal-oriented generation more gracefully than most competitors. Across the nine difficulty cells, the variance of SR values is the lowest in the benchmark, reflecting highly uniform performance regardless of the specific cognitive demand.

\textbf{Strengths and Weaknesses.} The defining strength of \textit{claude-sonnet-4.5} is its exceptional risk-adjusted efficiency: it achieves the highest Calmar Ratio (CR~=~1.650 at $\tau{=}0$) among all models, combining moderate returns (ARR~=~16.4\%) with tightly controlled drawdowns (MDD~=~0.150) and volatility (VOL~=~0.205). Crucially, the model also exhibits the narrowest run-to-run confidence bands, meaning that its 5 independent generations per query produce the most tightly clustered outcomes, a property highly desirable for production deployment where predictability is paramount. However, its overall SR (0.513) and ARR rank below \textit{gemini-3-pro-preview}, indicating that the model trades peak upside for consistency. On high-volatility assets (TSLA, BTCUSDT), it underperforms the aggressive-creative archetype by a wider margin, suggesting that its balanced signal logic is less effective in extreme market environments.

\textbf{Notable Patterns.}
\textit{claude-sonnet-4.5} stands out for a distinctive ``recovery'' pattern at Level~3: while most models degrade monotonically from L1 to L3, Claude's SR dips at L2 but partially recovers at L3, hinting that the model benefits from the additional degrees of freedom in open-ended tasks once it no longer needs to infer specific missing parameters. This non-monotonic profile is temperature-invariant and consistent across assets, suggesting it reflects an intrinsic reasoning strategy. Furthermore, the model generates the most diversified factor usage among all LLMs, drawing on a broader set of technical indicators rather than relying on a few dominant signals, a pattern that likely contributes to its low drawdown and high Calmar Ratio.

\begin{figure*}[h]
\centering
\begin{subfigure}[t]{0.48\textwidth}
\includegraphics[width=\textwidth]{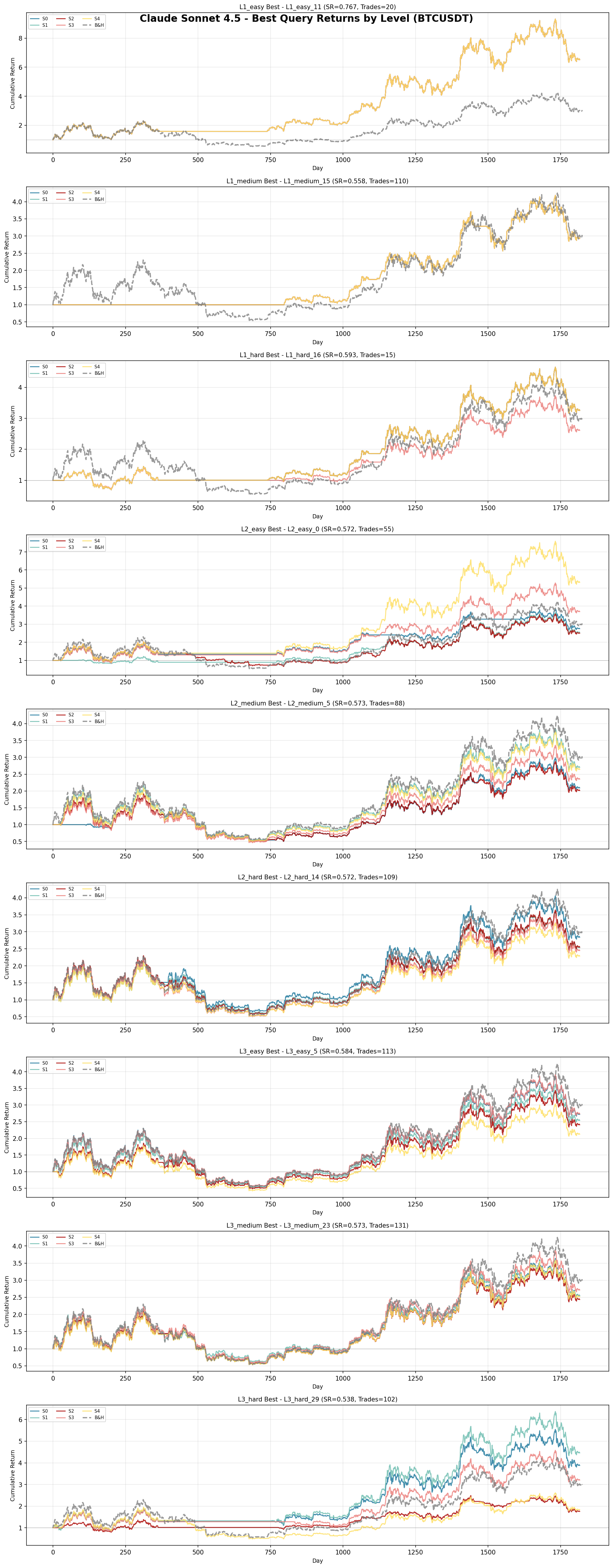}
\caption{T=0.0}
\end{subfigure}
\hfill
\begin{subfigure}[t]{0.48\textwidth}
\includegraphics[width=\textwidth]{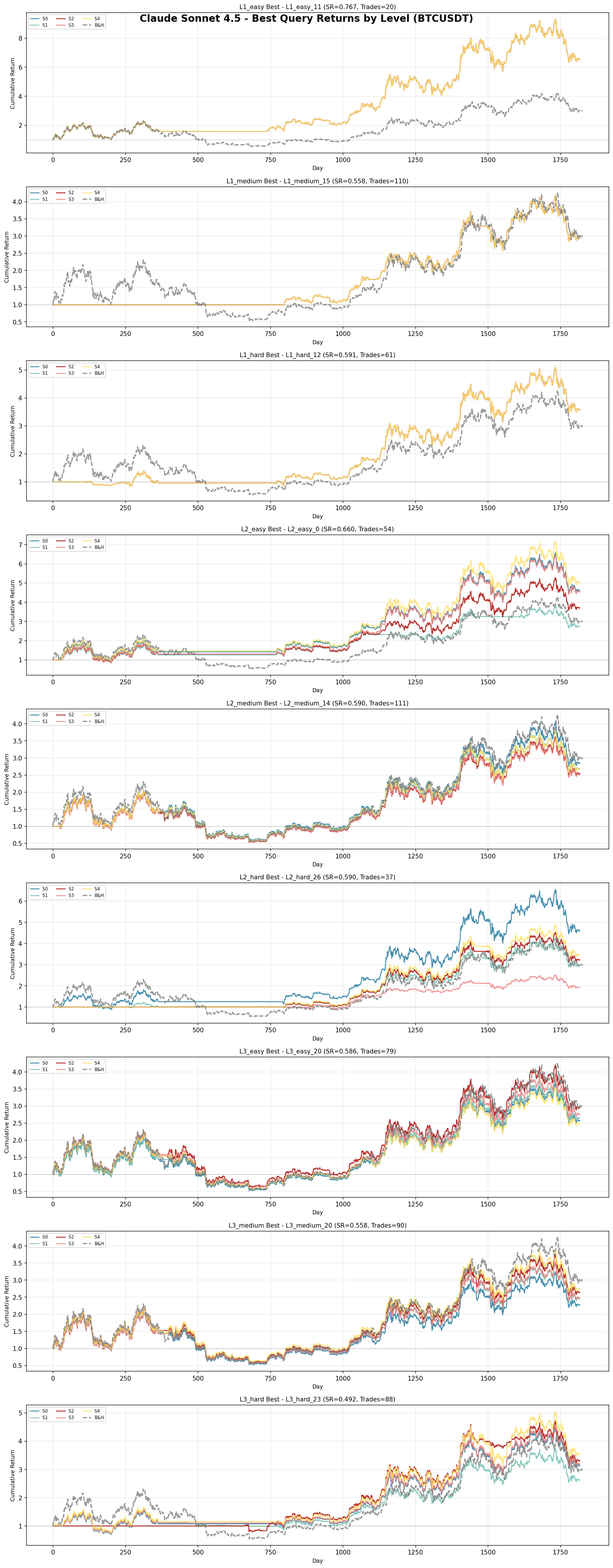}
\caption{T=0.7}
\end{subfigure}
\caption{Best-performing strategies generated by \textit{claude-sonnet-4.5}.}
\label{appx_fig:claude_best_returns}
\end{figure*}

\clearpage

\subsubsection{\textit{gemini-3-flash-preview}}

\begin{figure*}[h]
\centering
\begin{subfigure}[t]{0.48\textwidth}
\includegraphics[width=\textwidth]{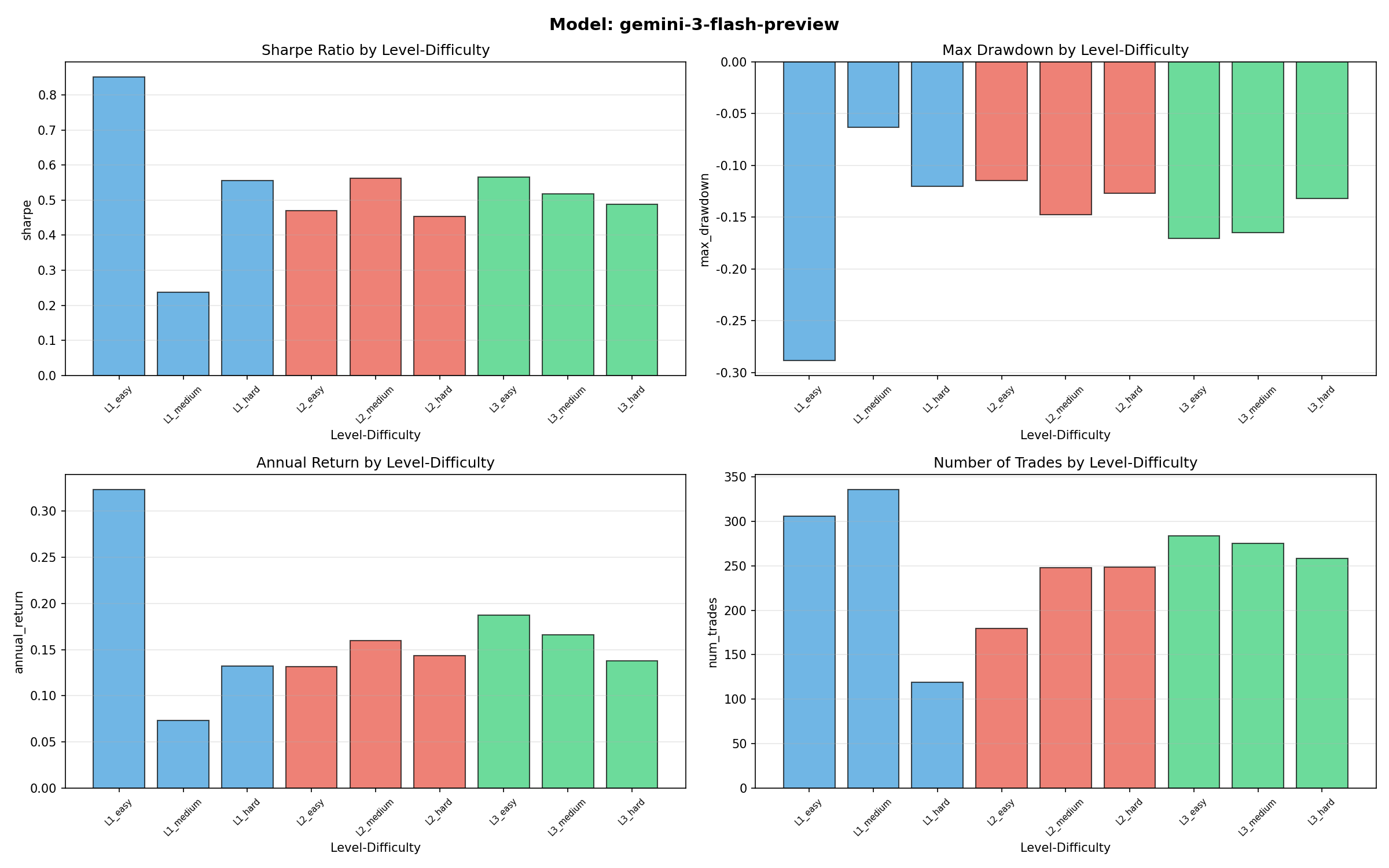}
\caption{T=0.0}
\end{subfigure}
\hfill
\begin{subfigure}[t]{0.48\textwidth}
\includegraphics[width=\textwidth]{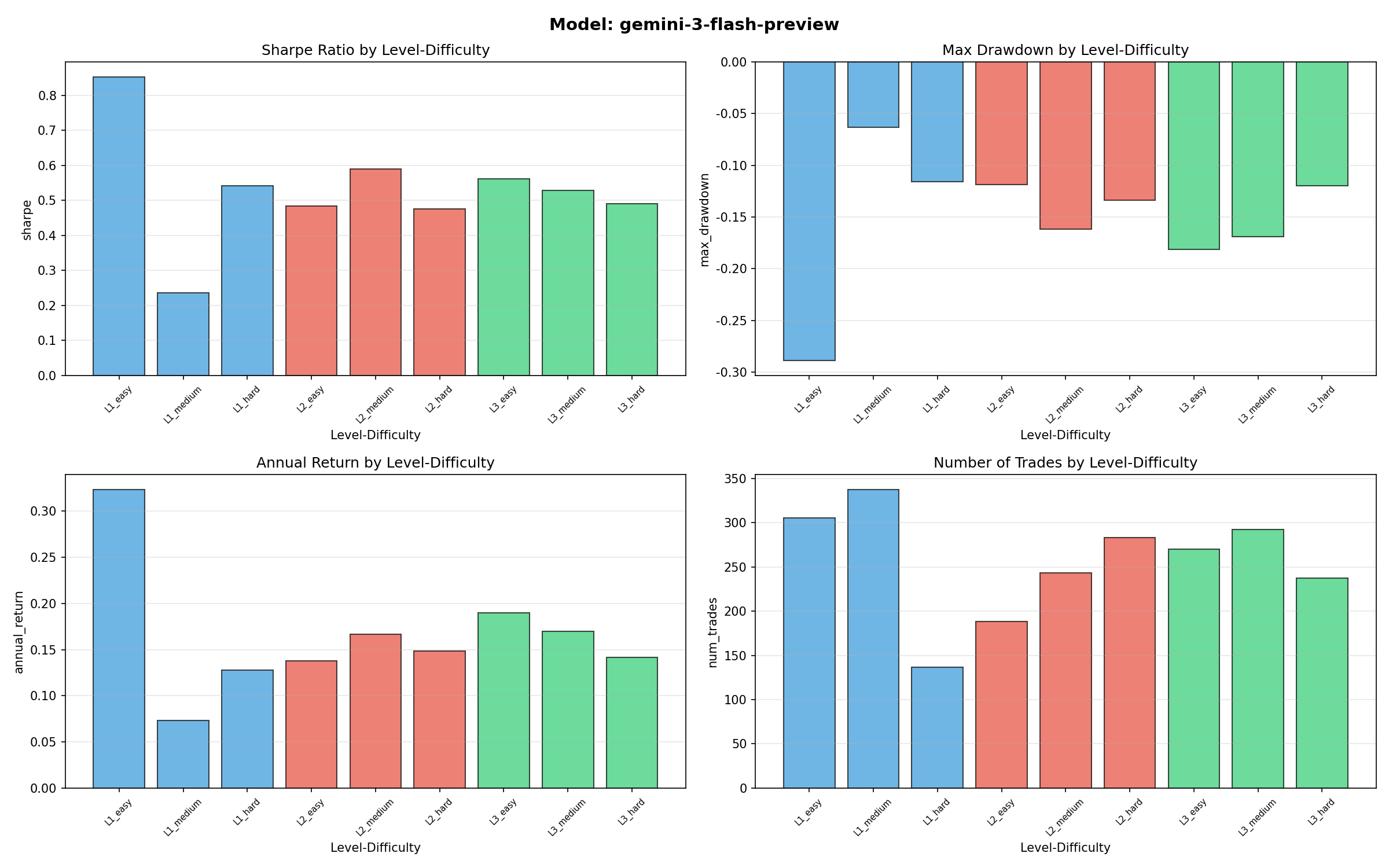}
\caption{T=0.7}
\end{subfigure}
\caption{\textit{gemini-3-flash-preview} performance across 9 difficulty levels.}
\label{appx_fig:gemini_level_performance}
\end{figure*}

\textbf{Overview.} \textit{gemini-3-flash-preview} is Google's efficient model optimized for speed and cost-effectiveness. Within \projectname it occupies the \emph{balanced-stable} archetype alongside \textit{claude-sonnet-4.5}, serving as a cost-effective alternative to its Pro-tier sibling with a remarkably flat difficulty profile.

\textbf{Analysis.} \textit{gemini-3-flash-preview} achieves the second-highest overall Sharpe Ratio (0.523 at $\tau{=}0$) in the benchmark, with per-level SR values of 0.543 (L1), 0.493 (L2), and 0.532 (L3). The L1-to-L3 spread is merely 0.011, the smallest among all models, yielding an almost perfectly flat difficulty curve in \Cref{appx_fig:gemini_level_performance}. This suggests that the model generalizes uniformly across the full cognitive-demand spectrum without exhibiting the sharp degradation seen in the conservative-rigid archetype or the ascending pattern of \textit{gemini-3-pro-preview}. The model's risk metrics (MDD~=~0.148, VOL~=~0.204) are moderate, positioning it between the aggressive \textit{gemini-3-pro-preview} and the conservative \textit{\textit{gpt-5.2}}.

\textbf{Strengths and Weaknesses.} The key strength of \textit{gemini-3-flash-preview} is its combination of solid performance with cost efficiency: it delivers SR values within 16\% of the top-performing \textit{gemini-3-pro-preview} at a fraction of the inference cost and latency, making it an attractive option for high-throughput or budget-constrained deployment. Its Level~3 SR (0.532) substantially exceeds that of \textit{deepseek-v3.2} (0.329), \textit{\textit{gpt-5.2}} (0.336), and \textit{grok-4.1-fast} (0.331), indicating solid open-ended strategy design capabilities. Run-to-run variance is low, placing it among the most stable models in the benchmark. However, the model does not reach the peak returns or SR of \textit{gemini-3-pro-preview} on any single asset or difficulty level, and its Calmar Ratio (1.476) trails \textit{claude-sonnet-4.5} (1.650), reflecting a slight disadvantage in return-to-drawdown efficiency.

\textbf{Notable Patterns.}
The flat difficulty profile of \textit{gemini-3-flash-preview} contrasts with the ascending profile of its Pro sibling, suggesting that the two Gemini variants encode qualitatively different strategy-generation behaviors despite sharing an architectural lineage. On per-asset analysis, Flash shows less sensitivity to high-volatility assets (TSLA, cryptocurrencies) than Pro, with a tighter spread between its best and worst asset-level SR values. This asset-agnostic behavior, combined with its temperature invariance (virtually identical results at $\tau{=}0$ and $\tau{=}0.7$), makes it the most predictable Gemini variant and a robust baseline for cost-performance trade-off evaluations.

\begin{figure*}[h]
\centering
\begin{subfigure}[t]{0.48\textwidth}
\includegraphics[width=\textwidth]{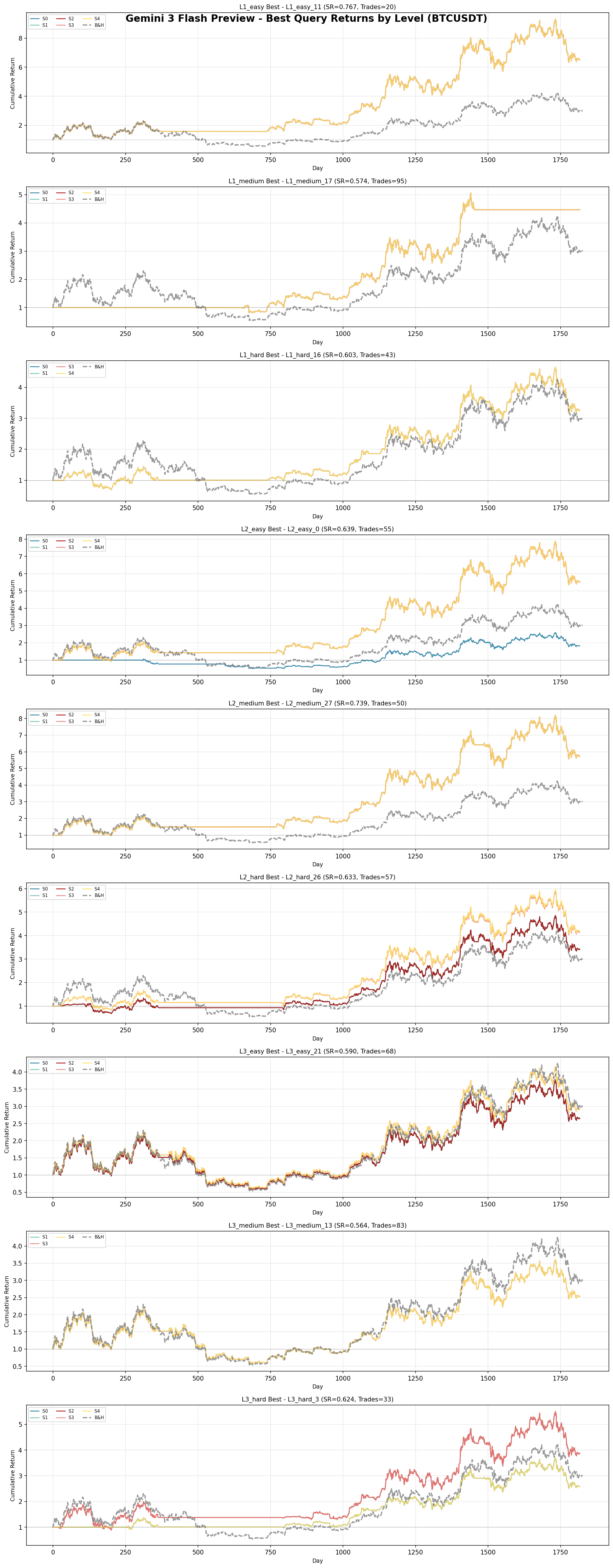}
\caption{T=0.0}
\end{subfigure}
\hfill
\begin{subfigure}[t]{0.48\textwidth}
\includegraphics[width=\textwidth]{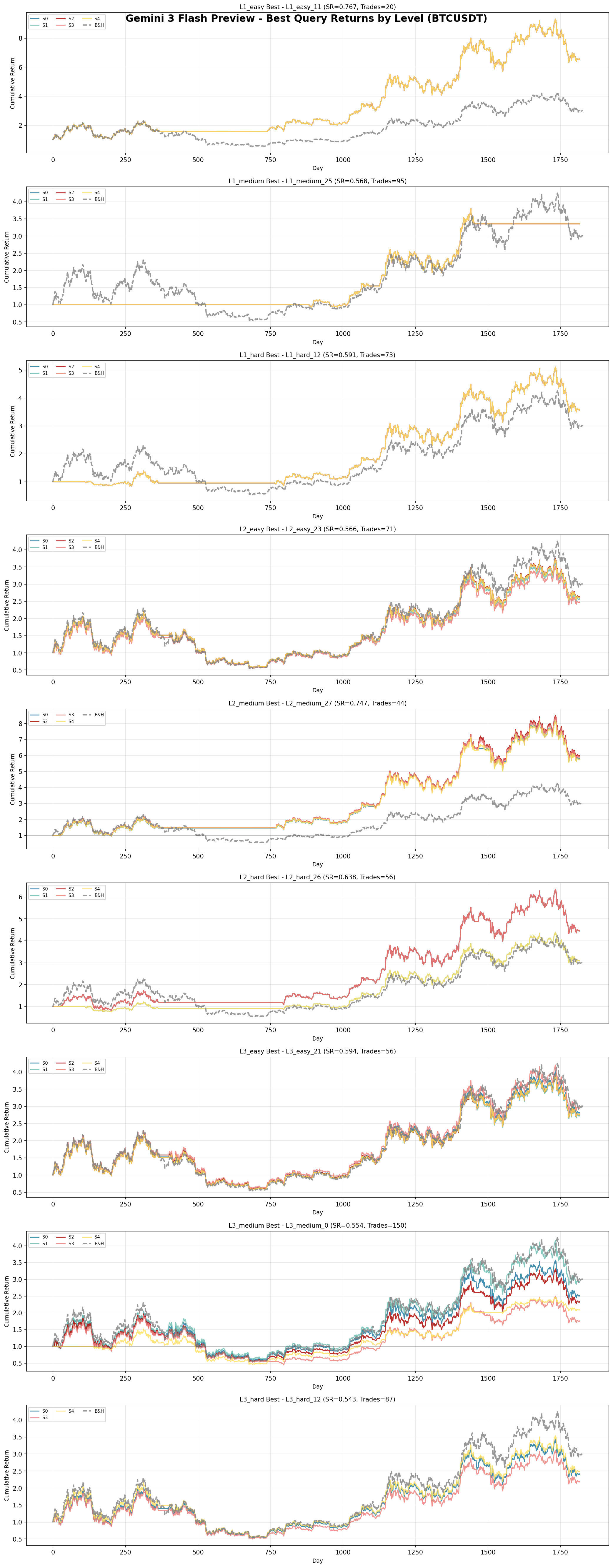}
\caption{T=0.7}
\end{subfigure}
\caption{Best-performing strategies generated by \textit{gemini-3-flash-preview}.}
\label{appx_fig:gemini_best_returns}
\end{figure*}

\clearpage

\subsubsection{\textit{deepseek-v3.2}}

\begin{figure*}[h]
\centering
\begin{subfigure}[t]{0.48\textwidth}
\includegraphics[width=\textwidth]{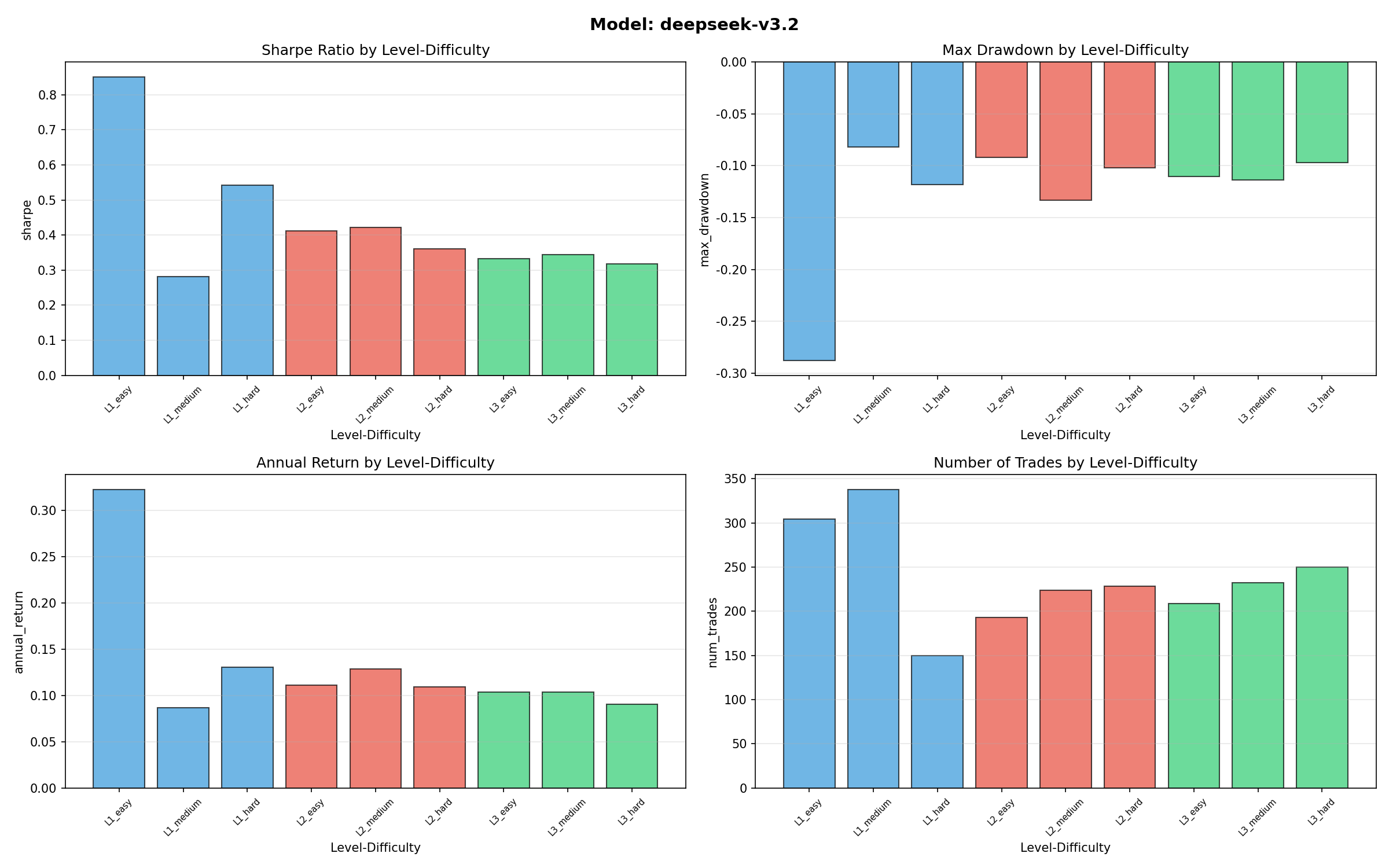}
\caption{T=0.0}
\end{subfigure}
\hfill
\begin{subfigure}[t]{0.48\textwidth}
\includegraphics[width=\textwidth]{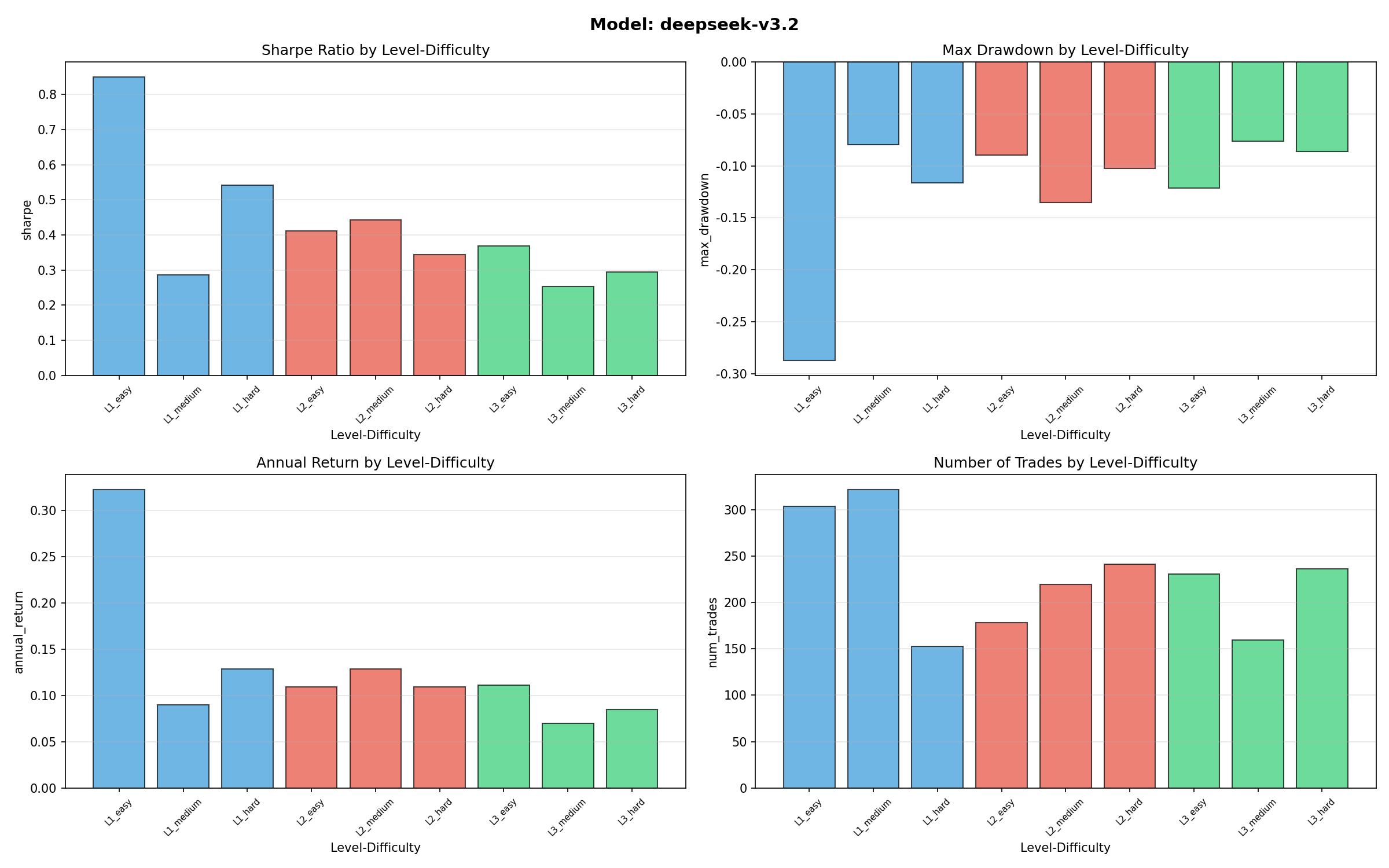}
\caption{T=0.7}
\end{subfigure}
\caption{\textit{deepseek-v3.2} performance across 9 difficulty levels.}
\label{appx_fig:deepseek_level_performance}
\end{figure*}

\textbf{Overview.} \textit{deepseek-v3.2} is an open-weights model with strong performance on coding and reasoning tasks. Within \projectname it falls under the \emph{conservative-rigid} archetype, exhibiting the most pronounced dissociation between code-translation competence and open-ended strategic reasoning among all evaluated models.

\textbf{Analysis.} \textit{deepseek-v3.2} achieves the highest Level~1 Sharpe Ratio (0.561 at $\tau{=}0$) in the entire benchmark, outperforming every other model on faithful if-then rule translation. However, as shown in \Cref{appx_fig:deepseek_level_performance}, performance degrades steeply as cognitive demands increase: SR drops to 0.366 at Level~2 and further to 0.329 at Level~3, producing the largest L1-to-L3 decline (0.232, or a 41\% relative drop) of any model. This steep gradient indicates that while \textit{deepseek-v3.2}'s code-generation machinery is highly competent at translating explicit specifications into executable Python, it struggles to fill in missing domain knowledge (Level~2) or design strategies from scratch (Level~3). Risk metrics are moderate (MDD~=~0.127, VOL~=~0.173), placing the model in a conservative band similar to \textit{\textit{gpt-5.2}}.

\textbf{Strengths and Weaknesses.} The primary strength of \textit{deepseek-v3.2} is its Level~1 dominance: on structured, fully specified queries, it generates the highest-quality trading code with low syntax error rates and the best risk-adjusted returns. This makes it an excellent choice for automated rule-translation pipelines where the strategy logic is pre-defined by a human quant. Its open-weights nature also offers deployment flexibility unavailable with proprietary models. However, the steep difficulty gradient is the model's most significant weakness: its Level~3 SR (0.329) is less than half that of \textit{gemini-3-pro-preview} (0.734), revealing a substantial gap in creative reasoning and domain grounding. Run-to-run variance increases notably at Level~3, with broader confidence bands emerging primarily on open-ended tasks, suggesting that the model's generation becomes less stable when guidance is sparse.

\textbf{Notable Patterns.}
The crossover between \textit{deepseek-v3.2} (leading at L1, trailing at L3) and \textit{gemini-3-pro-preview} (average at L1, dominant at L3) is the most striking ranking reversal in the benchmark, providing direct evidence that code-translation skill and strategic reasoning ability are dissociable cognitive capabilities. On a per-asset basis, \textit{deepseek-v3.2}'s degradation is most severe on high-volatility assets (TSLA, cryptocurrencies), where open-ended queries demand adaptive signal logic that the model struggles to produce. The pattern is temperature-invariant, confirming that the limitation is structural rather than sampling-related.

\begin{figure*}[h]
\centering
\begin{subfigure}[t]{0.48\textwidth}
\includegraphics[width=\textwidth]{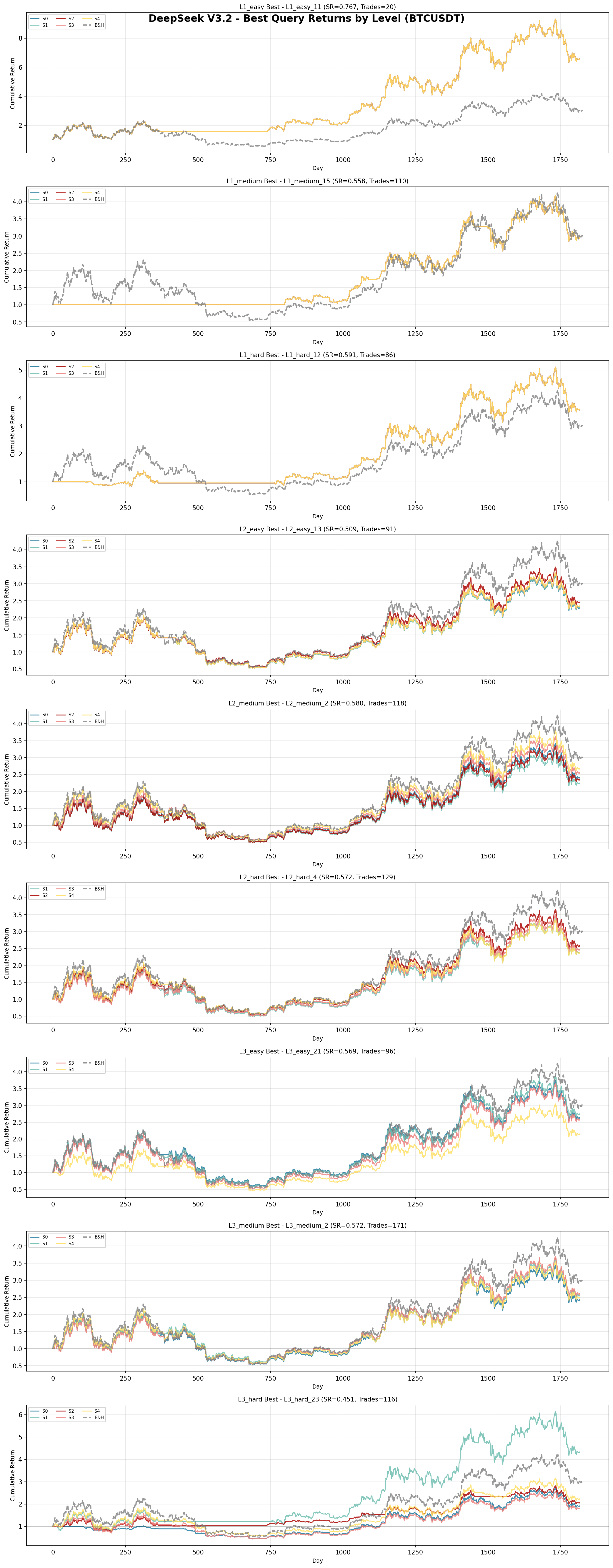}
\caption{T=0.0}
\end{subfigure}
\hfill
\begin{subfigure}[t]{0.48\textwidth}
\includegraphics[width=\textwidth]{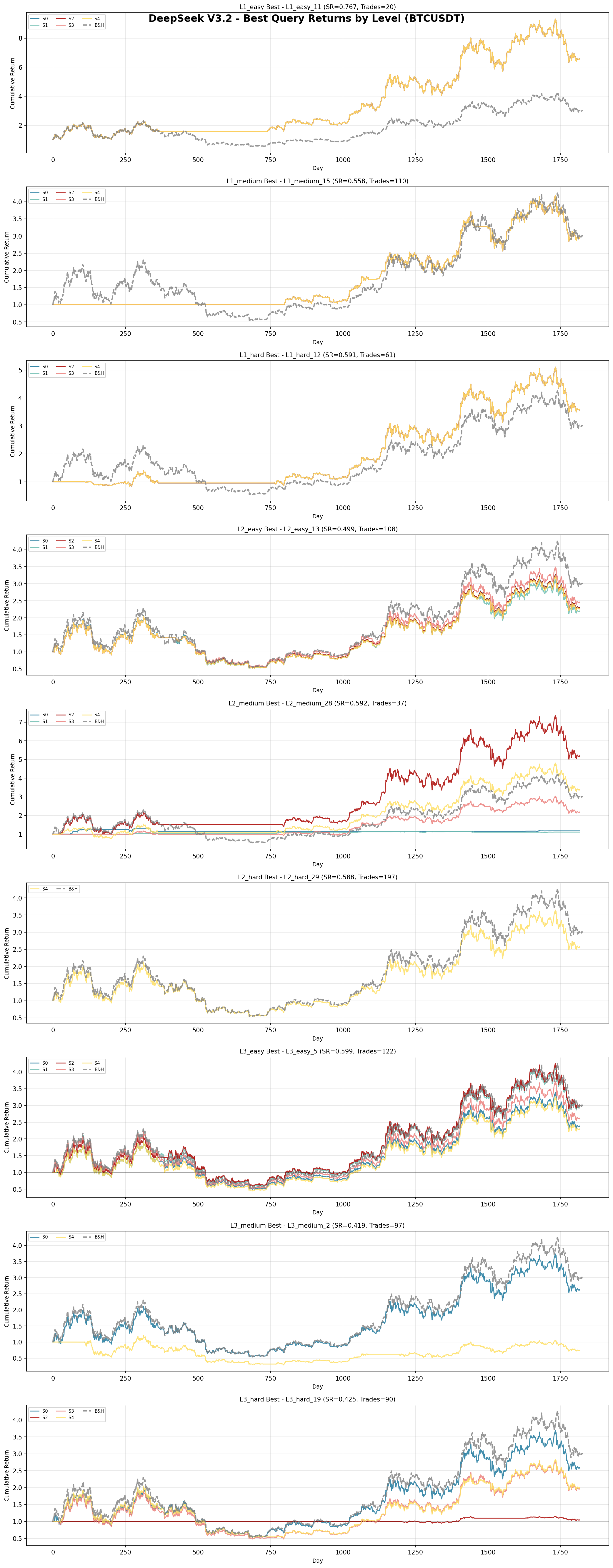}
\caption{T=0.7}
\end{subfigure}
\caption{Best-performing strategies generated by \textit{deepseek-v3.2}.}
\label{appx_fig:deepseek_best_returns}
\end{figure*}

\clearpage

\subsubsection{\textit{grok-4.1-fast}}

\begin{figure*}[h]
\centering
\begin{subfigure}[t]{0.48\textwidth}
\includegraphics[width=\textwidth]{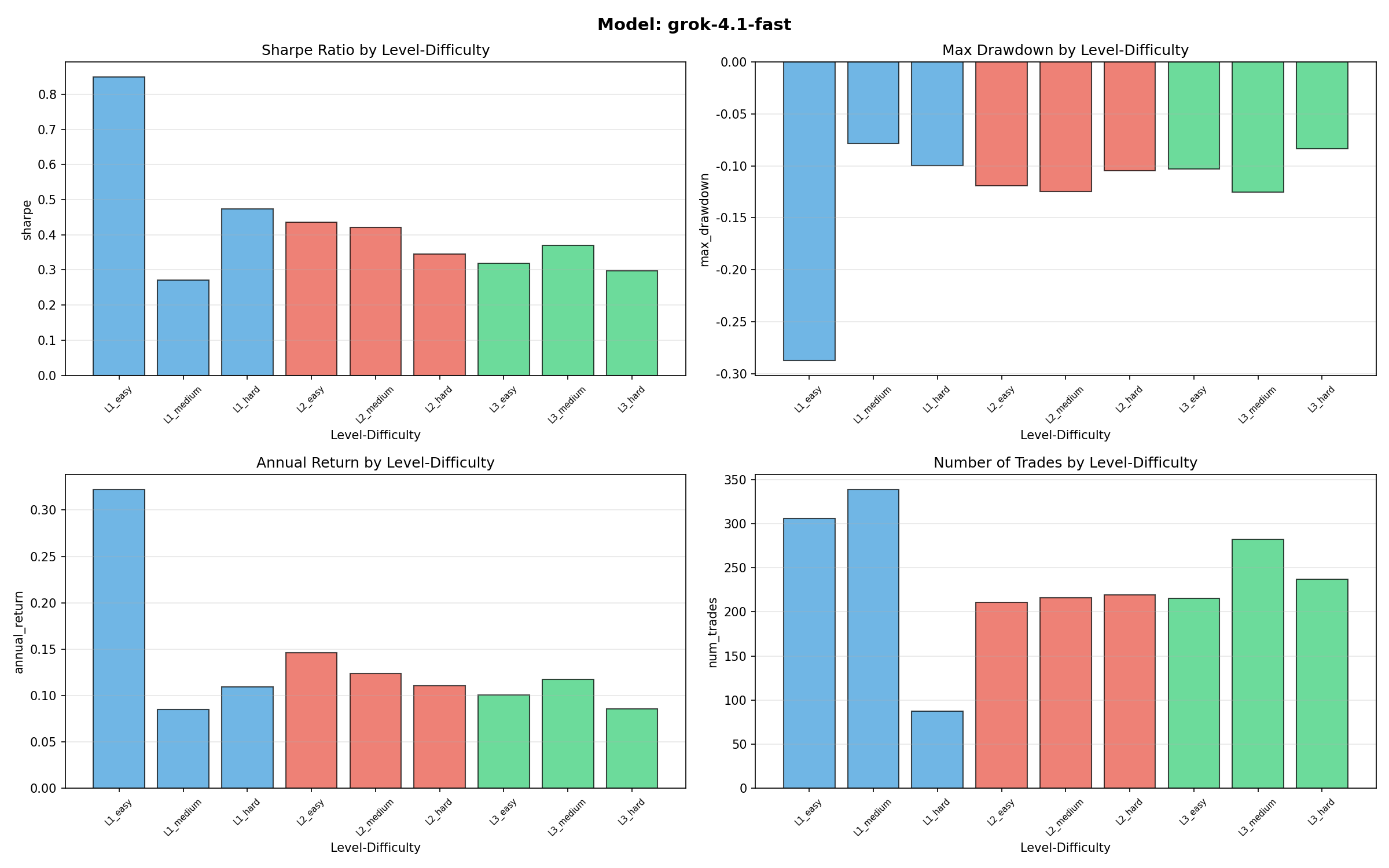}
\caption{T=0.0}
\end{subfigure}
\hfill
\begin{subfigure}[t]{0.48\textwidth}
\includegraphics[width=\textwidth]{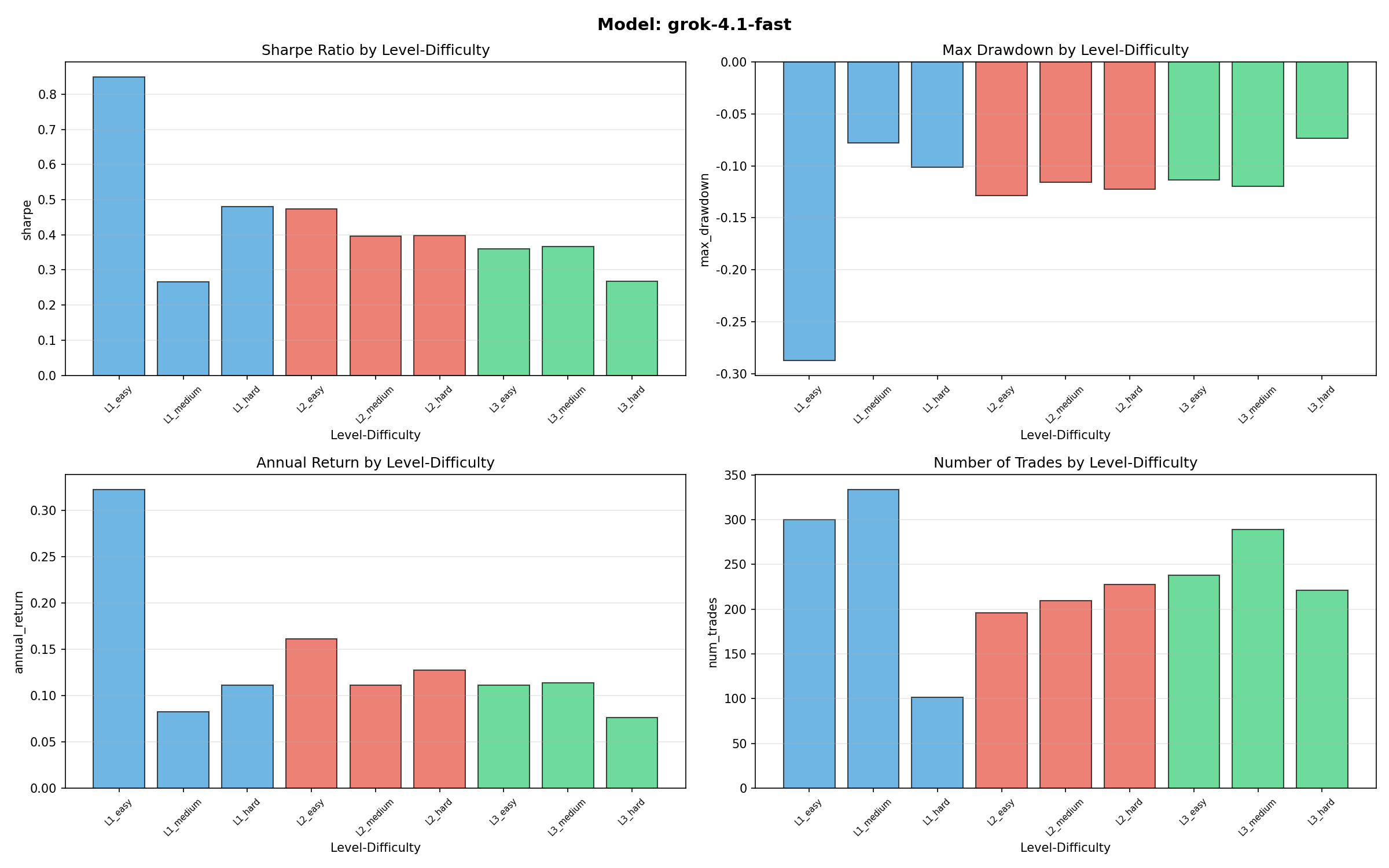}
\caption{T=0.7}
\end{subfigure}
\caption{\textit{grok-4.1-fast} performance across 9 difficulty levels.}
\label{appx_fig:grok_level_performance}
\end{figure*}

\textbf{Overview.} \textit{grok-4.1-fast} is xAI's model evaluated in this benchmark. It falls under the \emph{conservative-rigid} archetype but is further distinguished by the highest run-to-run variance among all models, making it the least predictable generator in \projectname.

\begin{figure*}[h]
\centering
\begin{subfigure}[t]{0.48\textwidth}
\includegraphics[width=\textwidth]{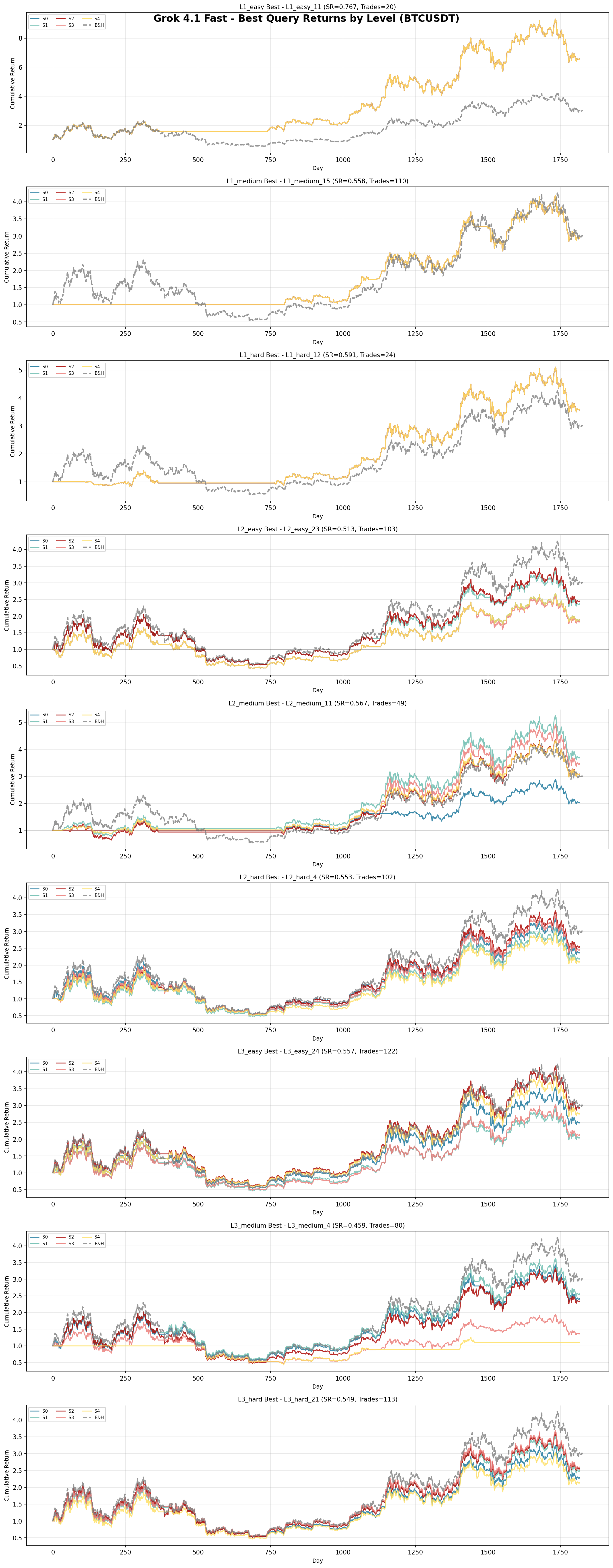}
\caption{T=0.0}
\end{subfigure}
\hfill
\begin{subfigure}[t]{0.48\textwidth}
\includegraphics[width=\textwidth]{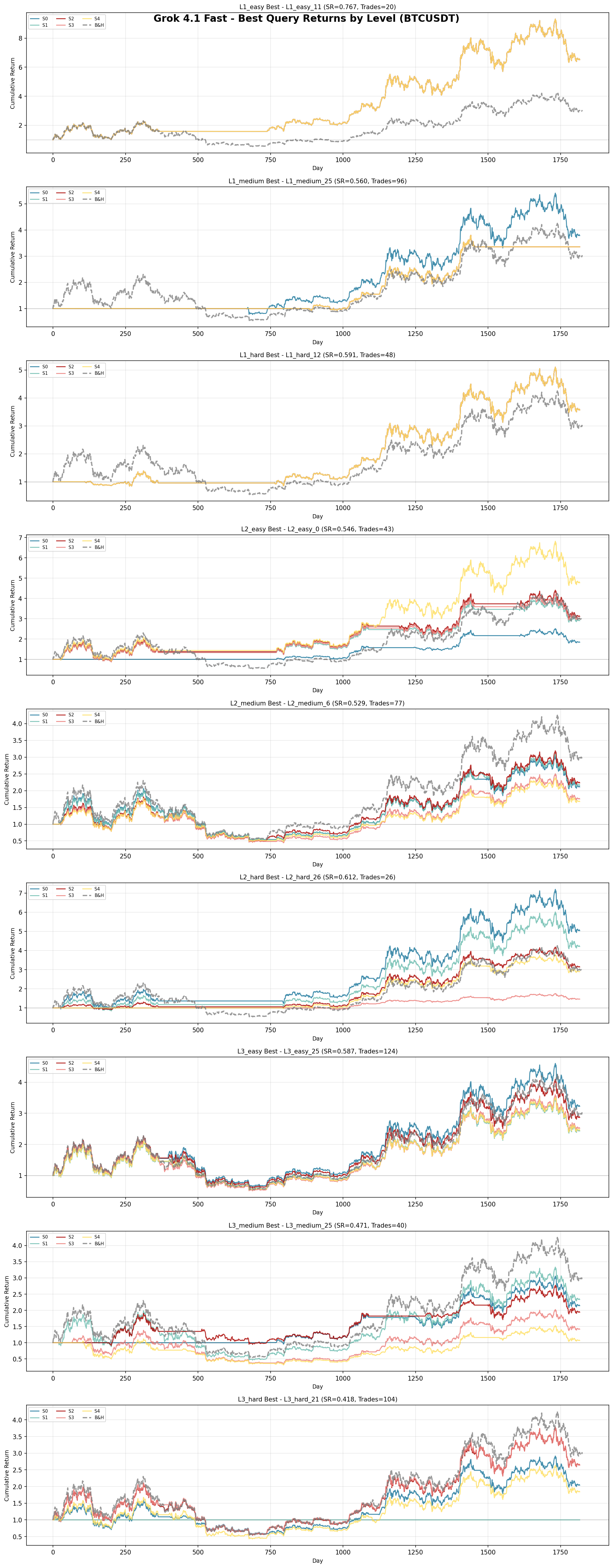}
\caption{T=0.7}
\end{subfigure}
\caption{Best-performing strategies generated by \textit{grok-4.1-fast}.}
\label{appx_fig:grok_best_returns}
\end{figure*}

\textbf{Analysis.} As shown in \Cref{appx_fig:grok_level_performance}, \textit{grok-4.1-fast} achieves an overall Sharpe Ratio of 0.421 at $\tau{=}0$, with per-level values of 0.532 (L1), 0.401 (L2), and 0.331 (L3). The L1-to-L3 decline (38\%) follows the conservative-rigid pattern, but what sets the model apart is the pronounced within-level variance: the 25th--75th percentile SR range across 5 runs is substantially wider than that of any other model, and erratic oscillations are visible in the aligned return curves (\Cref{appx_fig:grok_best_returns}). Risk metrics are superficially favorable (MDD~=~0.125, VOL~=~0.171), ranking second only to \textit{\textit{gpt-5.2}}, but this low average risk masks occasional high-drawdown outlier strategies that elevate the tail risk.

\textbf{Strengths and Weaknesses.} The model's primary strength is its competitive risk-adjusted efficiency under stochastic decoding: at $\tau{=}0.7$ it achieves the highest Calmar Ratio (1.692) in the benchmark, suggesting that sampling diversity occasionally helps it discover favorable parameter configurations. Its Level~1 performance (SR~=~0.532) is comparable to the middle tier, confirming adequate code-translation competence. However, the high cross-run variance is a significant liability: while upper-percentile runs can reach SR values near 0.7, lower-percentile runs drop below 0.2, producing a wide dispersion that undermines reliability. The model also shows the sharpest performance degradation on Level~2 and Level~3 tasks among non-DeepSeek models, indicating limited domain-knowledge grounding and creative strategy design capabilities.

\textbf{Notable Patterns.}
\textit{grok-4.1-fast} is the only model whose confidence bands in the aligned return curves occasionally cross those of higher-ranked models, meaning that on a per-run basis it can outperform \textit{claude-sonnet-4.5} or \textit{gemini-3-flash-preview}, but it can equally produce substantially worse outcomes. This high-variance, high-tail behavior resembles a ``lottery'' generation pattern: the model appears to sample from a wider distribution of strategy templates with less consistent quality filtering. The pattern is asset-dependent, with the widest variance on cryptocurrency assets where market noise amplifies the impact of inconsistent signal logic. For practitioners, this profile suggests that \textit{grok-4.1-fast} may benefit most from best-of-$k$ selection strategies, where multiple generations are sampled and the best-performing strategy is retained.

\clearpage

\subsection{Analysis and Discussion}

This subsection synthesizes the findings from the preceding quantitative analysis, per-level examination, per-asset comparison, comparative model profiling, and per-model detailed analysis. The Stage~2 evaluation, conducted on 270 difficulty-stratified queries with a temperature ablation ($\tau = 0$ vs.\ $\tau = 0.7$), complements the ecological validity of Stage~1 with controlled, diagnostic precision. We organize the synthesis around six key themes: temperature stability, difficulty-level discriminative power, cross-level ranking reversals, model risk personalities, cross-asset robustness, and code generation reliability.

The temperature ablation reveals near-perfect invariance of all metrics and model rankings across decoding regimes (\Cref{appx_tab:overall_performance,appx_fig:model_radar}), providing the strongest evidence to date that the code-generation evaluation paradigm is robust to the specific decoding strategy. The $3\times 3$ difficulty taxonomy produces a monotonically widening inter-model spread from Level~1 to Level~3 (\Cref{appx_fig:model_level_comparison,appx_tab:per_level_performance}), confirming that the benchmark design effectively separates models along a controlled cognitive-demand axis. Cross-level analysis uncovers ranking reversals that would be invisible in aggregate metrics, while per-asset analysis (\Cref{appx_tab:per_asset_stage2,appx_fig:symbol_grouped_bar,appx_fig:cross_asset_robustness}) confirms that model rankings generalize across both cryptocurrency and US equity markets. Together, these analyses demonstrate that \projectname provides a multi-dimensional, reproducible, and highly informative evaluation framework for LLM-based strategy generation.

\subsubsection{Findings and Conclusions}

\paragraph{Finding 1: Temperature-invariant evaluation.}
The side-by-side comparison of greedy ($\tau = 0$) and stochastic ($\tau = 0.7$) decoding is the defining feature of the Stage~2 evaluation. Across all six models and all six metrics, the absolute Sharpe Ratio difference between the two temperature settings is at most 0.008 (\textit{gemini-3-flash-preview}: 0.523 vs.\ 0.530), and the model ranking is fully preserved at both settings (\Cref{appx_tab:overall_performance}). The average standard-deviation difference across all nine difficulty levels is only 2.97\% (\Cref{appx_fig:temperature_stability}), and the radar-chart polygon shapes at $\tau = 0$ and $\tau = 0.7$ are virtually identical (\Cref{appx_fig:model_radar}). This near-invariance provides strong evidence that the fundamental structure of generated strategy code---signal logic, entry/exit conditions, and risk management rules---is largely determined by the model's learned representations rather than by the randomness of the sampling process. In stark contrast, direct-trading evaluations are notoriously sensitive to temperature, with even small changes producing dramatically different action sequences and portfolio outcomes. The temperature stability documented here establishes a unique and practically significant advantage of the code-generation evaluation paradigm.

\paragraph{Finding 2: Systematic difficulty progression with maximum discriminative power at Level~3.}
The $3\times 3$ difficulty taxonomy produces a monotonically widening inter-model Sharpe Ratio spread across the three levels. At Level~1 (logic translation), all six models achieve nearly identical performance, with the inter-model SR range of merely 0.029 (0.532--0.561). At Level~2 (parameter inference), the spread widens to 0.238 (0.366--0.604), more than $8\times$ the Level~1 range. At Level~3 (goal-oriented generation), the spread reaches 0.405 (0.329--0.734), nearly $14\times$ the Level~1 range (\Cref{appx_fig:model_level_comparison,appx_tab:per_level_performance}). This systematic amplification confirms that the benchmark's difficulty hierarchy is effective: Level~1 tasks primarily test code generation competence (a near-solved problem for frontier LLMs), while Level~2 and Level~3 tasks progressively probe domain knowledge and creative strategy design, which remain strongly differentiating capabilities. The fine-grained $3\times 3$ breakdown (\Cref{appx_fig:cross_level_detailed,appx_fig:boxplot_by_level_difficulty}) further reveals that the primary cognitive leap occurs between Level~1 and Level~2 (a 16\% SR decline), with the transition from parameter inference to goal-oriented design introducing additional variance rather than a sharp mean decline.

\paragraph{Finding 3: Cross-level ranking reversals reveal complementary cognitive capabilities.}
The per-level analysis exposes a striking ranking reversal that aggregate metrics would entirely obscure. \textit{deepseek-v3.2} achieves the highest Level~1 Sharpe Ratio (0.561), outperforming all other models on faithful code translation, yet drops to the bottom tier at Level~3 (SR = 0.329). Conversely, \textit{gemini-3-pro-preview} is unremarkable at Level~1 (SR = 0.545, nearly identical to all competitors) but dominates Level~3 (SR = 0.734), far exceeding every other model. This crossover demonstrates that the three difficulty levels measure fundamentally different cognitive capabilities: Level~1 rewards syntactic fidelity, Level~2 rewards domain-grounded parameter inference, and Level~3 rewards end-to-end strategic reasoning. No single model dominates across all levels, and the benchmark's multi-level design is essential for exposing these complementary strengths and weaknesses. Practitioners can use this diagnostic information to select models tailored to their specific use case: \textit{deepseek-v3.2} for rule-translation tasks, \textit{gemini-3-pro-preview} for open-ended strategy discovery.

\paragraph{Finding 4: Persistent model risk personalities across real-world and structured queries.}
The ``risk personality'' profiles identified in Stage~1 are fully reproduced in Stage~2, confirming that they are intrinsic properties of each LLM's strategy generation behavior rather than artifacts of a particular query distribution. \textit{gemini-3-pro-preview} consistently favors aggressive, high-conviction signal logic, achieving the highest SR (0.628), ARR (0.208), and SoR (1.004) at $\tau = 0$ while also incurring the highest MDD (0.191) and VOL (0.262). \textit{\textit{gpt-5.2}} produces the most conservative strategies with the lowest MDD (0.119) and VOL (0.163), prioritizing capital preservation over return maximization. \textit{claude-sonnet-4.5} occupies a balanced position, achieving the best Calmar Ratio (CR = 1.650 at $\tau = 0$) with a favorable return-to-drawdown trade-off. These characteristic profiles are stable across both temperature settings, across all seven backtest assets, and across all three difficulty levels (\Cref{appx_fig:model_radar,appx_fig:model_grouped_bar,appx_fig:model_boxplot}). The persistence of these profiles from Stage~1 (real-world queries) to Stage~2 (structured queries) demonstrates that model risk personalities are robust, reproducible properties that practitioners can rely on for model selection based on deployment-specific risk tolerances.

\paragraph{Finding 5: Cross-asset robustness.}
The per-asset analysis (\Cref{appx_tab:per_asset_stage2,appx_fig:symbol_grouped_bar,appx_fig:cross_asset_robustness}) demonstrates that model rankings are preserved across all seven backtest assets spanning cryptocurrency (BTCUSDT, ETHUSDT) and US equity (AAPL, GOOGL, MSFT, NVDA, TSLA) markets. \textit{gemini-3-pro-preview} consistently leads on return-oriented metrics and \textit{\textit{gpt-5.2}} consistently leads on risk metrics, regardless of asset class. A shared difficulty gradient is also evident: AAPL and GOOGL are the easiest assets (SR frequently exceeding 0.7), MSFT is the hardest (SR $\approx$ 0.35--0.58), cryptocurrency assets occupy an intermediate position (SR $\approx$ 0.23--0.42), and TSLA exhibits the highest absolute returns coupled with the widest variance. This difficulty gradient is reproducible across all models and at both temperature settings, confirming that it reflects genuine market characteristics rather than model-specific or sampling artifacts. The consistency of rankings across such diverse market environments provides strong evidence that the benchmark captures fundamental differences in strategy generation capability.

\paragraph{Finding 6: High code generation reliability.}
All six frontier LLMs achieve high code generation success rates on the Stage~2 benchmark, with an overall Pass@1 of 97.9\% at $\tau = 0$ and 96.7\% at $\tau = 0.7$ (\Cref{appx_tab:pass_rate_comparison}). Pass@5 rates approach 100\% (99.69\% at $\tau = 0$, 99.94\% at $\tau = 0.7$), indicating that syntax failures are stochastic rather than systematic. Importantly, pass rates remain consistently high across all three difficulty levels (L1: 98\%+, L2: 97\%+, L3: 96\%+), with only minimal degradation as task complexity increases, demonstrating that syntax correctness is largely independent of strategic complexity. Among the small fraction of failures, runtime errors outnumber pure syntax errors by approximately 15:1 (\Cref{appx_tab:error_statistics}), indicating that failures stem from edge-case handling (division by zero, index bounds) rather than fundamental coding deficiencies. The narrow range of syntax performance across models (93--100\%) contrasts sharply with the wide range of strategic performance (SR: 0.415--0.628), confirming that code generation competence is no longer a differentiating factor among frontier LLMs; the competitive advantage now lies in strategic reasoning and domain knowledge.

\paragraph{Conclusion.}
The Stage~2 evaluation comprehensively validates the $3\times 3$ difficulty taxonomy as an effective diagnostic tool that reveals capability differences invisible to aggregate metrics. The controlled query design amplifies inter-model separation relative to Stage~1 (7.8pp SR spread vs.\ 5.5pp), while the temperature ablation demonstrates that the code-generation paradigm produces evaluation outcomes that are virtually invariant to the decoding strategy---a property absent in direct-trading benchmarks. The six findings above collectively establish that \projectname provides a multi-dimensional, reproducible, and highly discriminative evaluation framework: it identifies distinct model risk personalities, exposes complementary cognitive strengths through cross-level ranking reversals, confirms cross-asset robustness, and maintains high code generation reliability across all conditions. Together with the ecological validity established by Stage~1, the Stage~2 results demonstrate that the two-stage design of \projectname offers a comprehensive and principled approach to benchmarking LLM capabilities in quantitative finance, combining the authenticity of real-world queries with the diagnostic precision of difficulty-stratified synthetic queries.